\documentclass[trackchanges]{aastex7}

\usepackage{color}
\begin{document}

\title{A Seven-Day Multi-Wavelength Flare Campaign on AU Mic III:
Quiescent and Flaring Properties of the X-ray Spectra and Chromospheric lines}

\author[0000-0002-0412-0849]{Yuta Notsu}
\affil{Laboratory for Atmospheric and Space Physics, University of Colorado Boulder, 3665 Discovery Drive, Boulder, CO 80303, USA}
\affil{National Solar Observatory, 3665 Discovery Drive, Boulder, CO 80303, USA}
\affil{Department of Astrophysical and Planetary Sciences, University of Colorado Boulder, 2000 Colorado Ave, Boulder, CO 80305, USA}
\email{Yuta.Notsu@colorado.edu}

\author[0000-0001-5974-4758]{Isaiah I.Tristan}
\affil{Rice University, Houston, TX 77005, USA}
\affil{Laboratory for Atmospheric and Space Physics, University of Colorado Boulder, 3665 Discovery Drive, Boulder, CO 80303, USA}
\affil{National Solar Observatory, 3665 Discovery Drive, Boulder, CO 80303, USA}
\affil{Department of Astrophysical and Planetary Sciences, University of Colorado Boulder, 2000 Colorado Ave, Boulder, CO 80305, USA}
\email{Isaiah.Tristan@colorado.edu}

\author[0000-0001-5643-8421]{Rachel A. Osten}
\affil{Space Telescope Science Institute, Baltimore, MD 21218, USA}
\email{osten@stsci.edu}

\author[0000-0003-2631-3905]{Alexander Brown}
\affiliation{Center for Astrophysics and Space Astronomy,
University of Colorado, 389 UCB, Boulder, CO 80309, USA}
\email{alexander.brown@Colorado.edu}

\author[0000-0001-7458-1176]{Adam F. Kowalski}
\affil{Laboratory for Atmospheric and Space Physics, University of Colorado Boulder, 3665 Discovery Drive, Boulder, CO 80303, USA}
\affil{National Solar Observatory, 3665 Discovery Drive, Boulder, CO 80303, USA}
\affil{Department of Astrophysical and Planetary Sciences, University of Colorado Boulder, 2000 Colorado Ave, Boulder, CO 80305, USA}
\email{Adam.F.Kowalski@colorado.edu}

\author[0000-0001-5440-1879]{Carol A. Grady}
\affil{Eureka Scientific, Oakland, CA 94602, USA}
\email{cagrady@comcast.net}

\correspondingauthor{Yuta Notsu}
\email{Yuta.Notsu@colorado.edu}

\begin{abstract}
We present the X-ray quiescent and flaring properties from a unique, 7-day multiwavelength observing campaign on the M1 flare star AU Mic.  Combining the XMM-Newton X-ray spectra with the chromospheric line and broadband NUV and optical continuum observations provides a dataset \color{black}
that is \color{black} one of the most comprehensive to date.
We analyze the sample of 38 X-ray flares and study in detail the  X-ray flare temperature ($T$) and emission measure (EM) evolutions of three largest flares with the X-ray flare energies of $>10^{33}$ erg. 
The $T-\mathrm{EM}$ evolution tracks and multi-wavelength emission evolutions
of the largest-amplitude Neupert-type flare 
reveal that the so-called ``Flare H-R diagram" is consistent with thermal coronal flare emission evolution.
The two other more gradual and longer duration X-ray flares 
are interpreted as having larger size scales.
None of the 17 H$\alpha$ and H$\beta$ flares show clear blue/red 
wing asymmetries, including the ones associated with the potential X-ray dimming event previously reported.
The above largest-amplitude Neupert flare shows clear symmetric H$\alpha$ and H$\beta$
broadenings with roughly $\pm$400 and $\pm$600 km s$^{-1}$, respectively, 
which are synchronized with the optical/NUV continuum emission evolution. 
Radiative hydrodynamic modeling results suggest that 
electron beam heating parameters 
that have been used to reproduce M-dwarf flare NUV/optical continuum emissions can reproduce these large broadenings of H$\alpha$ and H$\beta$ lines.
These results suggest that these most energetic M-dwarf flares are associated with
stronger magnetic field flux densities and larger size scales than solar flares but 
can be interpreted in terms of the standard flare model.
\end{abstract}

\keywords{Stellar flares (1603); Optical flares (1166); Stellar x-ray flares (1637); M dwarf stars (982); Flare stars (540); Red dwarf flare stars (1367); Stellar chromospheres (230); Stellar coronae (305); Stellar coronal mass ejections (1881)}

\section{Introduction} \label{sec:intro}

Stellar flares are sudden and intense releases of magnetic energy in the stellar atmosphere, showing increases in brightness across the entire electromagnetic spectrum from radio to X-ray  \color{black}  (e.g., \citealt{Pettersen_1989_SoPh,Guedel2004_A&ARv,Reid+2005,Shibata+2011,Benz_2017_LRSP,Kowalski+2024_LRSP} for reviews). \color{black}
Recent Kepler \& TESS optical photometric survey data have brought us various statistical results on how flare activities relate with stellar temperature, rotation, and age \color{black} (e.g., \citealt{Maehara+2012,Shibayama+2013,Notsu+2013_ApJ,Notsu+2019,Hawley+2014,Candelaresi+2014,Davenport+2016,Howard+2020_ApJ,Okamoto+2021,Feinstein+2020,Feinstein+2024_AJ,Vasilyev+2024_Sci}). 
Young- \color{black} to mid-age M-dwarfs are generally known to be magnetically active and show frequent flares \color{black}(e.g., \citealt{Lacy+1976,Reid+2005,Kowalski+2009_AJ,Hawley+2014,Maehara+2021,Tristan+2023_ApJ,Feinstein+2024_AJ}).  \color{black}
With renewed emphasis on finding and characterizing habitable zone planets around M dwarf stars, characterizing the flare activity of M dwarfs is a key ingredient to understand the impact that the star can have on its near stellar environment \color{black}(e.g., \citealt{Segura+2010,Segura_2018haex.book,Tilley+2019,Linsky2019,Yamashiki+2019,Airapetian+2020,France+2020_AJ,do-Amaral+2022_ApJ}). \color{black}

In the standard solar flare model  (e.g., \citealt{Shibata+2011,Kowalski+2024_LRSP} for reviews), the accelerated particles propagate through the corona and deposit their energy in the \color{black} 
lower chromosphere and upper photosphere \color{black} producing non-thermal hard X-ray (HXR; $\gtrsim$25 keV) and optical/near-UV(NUV) continuum (white-light) emission. 
The overpressure ablates chromospheric material into the corona,
which increases the coronal density and temperature and is known as chromospheric evaporation (\color{black}e.g., \color{black}\citealt{Fisher+1985_ApJ_289_414,Fisher+1985_ApJ_289_425,Allred+2005,Allred+2006}).
Chromospheric evaporation is traced as the temporal lag of the gradual soft X-ray ($\lesssim$15 keV) response relative to the impulsive non-thermal emission (e.g., HXR emission, optical/NUV continuum emission), which is known as the empirical Neupert effect (e.g., \citealt{Dennis_Zerro_1993_SoPh,Neidig_Kane_1993_SoPh,McTiernan+1999_ApJ,Veronig+2002,Namekata+2017_ApJ}). It is noted that \color{black} the \color{black} Neupert effect was first observed in solar flares as the correspondence between the derivative of thermal, soft X-ray emission and the impulsive non-thermal gyrosynchrotron \color{black} radio \color{black} centimeter-wave flux \citep{Neupert+1968}. The Neupert effect has also been reported in \color{black} many stellar flare observations 
(e.g., \citealt{Hawley+1995,Guedel+1996,Guedel+2002_ApJL,Osten+2004_ApJS,Mitra-Kraev+2005a,Wargelin+2008_ApJ,Fuhrmeister+2011_A+A,Lalitha+2013_A+A,Caballero-Garcia+2015_MNRAS,Stelzer+2022_A&A}), \color{black}  and it is generally interpreted as an evidence of the “standard” model of solar and stellar flares (\color{black}cf.\color{black} \citealt{Shibata+2011,Kowalski+2024_LRSP}). However, there are also solar and stellar flares that do not follow Neupert effect (\color{black}e.g., \color{black} \citealt{Warmuth+2009_ApJ,Fleishman+2016_ApJ,Osten+2005_ApJ}), \color{black} which require \color{black} better understanding of the energy partition and multi-wavelength timing in stellar flares (For the details, see \citealt{Tristan+2023_ApJ}, Section 7.7 of \citealt{Kowalski+2024_LRSP}, and references therein).

Line profile dynamics of chromospheric lines \color{black} has been an important topic of stellar flares for years 
(e.g., see also 
\citealt{Kowalski+2024_LRSP,Notsu+2024_ApJ} and references therein). 
Optical spectroscopic observations of M-dwarf flares, especially recent extensive monitoring observations, have reported many blue wing asymmetries (enhancements of ``blue" wing of lines) in chromospheric lines (e.g., H$\alpha$, H$\beta$, Ca II line observations in 
\citealt{Houdebine+1990,Eason+1992,Gunn+1994,Crespo-Chacon+2006,Fuhrmeister+2008,Fuhrmeister+2011_A+A,Fuhrmeister+2018,Vida+2016,Vida+2019,Honda+2018,Muheki+2020_EVLac,Maehara+2021,Notsu+2024_ApJ,Inoue+2024_PASJ,Kajikiya+2025_ApJ_PaperII,Kajikiya+2025_ApJ_PaperI}). 
\color{black}
In analogy with solar eruption events, these blue wing asymmetries during M-dwarf flares are often interpreted as the Doppler shifts of filament/prominence eruptions (\citealt{Leitzinger+2022,Otsu+2022}).
They are also discussed as candidates of stellar CMEs (coronal mass ejections), but there exists uncertainty of interpretations especially distinguishing between filament/prominence eruptions and potential flare-related blueshifts
\color{black}
(e.g., \citealt{Fuhrmeister+2018,Vida+2019,Notsu+2024_ApJ,Kajikiya+2025_ApJ_PaperI}). 
\color{black}
In addition to blue wing asymmetries, 
red wing asymmetries (enhancements of ``red" wing of lines) and symmetric line broadenings have been also observed during M-dwarf stellar flares \color{black} (e.g., \citealt{Houdebine+1990,Fuhrmeister+2018,Wu+2022,Namizaki+2023_ApJ,Wollmann+2023_A+A,Notsu+2024_ApJ,Kajikiya+2025_ApJ_PaperII})\color{black}. 
One possible cause of the red wing asymmetries is the process called chromospheric condensations, which is the downward flow of cool plasma in the chromosphere (\citealt{Ichimoto+1984,Graham+2015}), while another possible cause is the flare-driven coronal rain or the post-flare loop (\citealt{Wu+2022,Wollmann+2023_A+A}). The symmetric broadenings, \color{black} which are interpreted as non-thermal broadening or
Stark (pressure) broadening, \color{black} can be caused by high-energy nonthermal electron beams penetrating into the lower atmosphere (\citealt{Oks+2016,Namekata+2020_PASJ,Kowalski+2022,Kowalski+2024_ApJ}).
These blue/red wing asymmetries and symmetric broadenings of chromospheric line profiles can provide hints on various physical aspects of stellar flares (e.g., evolution of flare loop dynamics, energy release process, potential mass ejections), 
but previous investigations have been mostly limited to optical wavelength (e.g., simultaneous optical spectroscopy and optical photometry in \citealt{Notsu+2024_ApJ,Kajikiya+2025_ApJ_PaperI,Odert+2025_MNRAS}). 
As a limited example, the NICER (Neutron star Interior Composition Explore) X-ray flare emissions have been simultaneously reported with the H$\alpha$ blue wing asymmetries for two flares in  \citet{Notsu+2024_ApJ,Inoue+2024_PASJ}, but their investigations are limited because of the sample size (\color{black} only two flares have simultaneous NICER X-ray data in their datasets\color{black}) and the time-resolution (observation gap) problem caused by NICER's orbital motion. 
More time-resolved, multi-wavelength (e.g., coronal X-ray emission) investigations are necessary to constrain how these chromospheric line emission and profile asymmetries/broadenings occur in association with flare energy release processes (e.g., the Neupert effect described above), and clarify the generation mechanism of the line asymmetries/broadenings.

Recently, we have executed a large observational campaign (hereafter, T23; \citealt{Tristan+2023_ApJ}) on a young M1 dwarf flare star, AU Mic, over 7 days with the X-ray Multi-Mirror Mission (XMM-Newton; \citealt{Jansen+2001_A&A}), the Jansky Very Large Array (JVLA), 
the Neil Gehrels Swift Observatory (Swift), 
the Las Cumbres Observatory Global Telescope network (LCOGT; \citealt{Brown+2013}), 
the Astrophysical Research Consortiums 3.5 m telescope at the Apache Point Observatory (APO), 
the Australia Telescope Compact Array (ATCA), 
and the Small and Moderate Aperture Research Telescope System 
(SMARTS; \citealt{Subasavage+2010_SPIE}) 0.9 and 1.5 m telescopes 
at the Cerro Tololo Inter-American Observatory (CTIO).
As summarized in Section 2.1 of T23, 
AU Mic is a nearby well-known M1-dwarf (Distance: 9.72 pc, Radius: 0.75$\pm$0.03 solar radius) with two observed planets (\citealt{Plavchan+2020_Nature,Martioli+2021_A&A,Gilbert+2022_AJ}),
resolved edge-on debris disk (\citealt{Kalas+2004_Science,Wisniewski+2019_ApJ}),
high magnetic activity level (\citealt{Duvvuri+2021_ApJ,Alvarado-Gome+2022_ApJ,Cohen+2022_ApJ,Klein+2022_MNRAS}), 
and a source of frequent and energetic flares in the multi-wavelength regimes (\citealt{Kunkel+1970_PASP,Robinson+2001_ApJ,Mitra-Kraev+2005a,Feinstein+2022_AJ,Ikuta+2023_ApJ,Odert+2025_MNRAS}), 
which may affect the exoplanet atmosphere (\citealt{do_Amaral_2025_ApJ}).

The first paper of this flare campaign (T23) have presented the overview of the whole campaign dataset, and an analysis of the thermal empirical Neupert effect between the XMM-Newton NUV photometry (OM UVW2; 1790--2890 \AA) and X-ray EPIC-pn photometry (European Photon Imaging Cameras; 0.2–12 keV). 
In T23, we found that 65\% of the observed AU Mic flares (30 of 46) do not follow the Neupert effect, which is 3 times \color{black} larger than the fraction from solar flare results \color{black}, and proposed a four-part Neupert effect classification (Neupert, quasi-Neupert, non-Neupert types I and II) to explain these multiwavelength responses. While the SXR emission generally lags behind the NUV as expected from the chromospheric evaporation flare models, the Neupert effect is more prevalent in larger, more impulsive flares.  
In T23, we also conducted preliminary flaring rate analysis with X-ray (from XXM-Newton) and $U$-band data (from LCOGT), which suggests that previously estimated energy ratios (cf. \citealt{Osten+2015}) hold for a collection of flares observed over the same time period, but not necessarily for an individual, multiwavelength flare.
These results imply that one model cannot explain all stellar flares and care should be taken when extrapolating to different wavelength regimes.

The second paper of the campaign 
(\citealt{Tristan+2025_ApJ}, T25) presented high-time-resolution radio light curves from the VLA Ku band (12--18 GHz) and the ATCA K band (16--25 GHz), 
which observe gyrosynchrotron radiation and directly probe the action of accelerated electrons within flaring loops.
The observation revealed 16 VLA and 3 ATCA flares, which include both optically thick and thin flares. We estimate the total kinetic energies of gyrating electrons in optically thin flares to be between 10$^{30}$ to 10$^{34}$ erg when the local magnetic field strength is above 700 G. These energies are able to explain the combined radiated energies from multi-wavelength observations. Estimations from optically thick radiation indicate higher loop-top magnetic field strengths ($\sim$1 kG) and sustained high electron densities ($\sim10^6$ cm$^{-3}$) compared to previous observations of large M-dwarf flares, which is in line with modern radiative-hydrodynamic simulations of these energetic events.

This is the third paper of the AU Mic multi-wavelength campaign, following these two papers (T23 and T25). Only the lightcurve data have been discussed for the X-ray data in T23 and T25, and this paper presents the analysis of the X-ray spectra from EPIC-pn, EPIC-MOS, and RGS (Reflection Grating Spectrometers) instruments of XMM-Newton, and optical high-dispersion spectroscopic data (e.g., H$\alpha$ and H$\beta$ lines) from SMARTS 1.5m telescope. 
Section \ref{sec:overview} overviews the multi-wavelength flare lightcurve dataset of the whole campaign, focusing on the flares detected in X-ray and H$\alpha$\&H$\beta$ data.
In Section \ref{subsec:X-ray_specana_quiescent}, we conduct spectral analysis of the quiescent phase X-ray spectra from the EPIC-pn, EPIC-MOS, and RGS instruments.
Then we do spectral analysis of the time-averaged flare components from EPIC-pn and EPIC-MOS in Section \ref{subsec:X-ray_specana_time-average}.
In Section \ref{sec:remarkable_flares}, 
we describe the time evolution of X-ray spectra and chromospheric line profiles of the remarkable flares with the Neupert classification and/or chromospheric line profile evolution.
In Section \ref{sec:discussions}, on the basis of these data analysis results, we discuss quiescent emissions, flare temperature evolution to further study the flare loop evolutions, and investigate chromospheric emission and line profile evolutions from multi-wavelength points of view. 

\section{Observations, Data, and Flare lightcurve overview} \label{sec:overview}
\subsection{Observation Data}\label{subsec:obs_data}

The basic stellar parameters of the target star AU Mic and the observation campaign overview are described in Section 2 of T23. 
Here we briefly describe XMM-Newton X-ray and SMARTS optical spectroscopic data, 
which are the focus of the detailed spectral analysis in this paper. The data and analysis details of the NUV photometry (XMM-Newton OM UVW2), optical photometry (LCOGT $U$-band and SMARTS $V$-band) and VLA radio data are described in T23 and T25.

Optical spectroscopy was conducted with the cross-dispersed, fiber-fed echelle CTIO HIgh ResolutiON (CHIRON) spectrograph (\citealt{Tokovinin+2013}) attached to the SMARTS 1.5 m telescope at CTIO (\citealt{Subasavage+2010_SPIE}). 
The wavelength range and wavelength resolution of our CHIRON data are 4500--8900 \AA~and $R=\lambda/\Delta\lambda\sim$25,000, respectively.
This wavelength range includes H$\alpha$, H$\beta$, Ca II 8542\AA, He I D3 5876\AA, and Na I D1 and D2 lines. 
The exposure time and the resultant time cadence were 60s and 65s, respectively (Table 2 of T23), and S/N values of $\sim$40--50 are achieved at the continuum level around the H$\alpha$ 6563\AA~line. The CHIRON spectra were reduced using the CHIRON pipleline described in \citet{Tokovinin+2013}, and the equivalent width and line profiles of the major lines are analyzed in the same way as in \citet{Notsu+2024_ApJ}.

The XMM-Newton X-ray observations were conducted in four visits (Obs-IDs: 0822740301, 0822740401, 0822740501, and 0822740601, PI: Adam Kowalski), which spanned from 2018 October 10 to 17 (401.54 ks = 130.04 hours in total). The ODF files downloaded from the XMM-Newton archive \footnote{\url{https://nxsa.esac.esa.int/nxsa-web/}} were reduced with \texttt{SAS}\footnote{\url{https://www.cosmos.esa.int/web/xmm-newton/sas}} version 21.0.0 (xmmsas\_20230412\_1735) to obtain event tables for two EPIC instruments (EPIC-pn and EPIC-MOS1; \citealt{Strueder+2001_A&A,Turner+2001_A&A}) and spectra for the two RGS high-resolution spectrometers (RGS1 and RGS2; \citealt{den-Herder+2001_A&A}). It is noted that the Pipeline Processing System (PPS) data were used for the RGS lightcurves in T23, but we use the ODF files and reduce them with SAS for the detailed analysis in this paper. 
As for the EPIC events data, 
we used standard processing tasks \texttt{epproc} and \texttt{emproc}, 
and we selected the events for having energies in the band 0.2 -- 12.0 keV and \texttt{Flag} $==$ 0 and \texttt{Pattern} $\leq$4 (for EPIC-pn) and \texttt{Pattern} $\leq$ 12 (for EPIC-MOS1), respectively, following the \texttt{SAS} reduction guide \footnote{\url{https://www.cosmos.esa.int/web/xmm-newton/documentation}} \footnote{
\color{black} \texttt{Pattern} values represent the number and pattern of the CCD pixels triggered for a given event. We simply followed the standard filtering criteria provided by the \texttt{SAS} reduction guide.
}. The background-subtracted lightcurves are used for the EPIC-pn and EPIC-MOS1 lightcurves in the following part of this paper. As for the time-divided EPIC spectra in this paper, the spectra of the source and the background were created with \texttt{SAS}, along with the corresponding redistribution matrix files (\texttt{RMF}) and ancillary response files (\texttt{ARF}). For extracting the RGS spectra (Energy range: 0.33--2.5keV = 5--38 \AA, Spectral resolving power ($E/\Delta E$) between 150 and 800), we used the standard tasks \texttt{rgsproc} and \texttt{rgscombine}. We only used the first order data of RGS1 and RGS2. As for the RGS lightcurves in the following part of this paper, the RGS1 and RGS2 data are combined with the task \texttt{rgslccorr}, simply for a better count statistics. On the other hand, as for the RGS spectral analysis in the following part of this paper, the RGS1 and RGS2 data are not co-added and they are separately fitted, in order to avoid potential errors caused by combining two spectra from different cameras (RGS1 and RGS2) with \texttt{rgscombine} \footnote{\url{https://xmm-tools.cosmos.esa.int/external/sas/current/doc/rgscombine/}}.

\subsection{Multi-wavelength flare lightcurve atlas}\label{subsec:flare_atlas}

 \begin{figure}[ht!]
   \begin{center}
      \gridline{
    \fig{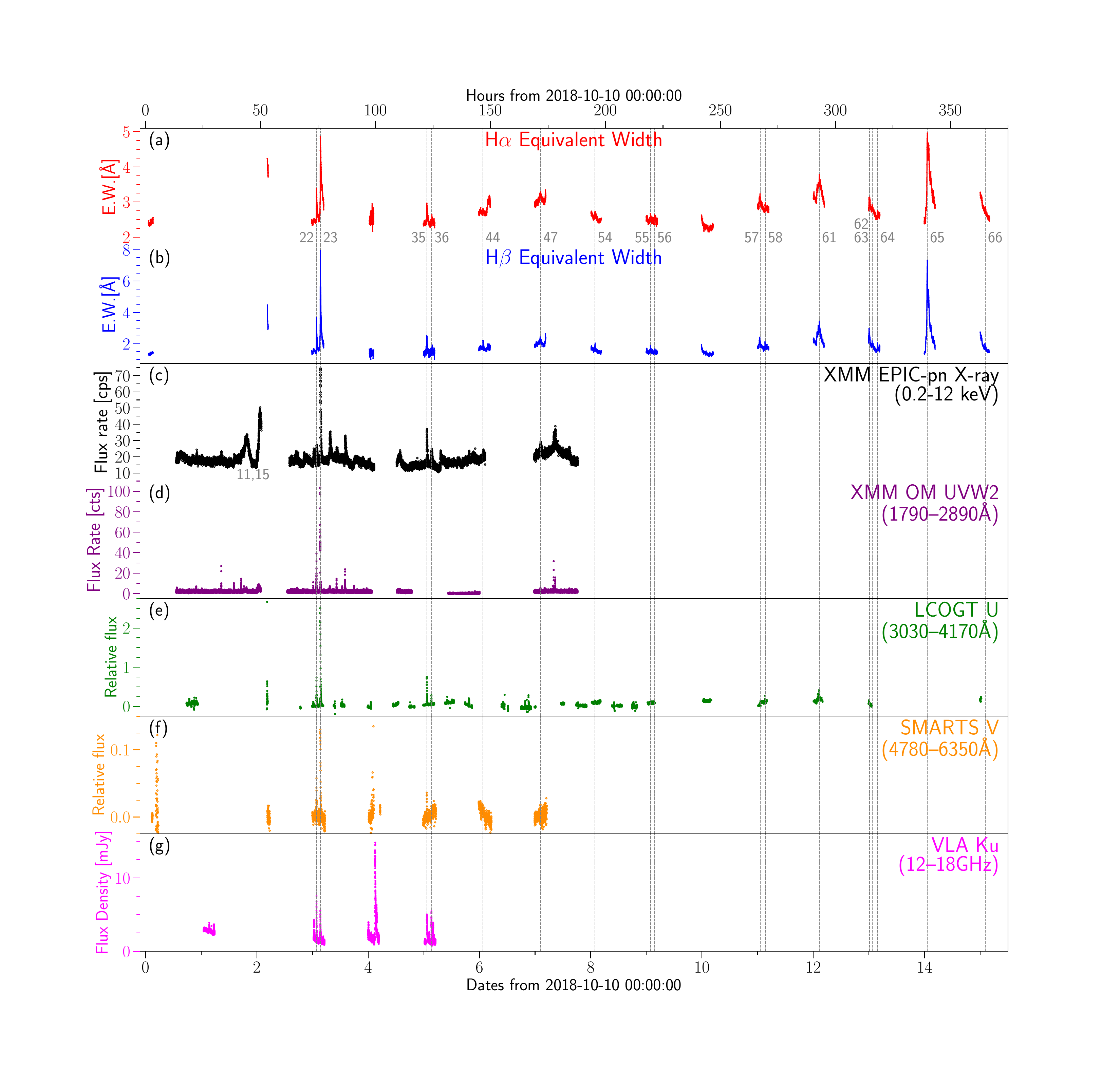}{1.0\textwidth}{\vspace{0mm}}
    }
     \vspace{-10mm}
     \caption{
Summary lightcurves of the AU Mic X-ray campaign that overlap with the SMARTS/CHIRON spectroscopic data (2018 October 10 -- 25).
(a)\&(b) H$\alpha$ and H$\beta$ equivalent width (E.W.) values from the SMARTS/CHIRON spectroscopic data  with the time cadence of 65 seconds. 
Gray numbers (in (a)) and gray dashed lines (in all the panels) show H$\alpha$ flares and their timings, respectively, which are identified in T23 (cf. Table 6 of T23). 
(c)\&(d) XMM EPIC-pn X-ray (0.2 -- 12 keV) and XMM OM UVW2 data are plotted in units of counts per second, and with 30 and 10 sec binnings, respectively. 
Flares 11 and 15, which are discussed in detail in Section \ref{subsec:ana_flare_11_and_15}, are marked with the gray numbers in (c). 
(e)\&(f) LCOGT U-band and SMARTS V-band photometric data are plotted in relative flux units, and with the time cadence of 46 and 47 seconds, respectively. 
(g) VLA Ku band data in the unit of flux density (mJy) and with the time binning of 10 seconds (from T25).
     }
   \label{fig:allEW_AUMic_XMM_opt}
   \end{center}
 \end{figure}

\begin{figure}[ht!]
   \begin{center}
      \gridline{
    \fig{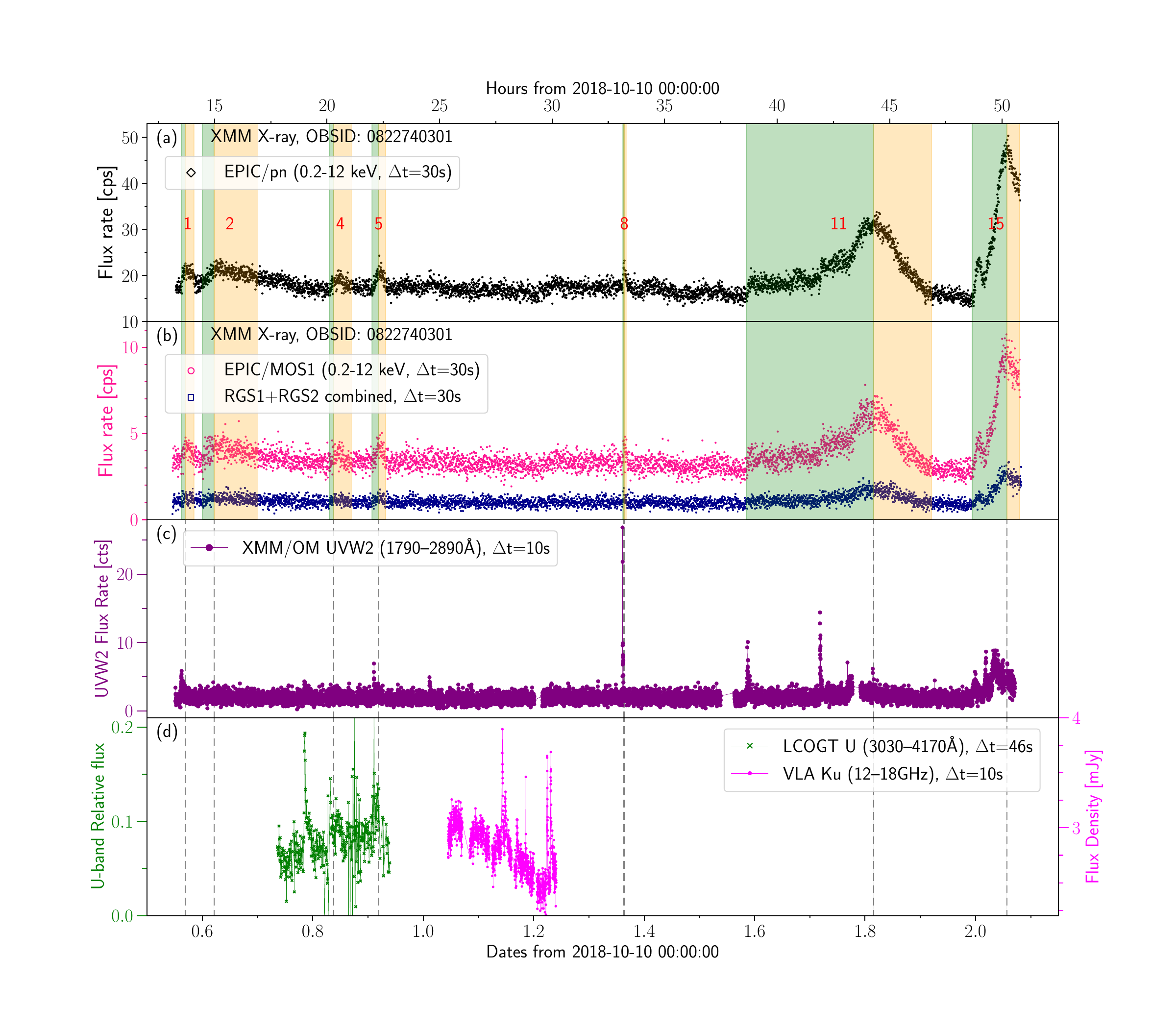}{0.9\textwidth}{\vspace{0mm}}
    }
     \vspace{-10mm}
     \caption{
(a) XMM EPIC-pn X-ray (0.2 -- 12 keV) lightcurve of AU Mic for the data period of Obs-ID 0822740301 (2018 October 10 13:13:59.760 -- 2018 October 12 01:42:19.177 UTC),
plotted in units of counts per second, and with 30 sec binning. 
The red numbers show flares identified by T23, and the start, peak, and end times of these flares (from Table 6 of T23) are shown with green and orange colored regions. 
(b) Same as (a) but for the EPIC-MOS1 (0.2 -- 12 keV) lightcurve and the combined one of RGS1 and RGS2 data.
(c)\&(d) XMM OM UVW2, LCOGT U-band, and VLA Ku band lightcurves as in Figure \ref{fig:allEW_AUMic_XMM_opt}(d)--(g), but for the data period of the X-ray data in (a)\&(b). The flare peak times in the XMM X-ray data (in (a)) are also plotted as gray dashed lines in (c)\&(d).
     }
   \label{fig:X-ray_Ha_obsid_0822740301_lc}
   \end{center}
 \end{figure}

 \begin{figure}[ht!]
   \begin{center}
      \gridline{
    \fig{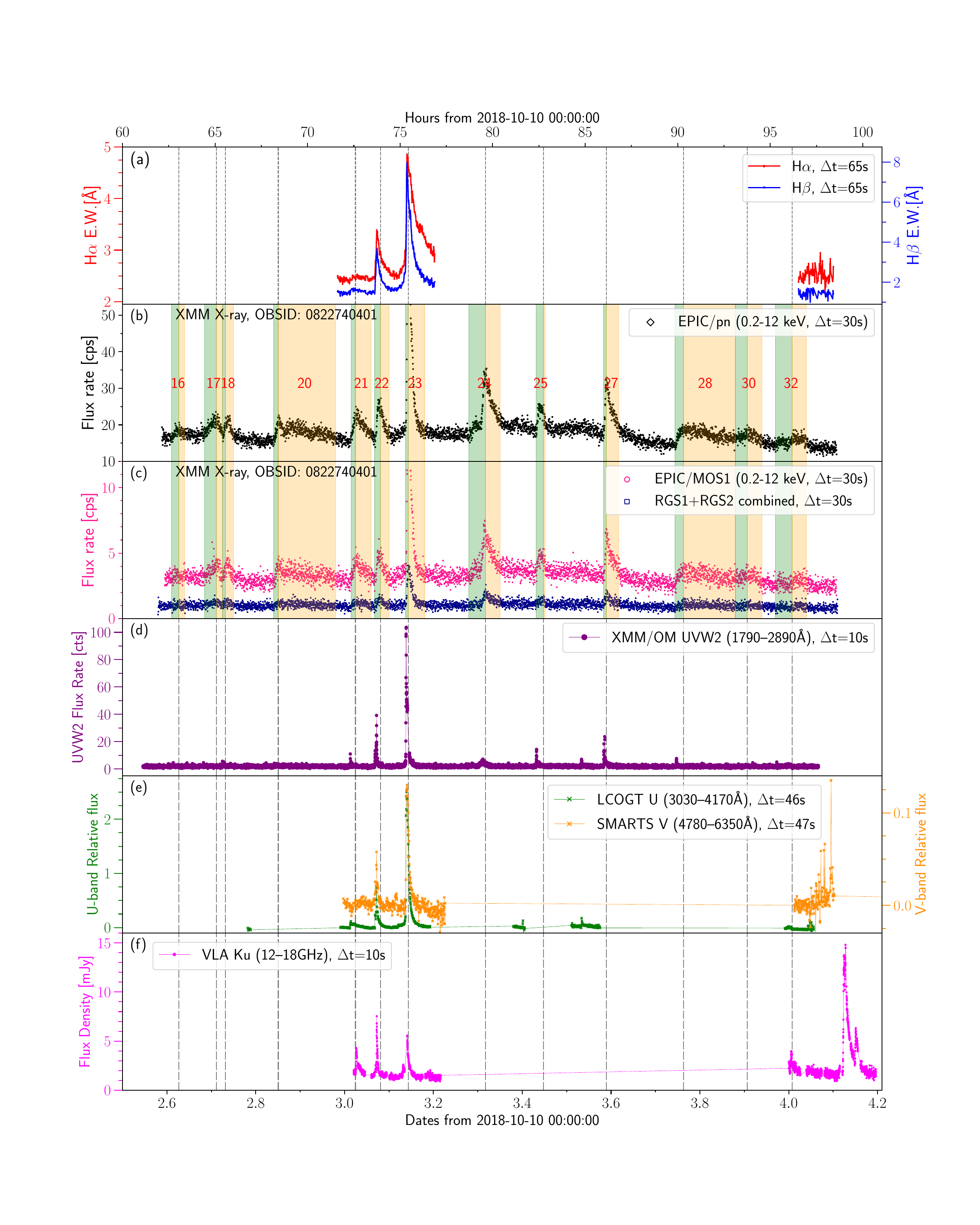}{0.9\textwidth}{\vspace{0mm}}
    }
     \vspace{-10mm}
     \caption{
(a) H$\alpha$ and H$\beta$ equivalent width (E.W.) lightcurves as in Figure \ref{fig:allEW_AUMic_XMM_opt}(a)\&(b), but for the data of 2018 October 13 and 14, which overlap with the X-ray data in (b)\&(c).
(b)\&(c) The X-ray lightcurve data as Figure \ref{fig:X-ray_Ha_obsid_0822740301_lc}(a)\&(b) but for Obs-ID 0822740401 (2018 October 12 13:06:33.808 to 2018 October 14 01:36:22.232 UTC). 
(d),(e),\&(f) XMM OM UVW2, LCOGT U-band, SMARTS V-band and VLA Ku band lightcurves as in Figure \ref{fig:allEW_AUMic_XMM_opt}(d)--(g) and Figure \ref{fig:X-ray_Ha_obsid_0822740301_lc}(c)\&(d), but for the data period of the X-ray data in (b)\&(c)
     }
   \label{fig:X-ray_Ha_obsid_0822740401_lc}
   \end{center}
 \end{figure}

 \begin{figure}[ht!]
   \begin{center}
      \gridline{
    \fig{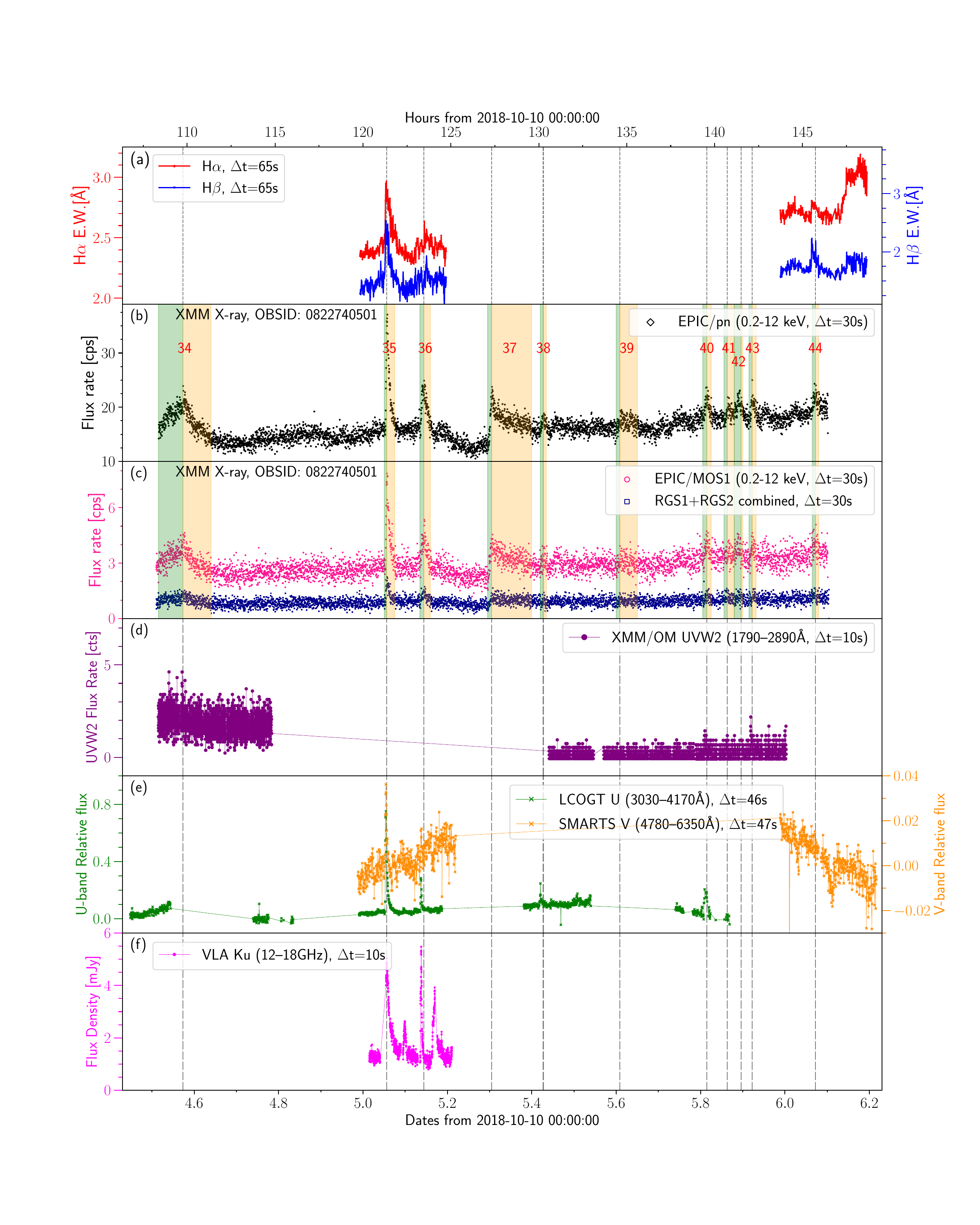}{0.9\textwidth}{\vspace{0mm}}
    }
     \vspace{-10mm}
     \caption{
Same as Figure \ref{fig:X-ray_Ha_obsid_0822740401_lc}, but for the period of the XMM X-ray data Obs-ID 0822740501 (2018 October 14 12:21:18.787 to 2018 October 16 00:04:00.273 UTC).
     }
   \label{fig:X-ray_Ha_obsid_0822740501_lc}
   \end{center}
 \end{figure}

\begin{figure}[ht!]
   \begin{center}
      \gridline{
    \fig{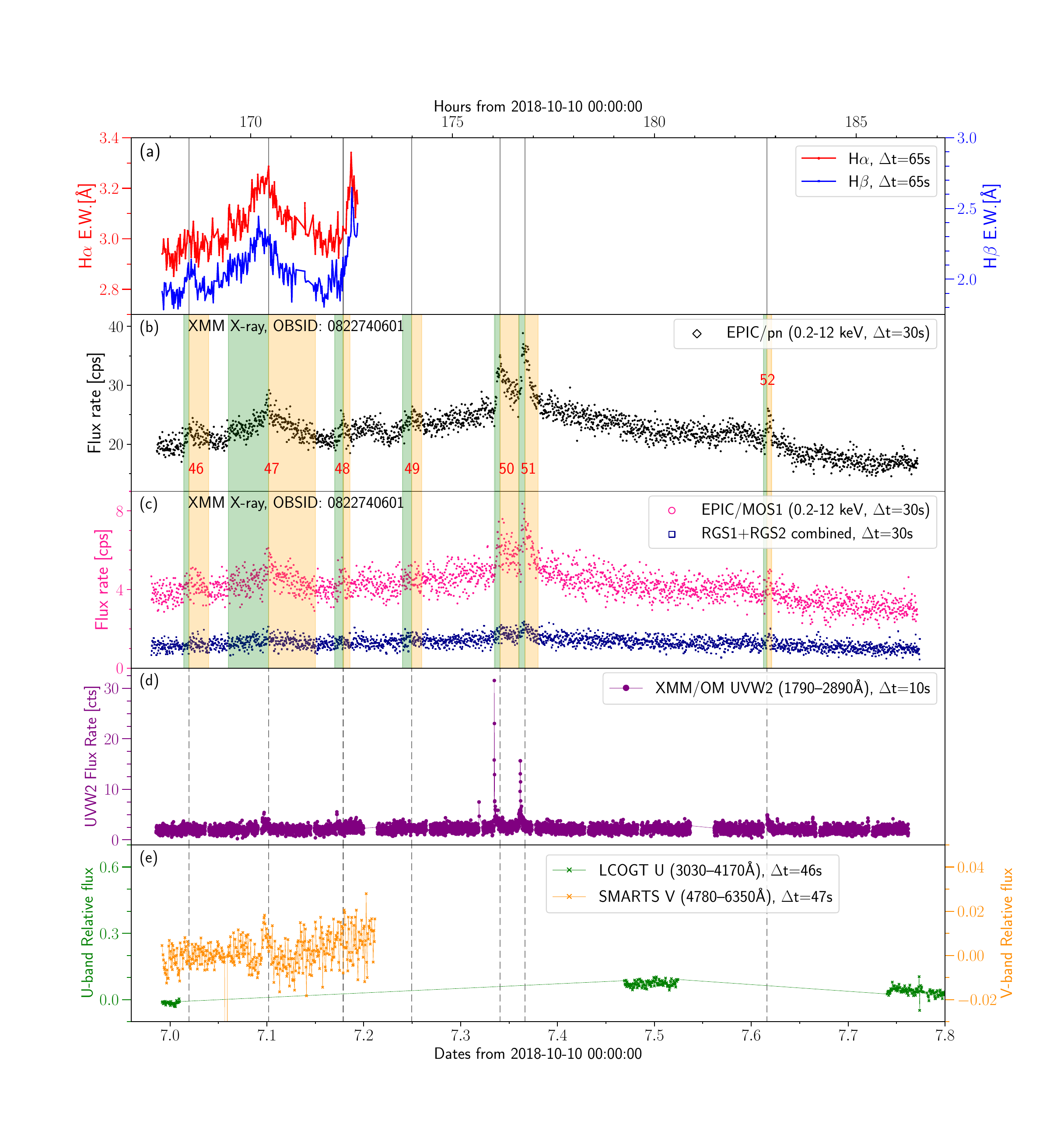}{0.9\textwidth}{\vspace{0mm}}
    }
     \vspace{-10mm}
     \caption{
Same as Figure \ref{fig:X-ray_Ha_obsid_0822740401_lc}, but for the period of the XMM X-ray data Obs-ID 0822740601 (2018 October 16 23:39:21.362 to 2018 October 17 18:17:33.004 UTC ). No overlapping VLA Ku band data exist here (cf. Figure \ref{fig:allEW_AUMic_XMM_opt}).
     }
   \label{fig:X-ray_Ha_obsid_0822740601_lc}
   \end{center}
 \end{figure}

\begin{figure}[ht!]
   \begin{center}
      \gridline{
    \fig{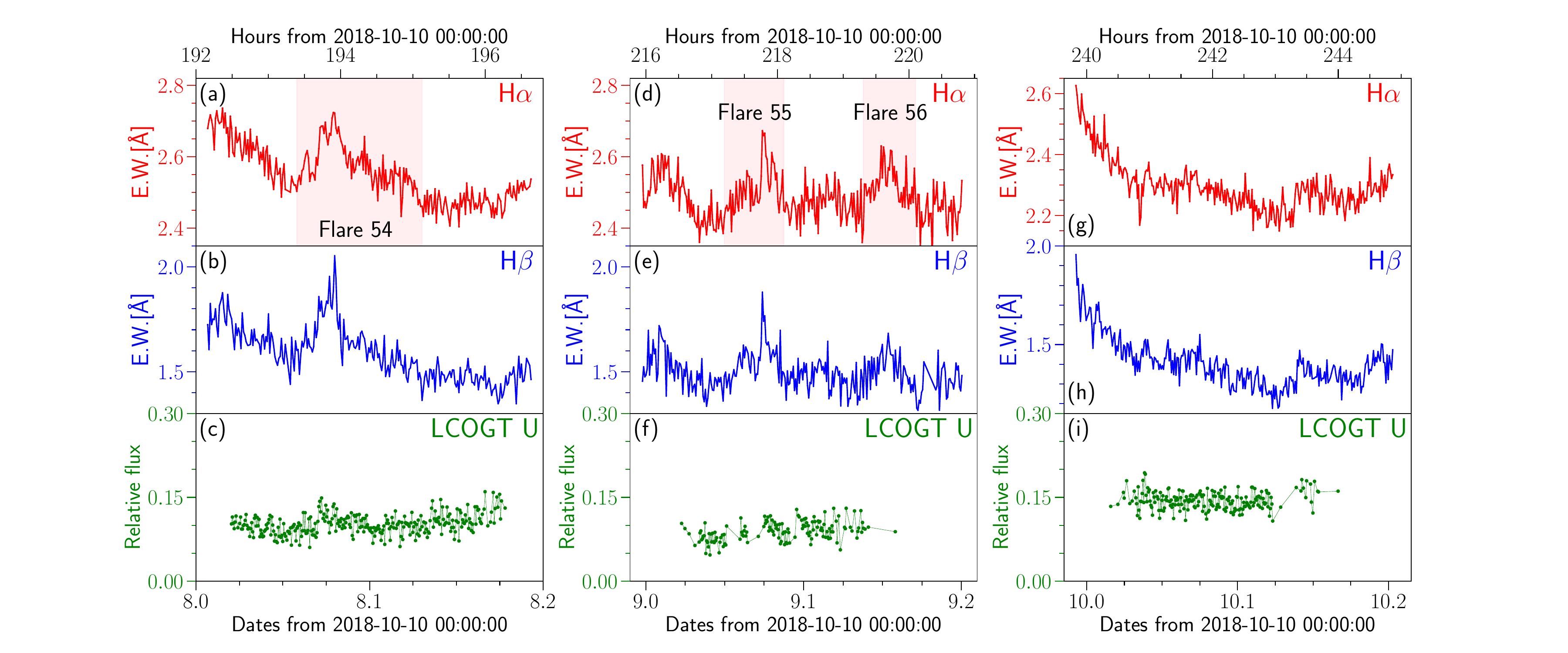}{1.0\textwidth}{\vspace{0mm}}
    }
     \vspace{-5mm}
     \caption{
H$\alpha$, H$\beta$, and LCOGT U-band lightcurves of 2018 October 18, 19, and 20. The flares identified in H$\alpha$ data by T23 are shown with pink colored regions, which correspond to the start and end times of these flares (from Table 6 of T23).
     }
   \label{fig:lc_HaHb_only_flares_Oct18-20}
   \end{center}
 \end{figure}

\begin{figure}[ht!]
   \begin{center}
      \gridline{
    \fig{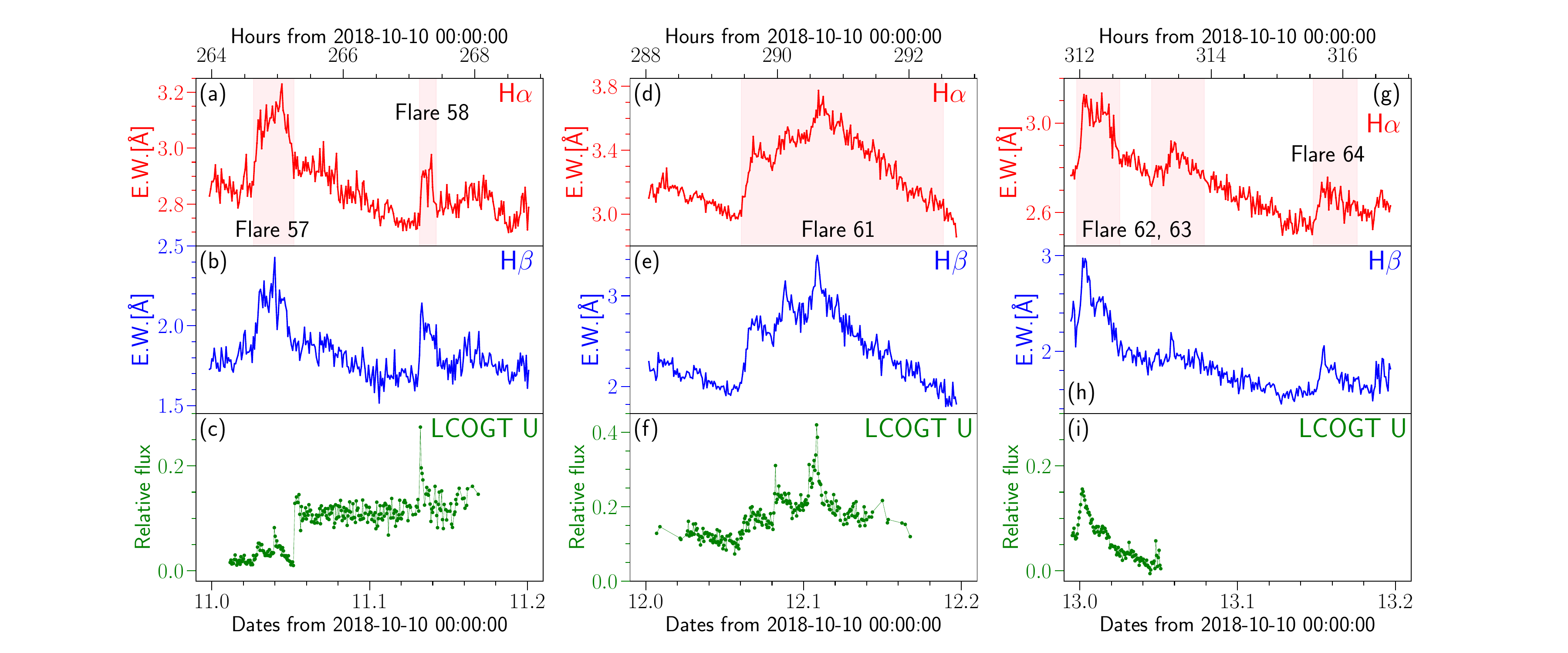}{1.0\textwidth}{\vspace{0mm}}
    }
     \vspace{-5mm}
     \caption{
Same as Figure \ref{fig:lc_HaHb_only_flares_Oct18-20}, but for the data of 2018 October 21, 22, and 23.
     }
   \label{fig:lc_HaHb_only_flares_Oct21-23}
   \end{center}
 \end{figure}

\begin{figure}[ht!]
   \begin{center}
      \gridline{
    \fig{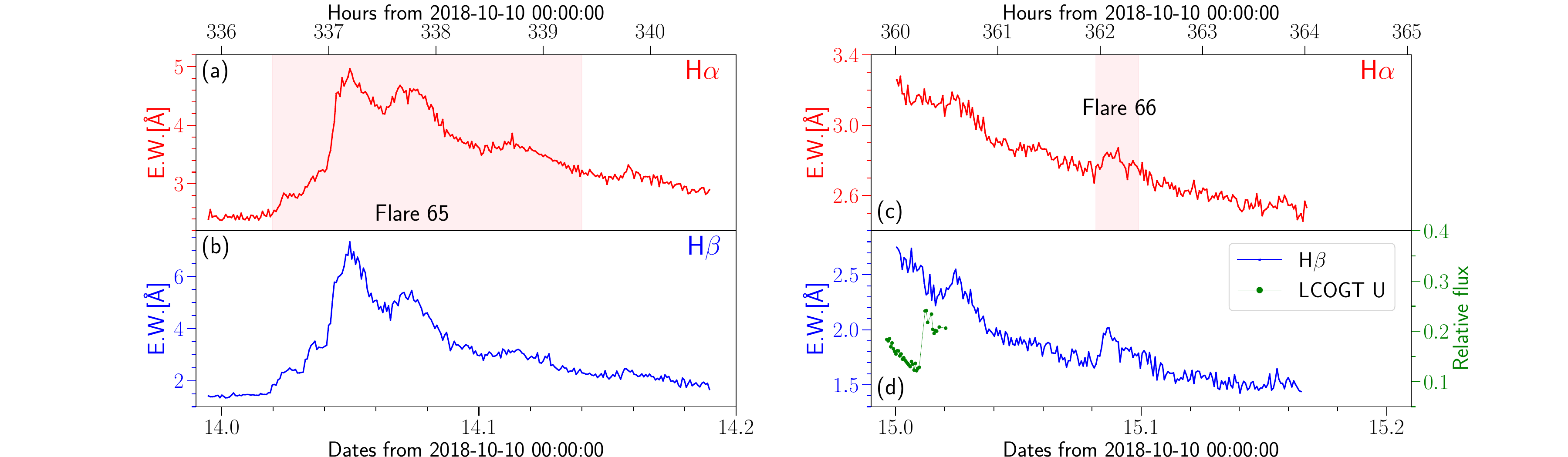}{1.0\textwidth}{\vspace{0mm}}
    }
     \vspace{-5mm}
     \caption{
Same as Figure \ref{fig:lc_HaHb_only_flares_Oct18-20}, but for the data of 2018 October 24 and 25.
     }
   \label{fig:lc_HaHb_only_flares_Oct24-25}
   \end{center}
 \end{figure}

Figure \ref{fig:allEW_AUMic_XMM_opt} shows the overall multi-wavelength lightcurves from 2018 October 10 to 25 for overlapping times with the SMARTS/CHIRON observations (see T23 for the complete campaign lightcurve data including the LCOGT U-band data on October 26-29 and the entire LCOGT V-band data). Enlarged multi-wavelength lightcurves are shown in  Figures \ref{fig:X-ray_Ha_obsid_0822740301_lc} -- \ref{fig:lc_HaHb_only_flares_Oct24-25}.
As already mentioned in T23, the H$\alpha$ and H$\beta$ show gradually varying non-flare components with H$\alpha$ equivalent width (E.W.) modulation amplitude of $\sim$0.5\AA, in addition to clear and sudden brightenings identified as flares.
The X-ray data may also show similar longer-term gradual variations, though it is not so clear as the H$\alpha$ modulation. These periods of gradual variation may correspond to rotational modulation (\citealt{Toriumi+2020,Maehara+2021,Schofer+2022})
considering AU Mic's rotation period of 4.863 days (\citealt{Ikuta+2023_ApJ}).
The X-ray and H$\alpha$ flares listed in Table 6 of T23 are shown in Figures \ref{fig:allEW_AUMic_XMM_opt} -- \ref{fig:lc_HaHb_only_flares_Oct24-25}, and the flare identification methods are presented in Section 3.1 of T23. 
\color{black}
It is noted some brightnings in the H$\alpha$ and H$\beta$ lightcurves are not defined as flares in T23 (e.g., large E.W. values but limited data duration on Oct 12, continuous decrease trend around Flare 66 on Oct 25), because of the difficulties caused by the limited length of the observation periods. 
As also described in \citet{Notsu+2024_ApJ}, 
overlapping flares can occur where a second flare
starts before the preceding
flare emission completely decays. There are also some partial flares and their observed flare properties (e.g., flare energies) could include various 
uncertainties since only portions of the flare phases were observed.
These points can cause some uncertainty of definitions of each flare, 
but the main purpose of the CHIRON data analysis in this paper 
is to discuss the line profile changes in obvious flares \footnote{\color{black} It is noted that the CHIRON data of Flares 21, 22, 23, 35, 36, and 65 are discussed in Section \ref{sec:remarkable_flares} but these flares are not strongly affected the flare identification uncertainty described here, since flare peak and start times of these flares can be defined clearly}. 
Because of this, some uncertainty of flare definitions were left,
as long as they are not expected to cause a serious problem for the main conclusion of this paper.  
\color{black}

38 flares are identified in EPIC-pn X-ray lightcurve data (Figures \ref{fig:X-ray_Ha_obsid_0822740301_lc} -- \ref{fig:X-ray_Ha_obsid_0822740601_lc}). The time-averaged spectral components of these 38 flares are discussed in Section \ref{subsec:X-ray_specana_time-average}.
17 flares are identified in H$\alpha$ and H$\beta$ lines (Figures \ref{fig:allEW_AUMic_XMM_opt} -- \ref{fig:lc_HaHb_only_flares_Oct24-25}). 
Among them, Flare 23 has  good multi-wavelength coverage and will be investigated in detail in Section \ref{subsec:ana_flare_23}. The clear line profile broadenings  during Flare 23 are emphasized in this paper.
Flares 11 and 15 are investigated  in Section \ref{subsec:ana_flare_11_and_15} since they have much longer durations and larger amplitudes than other X-ray flares, though there are no SMARTS/CHIRON optical spectra for these two flares. Flares 35 and 36 are highlighted in Section \ref{subsec:ana_flare_35_and_36} since there is a potential X-ray dimming event following these flares, which was originally reported by \citet{Veronig+2021} 
as a stellar CME candidate 
using the same XMM-Newton dataset. 
Flare 65 is additionally investigated  in Section \ref{subsec:ana_flare_65}, since this flare shows clear line profile changes, though there are no datasets other 
than the SMARTS/CHIRON optical spectra.  
Profiles of the H$\alpha$ and H$\beta$ lines of the other smaller flares  are investigated as done for other M-dwarfs in \citet{Notsu+2024_ApJ}, but the other flares did not show clear line profile blue/red wing asymmetries or 
\color{black} clear notable line broadenings compared with quiescent line profile widths of H$\alpha$ \& H$\beta$ lines ($\sim$100 km s$^{-1}$)\color{black}.

\clearpage

\section{X-ray spectral analysis} \label{sec:X-ray_spana1}
\subsection{Quiescent phase X-ray spectra}\label{subsec:X-ray_specana_quiescent}

\begin{figure}[ht!]
   \begin{center}
      \gridline{
\fig{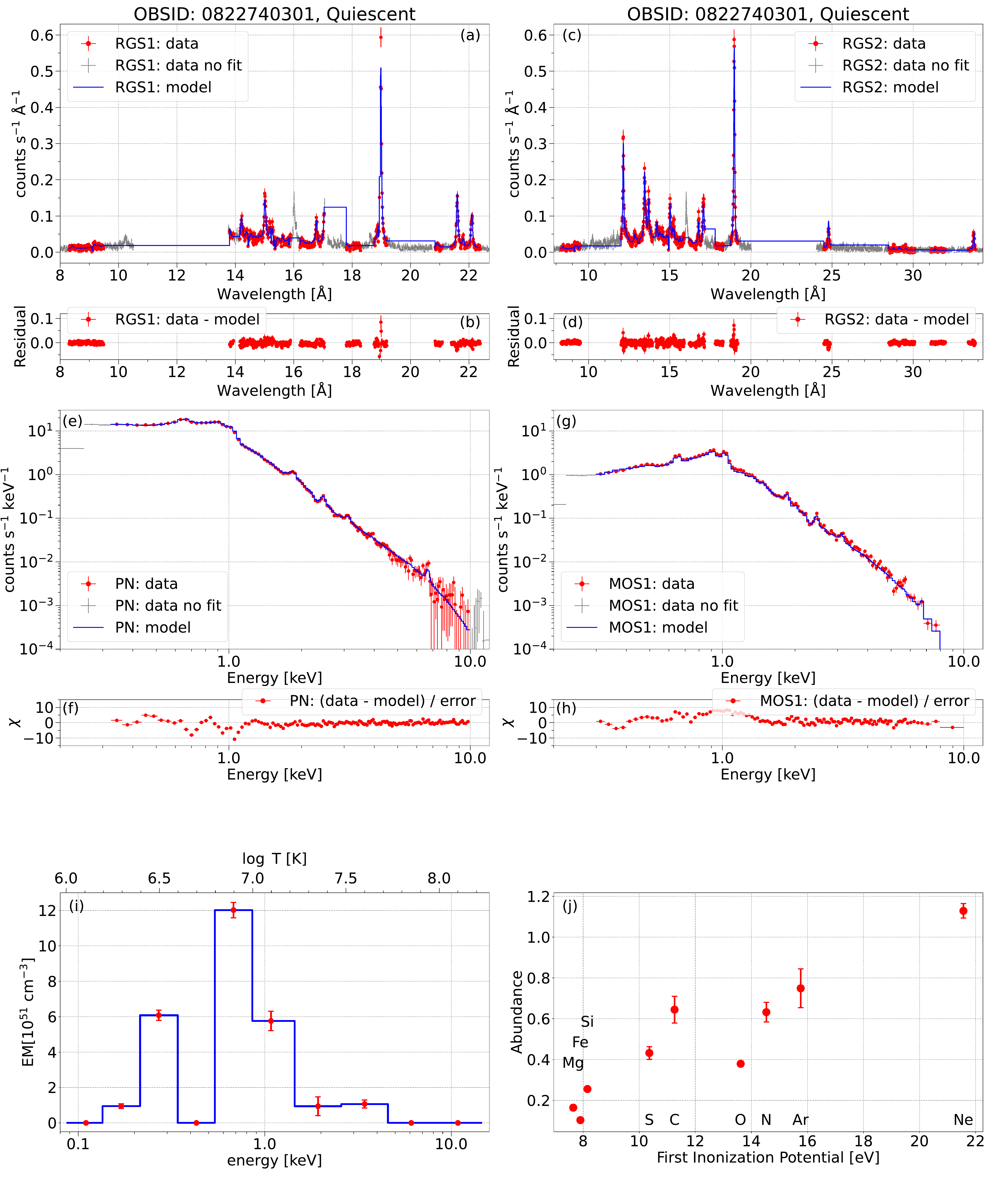}{1.0\textwidth}{\vspace{0mm}}
    }
     \vspace{-5mm}
     \caption{
The quiescent phase X-ray spectra of the Obs-ID 0822740301 (cf. Figure \ref{fig:X-ray_Ha_obsid_0822740301_lc}) with the 10-temperature fit with fixed temperature bins (Table \ref{table:X-ray_quie_fit_results}). 
(a), (c), (e), \& (g) The data (red circles) and model fits (blue lines) of RGS1, RGS2, PN, and MOS1 are plotted, respectively.
Gray points show the wavelength ranges excluded from the fittings (Table \ref{table:X-ray_quie_fit_wavelength}).  The difference between data and model fits are plotted in (b) and (d), and $\chi=(\rm{data}-\rm{model}) / \rm{error}$ are plotted in (f) and (h).
(i) Emission Measure (EM) distribution from the fitting. 
(j) Abundances versus the First Ionization Potential (FIP). 
The FIP values (eV) are from \citet{FIP_table}.}
   \label{fig:specfit_QuieALL1_0822740301}
   \end{center}
 \end{figure}

We first conduct spectral analysis of the quiescent component data for each Obs-ID, 
as a basis for the flare component analyses in later sections of this paper. The quiescent component data for each Obs-ID are extracted by integrating the PN, MOS1, RGS1, and RGS2 spectroscopic data over the time periods that were not identified as flares in EPIC-PN lightcurve data by T23 (= the regions that are not marked with green \& yellow colors in Figures \ref{fig:X-ray_Ha_obsid_0822740301_lc} -- \ref{fig:X-ray_Ha_obsid_0822740601_lc}). 
For example, the quiescent component average count rates of PN lightcurves (0.2 -- 12 keV) are 
17.07 (Obs-ID: 0822740301), 16.95 (Obs-ID: 0822740401), 15.71 (Obs-ID: 0822740501), 
and 21.42 (Obs-ID: 0822740601), respectively ($QC_{\rm{PN}}^{\rm{OBSID}}$ in Table \ref{table:X-ray_quie_fit_results}).
The extracted spectroscopic data are plotted in 
\color{black}
Figure \ref{fig:specfit_QuieALL1_0822740301} in this section 
and Figures \ref{fig:specfit_QuieALL1_0822740401} -- 
\ref{fig:specfit_QuieALL1_0822740601} in Appendix \ref{appen_sec:additional_figures}.
\color{black}
We then apply simultaneous X-ray spectral fitting to these RGS1, RGS2, PN, and MOS1 spectra with the 10 fixed temperature collisionally ionized equilibrium components (\texttt{vapec}) with interstellar absorption (\texttt{tbabs}). 
The approach of the 10 fixed temperature fitting is used, following \citet{Raassen+2007}, in order to avoid a priori assumptions about the number of free temperature bins, while the total number of free parameters stays approximately the same. 
In this study, the temperature bins are fixed as $T_{1}$ -- $T_{10}$ in Table \ref{table:X-ray_quie_fit_results}, with band widths of 0.2 dex in the range of $\log T\mathrm{[K]}=6.1-7.1$ (0.11--1.08 keV) and 0.25 dex in the range of 
$\log T\mathrm{[K]}=7.1-8.1$ (1.08 -- 10.85 keV). 
These bins are slightly changed from \citet{Raassen+2007} in two ways. First, 
the temperature bins in low temperature ranges ($\lesssim$0.1 keV, $\log T\mathrm{[K]}\lesssim 6$) where there are no emission measure (EM) contributions (see also Figure 4 of \citealt{Raassen+2007}) are covered by only one temperature bin at $\log T\mathrm{[K]}=6.1$. Second, the higher temperature ranges with more bins. We fix the hydrogen column density at the literature value $N_{\rm{H}}$ = 2.29$\times$10$^{18}$ cm$^{-2}$ (\citealt{Wood+2005_ApJS}). The fitting analysis is conducted using \texttt{Xspec} v12.14.1 and the associated \texttt{PyXspec} software (\citealt{Arnaud+1996_ASPC,Gordon_Arnaud_2021_ascl}).  
As for PN and MOS1 data, the range of 0.3--10 keV is used for fitting, following \citet{Pillitteri+2022}. 
Table \ref{table:X-ray_quie_fit_wavelength} summarizes the wavelength ranges 
of RGS1 and RGS2 spectra excluded from the fitting procedure, following Table 3 of \citet{Audard+2003_A+A}. 

\begin{deluxetable*}{l|c|c|c|c}[htbp]
   \tablecaption{Multi-temperature fitting parameters for the quiescent component of each Obs-ID, obtained from a model with fixed ten temperatures (see the text). Elemental abundances ($Z$) are relative to the solar photospheric values of \citet{Anders+1989_GeCoA}
   (the default setting of \texttt{Xspec}).
   Statistical 90\% confidence region errors are shown, and the values without error ranges are fixed.}
   \tablewidth{0pt}
   \tablehead{
     \colhead{Parameters} &  \multicolumn{4}{c}{Obs-IDs}  \\
   \colhead{}  &  \colhead{0822740301} &  \colhead{0822740401} &  \colhead{0822740501} &  \colhead{0822740601}    
}
   \startdata
      $QC_{\rm{PN}}^{\rm{OBSID}}$ (cps, 0.2 -- 12 keV) \tablenotemark{\rm \dag} & 17.07 & 16.95 & 15.71 & 21.42 \\
      $QC_{\rm{MOS1}}^{\rm{OBSID}}$ (cps, 0.2 -- 12 keV)\tablenotemark{\rm \dag} & 3.25 & 3.13 & 2.82 &  4.02 \\
      $QC_{\rm{RGS1}+\rm{RGS2}}^{\rm{OBSID}}$ (cps) \tablenotemark{\rm \dag} & 1.00 & 0.99 & 0.90 & 1.25 \\
\cline{1-5} 
   $N_{\rm{H}}$[cm$^{-2}$] \tablenotemark{\rm \ddag} &  \multicolumn{4}{c}{2.29$\times$10$^{18}$} \\ 
   $T_{1}$(keV), $\log T_{1}$(K) &  \multicolumn{4}{c}{0.11, 6.10} \\
   $T_{2}$(keV), $\log T_{2}$(K) &  \multicolumn{4}{c}{0.17, 6.30} \\
   $T_{3}$(keV), $\log T_{3}$(K) &  \multicolumn{4}{c}{0.27, 6.50} \\
   $T_{4}$(keV), $\log T_{4}$(K) &  \multicolumn{4}{c}{0.43, 6.70} \\
   $T_{5}$(keV), $\log T_{5}$(K) &  \multicolumn{4}{c}{0.68, 6.90} \\
   $T_{6}$(keV), $\log T_{6}$(K) &  \multicolumn{4}{c}{1.08, 7.10} \\
   $T_{7}$(keV), $\log T_{7}$(K) &  \multicolumn{4}{c}{1.93, 7.35} \\
   $T_{8}$(keV), $\log T_{8}$(K) &  \multicolumn{4}{c}{3.43, 7.60} \\
   $T_{9}$(keV), $\log T_{9}$(K) &  \multicolumn{4}{c}{6.10, 7.85} \\
   $T_{10}$(keV), $\log T_{10}$(K) &  \multicolumn{4}{c}{10.85, 8.10} \\
     \cline{1-5}   
      EM$_{1}$($10^{51}$ cm$^{-3}$) & 0.00 & 0.00 & 0.00 & 0.00 \\
      EM$_{2}$($10^{51}$ cm$^{-3}$) & 0.95 $\pm$ 0.13 & 0.80 $\pm$ 0.15 & 1.33 $\pm$ 0.14 & 1.22 $\pm$ 0.22 \\
      EM$_{3}$($10^{51}$ cm$^{-3}$) & 6.08 $\pm$ 0.29 & 5.85 $\pm$ 0.36 & 6.01 $\pm$ 0.32 & 6.58 $\pm$ 0.51 \\
      EM$_{4}$($10^{51}$ cm$^{-3}$) & 0.00 & 0.53 $\pm$ 0.48 & 0.86 $\pm$ 0.43 &  1.36 $\pm$ 0.73 \\
      EM$_{5}$($10^{51}$ cm$^{-3}$) & 12.02 $\pm$ 0.43 & 9.86 $\pm$ 0.56 & 10.68 $\pm$ 0.52 & 12.39 $\pm$ 0.80 \\
      EM$_{6}$($10^{51}$ cm$^{-3}$) & 5.76 $\pm$ 0.55 & 6.66 $\pm$ 0.55 & 4.70 $\pm$ 0.49 & 10.60 $\pm$ 0.81 \\
      EM$_{7}$($10^{51}$ cm$^{-3}$) & 0.94 $\pm$ 0.53 & 0.65 $\pm$ 0.52 & 0.86 $\pm$ 0.47 & 2.09 $\pm$ 0.77 \\
      EM$_{8}$($10^{51}$ cm$^{-3}$) & 1.07 $\pm$ 0.23 & 1.04 $\pm$ 0.22 & 1.09 $\pm$ 0.20 & 1.86 $\pm$ 0.31 \\
      EM$_{9}$($10^{51}$ cm$^{-3}$) & 0.00 & 0.00 & 0.00 & 0.00  \\
      EM$_{10}$($10^{51}$ cm$^{-3}$) & 0.00 & 0.00 & 0.00 & 0.00 \\
     \cline{1-5} 
      EM$_{\rm{tot}}$($10^{51}$ cm$^{-3}$) & 26.82 $\pm$ 0.96 & 25.39 $\pm$ 1.14 & 25.53 $\pm$ 1.04 & 36.09 $\pm$ 1.68 \\
        $T_{\rm{ave}}$(keV) & 0.81 $\pm$ 0.05 & 0.81 $\pm$ 0.06  & 0.78 $\pm$ 0.05 &  0.91 $\pm$ 0.06 \\
        $\log T_{\rm{ave}}$(K) & 6.97 $\pm$ 0.03 & 6.98 $\pm$ 0.03 & 6.96 $\pm$ 0.03 &  7.02 $\pm$ 0.03 \\
      $F_{\rm{X}}$(0.2 -- 12 keV) (10$^{-11}$ erg cm$^{-2}$ s$^{-1}$) & 2.57 $\pm$ 0.01 & 2.53 $\pm$ 0.01 & 2.36 $\pm$ 0.01 & 3.29 $\pm$ 0.01 \\
      $L_{\rm{X}}$(0.2 -- 12 keV) (10$^{29}$ erg s$^{-1}$) & 2.91 $\pm$ 0.01 & 2.86 $\pm$ 0.01  & 2.67 $\pm$ 0.01 & 3.72 $\pm$ 0.01 \\
     \cline{1-5}  
 $Z$(He) & 1.00 & 1.00 & 1.00 & 1.00 \\
 $Z$(C)  & 0.64 $\pm$ 0.07 & 0.69 $\pm$ 0.08 & 0.61 $\pm$ 0.06 & 0.60 $\pm$ 0.07 \\
 $Z$(N) & 0.63 $\pm$ 0.05 & 0.71 $\pm$ 0.06 & 0.56 $\pm$ 0.04 & 0.58 $\pm$ 0.05 \\
 $Z$(O) & 0.38 $\pm$ 0.01 & 0.42 $\pm$ 0.02 & 0.35 $\pm$ 0.01 & 0.33 $\pm$ 0.01 \\
 $Z$(Ne) & 1.13 $\pm$ 0.04 & 1.15 $\pm$ 0.04 & 1.04 $\pm$ 0.03 & 1.01 $\pm$ 0.04 \\
 $Z$(Mg) & 0.16 $\pm$ 0.01 & 0.17 $\pm$ 0.02 & 0.16 $\pm$ 0.01 & 0.13 $\pm$ 0.02 \\
 $Z$(Al) & 0.50 & 0.50 & 0.50 & 0.50 \\
 $Z$(Si) & 0.26 $\pm$ 0.01 & 0.29 $\pm$ 0.02 & 0.27 $\pm$ 0.01 & 0.23 $\pm$ 0.01 \\
 $Z$(S) & 0.43 $\pm$ 0.03 & 0.42 $\pm$ 0.03 & 0.37 $\pm$ 0.03 &  0.34 $\pm$ 0.03 \\
 $Z$(Ar) & 0.75 $\pm$ 0.10 & 0.89 $\pm$ 0.11 & 0.65 $\pm$ 0.09 & 0.58 $\pm$ 0.10 \\
 $Z$(Ca) \tablenotemark{\rm \ddag}  & = Fe & = Fe & = Fe &  = Fe\\
 $Z$(Fe) & 0.10 $\pm$ 0.01 & 0.11 $\pm$ 0.01 & 0.10 $\pm$ 0.01 & 0.10 $\pm$ 0.01 \\
 $Z$(Ni) \tablenotemark{\rm \ddag}  & = Fe & = Fe & = Fe & = Fe \\
      \cline{1-5}  
 $\chi^{2}/\rm{d.o.f}$ & 4337/1809 & 3477/1793 & 4046/1820 & 3407/1804 \\
 $\chi^{2}_{\rm{black}}$ & 2.40 & 1.94 & 2.22 & 1.89 \\
   \enddata
    \tablenotetext{\dag}{
Average count rates of PN, MOS1, and RGS1+RGS2 lightcurves in the quiescent phase of each Obs-ID data.
   }   
 \tablenotetext{\ddag}{
$N_{\rm{H}}$ is fixed to the literature value from \citet{Wood+2005_ApJS}.
Ca and Ni abundance values are fixed to that of Fe.
   }
   \label{table:X-ray_quie_fit_results}
 \end{deluxetable*}

 \begin{deluxetable*}{cc}[ht!]
   \tablecaption{RGS spectral wavelength ranges \textit{excluded} from the fitting procedure.}
   \tablewidth{0pt}
   \tablehead{
\colhead{Instrument} &  \colhead{Wavelength range (\AA)}    
}
   \startdata
RGS1 & $\leq$8.30 \\
RGS1 & 9.50--13.80  \tablenotemark{\rm \dag} \\
RGS1 & 13.95--14.15   \\
RGS1 & 15.90--16.20   \\
RGS1 & 17.15--17.80   \\
RGS1 & 18.30--18.75   \\
RGS1 & 19.20--20.80   \\
RGS1 & 21.10--21.40   \\
RGS1 & $\geq$22.40 \\
 \cline{1-2}
RGS2 & $\leq$8.30  \\
RGS2 & 9.50--12.00   \\
RGS2 & 13.95--14.15   \\
RGS2 & 15.90--16.20   \\
RGS2 & 17.15--17.80   \\
RGS2 & 18.30--18.75   \\
RGS2 & 19.20--24.50  \tablenotemark{\rm \dag}  \\
RGS2 & 24.90--28.50   \\
RGS2 & 30.10--31.10   \\
RGS2 & 32.00--33.40   \\
RGS2 & $\geq$33.85   \\
   \enddata
  % \tablecomments{
 %  }
 \tablenotetext{\dag}{
The two wavelength ranges of 10.6 -- 13.8\AA~and 20.0 -- 24.1\AA~are not available in RGS1 and RGS2, respectively, because of the CCD failures (\citealt{de-Vries_2015_A+A}). These two ranges are not used for fitting, in addition to the ranges from \citet{Audard+2003_A+A}.
   }
   \label{table:X-ray_quie_fit_wavelength}
 \end{deluxetable*}

The fitted RGS1, RGS2, PN, and MOS1 spectra in quiescent phase of each ObsID are shown in 
\color{black} 
Figures \ref{fig:specfit_QuieALL1_0822740301} \&  
\ref{fig:specfit_QuieALL1_0822740401} -- 
\ref{fig:specfit_QuieALL1_0822740601}, 
\color{black} 
and the best fitting parameters with statistical 90\% confidence region errors are summarized in Table \ref{table:X-ray_quie_fit_results}.
This table includes the total count rate of each quiescent-phase spectra ($QC_{\rm{PN}}^{\rm{OBSID}}$, $QC_{\rm{MOS1}}^{\rm{OBSID}}$, $QC_{\rm{RGS1}+\rm{RGS2}}^{\rm{OBSID}}$), the 10 temperature bins ($T_{1}$--$T_{10}$), 
the EMs (EM$_{1}$--EM$_{10}$), the total emission measure EM$_{\rm{tot}}=\rm{EM}_{1} + .... +  \rm{EM}_{10}$, and 
the EM-weighted average temperature $T_{\rm{ave}}$, the total X-ray fluxes ($F_{\rm{X}}$) and luminosities ($L_{\rm{X}}$) calculated over the range of 0.2 -- 12 keV based on the fitted model parameters, and abundances ($Z$). 
The abundances are relative to solar photospheric values (the default 
values in \texttt{Xspec}: \citealt{Anders+1989_GeCoA}). 
If the EM value turns out to be zero after the first fitting attempt, 
the value is simply fixed to be zero, since the error calculations around zero often 
do not converge well. As a result, 
$EM_{1}$, $EM_{9}$, and $EM_{10}$ are fixed to be zero in the final results shown in Table \ref{table:X-ray_quie_fit_results}. $EM_{4}$ is also fixed to zero
in the case of OBSID=0822740301.
The He and Al abundances are fixed to 1.00 and 0.50, and Ca and Ni abundances are fixed to that of Fe. 
These elements are included in \texttt{vapec}, \color{black}
but there are no major spectral lines of these elements (He, Al, Ca, Ni)
in the M-dwarf XMM X-ray spectra 
(cf. \citealt{Raassen+2007,Pillitteri+2022}) and 
because of this, these assumed values do not have meaningful
effects on the following results of temperature and emission measure values. We confirmed this point by changing these 4 element values manually in the range of 0.1--1.0. 
\color{black}
The inferred emission measure distribution (EMD) against temperatures in 
\color{black}
Figures \ref{fig:specfit_QuieALL1_0822740301} (i) \& 
\ref{fig:specfit_QuieALL1_0822740401} (i) -- 
\ref{fig:specfit_QuieALL1_0822740601} (i)
\color{black}
shows that the quiescent phase X-ray spectra have EM contributions over $\log T  \mathrm{[K]} \sim6.3-7.6$ ($\sim0.17-3.43$ keV) bins with the highest peaks at $\log T\mathrm{[K]}\sim6.9$ ($\sim$0.68 keV) and with the EM-weighted average temperature $\log T_{\rm{ave}} \mathrm{[K]}=6.96-7.02$  ($\sim$0.78--0.91 keV). 
The abundance of the quiescent phase spectra are shown in 
\color{black}
Figures \ref{fig:specfit_QuieALL1_0822740301} (j) \& 
\ref{fig:specfit_QuieALL1_0822740401} (j) -- 
\ref{fig:specfit_QuieALL1_0822740601} (j).
\color{black}

\begin{figure}[ht!]
   \begin{center}
      \gridline{
\fig{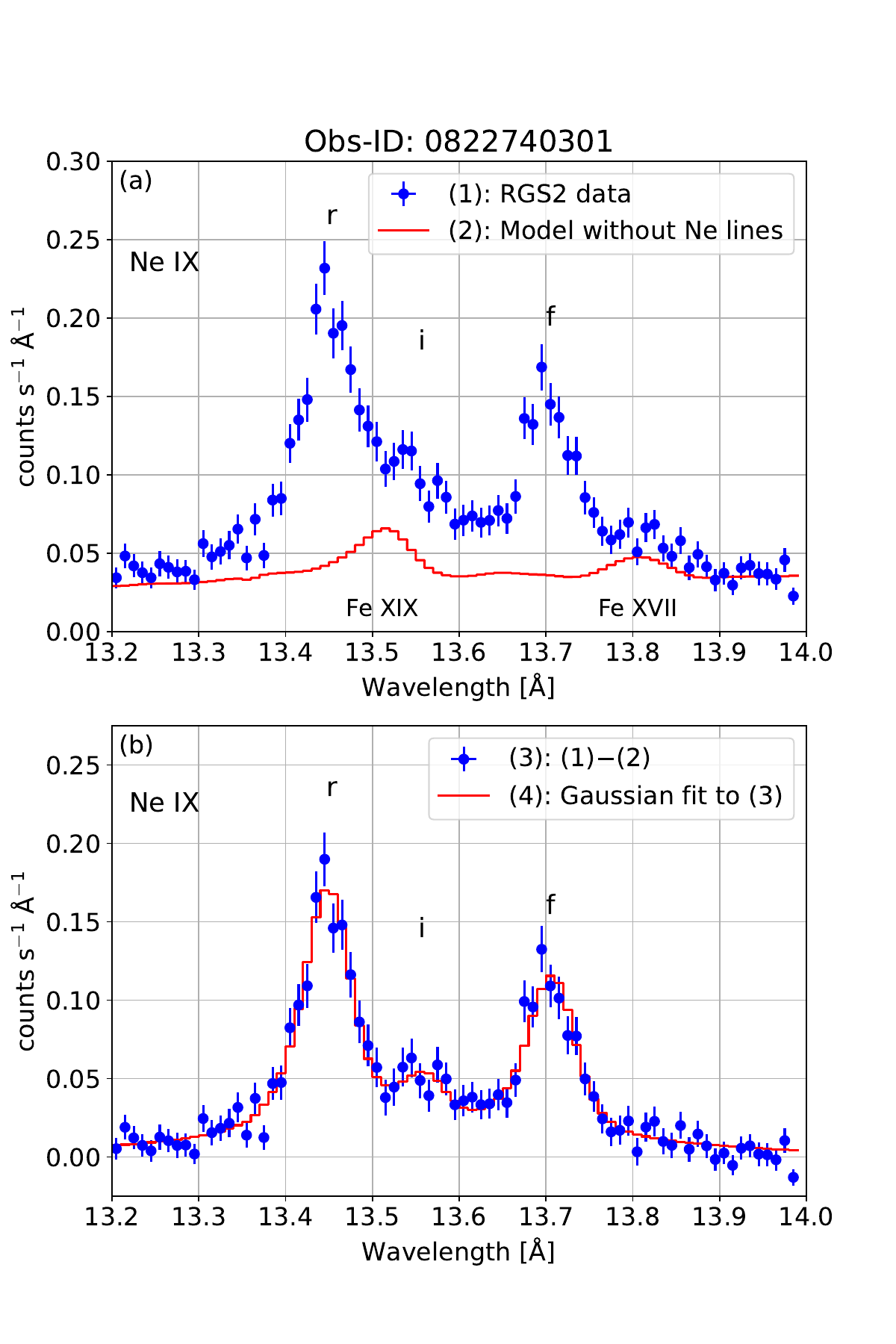}{0.5\textwidth}{\vspace{0mm}}
\fig{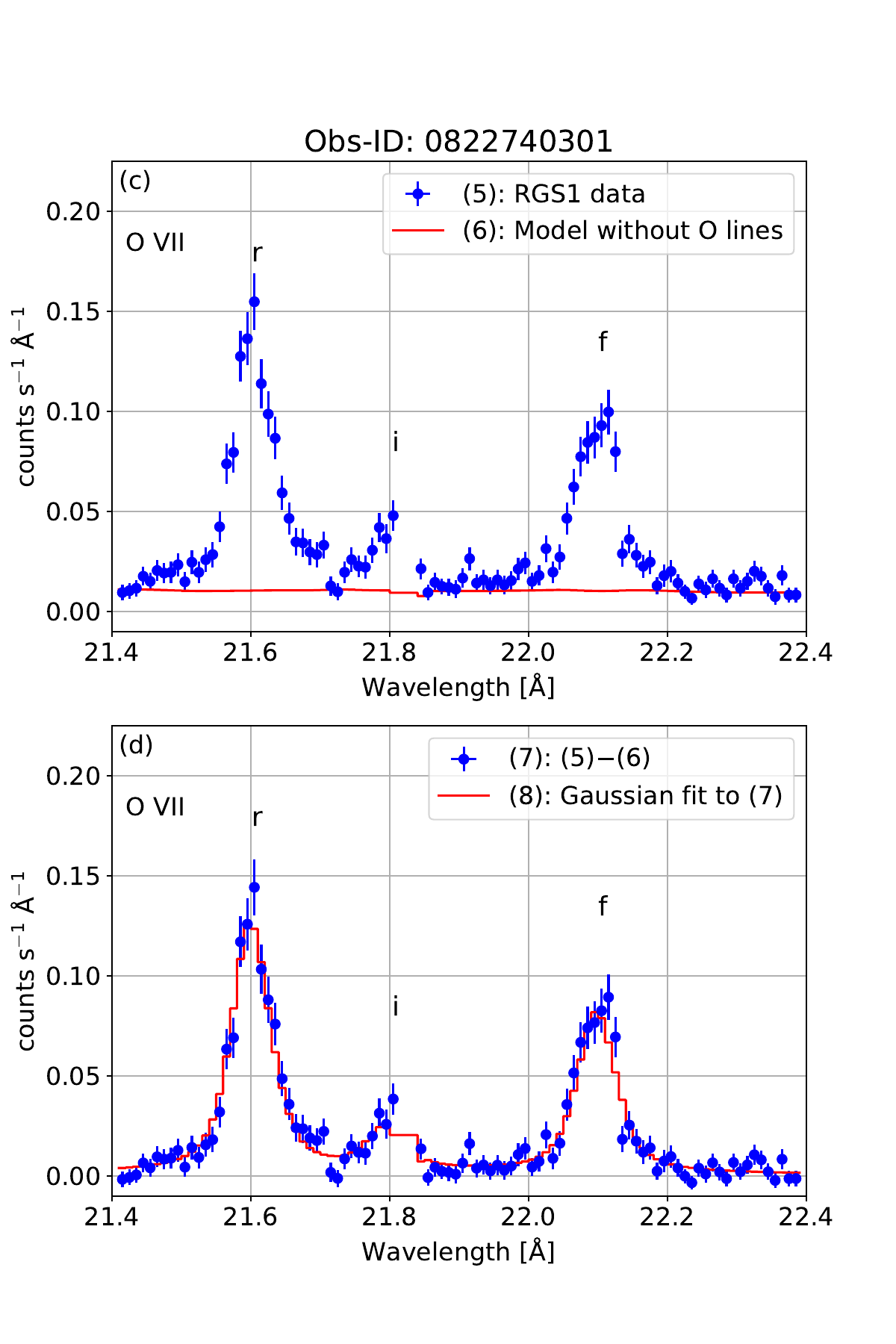}{0.5\textwidth}{\vspace{0mm}}
    }
     \vspace{-5mm}
     \caption{
(a) 
(1) The quiescent phase RGS2 spectra of the Obs-ID 0822740301 around Ne IX lines 
in 13.2--14.0\AA~
and (2) the 10-temperate model fit from Figure \ref{fig:specfit_QuieALL1_0822740301}  
(cf. Table \ref{table:X-ray_quie_fit_results} but only Ne abundance ($Z$(Ne)) is set to zero. 
(b) (3) The difference of data (1) and model fit (2). (4) Gaussian fit to (3) (See the main text for the details).
(c) \& (d) Same as (a) \& (b) but for the RGS1 spectra O VII lines in 21.4--22.4\AA.
}
   \label{fig:quieden_0822740301}
   \end{center}
 \end{figure}

The electron densities $n_{\rm{e}}$ can be investigated with 
the line fluxes of the line triplets in the He-like ions, 
consisting of a resonance (r), an intercombination (i) 
and a forbidden line (f) (\citealt{Gabriel_Jordan_1969_MNRAS}).
Here we focus on the ratios of the fluxes of the forbidden line to the intercombination line (f/i) of Ne IX and O VII lines in the wavelength range of RGS (5 -- 38\AA), by referring to the analysis methods of \citet{Osten+2006,Raassen+2007}.
Three Gaussian profiles were folded through the response matrix and fitted to the observed r, i, and f line profiles, establishing the positions, fluxes as well as the widths of the lines (\color{black}Figure \ref{fig:quieden_0822740301} in this section and Figures \ref{fig:quieden_0822740401} -- \ref{fig:quieden_0822740601} in 
Appendix \ref{appen_sec:additional_figures} \color{black}).
Apart from the instrumental broadening, no additional line broadening was noted. The model spectral components with Ne or O abundances being fixed to zero are subtracted before Gaussian fitting process in order to remove the contributions from other lines (e.g., Fe XIX and Fe XVII lines around Ne IX lines in 
\color{black}
Figures \ref{fig:quieden_0822740301}(a) \& \ref{fig:quieden_0822740401} (a) -- \ref{fig:quieden_0822740601} (a)
\color{black}
) into the fitting. 
The measured wavelengths, line fluxes, and the resultant f/i ratios are listed in Table \ref{table:quieden_NeIX_OVII}. Some minor wavelength deviations of a few m\AA~seem to be present, and they are comparable with the instrumental wavelength accuracy. 
As for Obs-IDs 0822740301, 0822740401, and 0822740501, there are some wavelength gaps at the O VII intercombination line (\color{black}Figures \ref{fig:quieden_0822740301}(d) \& \ref{fig:quieden_0822740401} (d) -- \ref{fig:quieden_0822740601} (d)\color{black}), 
and this can cause the fitted flux value to be systematically lower and the resultant f/i (O VII) values are thus systematically overestimated.
Then we compare the estimated  f/i ratio values (for Ne IX and O VII) with 
theoretical curves presented in Figure 8 of \citet{Osten+2006}.
These comparisons provide the quiescent-state $n_{\rm{e}}$ ranges listed in Table \ref{table:quieden_NeIX_OVII}: 
$\log n_{\rm{e}}\mathrm{[cm}^{-3}\,\mathrm{]}\lesssim11.5-11.9$ 
from Ne IX lines, and  
$\log n_{\rm{e}}\mathrm{[cm}^{-3}\,\mathrm{]}\lesssim10.0-10.2$
or $\log n_{\rm{e}}\mathrm{[cm}^{-3}\,\mathrm{]}\sim9.3-10.5$
from O VII lines. 
The $n_{\rm{e}}$ values estimated from O VII lines can be systematically lower because of the wavelength gap described above. Thus only $n_{\rm{e}}$ values from Ne IX lines are used in the later discussion (Section \ref{sec:discuss_Xray_flare}).

 \begin{deluxetable*}{lcccc}[ht!]
   \tablecaption{Quiescent component fluxes of Ne IX and O VII lines used for electron density discussions, measured with RGS spectra. Statistical 1$\sigma$ errors are shown.}
   \tablewidth{0pt}
   \tablehead{
\colhead{Obs-ID} & 
\colhead{Ion} &  
\colhead{$\lambda_{0}$(\AA)}\tablenotemark{\rm \dag}   & \colhead{$\lambda_{\rm{obs}}$(\AA)}\tablenotemark{\rm \dag}  &\colhead{Flux} \tablenotemark{\rm \dag}   
}
   \startdata
0822740301 & Ne IX (r) & 13.417 & 13.449 $\pm$ 0.004  & 3.23 $\pm$ 0.16 \\
 & Ne IX (i) & 13.553  &  13.560 $\pm$  0.010  & 0.66 $\pm$ 0.12  \\
 & Ne IX (f) & 13.700 & 13.701 $\pm$  0.004 & 2.12 $\pm$ 0.13 \\
 & \multicolumn{4}{c}{f/i (Ne IX) = 3.2 $\pm$ 0.6, \ \ 
 $\log n_{\rm{e}}\mathrm{[cm}^{-3}\mathrm{]}\lesssim$11.5 }\\ 
     \cline{2-5}
 & O VII (r) & 21.602 & 21.597 $\pm$ 0.007 &  3.65 $\pm$ 0.18 \\
 & O VII (i) & 21.804 & 21.790 $\pm$ 0.015 &  0.62 $\pm$ 0.12 \\
 & O VII (f) & 22.101 & 22.095 $\pm$ 0.008 & 2.47 $\pm$ 0.15 \\
 & \multicolumn{4}{c}{f/i (O VII) = 4.0 $\pm$ 0.8, \ \ 
 $\log n_{\rm{e}}\mathrm{[cm}^{-3}\mathrm{]}\lesssim$10.1 }\\ 
    \cline{1-5}
0822740401 & Ne IX (r) & 13.417 & 13.444 $\pm$ 0.003 &  3.25 $\pm$ 0.18 \\
 & Ne IX (i) & 13.553  & 13.550 $\pm$  0.012 &  0.50 $\pm$ 0.14 \\
 & Ne IX (f) & 13.700 & 13.690 $\pm$ 0.011 & 1.83 $\pm$ 0.14 \\
 & \multicolumn{4}{c}{f/i (Ne IX) = 3.7  $\pm$ 1.0, \ \ 
 $\log n_{\rm{e}}\mathrm{[cm}^{-3}\mathrm{]}\lesssim$11.5 }\\ 
     \cline{2-5}
 & O VII (r) & 21.602 & 21.604 $\pm$ 0.005 & 4.10 $\pm$ 0.21 \\
 & O VII (i) & 21.804 & 21.803 $\pm$ 0.010 & 0.55 $\pm$ 0.14 \\
 & O VII (f) & 22.101 & 21.094 $\pm$ 0.001 & 2.45 $\pm$ 0.17 \\
 & \multicolumn{4}{c}{f/i (O VII) = 4.5  $\pm$ 1.2, \ \ 
 $\log n_{\rm{e}}\mathrm{[cm}^{-3}\mathrm{]}\lesssim$10.0 }\\ 
    \cline{1-5}
0822740501 & Ne IX (r) & 13.417 & 13.439 $\pm$ 0.006  & 2.91 $\pm$ 0.13 \\
 & Ne IX (i) & 13.553  &  13.544 $\pm$ 0.006 & 0.70 $\pm$ 0.11 \\
 & Ne IX (f) & 13.700 &  13.690 $\pm$ 0.013  & 1.96 $\pm$ 0.12 \\
 & \multicolumn{4}{c}{f/i (Ne IX) = 2.8 $\pm$ 0.5, \ \ 
 $\log n_{\rm{e}}\mathrm{[cm}^{-3}\mathrm{]}\lesssim$11.7 }\\ 
     \cline{2-5}
 & O VII (r) & 21.602 & 21.602 $\pm$ 0.006 &  3.68 $\pm$ 0.19 \\
 & O VII (i) & 21.804 & 21.796 $\pm$ 0.013 &  0.55 $\pm$ 0.14\\
 & O VII (f) & 22.101 & 22.103 $\pm$ 0.003  &  2.26 $\pm$ 0.13 \\
 & \multicolumn{4}{c}{f/i (O VII) = 4.1 $\pm$ 1.1, \ \ 
 $\log n_{\rm{e}}\mathrm{[cm}^{-3}\mathrm{]}\lesssim$10.2 }\\ 
    \cline{1-5}
0822740601 & Ne IX (r) & 13.417 & 13.446 $\pm$ 0.006 & 3.22 $\pm$ 0.21 \\
 & Ne IX (i) & 13.553  & 13.552 $\pm$ 0.008 & 0.76 $\pm$ 0.16 \\
 & Ne IX (f) & 13.700 & 13.699 $\pm$ 0.006 & 2.09 $\pm$ 0.24 \\
 & \multicolumn{4}{c}{f/i (Ne IX) = 2.8 $\pm$ 0.7, \ \ 
 $\log n_{\rm{e}}\mathrm{[cm}^{-3}\mathrm{]}\lesssim$11.9 }\\ 
     \cline{2-5}
 & O VII (r) & 21.602 & 21.607 $\pm$ 0.005 & 3.62 $\pm$ 0.22 \\
 & O VII (i) & 21.804 & 21.808 $\pm$ 0.010 & 0.83 $\pm$ 0.15 \\
 & O VII (f) & 22.101 & 22.110 $\pm$ 0.007 & 2.30 $\pm$ 0.21 \\
 & \multicolumn{4}{c}{f/i (O VII) = 2.8 $\pm$ 0.6, \ \ 
 $\log n_{\rm{e}}\mathrm{[cm}^{-3}\mathrm{]}\sim$9.3--10.5 }\\ 
   \enddata
  % \tablecomments{
 %  }
 \tablenotetext{\dag}{
$\lambda_{0}$ is the theoretical energy and wavelength from \citet{Kelly_1987} (see also Table 4 of \citealt{Raassen+2007}). $\lambda_{\rm{obs}}$ is the observed wavelength from the Gaussian fitting. Flux at the Earth is also from the Gaussian fitting, and shown in units of photons cm$^{-2}$ s$^{-1}$.} 
   \label{table:quieden_NeIX_OVII}
 \end{deluxetable*}

\subsection{Time-averaged X-ray flare spectra}\label{subsec:X-ray_specana_time-average}

We conduct spectral fitting of flare components by \color{black}
combining \color{black}
the quiescent state parameters in Section \ref{subsec:X-ray_specana_quiescent} and PN \& MOS1 spectra during flares. It is noted that the RGS data are not used for the spectral analysis of the flare phases in this study since most of them have too low counts for the simple spectral analysis of each flare, but the flare RGS spectra will be discussed in a future paper with more detailed analysis (e.g., combining RGS data of multiple flares). 
In this Section \ref{subsec:X-ray_specana_time-average}, we conduct spectral analysis for time-averaged components of rise and decay phase of all the X-ray flares, in order to characterize overall spectral properties of all small-to-large X-ray flares. The rise and decay phases of each flare are shown as green and orange colored regions in Figures \ref{fig:X-ray_Ha_obsid_0822740301_lc} - \ref{fig:X-ray_Ha_obsid_0822740601_lc}, on the basis of the time values listed in Table 6 of T23 (cf. $t_{\rm{rise}}$, $t_{\rm{decay}}$, and $t_{\rm{total}}$ in Table \ref{table:X-ray_flarefit_timeave} \color{black} 
in Appendix \ref{appen_sec:additional_figures}\color{black}). 
The spectral analysis of the rise and decay spectra is performed as follows. We use the standard fitting technique with \texttt{Xspec} and \texttt{PyXspec}, assuming collisionally ionized equilibrium components (\texttt{vapec}) with interstellar absorption (\texttt{tbabs}), as done in Section \ref{subsec:X-ray_specana_quiescent}.

    \begin{figure}[ht!]
   \begin{center}
      \gridline{
\fig{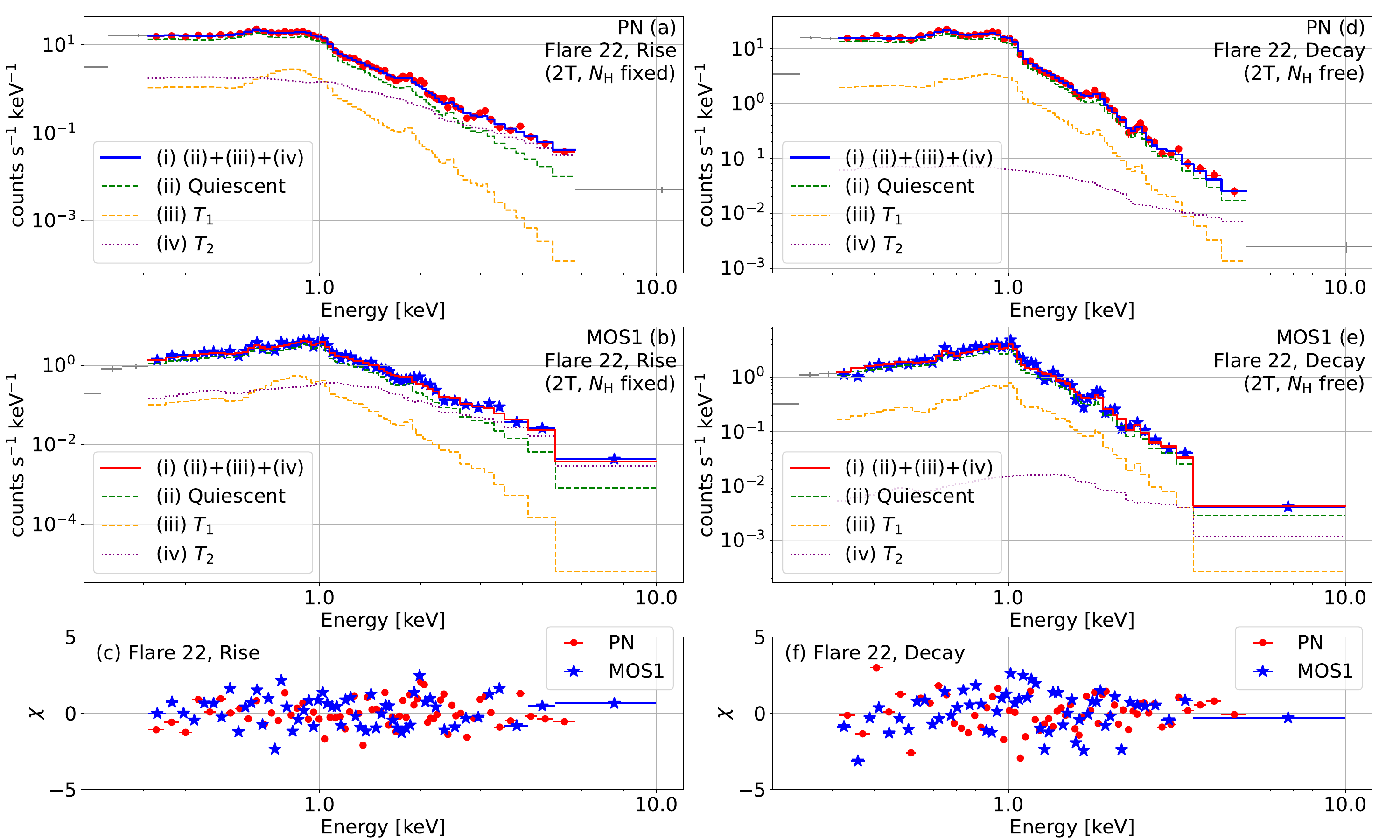}{0.95\textwidth}{\vspace{0mm}}
    }
     \vspace{-5mm}
     \caption{
The PN and MOS1 spectra and best-fit model results for rise and decay phase data of each flare. The figures of Flare 22 are shown here as an example, and those of all the X-ray flares (except for Flares 11 and 15 without the appropriate time-averaged fitting results) are available in \textbf{Online Figure Set 1}. It is also noted the figures of Flares 35 and 36 are also shown later in Figure \ref{fig:RiseDecayFit_fig1_Flare35and36} \color{black} in Appendix \ref{appen_sec:additional_figures}. \color{black} 
(a) The PN spectral data and fit of the rise phase. Red circles are the data used for fitting, while gray points show the ranges not used for fitting. The blue solid, green dashed, orange dashed, and purple dotted lines are the total, quiescent, low-temperature ($T_{1}$), and high temperature ($T_{2}$) model components, respectively. 
If the 1-temperature fitting is applied (e.g., Flare 1), the orange dashed line is not plotted. 
(b) Same as (a), but for MOS1 data. Blue star marks and red solid line are used instead of red circles and blue solid line, respectively. 
(c) Residuals between data and total model components divided by errors are plotted. Red circles and blue star marks correspond to the PN and MOS1 data, respectively.
(d)--(f) Same as (a)--(c), but for the decay phase spectra.
}
   \label{fig:RiseDecayFit_fig1_Flare22}
   \end{center}
 \end{figure}

\begin{enumerate} 
\renewcommand{\labelenumi}{(\arabic{enumi})}

\item 
We measure the quiescent component count rate of each flare ($QC_{\rm{PN}}^{\rm{Flare}}$ in Table \ref{table:X-ray_flarefit_timeave} \color{black} in Appendix \ref{appen_sec:additional_figures}\color{black}) by averaging the PN count rates of the preflare and post-flare phases. The fraction between $QC_{\rm{PN}}^{\rm{Flare}}$ and the average over the OBSID ($QC_{\rm{PN}}^{\rm{OBSID}}$ in Table \ref{table:X-ray_quie_fit_results}) is then calculated as $f_{\rm{QCPN}}^{\rm{Flare}}=QC_{\rm{PN}}^{\rm{Flare}}/QC_{\rm{PN}}^{\rm{OBSID}}$.

\item 
We determine the quiescent component spectra of each flare (``(ii) Quiescent" in Figure \ref{fig:RiseDecayFit_fig1_Flare22}), using $f_{\rm{QCPN}}^{\rm{Flare}}$ and the quiescent component fitting results of each Obs-ID in Table \ref{table:X-ray_quie_fit_results}. 
The emission measures are determined by multiplying  
each Obs-ID's quiescent component emission measures by $f_{\rm{QCPN}}^{\rm{Flare}}$: $f_{\rm{QCPN}}^{\rm{Flare}} EM_{1}$, $f_{\rm{QCPN}}^{\rm{Flare}} EM_{2}$, ....., $f_{\rm{QCPN}}^{\rm{Flare}} EM_{10}$. The other parameters ($N_{\rm{H}}$, $T_{1}$--$T_{10}$, and abundances $Z$) are fixed at the values of each Obs-ID (= the values in Table \ref{table:X-ray_quie_fit_results}).

\item 
We fix this quiescent component as a steady component (cf. ``(ii) Quiescent" in Figure \ref{fig:RiseDecayFit_fig1_Flare22}) and fit the PN and MOS1 spectra by adding an additional 1-temperature component (cf. ``(iii) $T_{1}$"  in Figure \ref{fig:RiseDecayFit_fig1_Flare22}). 
As for this additional $T_{1}$ component, the abundance of Fe ($Z$(Fe)) is set as a free parameter, while $N_{\rm{H}}$ and abundance of the other elements are fixed to the same values as the quiescent component (Table \ref{table:X-ray_quie_fit_results}). 

\item
Next, we also calculate the 2-temperature flare component fitting ($T_{1}$ and $T_{2}$). As in the 1-temperature flare component fitting in the above, the quiescent component is fixed, the abundance of Fe ($Z$(Fe)) is set as a free parameter, and abundance values of the other elements are fixed to the same values as the quiescent component. As for $N_{\rm{H}}$ values, we try two cases: $N_{\rm{H}}$ fixed (Table \ref{table:X-ray_quie_fit_results}) and $N_{\rm{H}}$ set as a free parameter. 

\item
As a result, we now have results of the three fitting cases: (a) ``1T, $N_{\rm{H}}$ fixed", (b) ``2T, $N_{\rm{H}}$ fixed", and (c) ``2T, $N_{\rm{H}}$ free".  We check the reduced chi-square value ($\chi^{2}_{\rm{black}}$) of each case, and we judge that the fitting is failed for the cases with $\chi^{2}_{\rm{black}}>2.0$. 
As for the cases with $\chi^{2}_{\rm{black}}\leq2.0$, we conduct \texttt{F-test}\footnote{\url{https://heasarc.gsfc.nasa.gov/xanadu/xspec/manual/node82.html}}, and decide whether we should fit the data with an additional model component.  In this study, if the F-test probability is $\lesssim$0.1, we select the fitting case with an additional component as the more appropriate fitting result ((a)$\rightarrow$(b), (b)$\rightarrow$(c)). 
 
\end{enumerate}

The finally selected fitting results are shown in Figure \ref{fig:RiseDecayFit_fig1_Flare22} and  \textbf{Online Figure Set 1}.
The resultant fitting parameters of the flare components ($N_{\rm{H}}$, $T_{1}$, $EM_{1}$, $T_{2}$, $EM_{2}$, $Z_{\rm{Fe}}$, $\chi^{2}_{\rm{black}}$, $L_{\rm{X}}$) are summarized in Table \ref{table:X-ray_flarefit_timeave} \color{black} in Appendix \ref{appen_sec:additional_figures} \color{black}. 
The average temperature ($T_{\rm{ave}}$), total emission measure ($EM_{\rm{tot}}=EM_{1}+EM_{2}$), and total X-ray flare radiated energy in the 0.2 -- 12 keV range ($E_{\rm{X}}$) of each flare are calculated by integrating or time-averaging the rise and decay phase fitting results, and they are also listed in Table \ref{table:X-ray_flarefit_timeave}. 
The two Flares 11 and 15 are not included in the time-averaged analysis of the Rise and Decay phase spectra in this subsection, 
since the fitting analyses failed ($\chi^{2}_{\rm{black}}>2.0$) 
for these two fitting cases.
The detailed time-resolved X-ray spectral analysis of these two flares (Flare 11 and 15) and Flare 23 (the large Neupert-type flare with the largest multi-wavelength coverage in this campaign) are conducted in the following Sections \ref{subsec:ana_flare_23} \& \ref{subsec:ana_flare_11_and_15}. 
The $T_{\rm{ave}}$, $EM_{\rm{tot}}$, and $E_{\rm{X}}$ values of these three flares listed in the ``Total:" rows of Table \ref{table:X-ray_flarefit_timeave} are from the time-resolved analysis results in these following sections.

\section{Multiwavelength time-evolution analysis of remarkable flares} \label{sec:remarkable_flares}

\begin{deluxetable*}{l|cccccccc}[ht!]
   \tablecaption{Multi-wavelength flare energies of the remarkable flares in this study.}
   \tablewidth{0pt}
   \tablehead{
\colhead{} &
\multicolumn{8}{c}{Flares} 
\\
\colhead{} &
\colhead{21} &
\colhead{22} &
\colhead{23} &
\colhead{11} &
\colhead{15} &
\colhead{35} &
\colhead{36} &
\colhead{65} 
}
   \startdata
$E_{\rm{X}}$ (10$^{31}$ erg) \tablenotemark{$\dag$} & 31.3$_{-8.4}^{+10.0}$ & 22.7$_{-3.2}^{+8.3}$ & 145.2$_{-3.2}^{+2.0}$ & 363.3$_{-6.8}^{+2.7}$ & 288.1$_{-2.5}^{+2.1}$ & 30.9$_{-1.2}^{+0.8}$ & 18.0$_{-1.0}^{+0.7}$ & no obs. \\ 
$E_{\rm{UVW2}}$ (10$^{30}$ erg) \tablenotemark{$\dag$} & 7.1$\pm$0.3 & 120.0$\pm$0.0 &  970.0$\pm$30.0 & -- & 500.0$\pm$40.0 & no obs. & no obs. & no obs. \\ 
$E_{\rm{U}}$ (10$^{32}$ erg) \tablenotemark{$\dag$} & 1.1$\pm$0.0 &  2.6$\pm$0.0 &  16$\pm$0 & no obs. & no obs. &  2.8$\pm$0.0 & 0.40$\pm$0.01 & no obs.\\ 
$E_{\rm{V}}$ (10$^{32}$ erg) \tablenotemark{$\dag$} & -- & 2.0$\pm$0.0 & 9.8$\pm$0.1 & no obs.& no obs. & 1.6$\pm$0.1 & -- & no obs. \\ 
$E_{\rm{Ku}}$ (10$^{26}$ erg) \tablenotemark{$\ddag$} & 6.65$\pm$0.44 & 6.92$\pm$0.04 & 7.26$\pm$3.35 \tablenotemark{$\ddag$} & no obs. & no obs. &  20.17$\pm$0.47 & 7.21$\pm$0.05 \tablenotemark{$\ddag$} & no obs. \\ 
$E_{\rm{H}\alpha}$ (10$^{31}$ erg) \tablenotemark{$\sharp$} & -- & 2.6--2.7 & 14.1--14.9 & no obs. & no obs. & 1.7--1.8 & 0.8--1.0 & 35.6 \\ 
$E_{\rm{H}\beta}$ (10$^{31}$ erg) \tablenotemark{$\sharp$} & -- & 1.9--2.0 & 9.1--9.5 & no obs. & no obs. & 1.3 & 0.8 & 25.3 \\ 
   \enddata
\tablecomments{
The data columns with ``no obs." mean that there were no observation data for the specific bands
}
 \tablenotetext{\dag}{
$E_{\rm{X}}$ is the total soft X-ray flare energy in the 0.2 -- 12 keV range, taken from Tables \ref{table:X-ray_quie_fit_results}, \ref{table:flare23_Xray_fit}, \& \ref{table:flare11_Xray_fit}.
$E_{\rm{UVW2}}$, $E_{\rm{U}}$, and $E_{\rm{V}}$ are the flare energies in UVW2, U, and V-bands reported in Table 6 of T23.
   }
    \tablenotetext{\ddag}{
$E_{\rm{Ku}}$ is the flare energy in VLA radio Ku-band reported in T25. Flares 23 and 36 have significant data gaps during the flares, and the listed flare energies are only minimum values.
}
    \tablenotetext{\sharp}{
$E_{\rm{H}\alpha}$ and $E_{\rm{H}\beta}$ are the flare energies in 
H$\alpha$ and H$\beta$ lines, which are calculated in this study
following the methodology described in Section 2.5 of \citet{Notsu+2024_ApJ}. 
The SMARTS $V$-band flux changes are used as a proxy for the continuum level around the H$\alpha$ and H$\beta$ lines.
}
   \label{table:energy_remarkable_flares}
 \end{deluxetable*}

\subsection{Flare 23: The large Neupert-type flare with the largest multi-wavelength coverage in this campaign}\label{subsec:ana_flare_23}

\begin{figure}[ht!]
   \begin{center}
      \gridline{
\fig{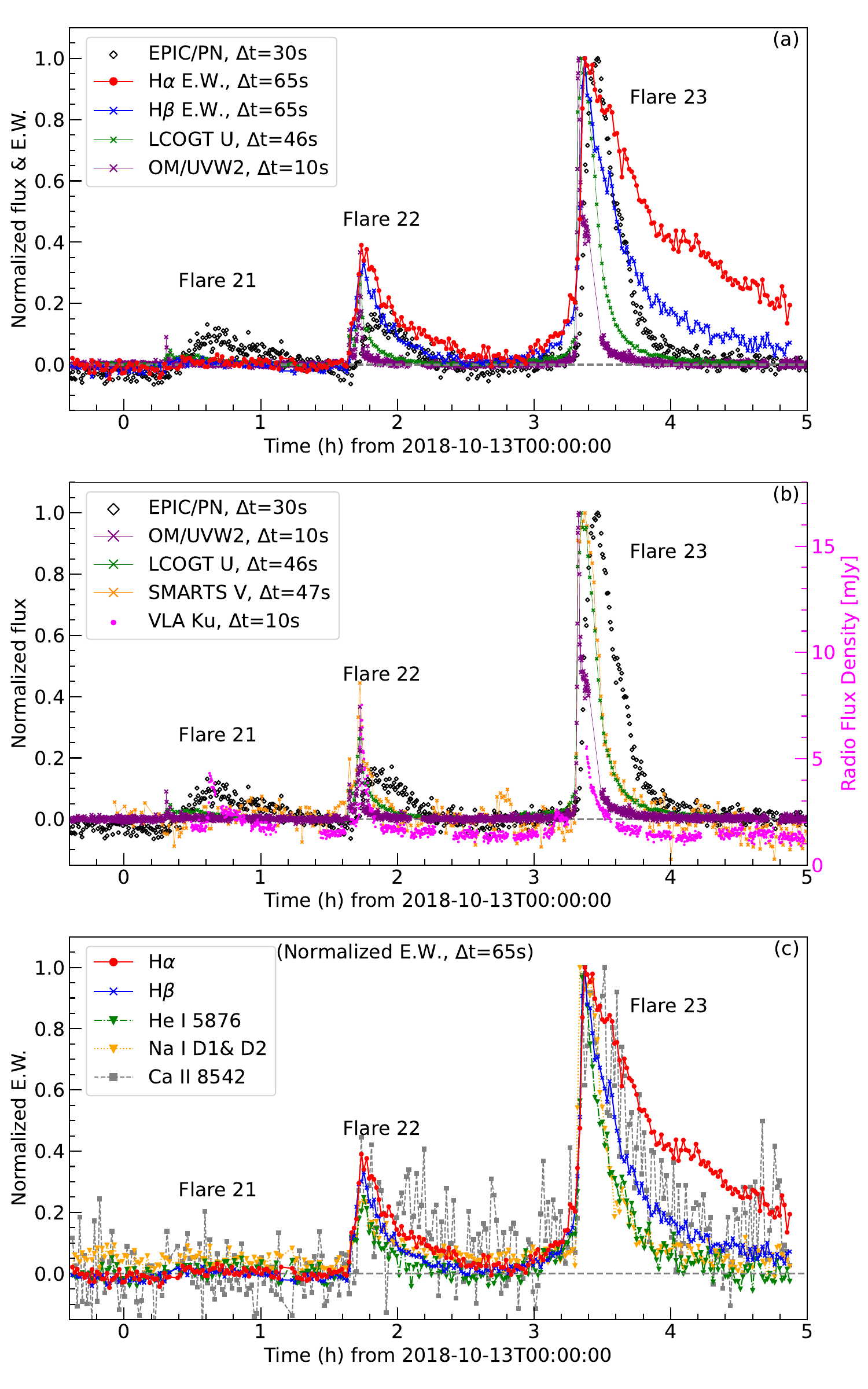}{0.55\textwidth}{\vspace{0mm}}
    }
     \vspace{-5mm}
     \caption{
Multi-wavelength lightcurves of Flares 21 -- 23 on 2018 October 13.
(a)\&(b) H$\alpha$ \& H$\beta$ equivalent width (E.W.), XMM EPIC-pn X-ray (0.2 -- 12 keV), XMM OM UVW2, LCOGT U-band, SMARTS V-band, and VLA Ku-band lightcurves as in Figure \ref{fig:X-ray_Ha_obsid_0822740401_lc}.
These fluxes and E.W. values are plotted as normalized with the peak and preflare values of Flare 23. The exact flux value is plotted only for the VLA Ku-band radio data. The peak H$\alpha$ and H$\beta$ E.W. values of Flare 23 are 4.86\AA~and 7.97\AA, while the preflare values are 2.45\AA~and 1.56\AA, respectively (cf. Figure \ref{fig:maps_flare23}).
(c) The E.W. lightcurves of H$\alpha$, H$\beta$, He I D3 5876\AA, Na I D1\&D2, and Ca II 8542\AA~lines from the CHIRON data, as normalized with the peak and preflare values of Flare 23 (as in (a)). The peak He I D3 5876\AA~E.W. value of Flare 23 is 1.03\AA, while the preflare value is 0.04\AA~(cf. Figure \ref{fig:maps_flare23}). As for the Na I D1\&D2, and Ca II 8542\AA~lines, 
the E.W. values are measured from the quiescent component subtracted spectra and as the excess equivalent widths (=the preflare E.W. values are zero), 
to identify excess emissions in the line core of the absorption lines (Figures \ref{fig:flare23_HeNaSpec}\&\ref{fig:flare23_Ca8542Spec}). The peak Na I D1\&D2, and Ca II 8542\AA~excess E.W. values are 0.71\AA~and 0.26\AA, respectively (cf. Figure \ref{fig:maps_flare23}).
}
   \label{fig:Oct13_TEM_multi_lc}
   \end{center}
 \end{figure}

\begin{figure}[ht!]
   \begin{center}
      \gridline{
\fig{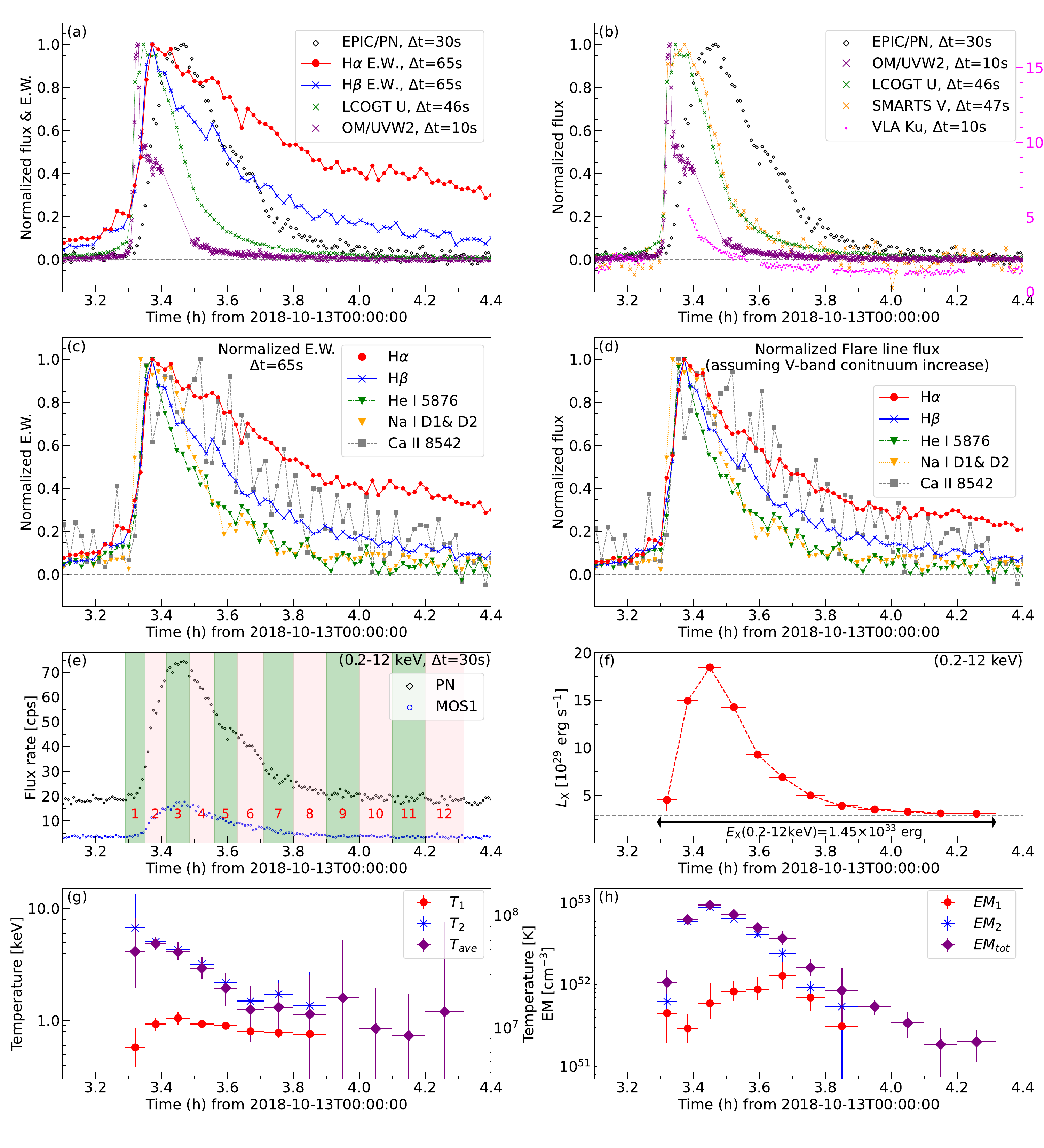}{1.0\textwidth}{\vspace{0mm}}
    }
     \vspace{-5mm}
     \caption{
(a)--(c) The same as Figure \ref{fig:Oct13_TEM_multi_lc} (a)--(c), but enlarged lightcurves of Flare 23.
(d) The normalized flare line fluxes, converted from the E.W. values in the panel (c), assuming $V$-band continuum increase.
(e) XMM EPIC-pn and EPIC-MOS1 lightcurves (0.2 -- 12 keV) of Flare 23.
The green and pink colored regions corresponds to the Phases 1--10, separated for the spectral analysis.
(f) -- (h) The spectral analysis results (X-ray luminosity in 0.2 -- 12 keV $L_{\rm{X}}$, temperature $T$, and emission measure EM) of Phases 1--12 of Flare 23, from Table \ref{table:flare23_Xray_fit}. 
}
   \label{fig:Flare23_TEM_multi_lc}
   \end{center}
 \end{figure}

Flare 23, which occurred on 2018 October 13, is the largest X-ray amplitude flare with the most comprehenisve multi-wavelength coverage, among all the flares in this campaign (Figures \ref{fig:allEW_AUMic_XMM_opt} \& \ref{fig:X-ray_Ha_obsid_0822740401_lc}). 
Figure \ref{fig:Oct13_TEM_multi_lc} shows the multi-wavelength lightcurves 
around Flare 23, and as seen in this figure, Flare 23 occurred successively after two smaller flares (Flares 21 and 22, see also Table \ref{table:energy_remarkable_flares}). Figure \ref{fig:Flare23_TEM_multi_lc} (a)--(c) are
the enlarged lightcurves of Figure \ref{fig:Oct13_TEM_multi_lc} (a)--(c), 
focusing on Flare 23.
Flares 22 and 23 are identified as ``Neupert" flares in T23, since they clearly show both UVW2 (NUV) and soft X-ray emissions, and the soft X-ray time-derivative peak is consistent with UVW2 response peak (see the timing criteria in T23 for the details). Flare 21 is identified as a ``Quasi-Neupert" flare in T23, since it exhibits a response in both UVW2 and soft X-ray emission, 
but the light-curve peaks are well separated in time, 
according to the criteria in T23. 
It is also noted that according to the VLA radio analysis in T25, 
Flares 21 and 23 show significant time intervals of optically thin gyrosynchrotron radio emissions, while Flare 22 shows that of optically thick emission. This suggests that the spectral peaks of the gyrosynchrotron radio emission can be different between these three 
successive flares, and more detailed physical discussions on the radio data are available in T25.

Figures \ref{fig:Oct13_TEM_multi_lc}(c) \& \ref{fig:Flare23_TEM_multi_lc}(c)
show the equivalent width (E.W.) time evolutions 
of H$\alpha$, H$\beta$, He I D3 5876\AA, 
Na I D1\&D2, and Ca II 8542\AA~optical chromospheric lines, 
during Flare 23. The E.W. values are normalized with the preflare and peak values of Flare 23, 
as described in the caption of Figure \ref{fig:Oct13_TEM_multi_lc}.  
In addition, the normalized line flux values incorporating the SMARTS $V$-band flux change (in Figure \ref{fig:Flare23_TEM_multi_lc}(b)) are plotted in Figure \ref{fig:Flare23_TEM_multi_lc}(d), in order to roughly estimate the line flux changes incorporating the continuum enhancement during Flare 23. 
It is noted that since the $V$-band effective wavelength is shorter (bluer) than those of these chromospheric lines, Figure \ref{fig:Flare23_TEM_multi_lc}(d) is just a rough proxy for the flux changes.
As seen in these figures, clear chromospheric emissions are observed in all of these lines (H$\alpha$, H$\beta$, He I D3 5876\AA, Na I D1\&D2, and Ca II 8542\AA) during Flares 22 and 23, while no clear chromospheric line emissions are identified during Flare 21. Especially for Flare 23, there are clear differences of the flare durations among the chromospheric lines: the H$\alpha$ duration is longer than those of H$\beta$ and Ca II 8542\AA~lines, while Na I D1\&D2 and He I 5876\AA~durations are even shorter than those of H$\beta$ and Ca II 8542\AA~lines (Figure \ref{fig:Flare23_TEM_multi_lc}(c)\&(d)). 
There is another notable difference between the 
H$\alpha$ and X-ray time evolutions:
the flare peak of the H$\alpha$ emission and also the other chromospheric lines 
are \color{black} occurring \color{black} earlier than that of the soft X-ray emission, but the H$\alpha$ and H$\beta$ emissions last longer than the soft X-ray emission 
(Figure \ref{fig:Flare23_TEM_multi_lc}(a)\&(c)).

The multi-wavelength flare \color{black} energies \color{black} of Flares 21--23 are summarized in Table \ref{table:energy_remarkable_flares}.
The H$\alpha$ and H$\beta$ \color{black} energies \color{black} ($E_{\rm{H}\alpha}$ and $E_{\rm{H}\beta}$) of Flares 22 and 23 are estimated 
by using the methodology described in Section 2.5 of \citet{Notsu+2024_ApJ}. 
In \color{black} the energy determination process\color{black}, the distance of AU Mic (9.72 pc) is used from Table 1 of T23.
The quiescent flux densities at the continuum levels around the
H$\alpha$ and H$\beta$ lines ($F_{\rm{H}\alpha, \rm{q}}^{\rm{cont}}$ 
and $F_{\rm{H}\beta, \rm{q}}^{\rm{cont}}$) are calculated based on the
flux-calibrated quiescent spectra of AU Mic shown in Figure 1 of T23,
which were originally observed with the 
HST/Faint Object Spectrograph (the data \color{black} are \color{black} available on the MAST archive \footnote{doi:\dataset[10.17909/6pe3-yp69]{https://doi.org/10.17909/6pe3-yp69}, \color{black} as already made available by T23.}): 
$F_{\rm{H}\alpha, \rm{q}}^{\rm{cont}}= 1.91\times 10^{-12}$ erg s$^{-1}$ cm$^{-2}$ \AA$^{-1}$ and
$F_{\rm{H}\beta, \rm{q}}^{\rm{cont}}= 8.26\times 10^{-13}$ erg s$^{-1}$ cm$^{-2}$ \AA$^{-1}$.
As also used in Figure \ref{fig:Flare23_TEM_multi_lc}(d), the SMARTS $V$-band flux changes are used in this study as a proxy for the continuum level around the H$\alpha$ and H$\beta$ lines ($\Delta f_{\rm{line}}^{\rm{cont}}$ in Eq. (9) of \citealt{Notsu+2024_ApJ}). 
Since the $V$-band flux changes can be larger than the
changes of the real local continuum levels especially around the H$\alpha$ line 
considering the quiescent AU Mic spectra (see Figure 1 of T23),
the resultant energy values are shown with \color{black} 
ranges: for example,  $E_{\rm{H}\alpha}$= 1.41 -- 1.49 $\times$10$^{32}$ erg for Flare 23 in Table \ref{table:energy_remarkable_flares} 
\color{black}.
The lower values of the \color{black}  energy range (e.g., 1.41 $\times$10$^{32}$ erg for the above Flare 23 case) 
\color{black} 
do not take into account any continuum flux changes ($\Delta f_{\rm{line}}^{\rm{cont}}=0$ in Eq. (9) of \citealt{Notsu+2024_ApJ}) and the upper values \color{black} (e.g., 1.49 $\times$10$^{32}$ erg for the above Flare 23 case) \color{black}
correspond to the values incorporating $V$-band continuum flux changes ($\Delta f_{\rm{line}}^{\rm{cont}}$
taken from SMARTS $V$-band light curves).

\color{black}
The results of the analysis of the rise and decay phase X-ray spectra of 
\color{black}
the three flares \color{black} are \color{black} already \color{black} presented \color{black} in Section \ref{subsec:X-ray_specana_time-average} and the resultant fitting values are summarized in Table \ref{table:X-ray_quie_fit_results}. 
The X-ray spectra and the best fit model results of Flare 22 are shown in Figure \ref{fig:RiseDecayFit_fig1_Flare22}, while those of Flares 21 and 23 are in \textbf{Online Figure Set 1}. 
As for Flare 23, we also conduct the spectral analysis of the time-resolved X-ray data (PN and MOS1 data) as follows. 
First, we divide the PN and MOS1 data into the 12 phases as 
shown in Figure \ref{fig:Flare23_TEM_multi_lc}(e) (No.1 -- No.12). 
These 12 phases were selected manually so that each phase data represent 
the rough evolution of the lightcurve changes, while the flare start and end times are fixed to the values reported in T23. Then for the PN and MOS1 spectral data of each phase, 
we conduct the standard spectral fitting analysis
with the same method described in Steps (1)--(5) of Section \ref{subsec:X-ray_specana_time-average}. Each phase spectral data and finally selected fitting results are shown in Figures \ref{fig:specfig_Flare23_TEM_quie_each_No.1-No.4}, \ref{fig:specfig_Flare23_TEM_quie_each_No.5-No.8} and, \ref{fig:specfig_Flare23_TEM_quie_each_No.9-No.12} in Appendix \ref{appen_sec:additional_figures}. 
The resultant fitting parameters of Phases 1--12 ($N_{\rm{H}}$ $T_{1}$, $EM_{1}$, $T_{2}$, $EM_{2}$, $Z_{\rm{Fe}}$, $\chi^{2}_{\rm{black}}$, the EM-weighted average temperature $T_{\rm{ave}}$, the total emission measure $EM_{\rm{tot}}=EM_{1}+EM_{2}$, the X-ray luminosity in the 0.2 -- 12 keV range $L_{\rm{X}}$) 
are summarized in Table \ref{table:flare23_Xray_fit}. 
As in this table, the 2-temperature fitting results are selected as more appropriate results (cf. Step (5) of Section \ref{subsec:X-ray_specana_time-average}) for the Phases 1--8 
(Figures \ref{fig:specfig_Flare23_TEM_quie_each_No.1-No.4}--\ref{fig:specfig_Flare23_TEM_quie_each_No.5-No.8}), 
while the 1-temperature results are selected for the Phases 9--12 (Figure \ref{fig:specfig_Flare23_TEM_quie_each_No.9-No.12}). 
Among the 2-temperature fitting cases (Phases 1--8), ``$N_{\rm{H}}$ free" is selected for the Phases 3, 5, and 6, while ``$N_{\rm{H}}$ fixed" is selected for the other phases.
The resultant $L_{\rm{X}}$, temperature ($T_{1}$, $T_{2}$, $T_{\rm{ave}}$), and emission measure ($EM_{1}$, $EM_{2}$, $EM_{\rm{tot}}$) values 
are plotted in Figure \ref{fig:Flare23_TEM_multi_lc}(f), (g), and (h).
The total X-ray energy of Flare 23 in the 0.2 -- 12 keV range is calculated as $E_{\rm{X}}=1.45^{+0.02}_{-0.03}\times 10^{33}$ erg, 
and this is $\sim$10 times larger than the H$\alpha$ energy ($E_{\rm{H}\alpha}$=1.41--1.49 $\times$10$^{32}$ erg, Table \ref{table:energy_remarkable_flares}).
This X-ray energy of Flare 23 is $\sim$5--6 times larger than that of Flares 21 and 22 (Table \ref{table:energy_remarkable_flares}).   

\begin{longrotatetable}
\begin{deluxetable*}{lcccccccccccc}
   \tablecaption{Best fitting \color{black} parameters \color{black} for the time-resolved X-ray spectra of Flare 23. Statistical
90\% confidence region errors are shown for the X-ray fitting parameters.}
   \tablewidth{0pt}
   \tablehead{
\colhead{Phase}\tablenotemark{\rm \dag}  &  
\colhead{Time}\tablenotemark{\rm \dag}   & 
\colhead{$N_{\rm{H}}$ \tablenotemark{\rm \ddag}} &
\colhead{$T_{1}$} &
\colhead{$EM_{1}$} &
\colhead{$T_{2}$} &
\colhead{$EM_{2}$} &
\colhead{$Z_{\rm{Fe}}$} &
\colhead{$\chi^{2}/\rm{d.o.f}$} &
\colhead{$\chi^{2}_{\rm{black}}$} &
\colhead{$T_{\rm{ave}}$ \tablenotemark{\rm *}} &
\colhead{$EM_{\rm{tot}}$ \tablenotemark{\rm *}} &
\colhead{$L_{\rm{X}}$(0.2 -- 12 keV)} 
\\
\colhead{} &
\colhead{(h)}  & 
\colhead{(10$^{20}$ cm$^{-2}$)}&
\colhead{(keV)} &
\colhead{(10$^{51}$ cm$^{-3}$)} &
\colhead{(keV)} &
\colhead{(10$^{51}$ cm$^{-3}$)} & 
\colhead{($Z_{\rm{Fe,}\odot}$)}& 
\colhead{}& 
\colhead{} &
\colhead{(keV)}&
\colhead{(10$^{51}$ cm$^{-3}$)} &
\colhead{(10$^{29}$ erg s $^{-1}$)} 
}
   \startdata
1 & 3.29--3.35 & fixed & 0.58$_{-0.19}^{+0.29}$ & 4.51$_{-2.54}^{+3.99}$ & 6.71$_{-3.08}^{+6.71}$ & 6.24$_{-1.74}^{+1.83}$ & 0.08$_{-0.08}^{+0.08}$ & 98$/$63 & 1.56 & 4.14$_{-2.17}^{+4.11}$ & 10.75$_{-4.39}^{+4.39}$ & 4.52$_{-1.15}^{+0.15}$\\ 
2 & 3.35--3.42 & fixed & 0.93$_{-0.12}^{+0.12}$ & 2.93$_{-0.97}^{+1.51}$ & 5.06$_{-0.41}^{+0.49}$ & 60.14$_{-2.20}^{+2.27}$ & 0.76$_{-0.76}^{+0.76}$ & 166$/$132 & 1.26 & 4.87$_{-0.59}^{+0.73}$ & 63.07$_{-2.73}^{+2.73}$ & 14.94$_{-0.39}^{+0.36}$\\ 
3 & 3.42--3.48 & 2.32$_{-0.81}^{+0.86}$ & 1.05$_{-0.12}^{+0.15}$ & 5.93$_{-2.15}^{+4.56}$ & 4.29$_{-0.34}^{+0.46}$ & 89.30$_{-4.56}^{+4.19}$ & 0.63$_{-0.63}^{+0.63}$ & 175$/$153 & 1.15 & 4.09$_{-0.60}^{+0.92}$ & 95.22$_{-6.19}^{+6.19}$ & 18.45$_{-0.38}^{+0.47}$\\ 
4 & 3.48--3.56 & fixed & 0.94$_{-0.06}^{+0.06}$ & 8.30$_{-1.92}^{+2.75}$ & 3.19$_{-0.21}^{+0.25}$ & 64.30$_{-2.73}^{+2.69}$ & 0.50$_{-0.50}^{+0.50}$ & 169$/$132 & 1.28 & 2.93$_{-0.60}^{+0.73}$ & 72.60$_{-3.85}^{+3.85}$ & 14.27$_{-0.31}^{+0.24}$\\ 
5 & 3.56--3.63 & 1.64$_{-1.24}^{+1.34}$ & 0.90$_{-0.06}^{+0.06}$ & 8.79$_{-2.35}^{+3.60}$ & 2.17$_{-0.20}^{+0.22}$ & 41.47$_{-3.24}^{+3.56}$ & 0.42$_{-0.42}^{+0.42}$ & 125$/$104 & 1.20 & 1.95$_{-0.59}^{+0.69}$ & 50.26$_{-5.06}^{+5.06}$ & 9.29$_{-0.27}^{+0.24}$\\ 
6 & 3.63--3.71 & 2.16$_{-1.83}^{+1.97}$ & 0.80$_{-0.10}^{+0.09}$ & 12.90$_{-4.12}^{+7.54}$ & 1.49$_{-0.17}^{+0.22}$ & 24.37$_{-4.98}^{+4.13}$ & 0.23$_{-0.23}^{+0.23}$ & 105$/$81 & 1.29 & 1.25$_{-0.60}^{+0.77}$ & 37.27$_{-8.59}^{+8.59}$ & 6.92$_{-0.20}^{+0.22}$\\ 
7 & 3.71--3.80 & fixed & 0.78$_{-0.08}^{+0.07}$ & 6.98$_{-2.21}^{+3.84}$ & 1.72$_{-0.26}^{+0.60}$ & 9.32$_{-2.95}^{+1.83}$ & 0.25$_{-0.25}^{+0.25}$ & 93$/$76 & 1.23 & 1.32$_{-0.61}^{+0.84}$ & 16.30$_{-4.25}^{+4.25}$ & 5.00$_{-0.09}^{+0.10}$\\ 
8 & 3.80--3.90 & fixed & 0.76$_{-0.41}^{+0.76}$ & 3.11$_{-2.32}^{+6.67}$ & 1.36$_{-0.33}^{+1.36}$ & 5.43$_{-5.22}^{+2.98}$ & 0.20$_{-0.20}^{+0.20}$ & 70$/$72 & 0.97 & 1.14$_{-0.97}^{+1.41}$ & 8.54$_{-7.30}^{+7.30}$ & 3.92$_{-0.12}^{+0.08}$\\ 
9 & 3.90--4.00 & fixed & 1.60$_{-0.55}^{+0.69}$ & 5.42$_{-1.15}^{+1.16}$ & -- & -- & 0.23$_{-0.23}^{+0.23}$ & 83$/$69 & 1.20 & 1.60$_{-3.11}^{+3.70}$ & 5.42$_{-1.16}^{+1.16}$ & 3.53$_{-0.09}^{+0.08}$\\ 
10 & 4.00--4.10 & fixed & 0.85$_{-0.20}^{+0.22}$ & 3.41$_{-1.17}^{+1.18}$ & -- & -- & 0.12$_{-0.12}^{+0.12}$ & 52$/$66 & 0.79 & 0.85$_{-1.09}^{+1.12}$ & 3.41$_{-1.18}^{+1.18}$ & 3.28$_{-0.10}^{+0.07}$\\ 
11 & 4.10--4.20 & fixed & 0.73$_{-0.25}^{+0.28}$ & 1.87$_{-1.12}^{+1.10}$ & -- & -- & 0.16$_{-0.16}^{+0.16}$ & 61$/$64 & 0.95 & 0.73$_{-0.99}^{+1.01}$ & 1.87$_{-1.10}^{+1.10}$ & 3.12$_{-0.12}^{+0.07}$\\ 
12 & 4.20--4.32 & fixed & 1.20$_{-0.86}^{+2.31}$ & 2.00$_{-0.88}^{+0.80}$ & -- & -- & 0.00$_{-0.00}^{+0.00}$ & 81$/$71 & 1.14 & 1.20$_{-2.51}^{+6.34}$ & 2.00$_{-0.80}^{+0.80}$ & 3.08$_{-0.10}^{+0.11}$\\ 
\cline{1-13} 
Total : & -- & -- & -- &  -- & -- & -- & -- & -- & -- & 3.00$_{-1.74}^{+2.03}$ \tablenotemark{$\sharp$} & 26.25$_{-1.01}^{+1.27}$ \tablenotemark{$\sharp$} & 145.2$_{-3.2}^{+2.0}$ \tablenotemark{$\sharp$} \\
 & -- & -- & -- &  -- & -- & -- & -- & -- & -- & (keV) & (10$^{51}$ cm$^{-3}$) & (10$^{31}$ erg)
   \enddata
  % \tablecomments{
 %  }
 \tablenotetext{\dag}{
Phase and Time ranges are shown in Figure \ref{fig:Flare23_TEM_multi_lc}(e).
   }
    \tablenotetext{\ddag}{
$N_{\rm{H}}$ = ``fixed" means that it is fixed to the literature value 
2.29$\times$10$^{18}$ cm$^{-2}$ from \citet{Wood+2005_ApJS}.
}
    \tablenotetext{\rm *}{
$EM_{\rm{tot}}=EM_{1}+EM_{2}$.
$T_{\rm{ave}}$ is the EM-weighted average of $T_{1}$ and $T_{2}$.
}
    \tablenotetext{\sharp}{
$T_{\rm{ave}}$, $EM_{\rm{tot}}$, and total flare energy $E_{\rm{X}}$(0.2 -- 12 keV) of the entire Flare 23.
}
   \label{table:flare23_Xray_fit}
 \end{deluxetable*}
\end{longrotatetable}

\clearpage 
    \begin{figure}[ht!]
   \begin{center}
      \gridline{
\fig{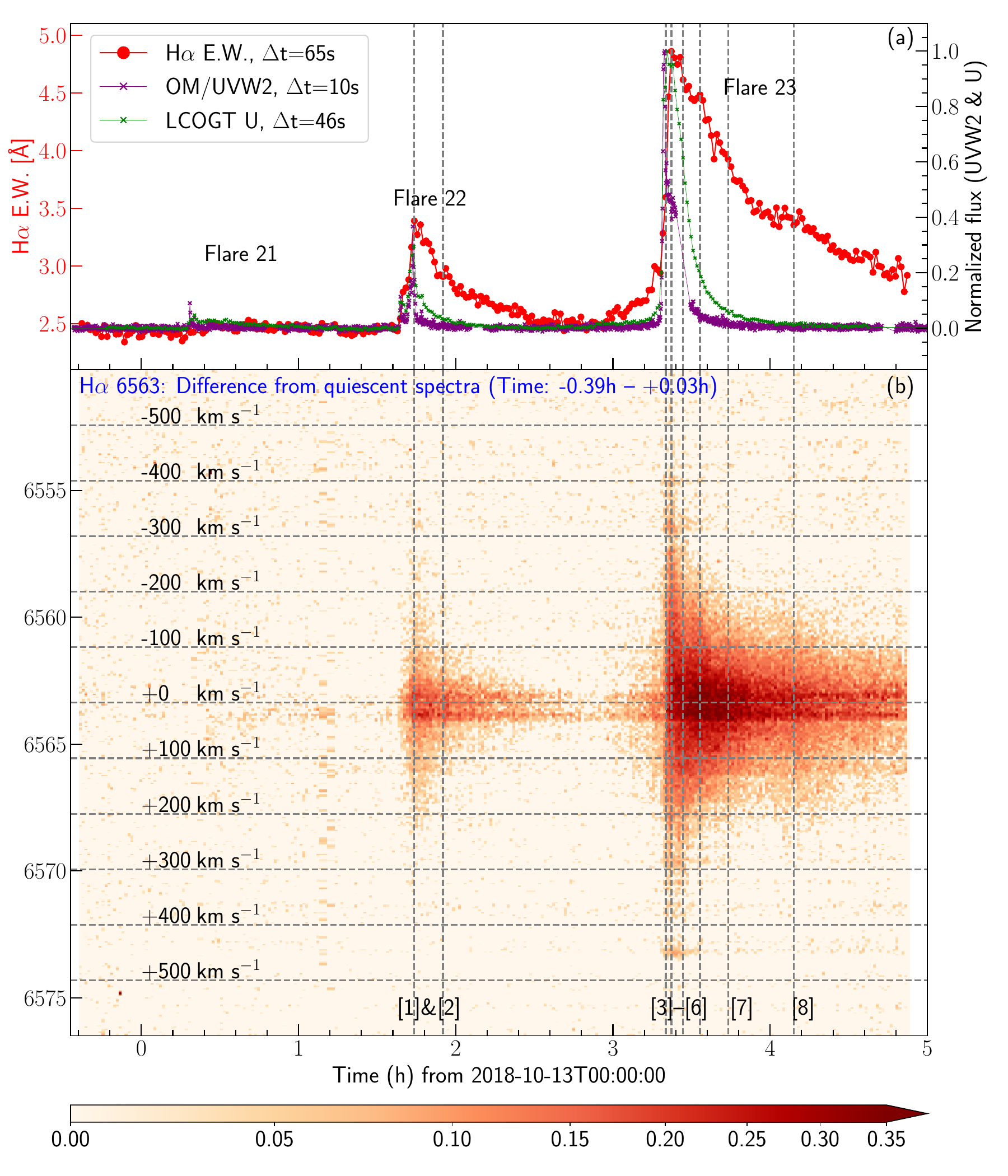}{0.45\textwidth}{\vspace{0mm}}
\fig{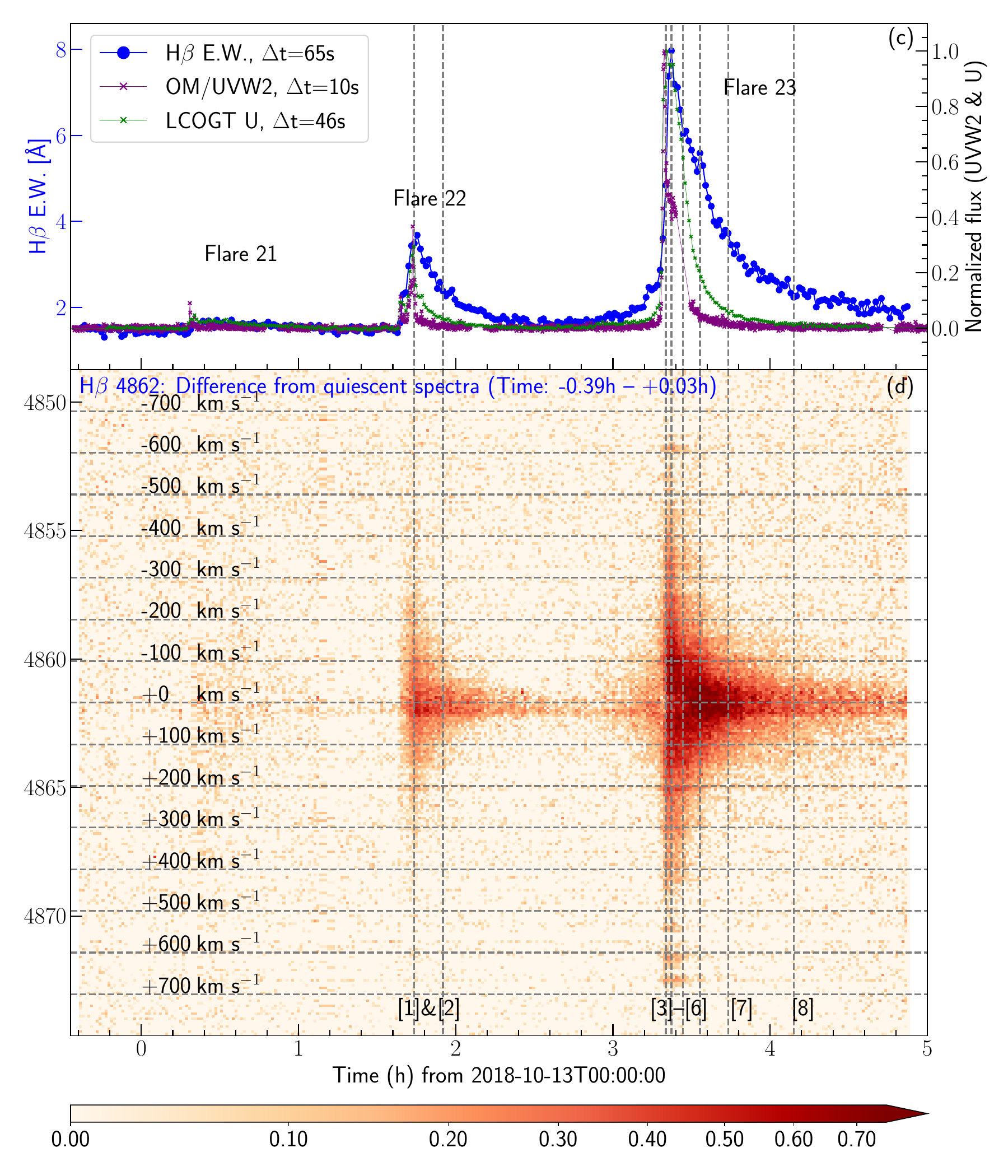}{0.45\textwidth}{\vspace{0mm}}
    }
     \vspace{-5mm}
      \gridline{
\fig{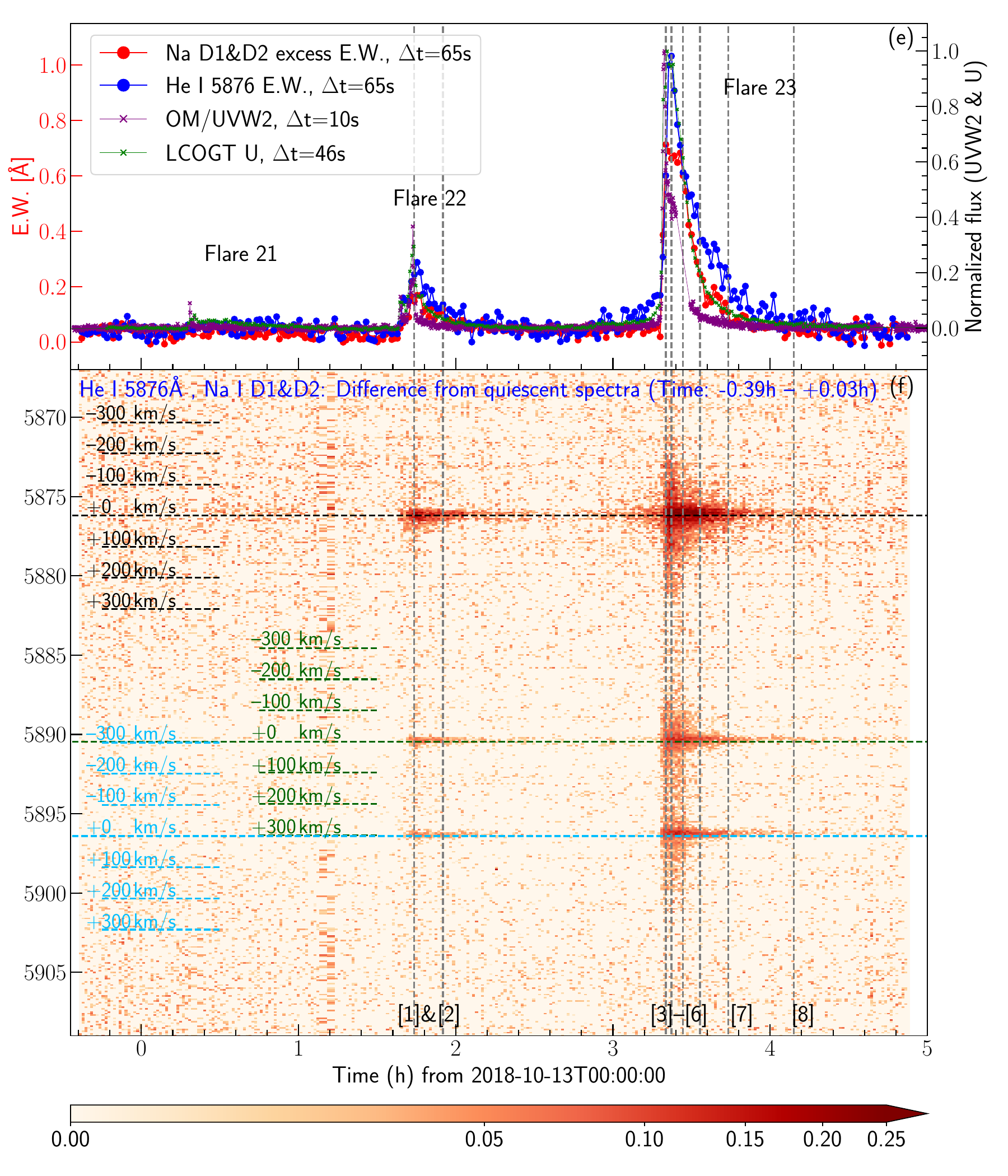}{0.45\textwidth}{\vspace{0mm}}
\fig{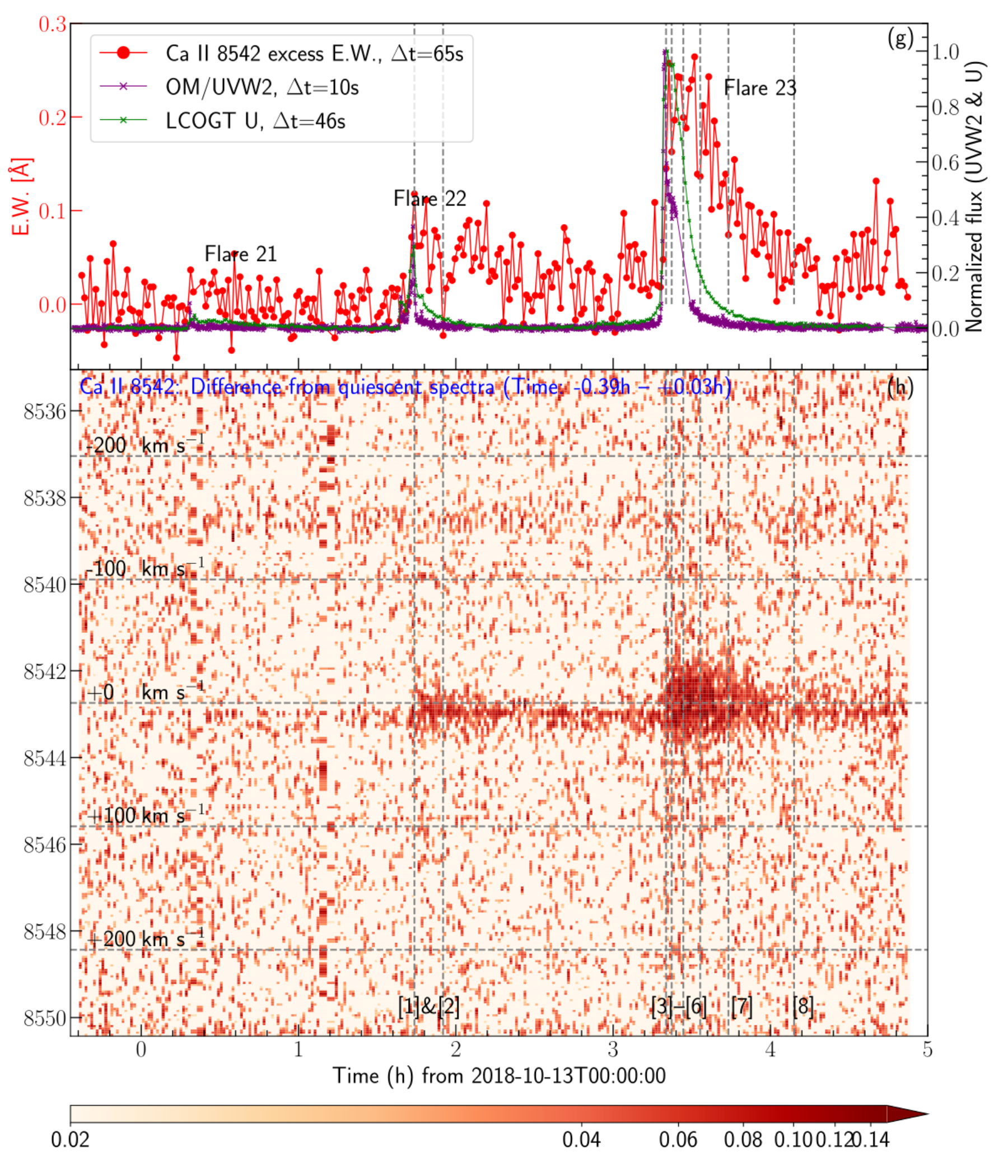}{0.45\textwidth}{\vspace{0mm}}
    }
     \vspace{-5mm}
     \caption{
(a) The H$\alpha$ equivalent width values covering Flares 21, 22, and 23 on 2018 October 13 are plotted with red circles, while the normalized OM UVW2 NUV \& LCOGT U-band fluxes are plotted with purple and green marks (cf. Figure \ref{fig:Oct13_TEM_multi_lc}). 
(b) Time evolution of the H$\alpha$ profile from the CHIRON spectra. 
The vertical axis represents the wavelength, while the gray horizontal dashed lines with velocity values represent the Doppler velocities from the H$\alpha$ line center. The gray vertical dashed lines indicate the time [1] -- \color{black} time \color{black} [8],
which are shown in (a) (lightcurve) and Figure \ref{fig:flare23_HaSpec} (spectra).
The color map represents the line profile changes from the quiescent profile 
(see Figure \ref{fig:flare23_HaSpec}(b), (d), (f), and (h)).
(c) \& (d) Same as panels (a) \& (b), but for the H$\beta$ line (cf. Figure \ref{fig:flare23_HbSpec}).
(e) \& (f) Same as panels (a) \& (b), but for He I D3 5876\AA~line and Na I D1\&D2 lines (cf. Figure \ref{fig:flare23_HeNaSpec}).
(g) \& (h) Same as panels (a) \& (b), but for the Ca II 8542 \AA~line 
 (cf. Figure \ref{fig:flare23_Ca8542Spec}).
The excess equivalent width values of Na I D1\&D2 and Ca II 8542 lines are plotted in (e) and (g) (cf. Figure \ref{fig:Oct13_TEM_multi_lc}).
}
   \label{fig:maps_flare23}
   \end{center}
 \end{figure}

\begin{figure}[ht!]
   \begin{center}
      \gridline{
\fig{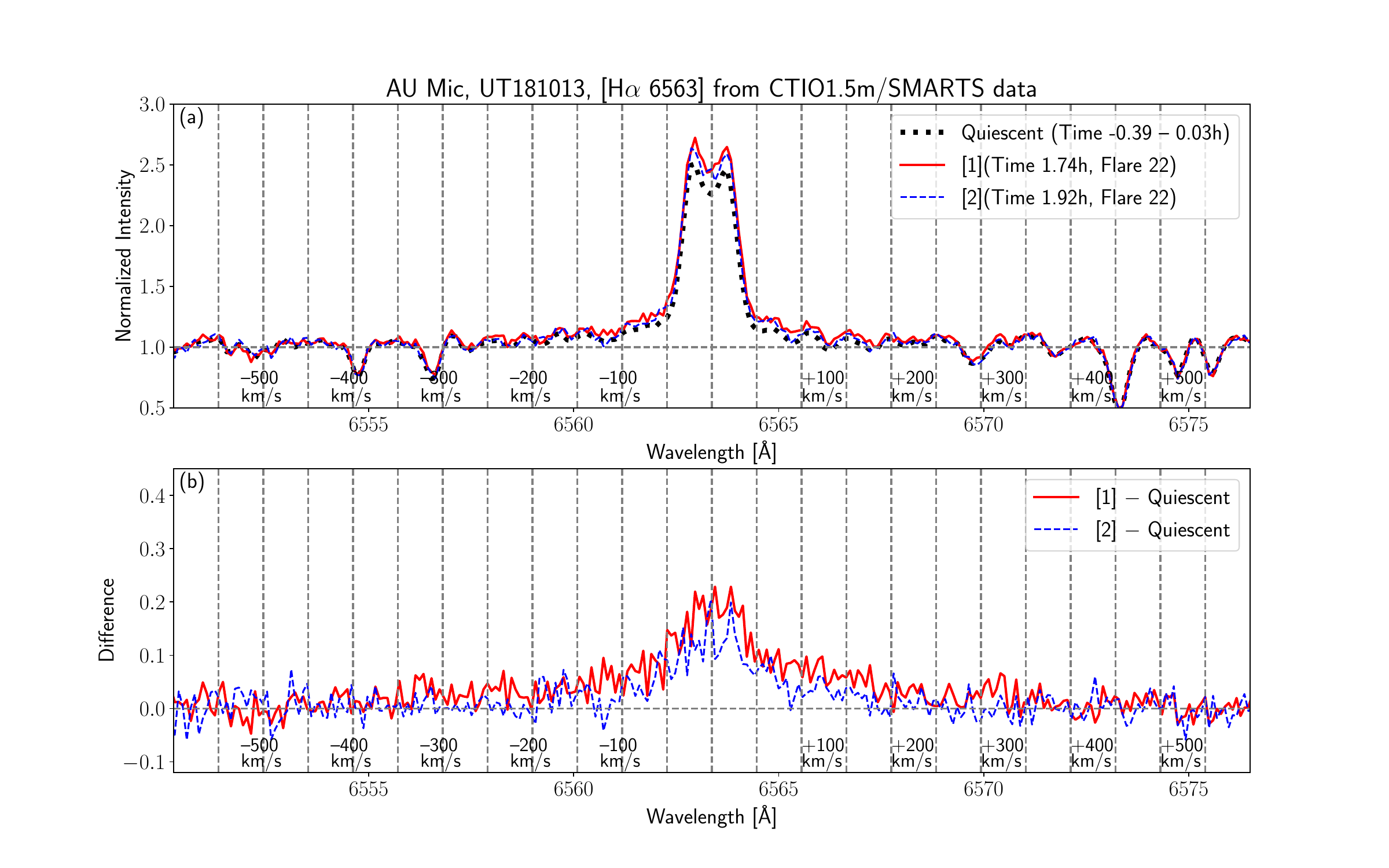}{0.5\textwidth}{\vspace{0mm}}
\fig{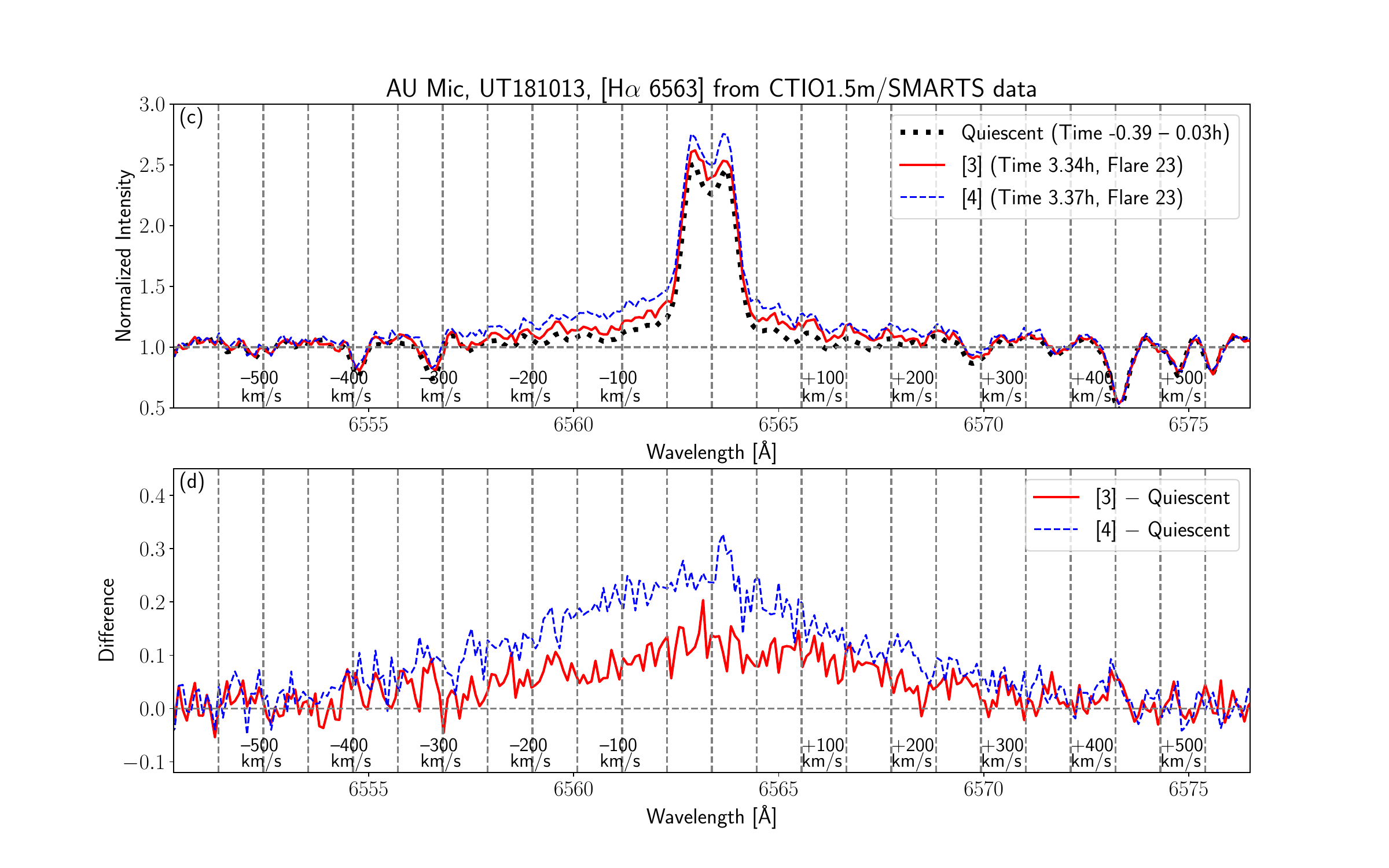}{0.5\textwidth}{\vspace{0mm}}
    }
     \vspace{-5mm}
      \gridline{
\fig{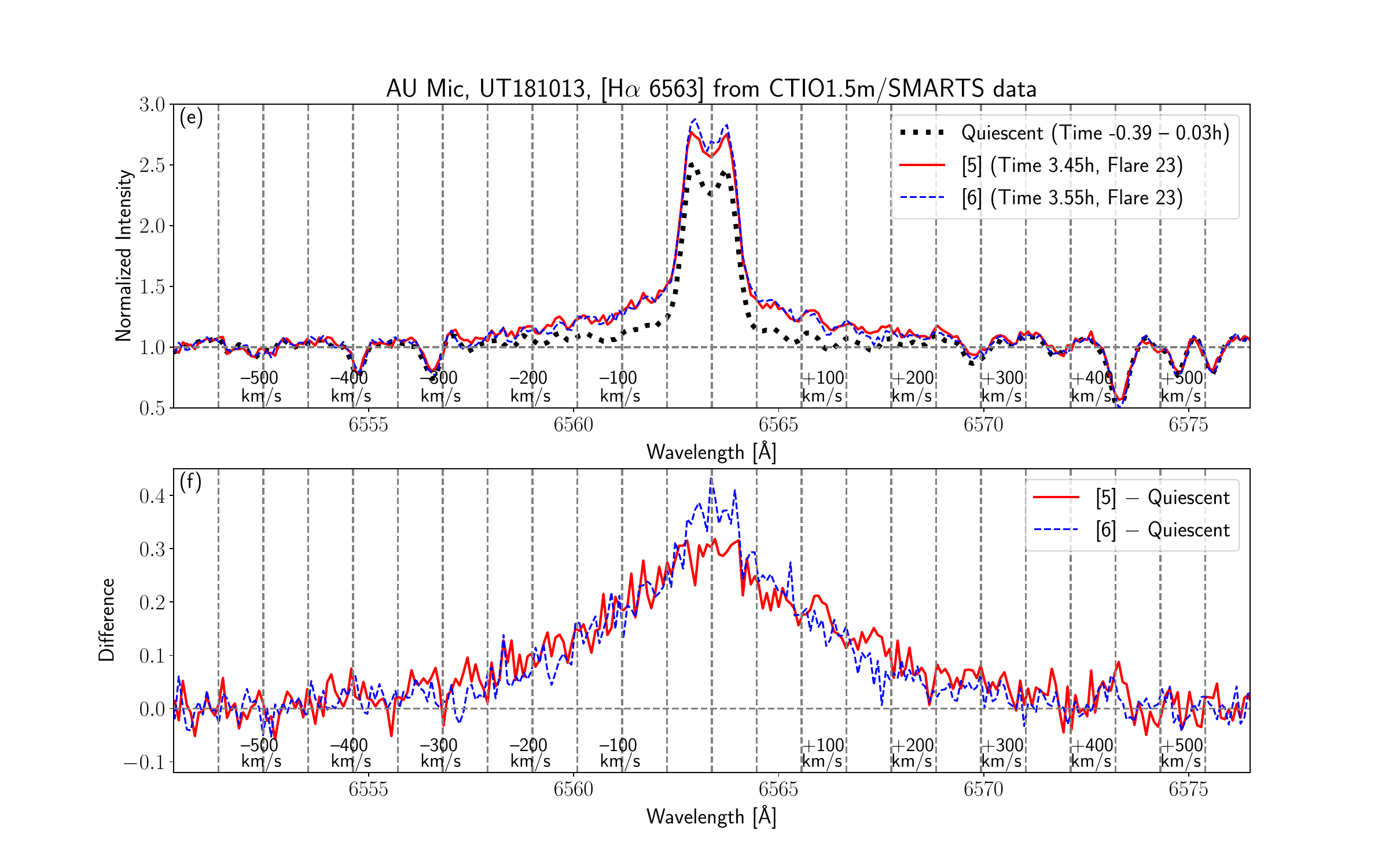}{0.5\textwidth}{\vspace{0mm}}
\fig{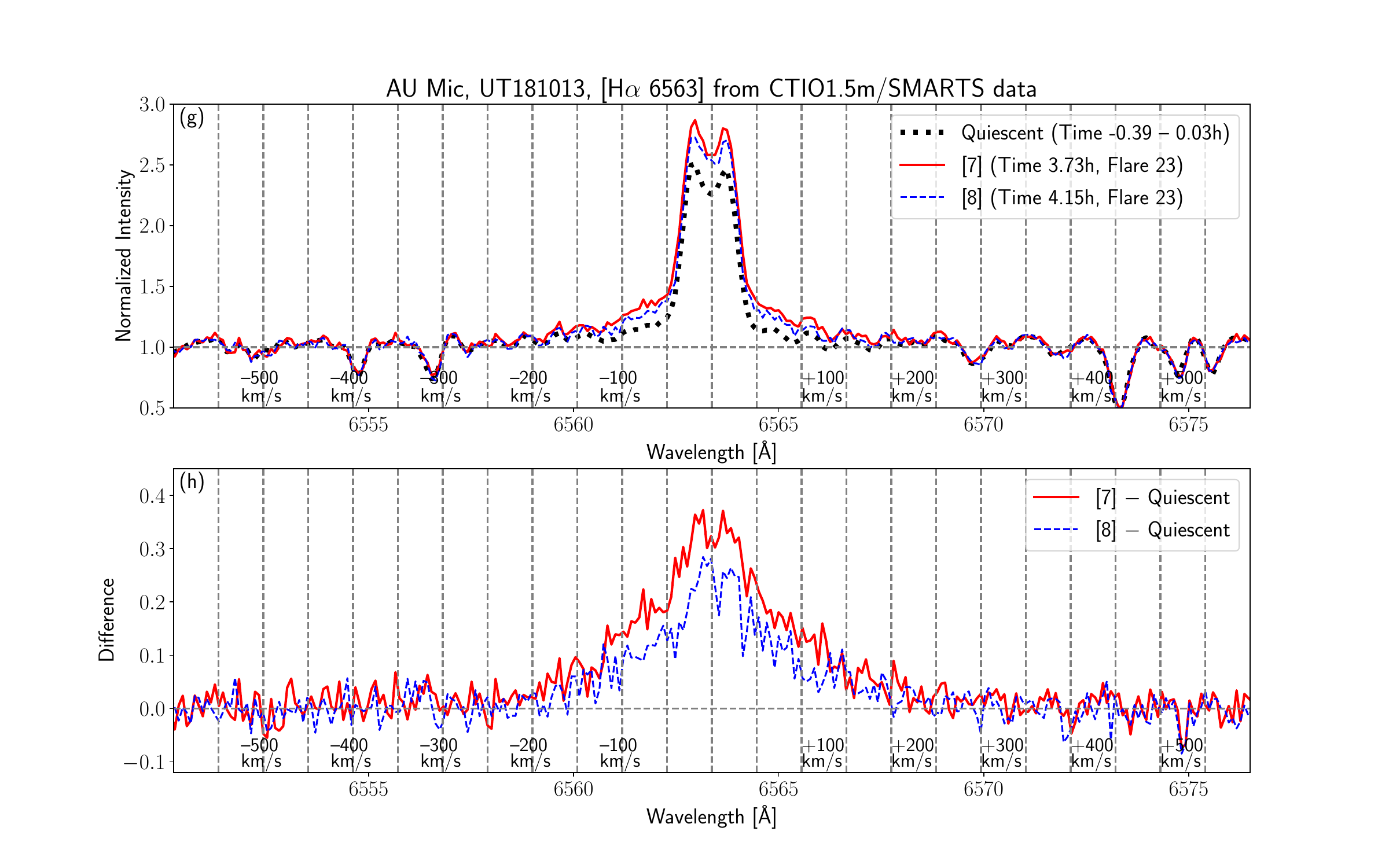}{0.5\textwidth}{\vspace{0mm}}
    }
     \caption{
(a) Line profiles of the H$\alpha$ emission line during Flare 22 on 2018 
October 13 from the CHIRON spectra. 
The horizontal and vertical axes represent the wavelength and flux normalized by the continuum. 
The gray vertical dashed lines with velocity values represent the Doppler velocities from the H$\alpha$ line center.
The red solid and blue dashed lines indicate the line profiles at the time [1] and \color{black} time \color{black} [2], respectively, which are indicated in 
Figure \ref{fig:maps_flare23} (a) \& (b) (lightcurve and intensity map)
and are during Flare 22. 
The black dotted line indicates the line profiles in the quiescent phase, which are the average profile during -0.39-- -0.03 hr on this date (see Figure \ref{fig:maps_flare23}(a)).
(c), (e),\& (g) Same as panel (a), but the line profiles at time [3] -- \color{black} time \color{black}  [8] during Flare 23.
(b), (d), (f), \& (h) Same as panels (a), (c), (e),\& (g), respectively,
but the line profile differences from the quiescent phase.
}
   \label{fig:flare23_HaSpec}
   \end{center}
 \end{figure}

\begin{figure}[ht!]
   \begin{center}
      \gridline{
\fig{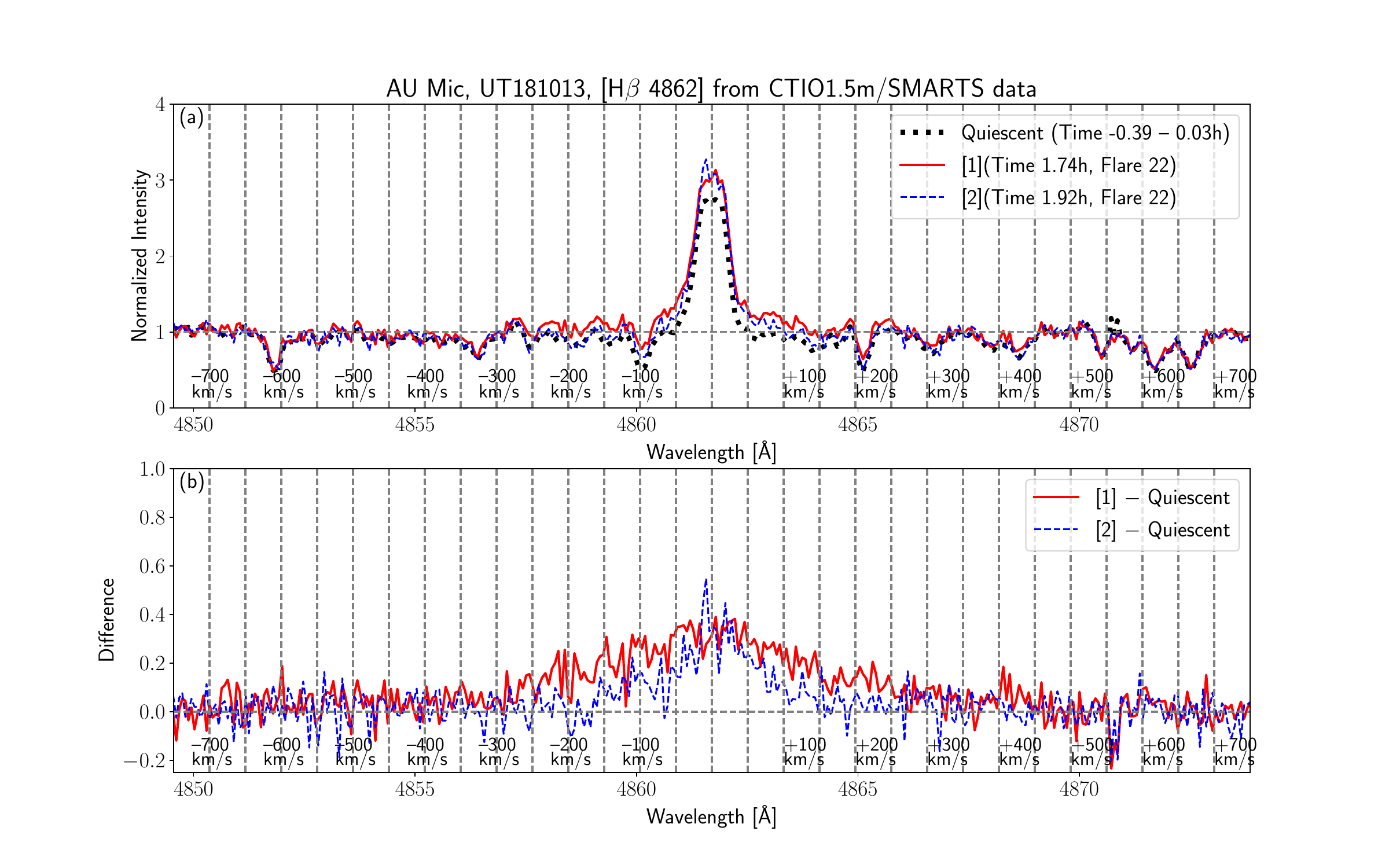}{0.5\textwidth}{\vspace{0mm}}
\fig{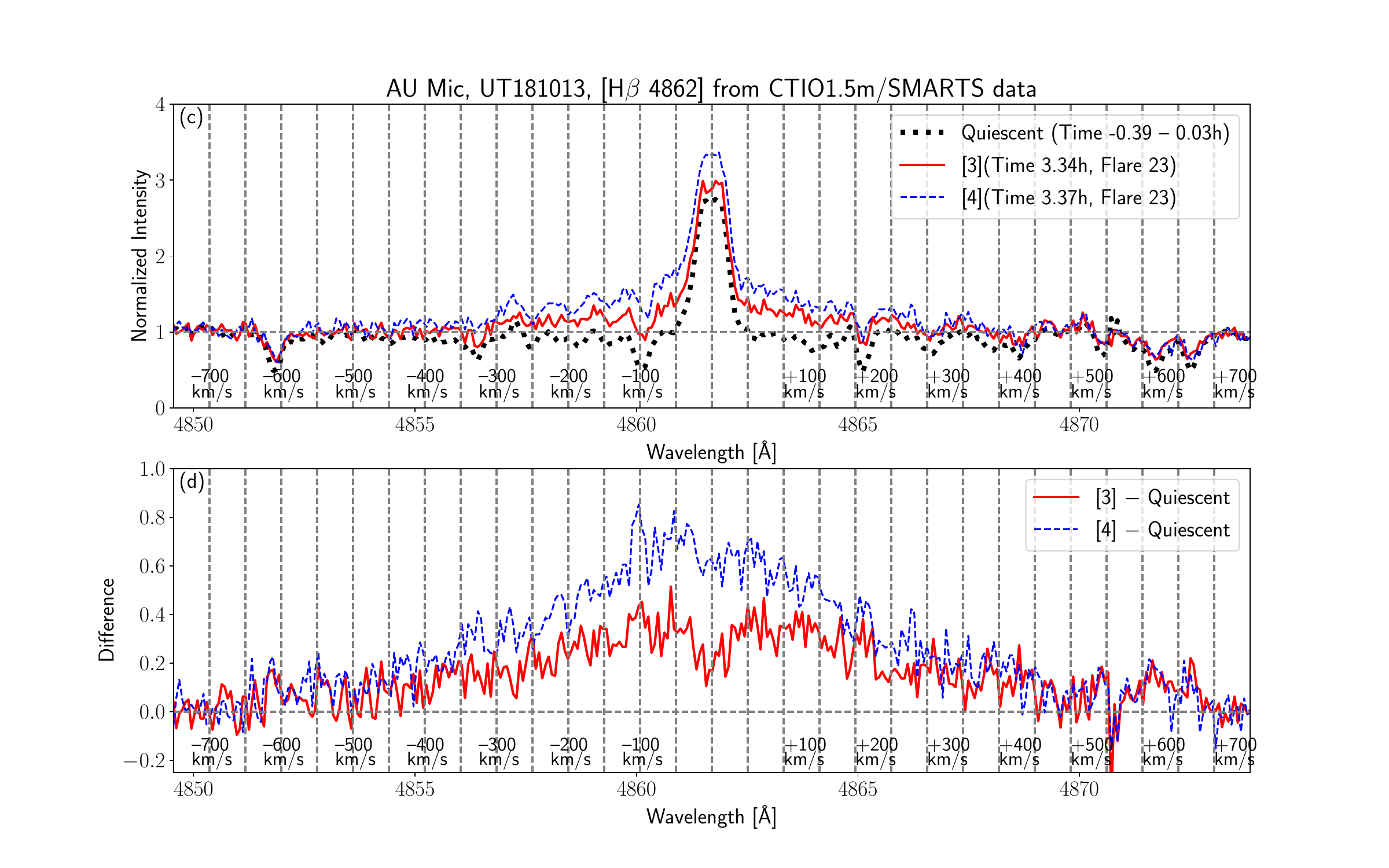}{0.5\textwidth}{\vspace{0mm}}
    }
     \vspace{-5mm}
      \gridline{
\fig{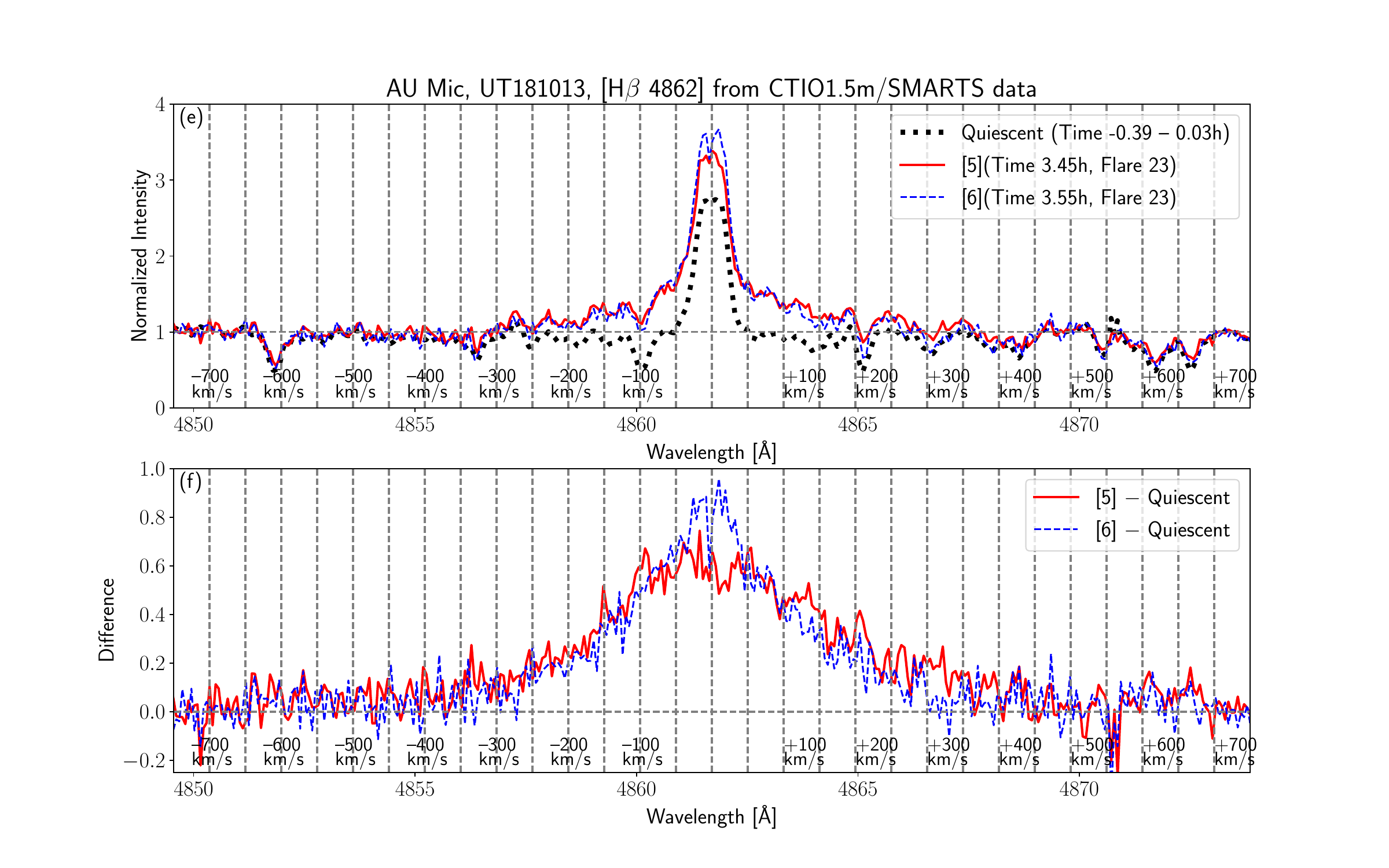}{0.5\textwidth}{\vspace{0mm}}
\fig{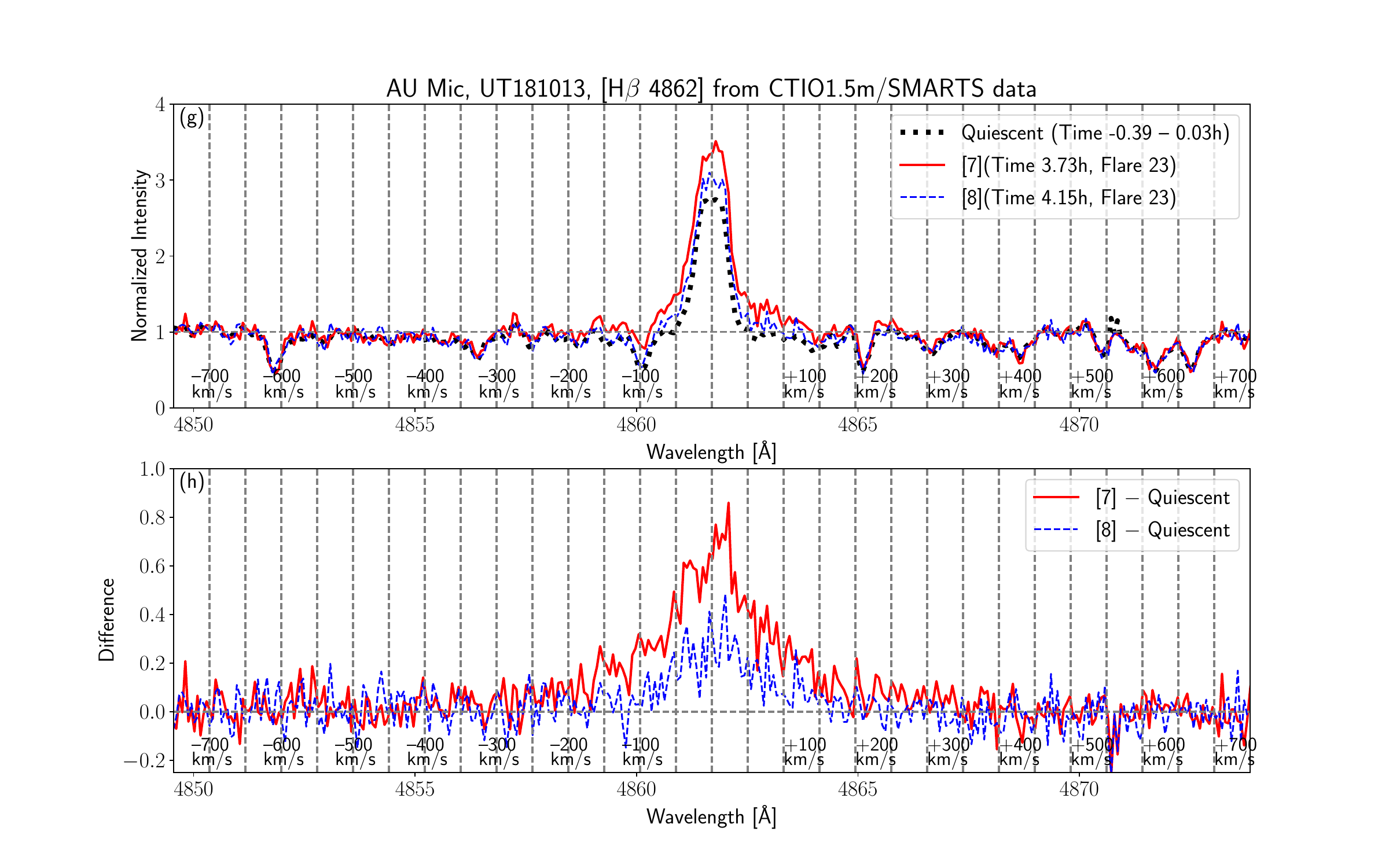}{0.5\textwidth}{\vspace{0mm}}
    }
     \caption{
Same as Figure \ref{fig:flare23_HaSpec} but for the H$\beta$ line.
}
   \label{fig:flare23_HbSpec}
   \end{center}
 \end{figure}

The line profiles of the H$\alpha$, H$\beta$, He I D3 5876\AA, 
Na I D1\&D2, and Ca II 8542 lines during Flares 22 and 23 
are shown in the color maps in Figure \ref{fig:maps_flare23} 
and the spectra in Figures \ref{fig:flare23_HaSpec},
\ref{fig:flare23_HbSpec}, \ref{fig:flare23_HeNaSpec}, and \ref{fig:flare23_Ca8542Spec} \color{black}(Figures \ref{fig:flare23_HeNaSpec}, and \ref{fig:flare23_Ca8542Spec} are in Appendix \ref{appen_sec:additional_figures}).\color{black} 
The E.W. lightcurves of these lines are also shown in Figure  \ref{fig:maps_flare23}. 
 The line broadening is nearly symmetric out to $\sim \pm$400 km s$^{-1}$
around the peak time of Flare 23 (the time [3] \footnote{\color{black} In the following part of the paper, the specific times focused in the lightcurves and spectra are shown with the numbers such as time [1], time [2], and so on (cf. Figures \ref{fig:maps_flare23} and \ref{fig:flare23_HaSpec}).} and \color{black} time \color{black} [4] in Figures \ref{fig:maps_flare23} and \ref{fig:flare23_HaSpec}).
This broadening rapidly decreases as the NUV/optical continuum (e.g., UVW2, U-band) emissions decay (see the line profile changes from the time [3] to \color{black} time \color{black} [7]), and the broadening is much smaller in the decay phase of the longer H$\alpha$ emission (see the line broadenings in the time [7] and \color{black} time \color{black} [8]). 
Comparing the line profile evolution in Figures \ref{fig:maps_flare23} -- \ref{fig:flare23_HbSpec}, 
the H$\beta$ line also shows the time evolution similar to the H$\alpha$ evolution  (the broadening around the flare peak, and the rapid decay of the broadening as the NUV/optical continuum emission decays) during Flare 23. The H$\beta$ line broadening occurs around the flare peak (= around the time [3] and \color{black} time \color{black} [4]) of Flare 23 is up to $\sim$600 km s$^{-1}$ and this is bigger than that of the H$\alpha$ line, but the decay phase line broadening of the H$\beta$ line ($\sim$100--150 km s $^{-1}$) is smaller than that of the H$\alpha$ line ($\sim$200 km s $^{-1}$) (see the spectra at the time [8]). Figure \ref{fig:flare23_HaSpec}(d) 
also suggests that the H$\alpha$ line profile at the time [3] may have 
slightly larger flux in the blue wing than in the red wing, 
but this potential asymmetry is too small to conclude that there is a clear blue wing enhancement as seen in previous studies of M-dwarf line asymmetries \color{black} (e.g., \citealt{Fuhrmeister+2018,Vida+2019,Notsu+2024_ApJ,Kajikiya+2025_ApJ_PaperI}). \color{black} 
No other notable line asymmetry signatures are seen in the H$\alpha$ and H$\beta$ line profiles during Flare 23.

The He I D3 5876\AA~and Na D1\&D2 lines also show smaller symmetric broadenings with $\sim \pm$200--250 km s$^{-1}$ and $\sim \pm$150--200 km s$^{-1}$, respectively, at around the time [3] and \color{black} time \color{black} [4] (= around the flare peak) of Flare 23, but 
their flare line emissions decay much faster than the H$\alpha$ and H$\beta$ lines, and the flare emissions end between the time [7] and \color{black} time \color{black} [8] (Figures \ref{fig:maps_flare23} and \ref{fig:flare23_HeNaSpec}). 
There are no notable line broadening changes in the Ca II 8542 emissions during Flare 23 
(Figures \ref{fig:maps_flare23} and \ref{fig:flare23_Ca8542Spec}), 
but the flare duration of Ca II 8542 are longer than the H$\beta$ line and shorter than the H$\alpha$ line (Figure \ref{fig:Flare23_TEM_multi_lc}(c) and (d)).
In addition to Flare 23, Flare 22 shows clear symmetric line broadening of the H$\alpha$ and H$\beta$ lines with $\sim \pm$200 km s$^{-1}$ and $\sim \pm$250--300 km s$^{-1}$, respectively, at around the flare peak (= around the time [1]), and these broadenings also decay rapidly as the NUV/optical continuum emissions (e.g., UVW2, U-band) decay 
(Figures \ref{fig:maps_flare23}, \ref{fig:flare23_HaSpec}, and \ref{fig:flare23_HbSpec}). 
Although the broadenings and flare size are smaller, 
these time evolution properties of Flare 22 (the symmetric broadenings of the line profiles at the flare peak and the rapid decay afterwards)
are roughly similar to those of Flare 23 described here, from rough qualitative points of view. We have more detailed discussions on the line broadening evolutions of Flare 23 in Section \ref{sec:discussions}.

\clearpage

\subsection{Flares 11\&15: The two long duration X-ray flares}\label{subsec:ana_flare_11_and_15}

\begin{figure}[ht!]
   \begin{center}
      \gridline{
\fig{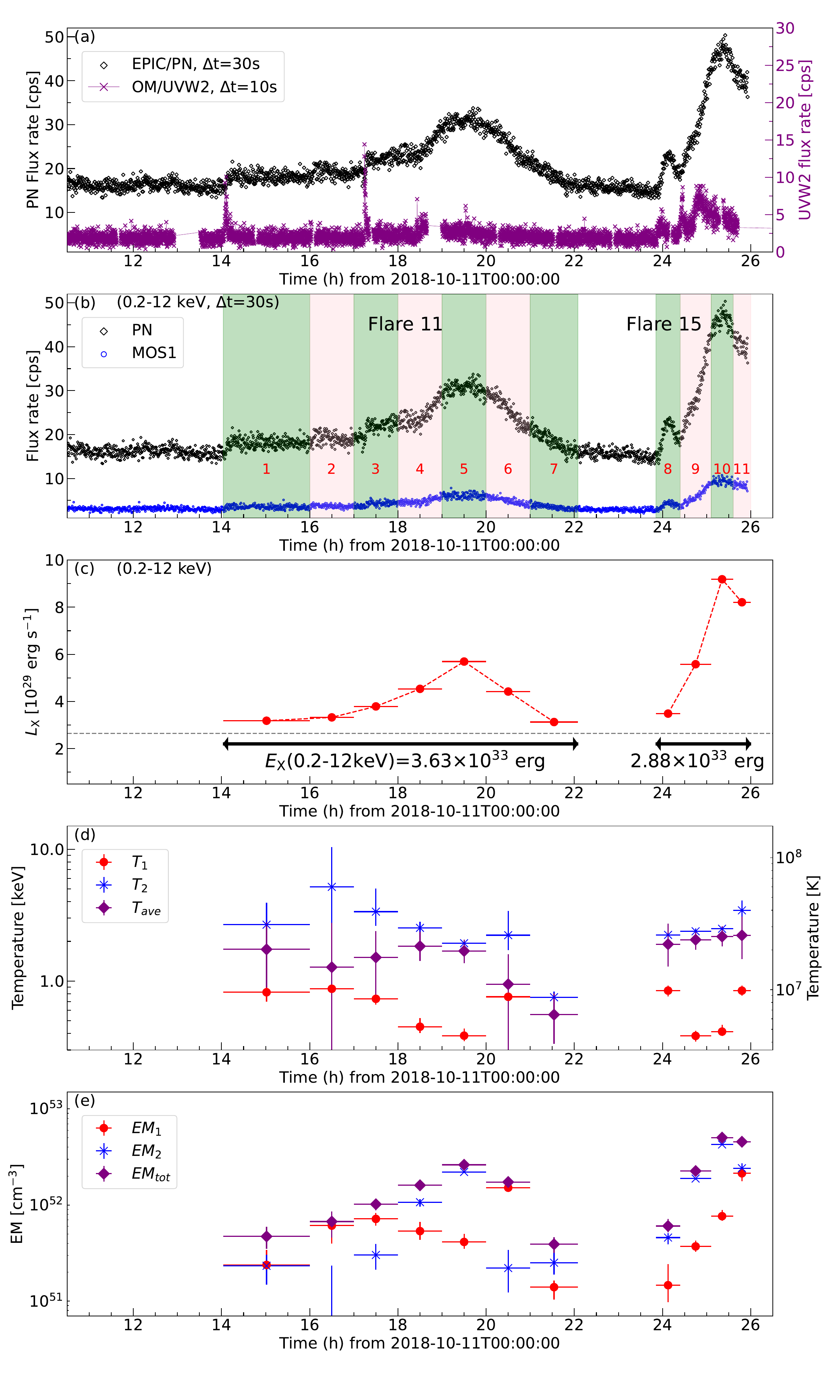}{0.7\textwidth}{\vspace{0mm}}
    }
     \vspace{-5mm}
     \caption{
(a) XMM EPIC-pn X-ray (0.2–12 keV) and XMM OM UVW2 lightcurves around Flares 11 and 15 (cf. Figure \ref{fig:X-ray_Ha_obsid_0822740301_lc}). 
(b) XMM EPIC-pn and EPIC-MOS1 X-ray lightcurves (0.2–12 keV) of Flares 11 and 15. 
The green and pink colored regions correspond to the Phases 1–10, separated for the spectral analysis.
(c)--(e) The spectral analysis results (X-ray luminosity in 0.2 -- 12 keV $L_{\rm{X}}$, temperature $T$, and emission measure EM) of Phases 1--11 of Flares 11 and 15 (from Table \ref{table:flare11_Xray_fit}). 
}
   \label{fig:Flare11and15_TEM_multi_lc}
   \end{center}
 \end{figure}

Figure \ref{fig:Flare11and15_TEM_multi_lc} shows the X-ray and UVW2 lightcurves around Flares 11 and 15, which occured on 2018 October 11.
Among all the flares in this campaign, 
Flares 11 has the longest X-ray flare duration ($\sim$8 hours), 
while Flare 15 is the flare with the second largest X-ray amplitude 
(Figures \ref{fig:allEW_AUMic_XMM_opt} 
and \ref{fig:X-ray_Ha_obsid_0822740301_lc}).
In T23, Flare 15 is identified as a ``Neupert/quasi-Neupert" flare since it passes one of the two timing criteria, while Flare 11 is identified as 
a ``Non-Neupert Type I" flare since there are no clear 
UVW2 response in association with the X-ray flare peak. 
It is noted there are some UVW2 responses at $\sim$14:10 UT and $\sim$17:16 UT in Figure \ref{fig:Flare11and15_TEM_multi_lc}(a), 
but T23 did not judge \color{black} that \color{black} this has a clear relation 
satisfying the Neupert criteria with the overall X-ray emission of Flare 11.

We conduct the spectral analysis of the time-resolved X-ray data (PN and MOS1 data) of these two flares, following the same method as Flare 23 in Section \ref{subsec:ana_flare_23}.
The data of Flare 11 and 15 are manually 
divided into Phases 1--7 and 8--11, respectively, as shown in
Figure \ref{fig:Flare11and15_TEM_multi_lc}(b).
\color{black} Each phase spectral data  and finally selected fitting results \color{black}  
are shown in 
Figures \ref{fig:specfig_Flare11_TEM_quie_each_No.1-No.4}, 
\ref{fig:specfig_Flare11_TEM_quie_each_No.5-No.7}, and 
\ref{fig:specfig_Flare15_TEM_quie_each_No.8-No.11} in Appendix \ref{appen_sec:additional_figures}. 
The resultant fitting parameters of Phases 1–11 
are summarized in Table \ref{table:flare11_Xray_fit}.
\color{black}
As shown in the table, the 2-temperature fitting results with ``$N_{\rm{H}}$ fixed" are selected in the fitting analysis of all the phases 
of Flares 11 and 15
(Figures \ref{fig:specfig_Flare11_TEM_quie_each_No.1-No.4} -- 
\ref{fig:specfig_Flare15_TEM_quie_each_No.8-No.11}).
\color{black}
The resultant $L_{\rm{X}}$, temperature ($T_{1}$, $T_{2}$, $T_{\rm{ave}}$), and emission measure ($EM_{1}$, $EM_{2}$, $EM_{\rm{tot}}$) values 
are plotted in Figure \ref{fig:Flare11and15_TEM_multi_lc}(c), (d), and (e).
The total X-ray energies of Flares 11 and 15 in the 0.2 -- 12 keV range are calculated as $E_{\rm{X}}=3.63^{+0.03}_{-0.07}\times 10^{33}$ erg 
and $E_{\rm{X}}=2.88^{+0.02}_{-0.03}\times 10^{33}$ erg, respectively 
(Table \ref{table:energy_remarkable_flares}).
As a result, these two flares (Flares 11 and 15) emit the first and second largest X-ray flare energies among all the flares in this campaign, while Flare 23  
is the third largest one
(Tables \ref{table:X-ray_flarefit_timeave} and \ref{table:energy_remarkable_flares}).

\begin{longrotatetable}
\begin{deluxetable*}{lcccccccccccc}
   \tablecaption{Best fitting \color{black} parameters \color{black} for the time-resolved X-ray spectra of Flares 11 \& 15.  Statistical
90\% confidence region errors are shown for the X-ray fitting parameters.}
   \tablewidth{0pt}
   \tablehead{
\colhead{Phase}\tablenotemark{\rm \dag}  &  
\colhead{Time}\tablenotemark{\rm \dag}   & 
\colhead{$N_{\rm{H}}$ \tablenotemark{\rm \ddag}} &
\colhead{$T_{1}$} &
\colhead{$EM_{1}$} &
\colhead{$T_{2}$} &
\colhead{$EM_{2}$} &
\colhead{$Z_{\rm{Fe}}$} &
\colhead{$\chi^{2}/\rm{d.o.f}$} &
\colhead{$\chi^{2}_{\rm{black}}$} &
\colhead{$T_{\rm{ave}}$}&
\colhead{$EM_{\rm{tot}}$} &
\colhead{$L_{\rm{X}}$(0.2 -- 12 keV)} 
\\
\colhead{} &
\colhead{(h)}  & 
\colhead{(10$^{20}$ cm$^{-2}$)}&
\colhead{(keV)} &
\colhead{(10$^{51}$ cm$^{-3}$)} &
\colhead{(keV)} &
\colhead{(10$^{51}$ cm$^{-3}$)} & 
\colhead{($Z_{\rm{Fe,}\odot}$)}& 
\colhead{}& 
\colhead{} &
\colhead{(keV)}&
\colhead{(10$^{51}$ cm$^{-3}$)} &
\colhead{(10$^{29}$ erg s $^{-1}$)} 
}
   \startdata
\multicolumn{13}{c}{Flare 11} \\
\cline{1-13}
1 & 14.04--16.00 & fixed & 0.82$_{-0.12}^{+0.12}$ & 2.38$_{-0.84}^{+1.03}$ & 2.68$_{-0.50}^{+1.25}$ & 2.33$_{-0.85}^{+0.71}$ & 0.05$_{-0.05}^{+0.05}$ & 304$/$177 & 1.72 & 1.74$_{-0.84}^{+1.04}$ & 4.71$_{-1.25}^{+1.25}$ & 3.18$_{-0.03}^{+0.02}$\\ 
2 & 16.00--17.00 & fixed & 0.87$_{-0.12}^{+0.10}$ & 6.10$_{-2.13}^{+0.76}$ & 5.18$_{-3.56}^{+5.18}$ & 0.62$_{-0.38}^{+1.72}$ & 0.04$_{-0.04}^{+0.04}$ & 247$/$143 & 1.72 & 1.27$_{-1.06}^{+1.49}$ & 6.73$_{-1.88}^{+1.88}$ & 3.33$_{-0.15}^{+0.02}$\\ 
3 & 17.00--18.00 & fixed & 0.73$_{-0.07}^{+0.06}$ & 7.17$_{-1.08}^{+1.13}$ & 3.36$_{-0.73}^{+1.69}$ & 3.02$_{-0.89}^{+0.91}$ & 0.06$_{-0.06}^{+0.06}$ & 234$/$154 & 1.52 & 1.51$_{-0.77}^{+0.88}$ & 10.19$_{-1.44}^{+1.44}$ & 3.79$_{-0.04}^{+0.03}$\\ 
4 & 18.00--19.00 & fixed & 0.45$_{-0.05}^{+0.07}$ & 5.35$_{-1.01}^{+1.30}$ & 2.53$_{-0.21}^{+0.29}$ & 10.70$_{-1.05}^{+0.82}$ & 0.05$_{-0.05}^{+0.05}$ & 314$/$182 & 1.73 & 1.84$_{-0.42}^{+0.52}$ & 16.05$_{-1.54}^{+1.54}$ & 4.54$_{-0.04}^{+0.03}$\\ 
5 & 19.00--20.00 & fixed & 0.39$_{-0.04}^{+0.05}$ & 4.13$_{-0.62}^{+0.90}$ & 1.94$_{-0.08}^{+0.07}$ & 22.09$_{-0.63}^{+0.71}$ & 0.13$_{-0.13}^{+0.13}$ & 341$/$193 & 1.77 & 1.69$_{-0.33}^{+0.37}$ & 26.21$_{-1.15}^{+1.15}$ & 5.69$_{-0.04}^{+0.03}$\\ 
6 & 20.00--21.00 & fixed & 0.76$_{-0.03}^{+0.03}$ & 15.13$_{-1.36}^{+1.16}$ & 2.23$_{-0.51}^{+1.17}$ & 2.21$_{-0.97}^{+1.19}$ & 0.06$_{-0.06}^{+0.06}$ & 216$/$150 & 1.44 & 0.95$_{-0.70}^{+0.65}$ & 17.34$_{-1.66}^{+1.66}$ & 4.42$_{-0.04}^{+0.02}$\\ 
7 & 21.00--22.08 & fixed & 0.20$_{-0.07}^{+0.10}$ & 1.40$_{-0.36}^{+0.25}$ & 0.75$_{-0.08}^{+0.08}$ & 2.51$_{-0.63}^{+0.69}$ & 0.20$_{-0.20}^{+0.20}$ & 243$/$130 & 1.87 & 0.56$_{-0.22}^{+0.25}$ & 3.91$_{-0.73}^{+0.73}$ & 3.13$_{-0.03}^{+0.02}$\\ 
\cline{1-13} 
Total: & -- & -- & -- &  -- & -- & -- & -- & -- & -- & 1.47$_{-0.96}^{+1.02}$ \tablenotemark{$\sharp$} & 11.19$_{-0.53}^{+0.54}$ \tablenotemark{$\sharp$} & 363.3$_{-6.7}^{+2.7}$ \tablenotemark{$\sharp$}  \\
 & -- & -- & -- &  -- & -- & -- & -- & -- & -- & (keV) & (10$^{51}$ cm$^{-3}$) & (10$^{31}$ erg)
\\
\cline{1-13}
\cline{1-13}
\multicolumn{13}{c}{Flare 15} \\
\cline{1-13}
8 & 23.85--24.40 & fixed & 0.85$_{-0.08}^{+0.08}$ & 1.47$_{-0.49}^{+0.96}$ & 2.24$_{-0.26}^{+0.35}$ & 4.58$_{-0.68}^{+0.53}$ & 0.37$_{-0.37}^{+0.37}$ & 137$/$127 & 1.08 & 1.90$_{-0.61}^{+0.83}$ & 6.05$_{-1.10}^{+1.10}$ & 3.49$_{-0.04}^{+0.03}$\\ 
9 & 24.40--25.10 & fixed & 0.38$_{-0.04}^{+0.04}$ & 3.71$_{-0.45}^{+0.56}$ & 2.39$_{-0.09}^{+0.12}$ & 18.86$_{-0.55}^{+0.51}$ & 0.26$_{-0.26}^{+0.26}$ & 319$/$186 & 1.72 & 2.06$_{-0.32}^{+0.35}$ & 22.57$_{-0.76}^{+0.76}$ & 5.58$_{-0.06}^{+0.05}$\\ 
10 & 25.10--25.60 & fixed & 0.41$_{-0.03}^{+0.05}$ & 7.65$_{-0.79}^{+1.24}$ & 2.49$_{-0.08}^{+0.08}$ & 42.65$_{-0.89}^{+0.86}$ & 0.21$_{-0.21}^{+0.21}$ & 343$/$206 & 1.66 & 2.18$_{-0.35}^{+0.44}$ & 50.30$_{-1.50}^{+1.50}$ & 9.18$_{-0.06}^{+0.07}$\\ 
11 & 25.60--26.00 & fixed & 0.85$_{-0.07}^{+0.07}$ & 21.37$_{-3.65}^{+4.35}$ & 3.44$_{-0.39}^{+0.64}$ & 24.14$_{-3.68}^{+3.10}$ & 0.05$_{-0.05}^{+0.05}$ & 208$/$170 & 1.22 & 2.22$_{-0.75}^{+0.86}$ & 45.50$_{-5.34}^{+5.34}$ & 8.21$_{-0.10}^{+0.08}$\\ 
\cline{1-13} 
Total: & -- & -- & -- &  -- & -- & -- & -- & -- & -- & 2.14$_{-1.01}^{+1.09}$ \tablenotemark{$\sharp$} & 30.43$_{-1.10}^{+1.17}$ \tablenotemark{$\sharp$} & 288.1$_{-2.5}^{+2.2}$ \tablenotemark{$\sharp$}  \\
 & -- & -- & -- &  -- & -- & -- & -- & -- & -- & (keV) & (10$^{51}$ cm$^{-3}$) & (10$^{31}$ erg)
   \enddata
  % \tablecomments{
 %  }
 \tablenotetext{\dag}{
Phase and Time ranges are shown in Figure \ref{fig:Flare11and15_TEM_multi_lc}(b).
   }
    \tablenotetext{\ddag}{
$N_{\rm{H}}$ = ``fixed" means that it is fixed to the literature value 
2.29$\times$10$^{18}$ cm$^{-2}$ from \citet{Wood+2005_ApJS}.
}
    \tablenotetext{\sharp}{
 $T_{\rm{ave}}$, $EM_{\rm{tot}}$, and total flare energy $E_{\rm{X}}$(0.2 -- 12 keV) of entire emissions of Flares 11 and 15.
}
   \label{table:flare11_Xray_fit}
 \end{deluxetable*}
\end{longrotatetable}

\clearpage
\subsection{Flares 35\&36: Two flares preceding with the potential X-ray dimming  event}\label{subsec:ana_flare_35_and_36}

\begin{figure}[ht!]
   \begin{center}
      \gridline{
\fig{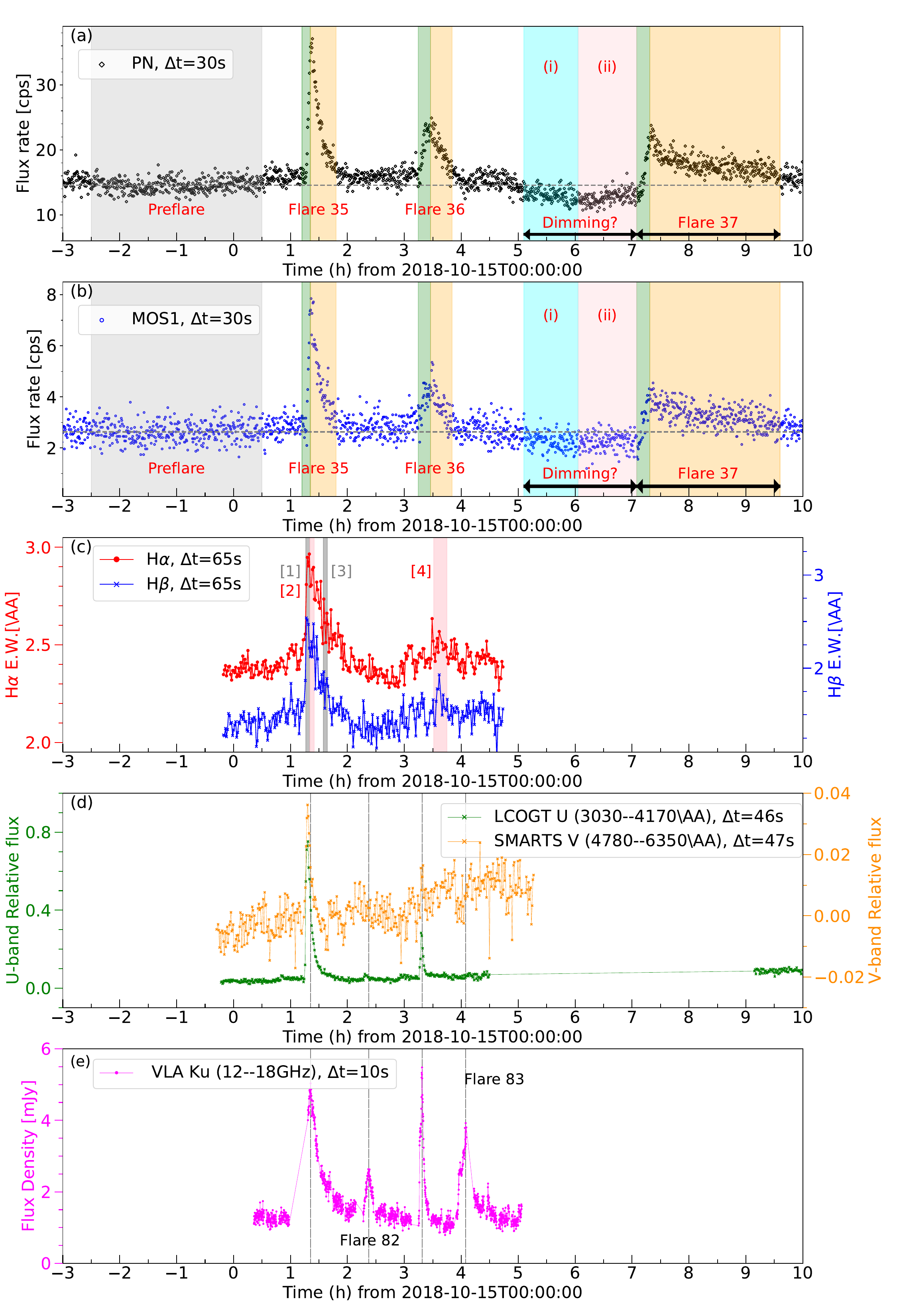}{0.75\textwidth}{\vspace{0mm}}
    }
     \vspace{-5mm}
     \caption{
Multi-wavelength lightcurves around Flares 35--37 (enlarged lightcurve of Figure \ref{fig:X-ray_Ha_obsid_0822740501_lc}).
(a) XMM EPIC-pn X-ray (0.2 -- 12 keV) lightcurve.
The rise and decay phases of Flares 35--37 are shown with green and yellow colored regions (see Figure \ref{fig:X-ray_Ha_obsid_0822740501_lc}). 
The preflare phase and the potential dimming phase (separated into (i) and (ii), cf. \citealt{Veronig+2021}) are also shown with gray, cyan, and pink colored regions. 
(b) XMM EPIC-MOS1 X-ray (0.2 -- 12 keV) lightcurve.
(c) H$\alpha$ and H$\beta$ equivalent width (E.W.) values. The gray and pink 
colored regions correspond to Time [1]--[4], whose spectra are shown 
in Figure \ref{fig:spec_HaHb_flare35and36}.
(d)\&(e) LCOGT U-band, SMARTS V-band, and VLA Ku-band lightcurves. The vertical gray dashed lines show peak times of VLA Ku-band radio flares identified in T25.
}
   \label{fig:Flare35and36_TEM_multi_lc1}
   \end{center}
 \end{figure}

\citet{Veronig+2021} analyzed the archival X-ray lightcurve data, and reported the dimming detections associated with flares on G, K, M dwarfs,
which are potential candidates of stellar CMEs. 
Among the reported 21 events, they used the same AU Mic data as this paper to identify the three events (Extended Data Figure 6 (Figure E6) and Supplementary Table 2 of \citealt{Veronig+2021}). The left panel of Figure E6 of \citet{Veronig+2021} shows that their reported first dimming of AU Mic exists between Flares 8 and 11 in Figure \ref{fig:X-ray_Ha_obsid_0822740301_lc} of this paper, 
while their second one (in the middle panel of Figure E6 of \citealt{Veronig+2021}) exists after Flare 27 and this is overlapped with Flares 28--32 in Figure \ref{fig:X-ray_Ha_obsid_0822740401_lc} of this paper.
We note that the X-ray dimming amplitudes of these two (in the left and middle panels of of Figure E6 of \citealt{Veronig+2021}) could be comparable to and could be hard to distinguish \color{black} from \color{black} the other quiescent X-ray level modulations nearby.
The right panel of Figure E6 of \citet{Veronig+2021} shows that their reported third dimming of AU Mic exist after Flares 35 and 36 in Figure \ref{fig:X-ray_Ha_obsid_0822740501_lc} of this paper. 
Figure \ref{fig:Flare35and36_TEM_multi_lc1} shows the enlarged multi-wavelength lightcurves around these two flares and the potential dimming event.
As seen in this figure, both of Flare 35 and 36 have clear flare emissions 
in X-ray, chromospheric lines (H$\alpha$, H$\beta$), optical continuum ($U$-band for the both flares, $V$-band for Flare 35), and VLA radio Ku-band observations.
The VLA Ku-band data from T25 reported two additional radio flares (Flares 82 and 83) in Figure \ref{fig:Flare35and36_TEM_multi_lc1}(e) between Flares 35, 36, and the potential dimming event. 
Interestingly, there are no flare emissions clearly identified in the wavelengths other than gyrosynchrotron radio emissions during Flares 82 and 83 (Figure \ref{fig:Flare35and36_TEM_multi_lc1}, see also T25 for more details).

The Rise and Decay phase X-ray spectra of Flare 35 and 36 from the PN and MOS1 data (cf. green and yellow colored regions in Figure \ref{fig:Flare35and36_TEM_multi_lc1}(a)\&(b)) and the flare model fitting results (analyzed in Section \ref{subsec:X-ray_specana_time-average}) are shown in Figure \ref{fig:RiseDecayFit_fig1_Flare35and36} \color{black} in Appendix \ref{appen_sec:additional_figures}.\color{black}
The X-ray flare energies are estimated from these fitting results as already listed in Table \ref{table:energy_remarkable_flares} in Section \ref{subsec:X-ray_specana_time-average}. The H$\alpha$ and H$\beta$ flare energies are estimated by using the methodology \color{black} for \color{black} Flare 23 in Section \ref{subsec:ana_flare_23}. The multi-wavelength flare energy values of Flares 35 and 36 are summarized in Table \ref{table:energy_remarkable_flares}. 

\begin{figure}[ht!]
   \begin{center}
      \gridline{
\fig{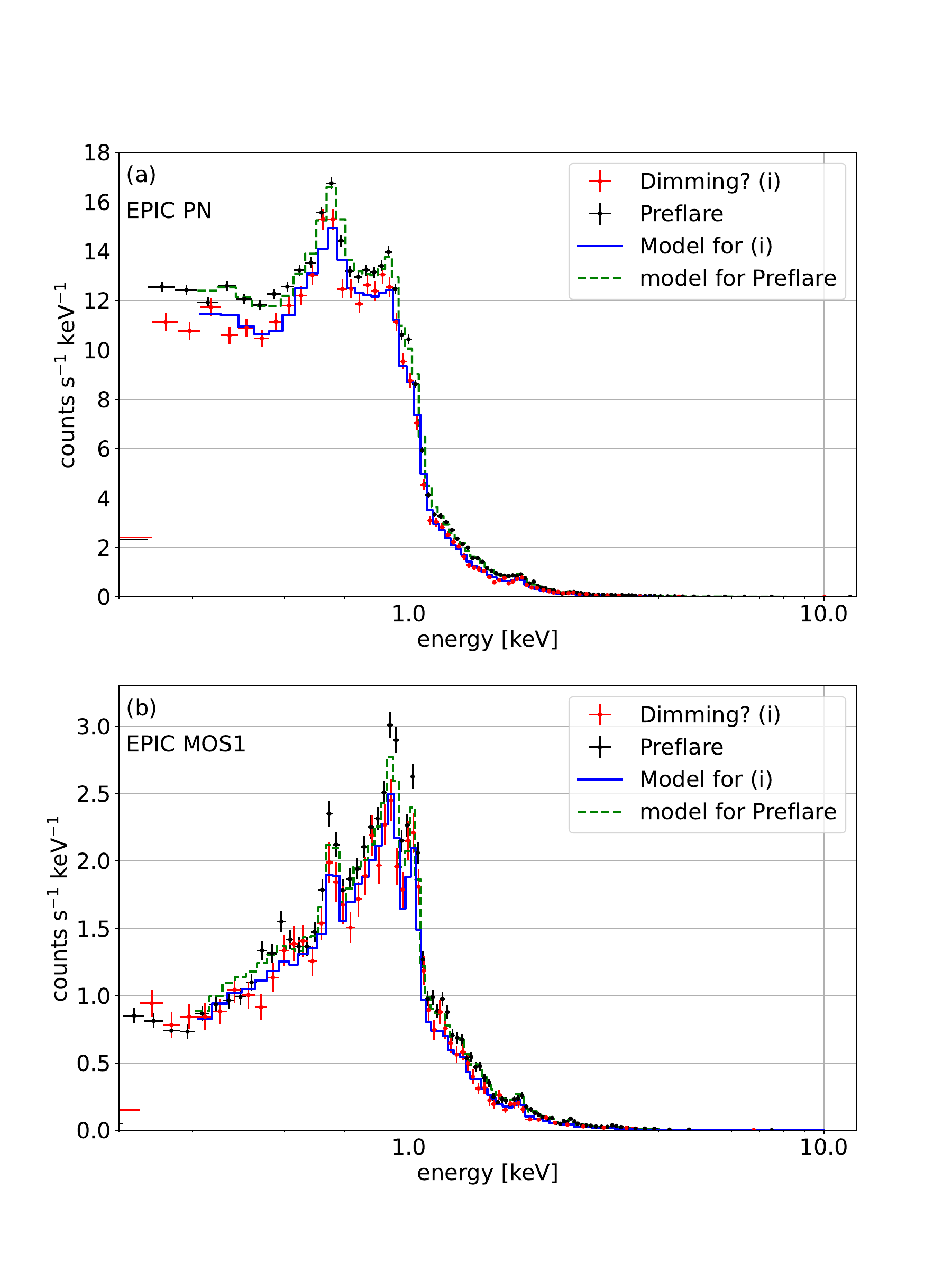}{0.45\textwidth}{\vspace{0mm}}
\fig{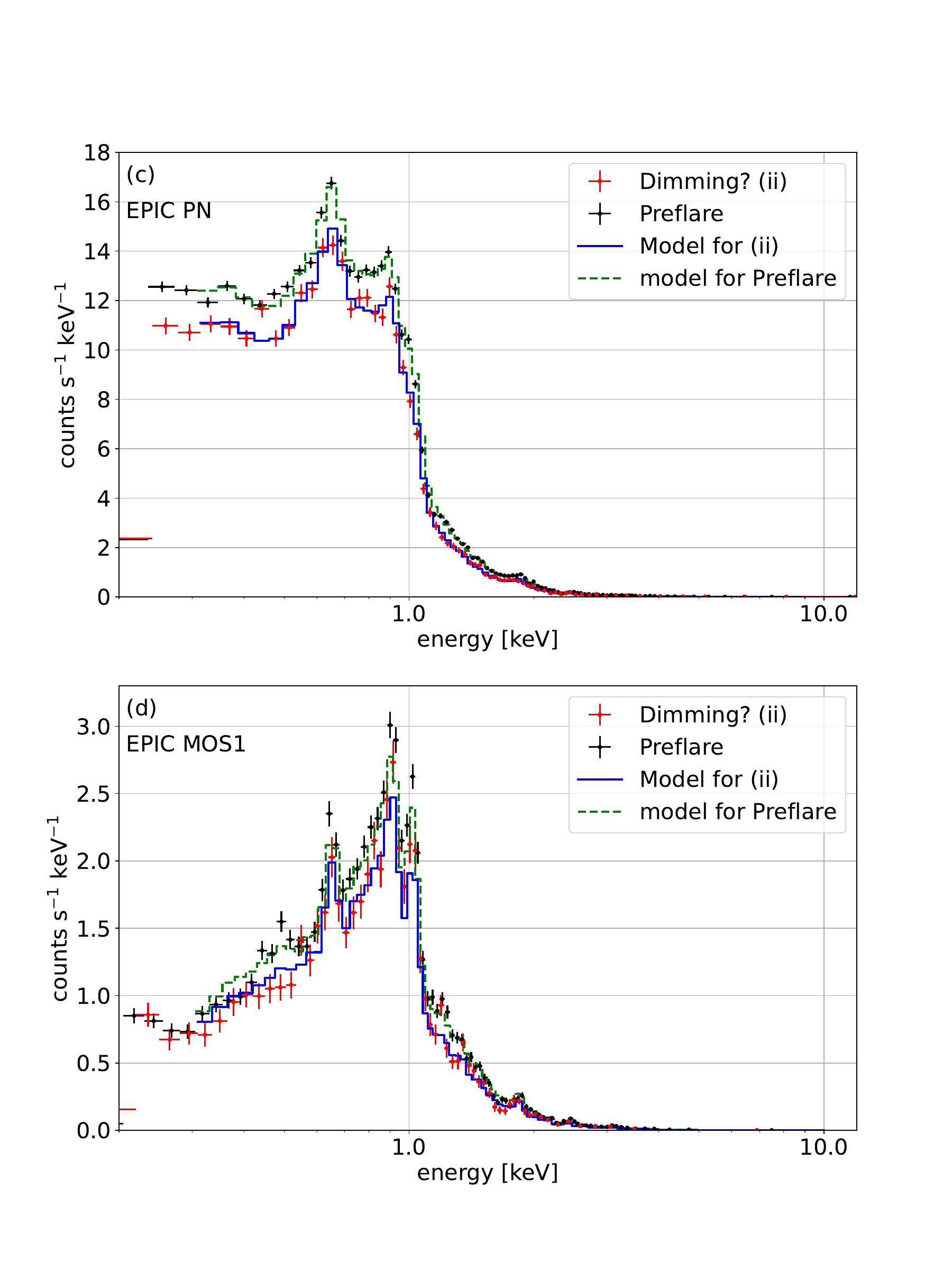}{0.45\textwidth}{\vspace{0mm}}
    }
     \vspace{-5mm}
      \gridline{
\fig{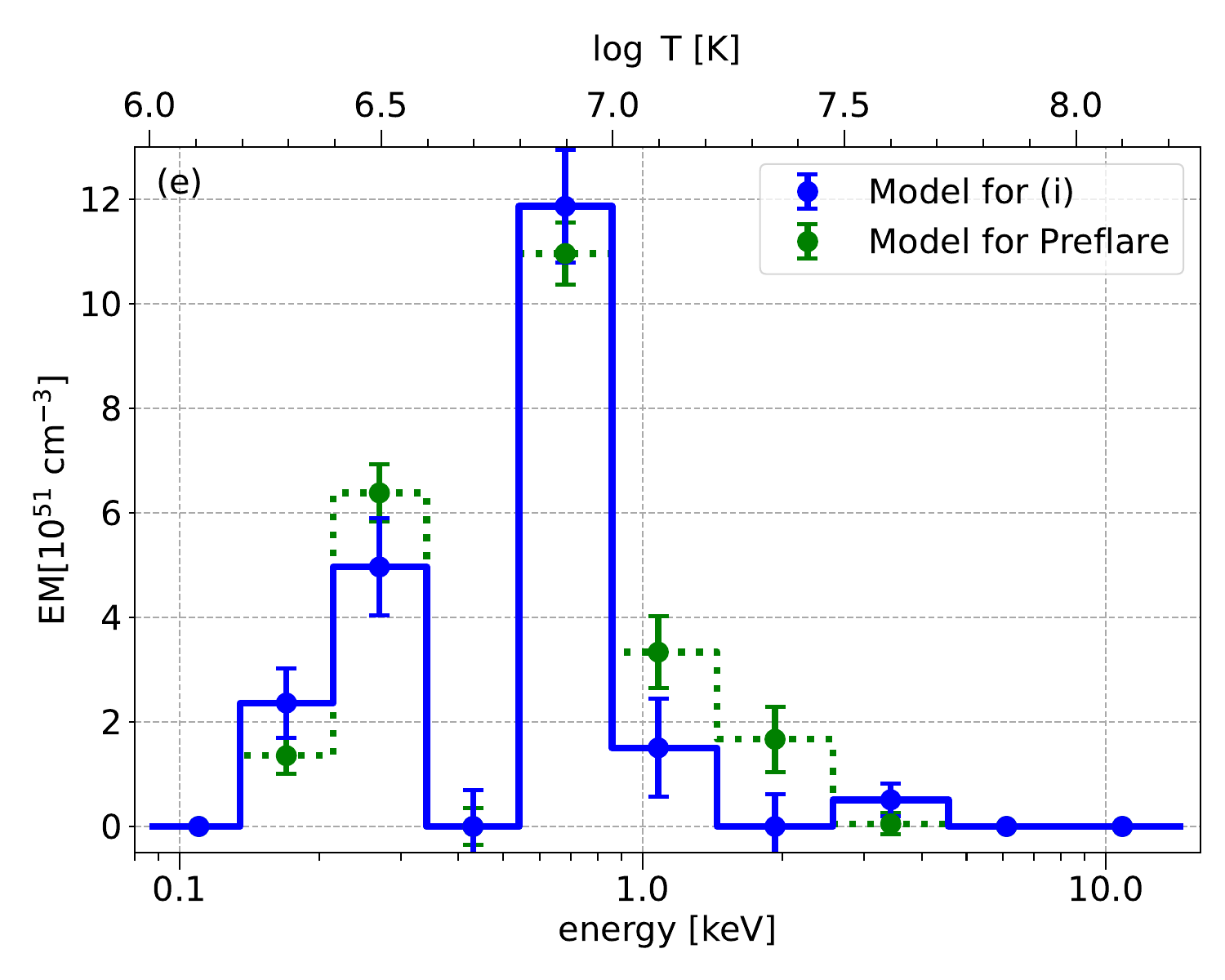}{0.4\textwidth}{\vspace{0mm}}
\fig{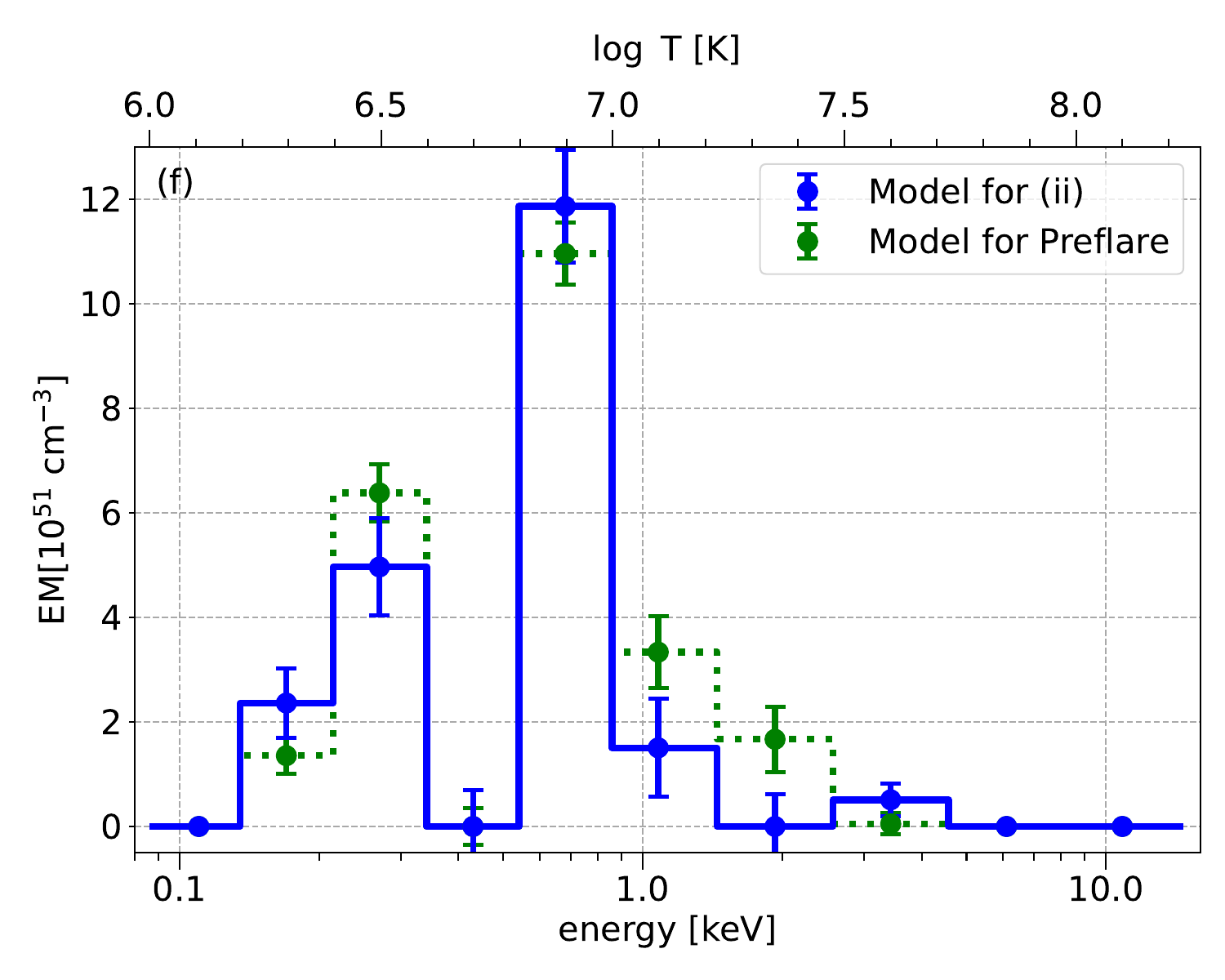}{0.4\textwidth}{\vspace{0mm}}
    }
     \vspace{-5mm}
     \caption{
(a) The red and black points are the PN spectra of the potential dimming phase (i) and preflare phase around Flares 35 and 36, shown in Figure \ref{fig:Flare35and36_TEM_multi_lc1}. 
The blue solid and green dashed lines are model fitting results to these two, respectively. 
Different from the other figures of PN spectra in this paper, the y-axis of this figure is plotted in the linear scale scale, following Figure 2 of \citet{Veronig+2021}.
(b) Same as (a), but for MOS1 data.
(c) \& (d) Same as (a) \& (b), but for the dimming phase (ii) data.
(e)Emission Measure (EM) distribution from the model fitting to the potential dimming phase (i) and Preflare phase. (f) Same as (e) but for the dimming phase (ii).   Statistical
90\% confidence region errors are shown in (e)\&(f).
}
   \label{fig:Oct15_dimming_spectra}
   \end{center}
 \end{figure}

We then conduct additional X-ray fitting analysis to estimate how the X-ray flux values are different between the preflare and potential dimming phases (the gray colored preflare region, (i), and (ii) in Figure \ref{fig:Flare35and36_TEM_multi_lc1}(a)\&(b)). First, we fix the 10-temperature bins, abundance, and $N_{\rm{H}}$ values as the quiescent values estimated in Section \ref{subsec:X-ray_specana_quiescent} (see the values with Obs-ID:0822740501 in Table \ref{table:X-ray_quie_fit_results} and Figure \ref{fig:specfit_QuieALL1_0822740501}). Then we conduct the standard fitting (as done in the above sections assuming \texttt{vapec} and \texttt{tbabs}) for the preflare phase and the two dimming spectra ((i) and (ii)) with the emission measure values ($EM_{1}$--$EM_{10}$) as free parameters. 
The spectra of preflare and potential dimming phases with the  fitting results are shown in Figure \ref{fig:Oct15_dimming_spectra}. 
Using these fitting results, the X-ray flux in the 0.2 -- 12 keV range in preflare and potential dimming phases ((i) and (ii)) are estimated as 
$F_{\rm{X}}^{\rm{preflare}}=2.16^{+0.05}_{-0.01}\times 10^{-11}$ erg cm$^{-2}$ s$^{-1}$,
$F_{\rm{X}}^{\rm{Phase (i)}}=1.93^{+0.24}_{-0.01}\times 10^{-11}$ erg cm$^{-2}$ s$^{-1}$, and
$F_{\rm{X}}^{\rm{Phase (ii)}}=1.89^{+0.28}_{-0.02}\times 10^{-11}$ erg cm$^{-2}$ s$^{-1}$.
The $F_{\rm{X}}^{\rm{Phase (i)}}$ and $F_{\rm{X}}^{\rm{Phase (ii)}}$ 
values roughly correspond to 89.4$^{+11.6}_{-0.02}$\% and 87.5$^{+13.4}_{-0.03}$\% of 
the $F_{\rm{X}}^{\rm{preflare}}$ value, respectively.

\begin{figure}[ht!]
   \begin{center}
      \gridline{
\fig{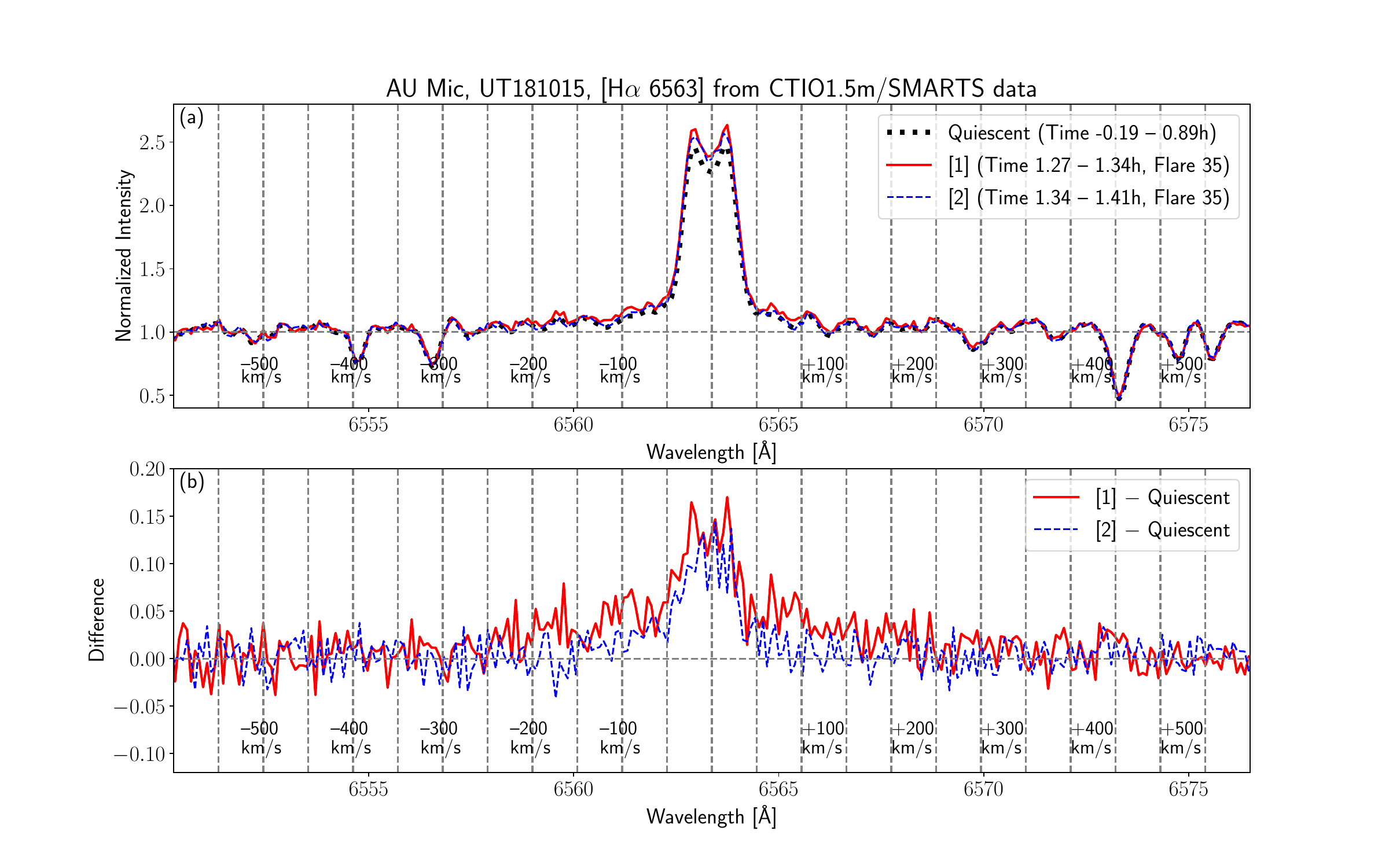}{0.5\textwidth}{\vspace{0mm}}
\fig{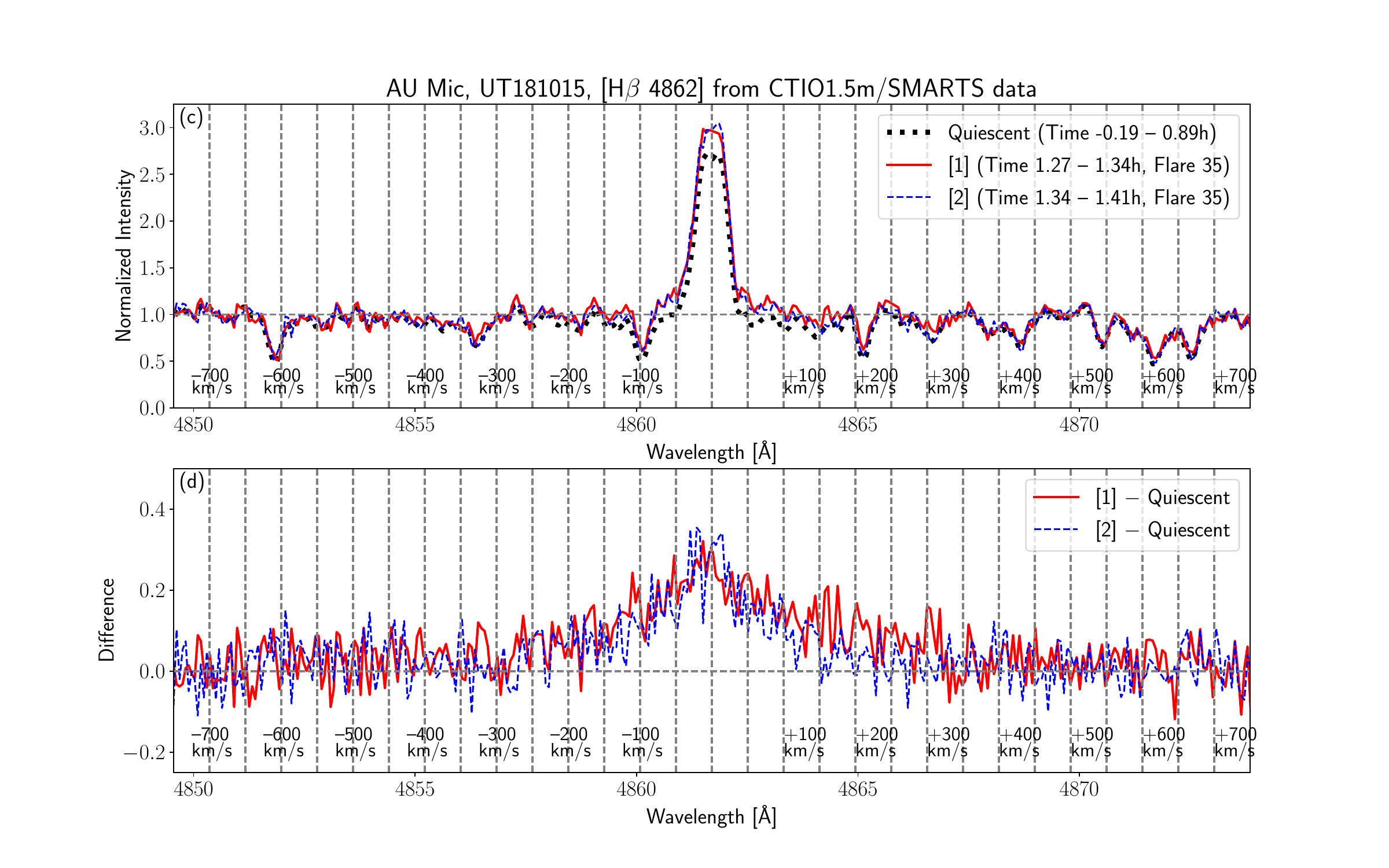}{0.5\textwidth}{\vspace{0mm}}
  }
     \vspace{-5mm}
      \gridline{
\fig{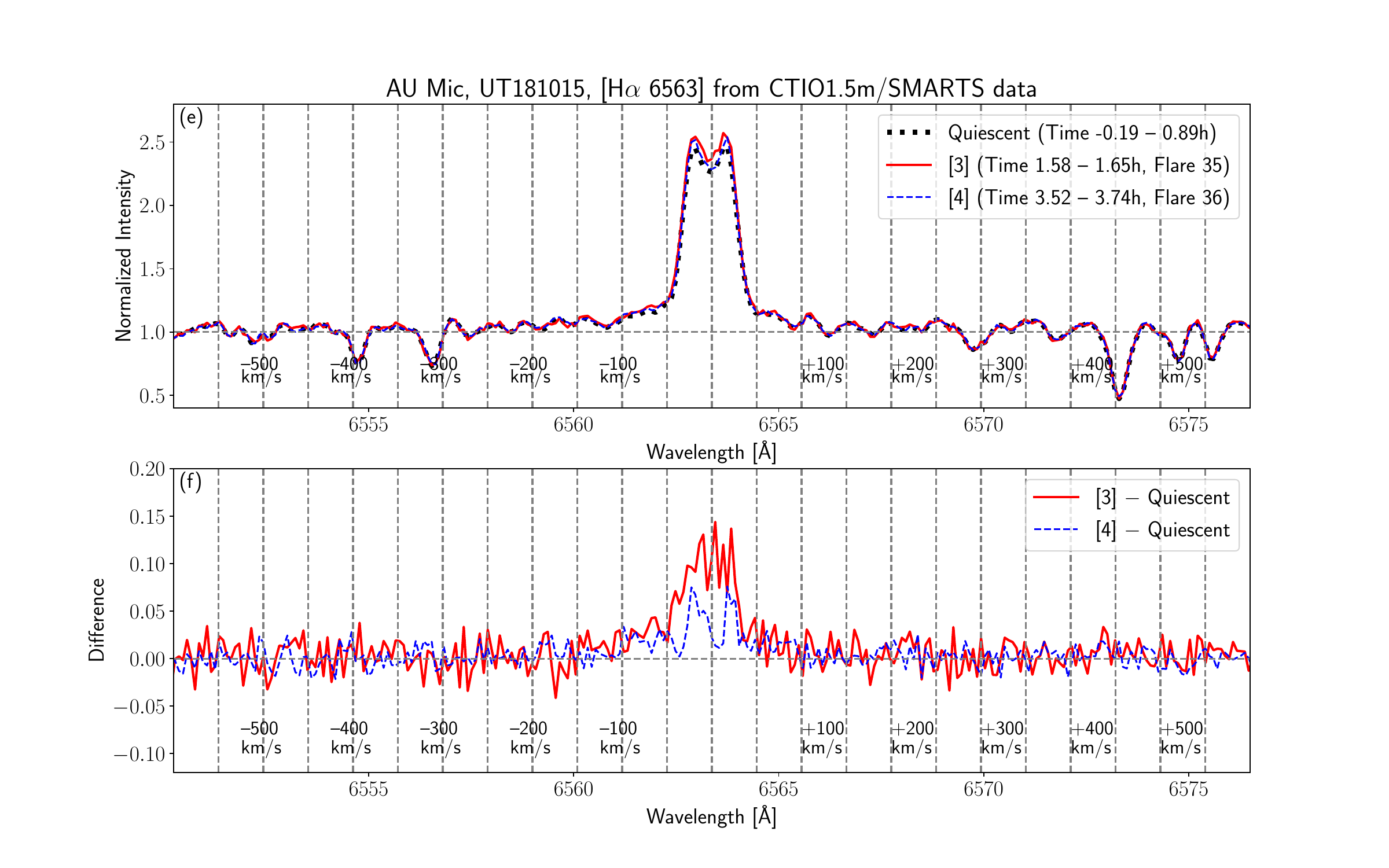}{0.5\textwidth}
{\vspace{0mm}}
\fig{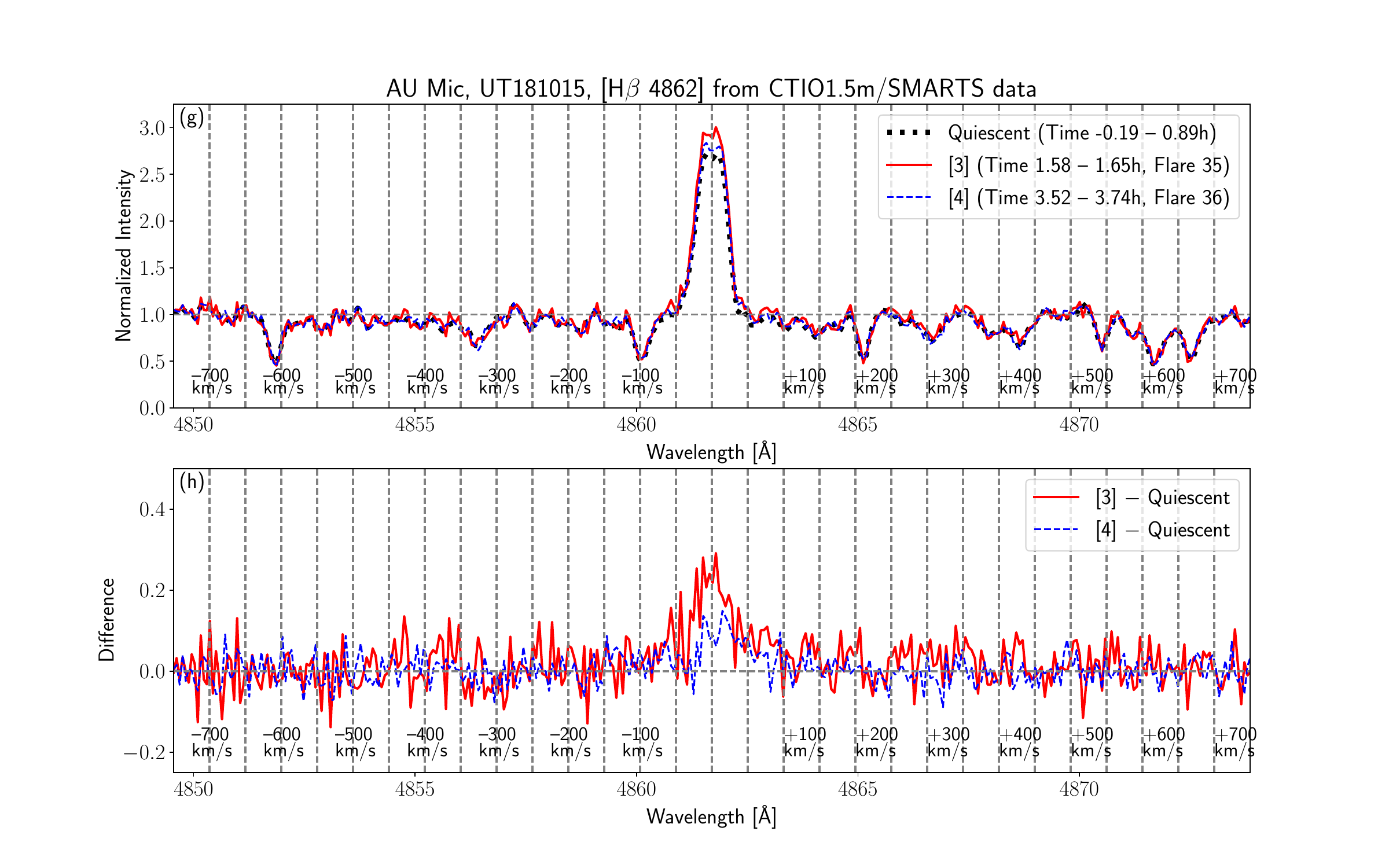}{0.5\textwidth}
{\vspace{0mm}}
}
     \vspace{-5mm}
     \caption{
(a) Line profiles of the H$\alpha$ emission line during Flare 35 on 2018 
October 15 from the CHIRON spectra.
The horizontal and vertical axes represent the wavelength and flux normalized by the continuum. The gray vertical dashed lines with velocity values represent the Doppler velocities from the H$\alpha$ line center.
The red solid and blue dashed lines indicate the line profiles at the time [1] and \color{black} time \color{black} [2], respectively, which are indicated in 
Figure \ref{fig:Flare35and36_TEM_multi_lc1} (c) (lightcurve)
and are during Flare 35. 
The black dotted line indicates the line profiles in quiescent phase, which are the average profile during -0.19-- +0.89 hr on this date (see Figure \ref{fig:Flare35and36_TEM_multi_lc1} (c)).
(e) Same as panel (a), but the line profiles at time [3] and \color{black} time \color{black} [4] during Flares 35 and 36, respectively.
(c) \& (g) Same as (a) \& (e), but for the H$\beta$ emission line.
(b), (d), (f), \& (h) Same as panels (a), (c), (e),\& (g), respectively,
but the line profile differences from the quiescent phase.
}
   \label{fig:spec_HaHb_flare35and36}
   \end{center}
 \end{figure}

The line profiles of the H$\alpha$ and H$\beta$ lines during Flares 35 and 36 
(the time [1] -- \color{black} time \color{black} [4] identified in Figure \ref{fig:Flare35and36_TEM_multi_lc1}(c))
are shown in Figure \ref{fig:spec_HaHb_flare35and36}. There are no clear 
significant line (blue or red) wing asymmetries similar to the ones in previous studies of M-dwarf line asymmetries \color{black}(e.g.,\citealt{Houdebine+1990,Fuhrmeister+2018,Vida+2019,Notsu+2024_ApJ,Kajikiya+2025_ApJ_PaperI}) \color{black}. This suggests that signatures of the filament/prominence eruptions \footnote{\color{black} It is noted that the blue or red wing asymmetries of M-dwarfs have been often interpreted as candidates of filament/prominence eruptions (and potentially evolved CMEs), as we have also described in Section \ref{sec:intro}. For more details, especially 
see the discussions of recent M-dwarf papers on asymmetry interpretations (e.g., \citealt{Fuhrmeister+2018,Vida+2019,Notsu+2024_ApJ,Kajikiya+2025_ApJ_PaperI}). 
} are at least not clearly confirmed with this potential X-ray dimming event, which has been discussed 
as a signature of a stellar CME candidate (\citealt{Veronig+2021}).

\clearpage

\subsection{Flare 65: Another energetic flare showing clear broadenings of chromospheric lines}\label{subsec:ana_flare_65}

\begin{figure}[ht!]
   \begin{center}
      \gridline{
\fig{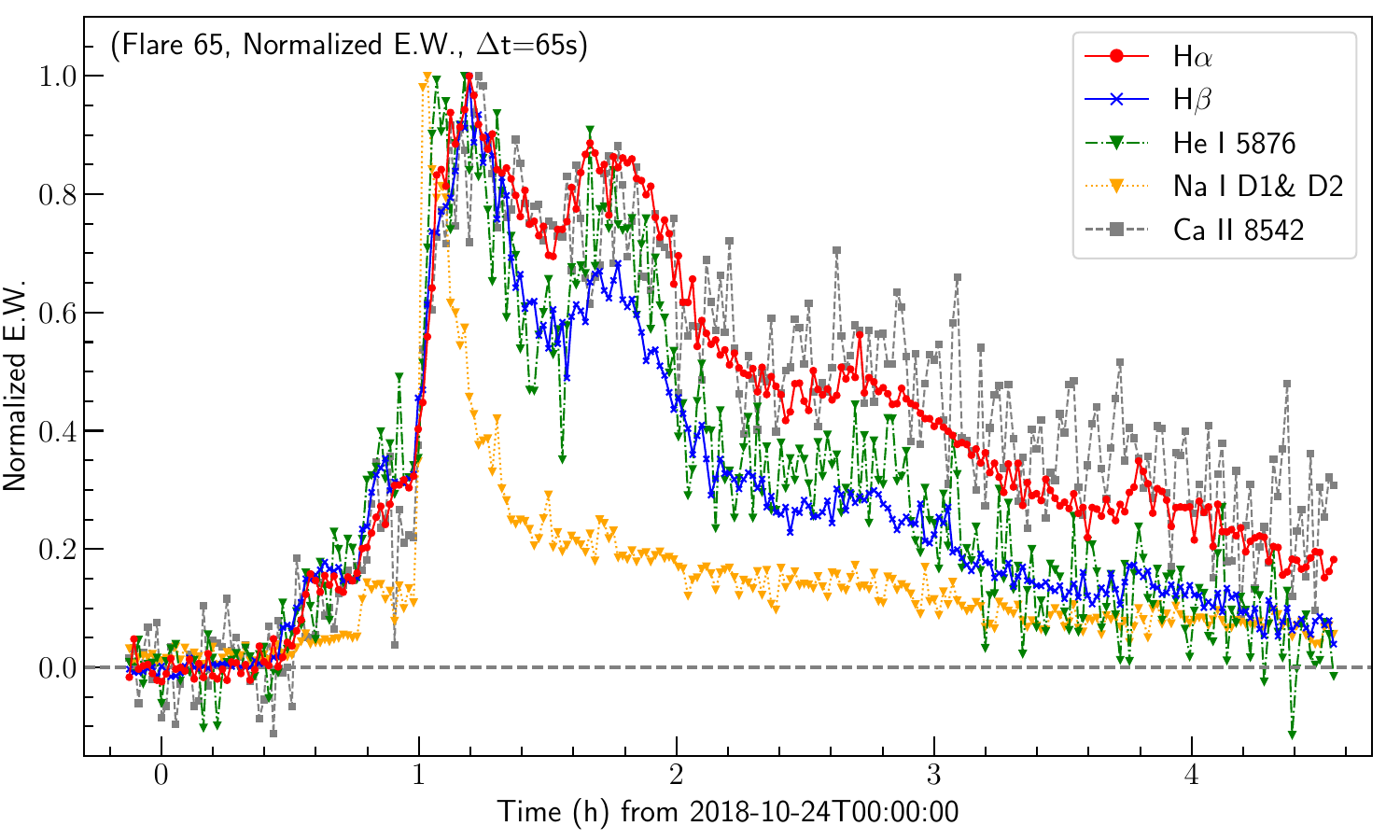}{0.65\textwidth}{\vspace{0mm}}
    }
     \vspace{-5mm}
     \caption{
The Equivalenth width (E.W.) lightcurves of H$\alpha$, H$\beta$, He I D3 5876\AA, Na I D1\&D2, and Ca II 8542\AA~lines from the CHIRON data, as normalized with the peak and preflare values of Flare 65 on 2018 October 24. 
The peak H$\alpha$ and H$\beta$ E.W. values of Flare 65 are 4.96\AA~and 7.33\AA, while the preflare values are 2.44\AA~and 1.43\AA, respectively (cf. Figure \ref{fig:maps_flare65}).
The peak He I D3 5876\AA~E.W. value of Flare 65 is 0.52\AA, while the preflare value is 0.03\AA~(cf. Figure \ref{fig:maps_flare65}). 
As for the Na I D1\&D2, and Ca II 8542\AA~lines, 
the E.W. values are measured as the excess equivalent widths as done for Flare 23 in Section \ref{subsec:ana_flare_23}. The peak Na I D1\&D2, and Ca II 8542\AA~excess E.W. values are 1.59\AA~and 0.41\AA, respectively (cf. Figure \ref{fig:maps_flare65}).
}
   \label{fig:lc1_flare65}
   \end{center}
 \end{figure}

Figure 65, which occurred on 2018 October 24, shows flare emissions in chromospheric lines for $\sim$4 hours (Figure \ref{fig:lc1_flare65}), and the flare peak H$\alpha$ E.W. value is almost comparable to that of Flare 23 (Figure \ref{fig:allEW_AUMic_XMM_opt}).
Unfortunately, there is no other multi-wavelength data coverage other than SMARTS chromospheric line observations (Figures \ref{fig:allEW_AUMic_XMM_opt}). 
As also seen for Flare 23 in Section \ref{subsec:ana_flare_23}, 
there are clear differences of the flare durations among the chromospheric lines (Figure \ref{fig:lc1_flare65}). 
The H$\alpha$ flare duration is longer than that of the H$\beta$ line, but as a notable difference from Flare 23, it is comparable to the Ca II 8542\AA~line. 
Another notable difference from Flare 23 is the He I 5876\AA~duration, which \color{black}is \color{black}  
much longer than that of the Na D1\&D2 lines and is roughly comparable to that of the H$\beta$ line.
The H$\alpha$ and H$\beta$ energy values of Flare 65 are estimated 
by using the methodology as Flare 23 in Section \ref{subsec:ana_flare_23} (Note that the $V$-band continuum flux changes cannot be incorporated since there are no $V$-band data): 
$E_{\rm{H}\alpha}$=3.56$\times$10$^{32}$ erg 
and 
$E_{\rm{H}\beta}$=2.53$\times$10$^{32}$ erg 
(Table \ref{table:energy_remarkable_flares}). As a result, the H$\alpha$ flare energy of Flare 65 is roughly 2.3--2.5 times larger than that of Flare 23 (Table \ref{table:energy_remarkable_flares}).

\begin{figure}[ht!]
   \begin{center}
      \gridline{
\fig{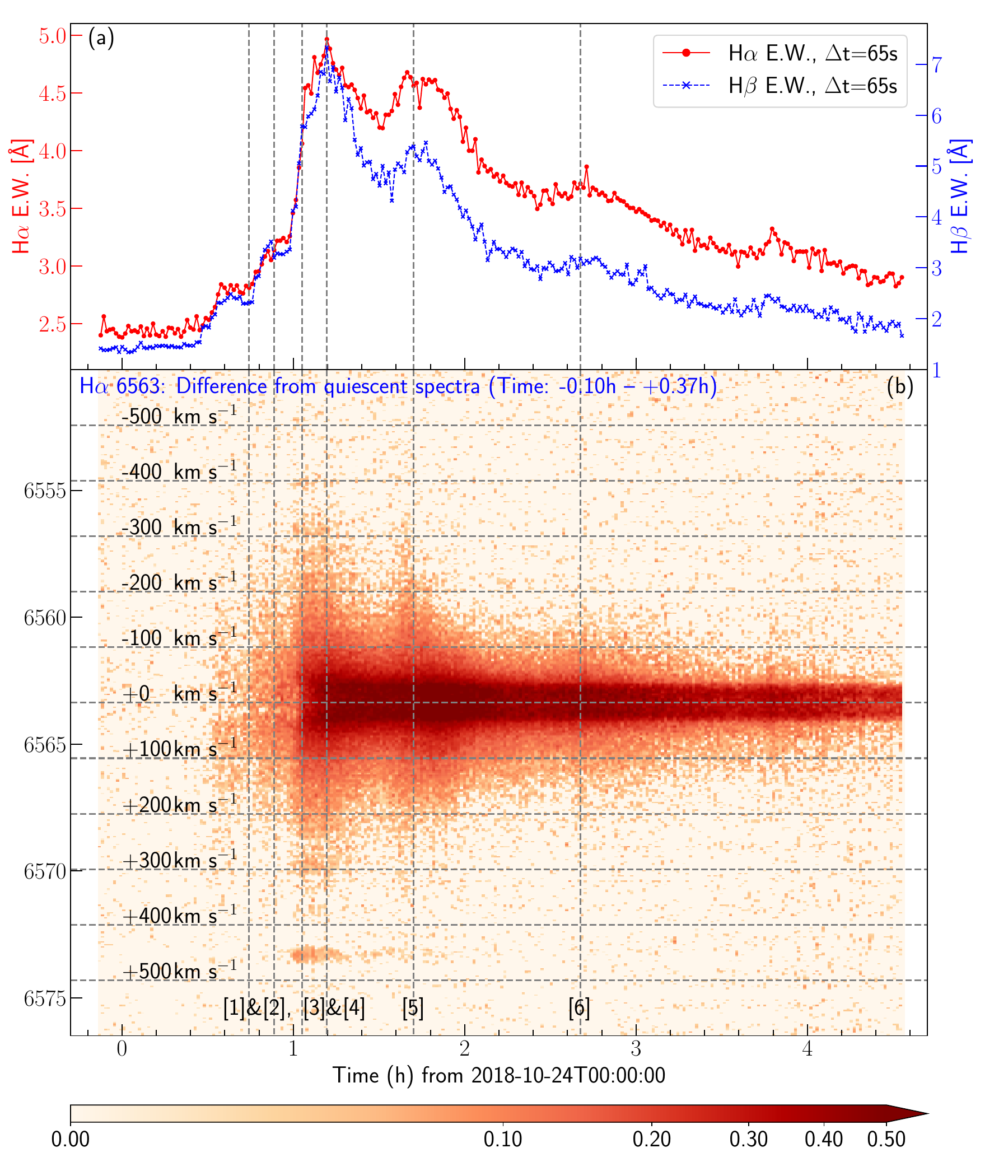}{0.47\textwidth}{\vspace{0mm}}
\fig{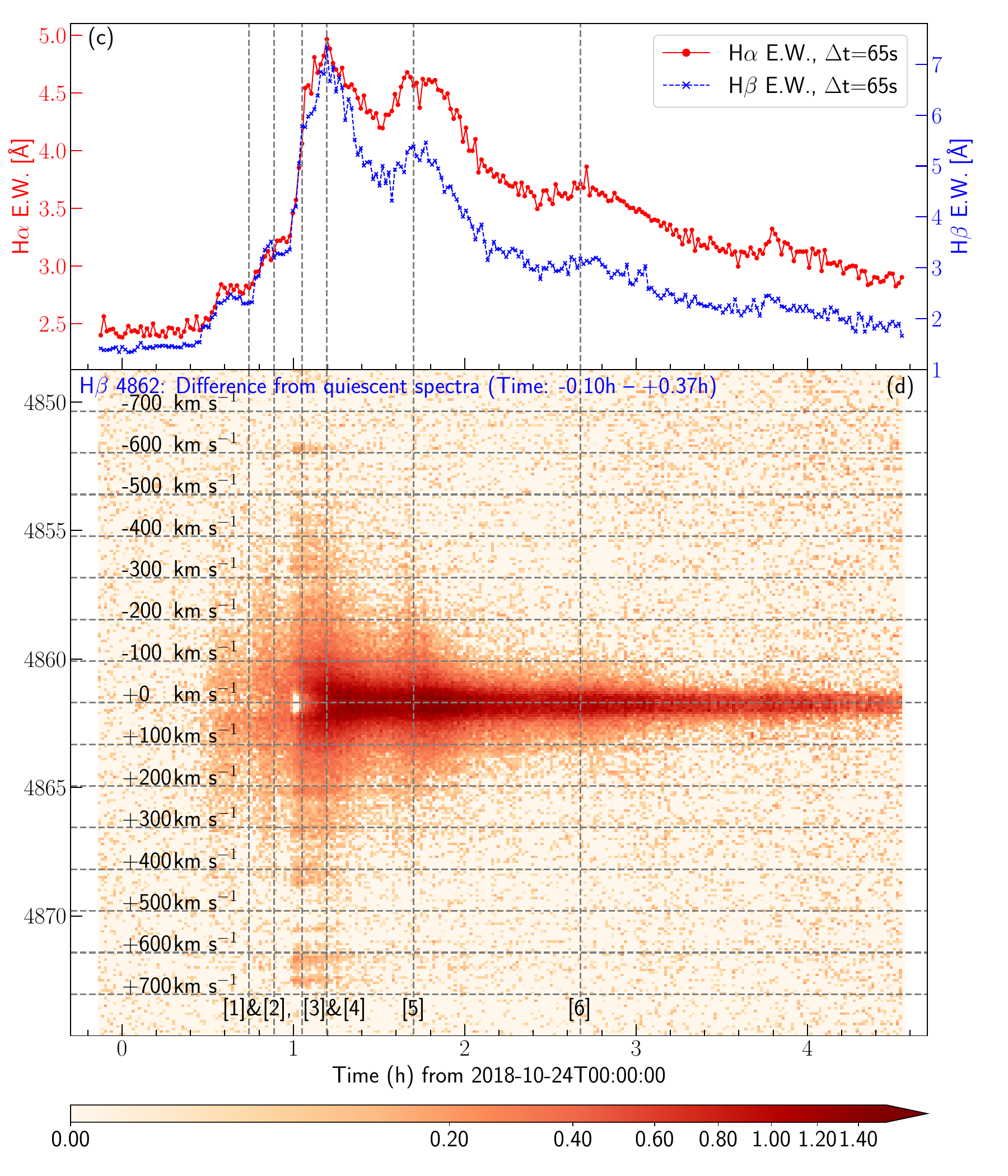}{0.47\textwidth}{\vspace{0mm}}
    }
     \vspace{-5mm}
      \gridline{
\fig{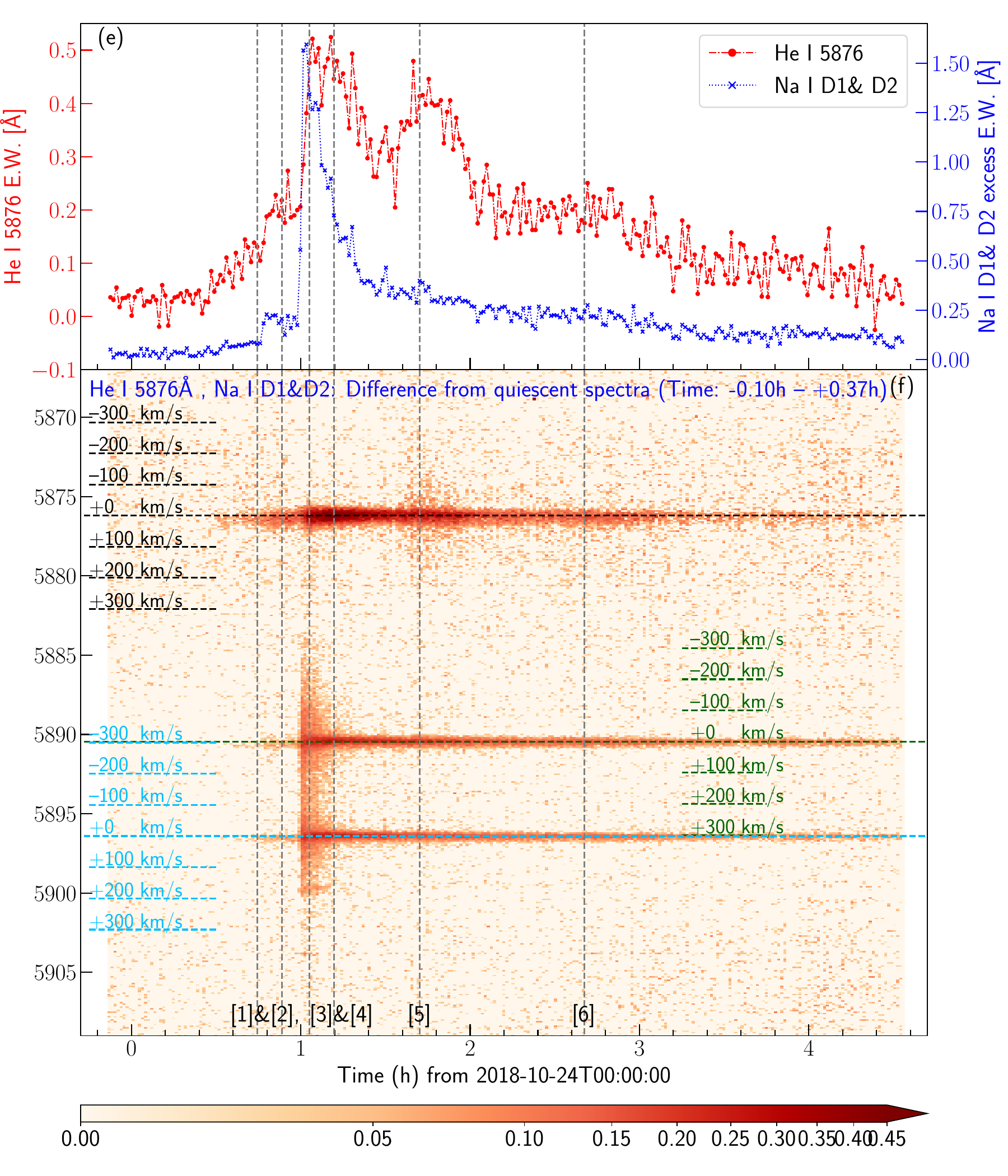}{0.47\textwidth}{\vspace{0mm}}
\fig{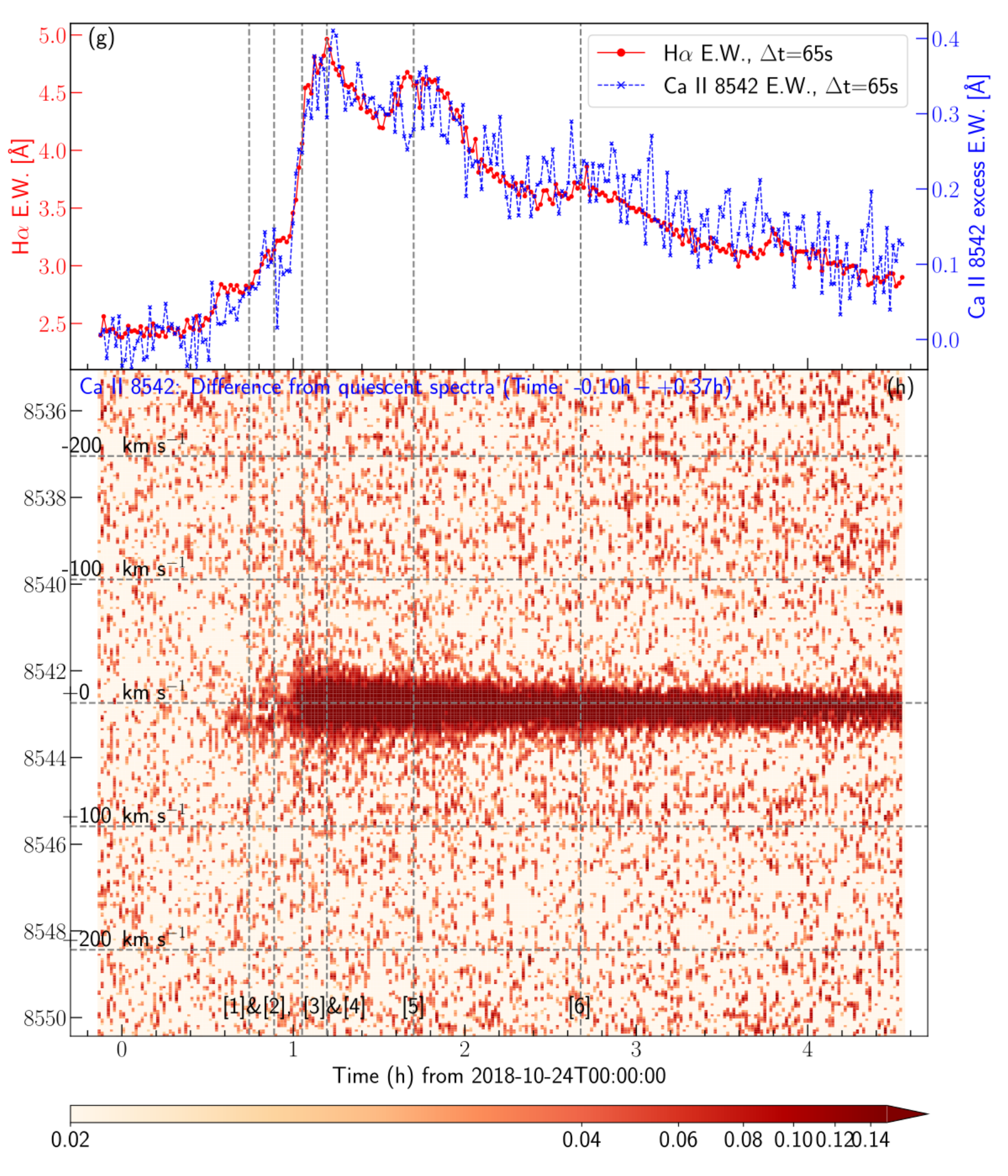}{0.47\textwidth}{\vspace{0mm}}
    }
     \vspace{-5mm}
     \caption{
(a)\&(c) The H$\alpha$ and H$\beta$ E.W. lightcurves of Flare 65 on 2018 October 24 are plotted with red circles and blue marks. 
(b) Time evolution of the H$\alpha$ profile from the CHIRON spectra. 
The vertical axis represents the wavelength, while the gray horizontal dashed lines with velocity values represent the Doppler velocities from the
H$\alpha$ line center. The gray vertical dashed lines indicate the time [1] --
\color{black} time \color{black} [6],
which are shown in (a) (lightcurve) and Figure \ref{fig:spec_HaHb_flare65}(spectra).
The color map represents the line profile changes from the quiescent profile 
(see Figure \ref{fig:spec_HaHb_flare65}(b), (f), and (j)).
(c) \& (d) Same as panels (a) \& (b), but for the H$\beta$ line.
(e) \& (f) Same as panels (a) \& (b), but for He I D3 5876\AA~line and Na I D1\&D2 lines.
(g) \& (h) Same as panels (a) \& (b), but for the Ca II 8542 \AA~line.
The excess equivalent width values of Na I D1\&D2 and Ca II 8542 lines are plotted in (e) and (g).
}
   \label{fig:maps_flare65}
   \end{center}
 \end{figure}

The line profiles of the H$\alpha$, H$\beta$, He I D3 5876\AA, 
Na I D1\&D2, and Ca II 8542\AA~lines during Flares 65
are shown in the color maps in Figure \ref{fig:maps_flare65} 
and the spectra in Figures \ref{fig:spec_HaHb_flare65} and \ref{fig:spec_HeNaCa8542_flare65} 
\color{black} in Appendix \ref{appen_sec:additional_figures}. \color{black}  
The E.W. lightcurves of these lines are also shown in Figure \ref{fig:maps_flare65}. 
The almost symmetric line broadenings with $\sim \pm$300--350 km s$^{-1}$ are clearly seen in the H$\alpha$ line at around the time [3], \color{black} time \color{black} [4], and \color{black} time \color{black} [5] (= around the two emission peaks) of Flare 65 (Figures \ref{fig:maps_flare65} and \ref{fig:spec_HaHb_flare65}). 
Comparing the line profile evolutions in Figures \ref{fig:maps_flare65} and  \ref{fig:spec_HaHb_flare65}, 
the H$\beta$ line also shows the time evolution similar to the H$\alpha$ evolution  (the broadenings around the flare emission peaks) during Flare 65. The H$\beta$ line broadenings are at around the flare peaks (= around the time [3], \color{black} time \color{black}  [4], and \color{black} time \color{black}  [5]) of Flare 65 are up to $\sim$400--500 km s$^{-1}$ and these are bigger than those of the H$\alpha$ line, but the decay phase line broadening of the H$\beta$ line ($\lesssim$100 km s $^{-1}$) is smaller than that of the H$\alpha$ line ($\sim$100--150 km s $^{-1}$) (see the spectra at the time [6]). 
These overall properties of the H$\alpha$ \& H$\beta$ broadenings around the peaks 
have similarities with Flare 23 (Figures \ref{fig:maps_flare23} -- \ref{fig:flare23_HbSpec}) although the exact broadening velocities are different.

The Na D1\&D2 lines also show smaller symmetric broadenings with $\sim \pm$200--250 km s$^{-1}$ at around the time [3] and \color{black} time \color{black} [4] (= around the first flare peak) of Flare 65, 
but this does not show clear broadenings at around the time [5] (= around the second flare peak) (Figures \ref{fig:maps_flare65} and \ref{fig:spec_HeNaCa8542_flare65}). 
In contrast, the He I D3 5876\AA~line does not show clear broadenings during Flare 65 (Figures \ref{fig:maps_flare65} and \ref{fig:spec_HeNaCa8542_flare65}), which is different from Flare 23 (Figures \ref{fig:maps_flare23} and \ref{fig:flare23_HeNaSpec}). 
There are no notable line broadening changes in the Ca II 8542 emissions during Flare 65
(Figures \ref{fig:maps_flare65} and \ref{fig:spec_HeNaCa8542_flare65}), which is the same as Flare 23 (Figures \ref{fig:maps_flare23} and \ref{fig:flare23_Ca8542Spec}).
This Flare 65 is briefly described in this subsection as a comparison with Flare 23, although they do not have other multi-wavelength observations.

\clearpage

\section{Discussions} \label{sec:discussions}
 
\subsection{Rotational modulations of the quiescent emissions}\label{subsec:discuss_quiescent}

   \begin{figure}[ht!]
   \begin{center}
      \gridline{
\fig{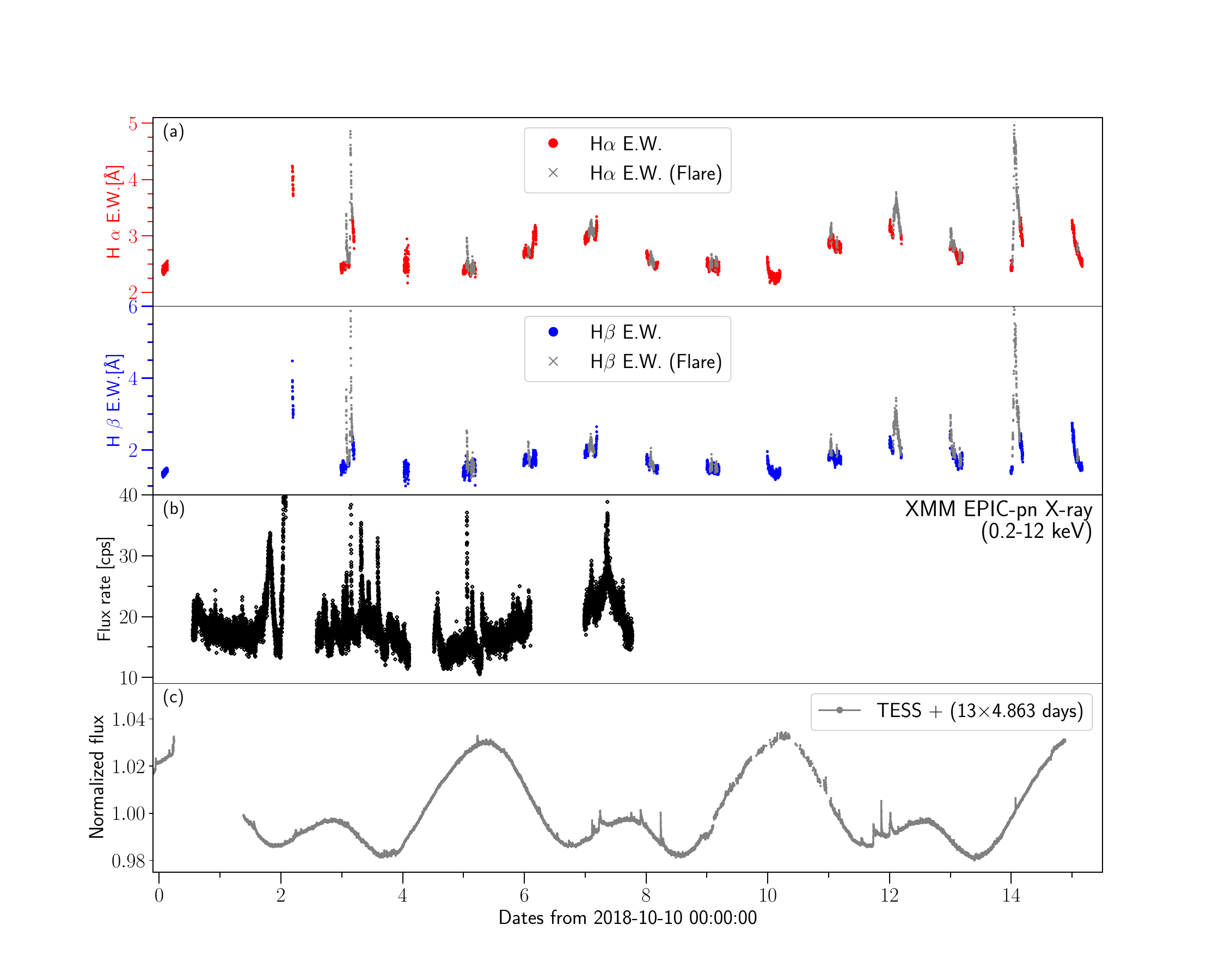}{1.0\textwidth}{\vspace{0mm}}
    }
     \vspace{-5mm}
     \caption{
\color{black} (a) \& (b) H$\alpha$ and H$\beta$ equivalent width (E.W.) lightcurves as plotted in Figure \ref{fig:allEW_AUMic_XMM_opt}.  
The flare times (based on the start and end times from T23) are shown with gray points to show quiescent phase modulations. \color{black}
\color{black} (b) \color{black} XMM EPIC-pn X-ray lightcurve as plotted in Figure \ref{fig:allEW_AUMic_XMM_opt}, but the enlarged one for the y-axis direction.
\color{black} (c)  \color{black}
TESS lightcurve of AU Mic Cycle 1 taken from Figure 1 (a) of \citealt{Ikuta+2023_ApJ}, 
but shifted for the x-axis direction with 13$\times$4.863 days (= 13 $\times$ AU Mic's rotation period).
}
   \label{fig:TESS_comparison_rotation}
   \end{center}
 \end{figure}

As mentioned in Section \ref{subsec:flare_atlas} (with Figures \ref{fig:allEW_AUMic_XMM_opt} -- \ref{fig:X-ray_Ha_obsid_0822740601_lc}), the X-ray, H$\alpha$,
and H$\beta$ lightcurves show some gradual modulations 
in addition to clear sudden brightenings identified as flares (Figure \ref{fig:TESS_comparison_rotation}(a)\&(b)).
\color{black}
TESS observed AU Mic in its Sector 1 period (2018 July 25 to August 22, cf. \citealt{Ikuta+2023_ApJ}), which was roughly 2 months before our AU Mic campaign (in 2018 October 10 to 25). \color{black}
In Figure \ref{fig:TESS_comparison_rotation}(c), we compare these with
the TESS lightcurve, by shifting with the AUMic's rotation period ($\sim$4.863 days, \citealt{Ikuta+2023_ApJ}). 
\citet{Ikuta+2023_ApJ} compared the TESS lightcurves of AU Mic in its Cycles 1 (Sector 1, July 25 to August 22 in 2018) and 
3 (Sector 27, July 4 to 30 in 2020), and found that the amplitudes and shapes of the lightcurve can change between the two years, as 
starspot sizes and distributions change. Because of this, we note that 
this comparison with the 2-month shifted TESS data 
could be only a rough proxy.
In Figure \ref{fig:TESS_comparison_rotation}, we can see that the quiescent (non-flare) H$\alpha$ and H$\beta$ emission components show quasi-periodic variations, which roughly corresponds to the rotational period shown in TESS data. 
\citet{Ikuta+2023_ApJ} also suggested more than two spot components are necessary to reproduce the TESS optical lightcurve, but we cannot judge whether there are two components in the H$\alpha$ and H$\beta$ lightcurves.
The H$\alpha$ and H$\beta$ E.W. quasi-periodic modulations have peaks at least around Oct 17-18 and Oct 22 in Figure \ref{fig:TESS_comparison_rotation}(a), 
which very roughly corresponds to the rotation phase when the total spot visible area become the minimum (Phase $\sim$0.4--0.5 in Figure 5(c) of \citealt{Ikuta+2023_ApJ}). 
In addition, the quiescent X-ray lightcurve roughly monotonously increases from October 14 to 18 (Figure \ref{fig:TESS_comparison_rotation}(b)), and 
the H$\alpha$ and H$\beta$ lightcurves also show the rough 
increase trend during the same period.
\color{black} This \color{black} might suggest the potential correlations and anti-correlations of the rotational modulations among the chromospheric line emissions (H$\alpha$\&H$\beta$ lines), coronal X-ray emissions (XMM EPIC-pn), and the optical white-light emission (TESS).
However, we cannot conclude this suggestion in detail only with this study, since the H$\alpha$ \& H$\beta$ and TESS observations were not simultaneously conducted, the XMM X-ray data only covered less than two rotation cycles, and the potential contribution of smaller 
flares in quiescent 
chromospheric lines and X-ray emissions cannot be ignored.

This topic itself (correlations/anti-correlations of the rotational modulations among the optical white-light, chromospheric line, and X-ray emissions) 
has been highlighted in recent studies with the Sun-as-a-star data (\citealt{Toriumi+2020}) and stellar data
(\citealt{Maehara+2021,Namekata+2022_ApJL,Schofer+2022,Notsu+2024_ApJ,Wargelin+2024_ApJ,Odert+2025_MNRAS}). 
\citet{Odert+2025_MNRAS} analyzed the simultaneous H$\alpha$ spectroscopy and $g$-band photometry data
of the same M-dwarf flare star AU Mic, taken on 56 nights spreading from 2022 October 31 until 2023 September 21.
They reported the potential anti-correlation between the H$\alpha$ emission 
and optical $g$-band photometry of the same star AU Mic,
which could be roughly consistent with the above suggestion of our study (potential anti-correlation of quasi-periodic modulations 
between the H$\alpha$ emission and TESS photometry). 
The amplitude of the quiescent H$\alpha$ E.W. modulations in Figure 7 of \citet{Odert+2025_MNRAS} are roughly comparable to that in Figure \ref{fig:TESS_comparison_rotation}(a) ($\lesssim$1 \AA).
We note that the datasets of our study and \citet{Odert+2025_MNRAS} show the similar results on the same early M-dwarf. 
Our H$\alpha$ data were taken in the intensive/continuous 15-day period (cf. Figure \ref{fig:TESS_comparison_rotation}(a)) but without the complete simultaneous optical photometry (TESS data are taken two months earlier, and ground-based photometry has a lot of gaps as seen in Figure \ref{fig:allEW_AUMic_XMM_opt}).
The H$\alpha$ data of \citet{Odert+2025_MNRAS} were taken intermittently over the 11-month period but with more observation period (56 days) and the simultaneous optical $g$-band photometry.
In addition, the potential contributions of flares in 
chromospheric line rotational modulations have also been mentioned in 
various studies \color{black} investigating \color{black} other M-dwarfs
(\citealt{Maehara+2021,Schofer+2022,Notsu+2024_ApJ}).
\citet{Schofer+2022} have suggested that the 
rotational signals in the chromospheric indicators 
are more common at lower activity levels, with the speculation that
an increasing flaring rate and/or an increasing number of plage regions
lead to a more homogeneous distribution of active regions 
in the chromosphere with increasing global activity.
More comprehensive observations of long-term chromospheric line and X-ray modulations with simultaneous TESS-like high-precision optical photometry are necessary in future studies, in order to discuss how commonly the modulation patterns as in Figure \ref{fig:TESS_comparison_rotation} can be seen in the case of various M-dwarfs.

\subsection{Coronal properties from quiescent phase X-ray data}\label{subsec:discuss_quiescent_X-ray}

Spectral analysis of quiescent component X-ray data are conducted in Section \ref{subsec:X-ray_specana_quiescent}. 
As seen in 
\color{black}
Figures \ref{fig:specfit_QuieALL1_0822740301}(j) \& 
\ref{fig:specfit_QuieALL1_0822740401} (j) -- 
\ref{fig:specfit_QuieALL1_0822740601} (j),
\color{black}
the abundances in the quiescent phase spectra tend to show a depletion of elements with a low First Ionization Potential (FIP), which correspond to the Inverse-FIP effect (\citealt{Brinkman+2001_A&A}).
The Inverse-FIP abundance trend is generally seen in the X-ray spectra of active stars in quescent and flare phases (e.g., \citealt{Audard+2003_A+A,Guedel2004_A&ARv,Raassen+2003_A+A,Raassen+2007,Pillitteri+2022,Chebly+2025_A+A,Didel+2024_MNRAS,Didel+2025_AJ}).

As listed in Table \ref{table:X-ray_quie_fit_results},
the total emission measure (EM$_{\rm{tot}}$) and EM-weighted average temperature 
($T_{\rm{ave}}$) of the AU Mic X-ray data are estimated to be 
EM$_{\rm{tot}}\sim2.5-3.6\times 10^{52}$ cm$^{-3}$ and
$T_{\rm{ave}}\sim 10^{6.96} - 10^{7.02}$ K ($\sim 0.78 - 0.91$ keV).
\citet{Takasao+2020_ApJ} investigated G-dwarf archival X-ray data 
and attempted to explain the quiescent phase EM vs. Temperature relation (EM$-T$ relation), on the basis of the present solar understanding: 
a steady corona model based on the so-called Rosner-Tucker-Vaiana (RTV) scaling laws (cf. \citealt{Rosner+1978_ApJ,Shibata_Yokoyama+2002}) and the observed power-law distribution function of surface magnetic features 
(e.g., \citealt{Harvey_Zwaan_1993_SoPh,Parnel+2009_ApJ}).
They derived a theoretical scaling law of the quiescent phase X-ray EM$-T$ relation for a star with single and multiple active region cases.
The relation for a single active region is originally derived by \citet{Shibata_Yokoyama+2002} (``Coronal EM$-T$ Scaling Law"):
\begin{eqnarray}
\rm{EM}_{\rm{sin}} 
&\approx& 10^{46} \left(\frac{f}{0.1}\right)
\left(\frac{T}{10^{6}~\rm{K}}\right)^{4}
\left(\frac{L}{10^{10}~\rm{cm}}\right)
\label{eq:EM-T_quie_single_L} \\ 
&\approx& 
10^{44} \left(\frac{f}{0.1}\right)
\left(\frac{T}{10^{6}~\rm{K}}\right)^{15/2}
\left(\frac{F_{\rm{h}}}{10^{7}~\rm{erg}~\rm{cm}^{-2}~\rm{s}^{-1}}\right)^{-1} 
\rm{cm}^{-3} 
\label{eq:EM-T_quie_single_Fh}
\end{eqnarray}
for $T<10^{7}$ K, where $L$ is the size of the active region, 
$f$ is the filling factor of the corona, and $F_{\rm{h}}$ 
is the coronal heating flux.
Then \citet{Takasao+2020_ApJ} 
assumed the size distribution ($dN/dA$) of magnetic features 
follows a single power law relation 
as the solar case (cf. \citealt{Harvey_Zwaan_1993_SoPh,Parnel+2009_ApJ}): 
  \begin{eqnarray}
\frac{dN}{dA}=N_{0}\gamma_{0}A^{-\alpha} \ \rm{cm}^{-2} \,
\label{eq:size_powerlaw}
    \end{eqnarray}
where $\gamma_{0}=A_{\odot}N_{\rm{f}}\bar{B}^{-\alpha+1}$,
$A_{\odot}=4\pi R_{\odot}^{2}\approx 6.2\times 10^{22}$ cm$^{2}$,
$N_{\rm{f}}=3 \times 10^{-4}$ in cgs units, 
$\bar{B}$ is the typical magnetic field strength of the active regions,
and $\alpha<2$ (e.g., $\alpha=1.85\pm$0.14 for the solar active regions, \citealt{Parnel+2009_ApJ}). 
Here it is hypothesized that the stellar size distribution functions 
have the same power-law index as the solar one, 
but the difference appears in \color{black} the coefficient $N_{0}$\color{black}.
$N_{0}=1$ corresponds to the Sun, while $N_{0}>1$ corresponds
to a star having more active regions with a give size than the Sun (cf. \citealt{Schrijver_2001_ApJ}).
Combining Equations (\ref{eq:EM-T_quie_single_Fh}) \& (\ref{eq:size_powerlaw}),
the $T-\mathrm{EM}$ relation of multiple active regions is
  \begin{eqnarray}
\rm{EM}_{\rm{tot}} &\approx& \frac{14N_{0}\gamma_{0}}{29 - 14\alpha}10^{60-16\alpha}
\left(\frac{f}{0.1}\right)
\left(\frac{F_{\rm{h}}}{10^{7}~\rm{erg}~\rm{cm}^{-2}~\rm{s}^{-1}}\right)^{-3+2\alpha}
\left(\frac{T_{\rm{max}}}{10^{6}~\rm{K}}\right)^{(29-14\alpha)/2}
\ \rm{cm}^{-3} \,
\label{eq:EM-T_quie_multi}
  \end{eqnarray}
where $T_{\rm{max}}$ is the maximum temperature of active regions, and 
the coronal temperature derived from observations $T_{\rm{obs}}$ can be
obtained as: 
  \begin{eqnarray}
T_{\rm{obs}} = \frac{29-14\alpha}{31-14\alpha}T_{\rm{max}} 
\approx 0.61 T_{\rm{max}}
\label{eq:Tobs_Tmax}
  \end{eqnarray}
for $\alpha=1.85$ (the present solar case).

   \begin{figure}[b!]
   \begin{center}
      \gridline{
\fig{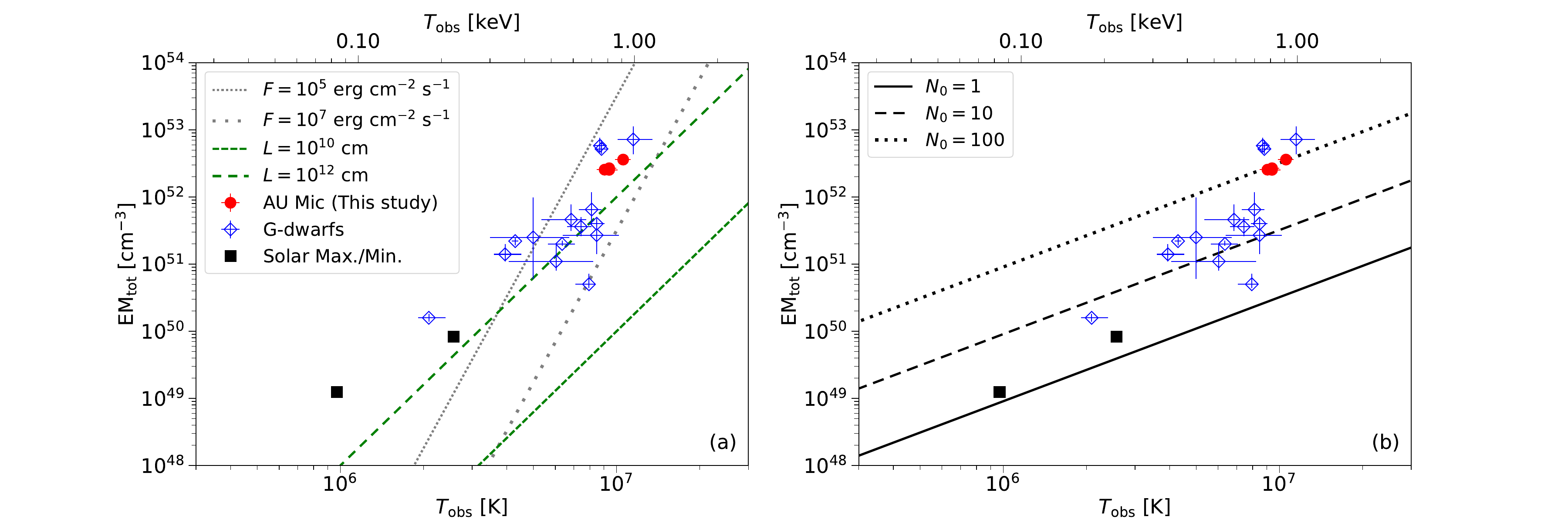}{1.0\textwidth}{\vspace{0mm}}
    }
     \vspace{-5mm}
     \caption{
$EM_{\rm{tot}}-T_{\rm{obs}}$ relation. 
Red filled circles show $EM_{\rm{tot}}$ and $T_{\rm{ave}}$ values
of AU Mic estimated in this study (Table \ref{table:X-ray_quie_fit_results}).
Blue open diamonds and black filled squares indicate the data of G-dwarfs 
and the solar maximum/minimum, respectively, 
and they are taken from Figure 5 of \citet{Takasao+2020_ApJ}. 
(a) Gray dotted and green dashed lines show Equations 
(\ref{eq:EM-T_quie_single_L}) \& (\ref{eq:EM-T_quie_single_Fh}) with different parameters.
(b) Data points with Equation (\ref{eq:EM-T_quie_multi}) 
with different values of $N_{0}$, $f=0.1$, $\bar{B}=100$ G, and $F_{\rm{h}}=10^{7}$erg cm$^{-2}$s$^{-1}$ (see the main text for the details).
}
   \label{fig:quie_T_EM}
   \end{center}
 \end{figure}

 As done in \citet{Takasao+2020_ApJ}, we compare these simple theoretical 
$T-\mathrm{EM}$ scaling laws with the observed values. 
In Figure \ref{fig:quie_T_EM}(a), the AU Mic data (in this study),
G-dwarf data (from Figure 5 of \citealt{Takasao+2020_ApJ}), and 
the solar maximum/minimum data (originally from \citealt{Peres+2000_ApJ} and also used in Figure 5 of \citealt{Takasao+2020_ApJ}) are compared with the single active region model (Equations (\ref{eq:EM-T_quie_single_L}) \& (\ref{eq:EM-T_quie_single_Fh})).
It is evident that the single active region model 
cannot explain
most of the data points with reasonable parameter sets, as already 
described in \citet{Takasao+2020_ApJ}.
For example, we can see 
the AU Mic data points locate above the line of $L=10^{12}$ cm.
Considering the AU Mic's radius of 0.75 $R_{\odot}$ $\approx$ 5.2$\times10^{10}$ cm
(cf. Table 1 of T23), this means the spatial scale of one X-ray active region of AU Mic is more than 20--30$\times$ larger than the stellar radius, 
which is \color{black} almost impossible \color{black}.
Then in Figure \ref{fig:quie_T_EM}(b), the data are compared with 
the multiple active region model (Equation (\ref{eq:EM-T_quie_multi})).
$\bar{B}=100$ G, $\alpha$=1.85, $f=0.1$, and $F_{\rm{h}}=10^{7}$erg cm$^{-2}$s$^{-1}$ are adapted in this figure (See \citealt{Takasao+2020_ApJ} for the limitations of assumptions). 
We can see that most of the datapoints are well above the line of 
$N_{0}=1$, and in particular, the AU Mic data in this study may agree with the case of $N_{0}\sim100$.
These results may suggest that X-ray active stars like AU Mic have much more (e.g., $\sim$100$\times$) active regions for a given size than the Sun,
though there are some limitations and uncertainty of the scaling-law assumptions as discussed in \citet{Takasao+2020_ApJ}. 
The assumed parameters (e.g., $f=0.1$, $\bar{B}=100$ G, and $F_{\rm{h}}=10^{7}$erg cm$^{-2}$s$^{-1}$) can also affect the discussions. 
\color{black}
For example, in the case of M-dwarfs, $\bar{B}$ values
can be larger at least by a few factors, if we speculate from 
the observational results that
the photospheric magnetic field strength of active M-dwarfs 
are reported to be a few times larger than the solar active region fields (e.g., \citealt{Kochukhov+2021} for review).
\color{black}
This \color{black} causes \color{black} a smaller $\gamma_{0}$ \color{black}
in the case of $1<\alpha<2$
and then \color{black} the $N_{0}$ value needs to become larger to compensate for the same $EM_{\rm{tot}}$ value \color{black} in Equation (\ref{eq:EM-T_quie_multi}).
In addition, as only the data of AU Mic, G-dwarfs, and the Sun are plotted in Figure \ref{fig:quie_T_EM}, it is important to compare with the X-ray data of other M-dwarfs as a next step study.

\subsection{Flare energetics and coronal parameters from X-ray flare spectral analysis}\label{sec:discuss_Xray_flare}

\begin{figure}[ht!]
   \begin{center}
      \gridline{
\fig{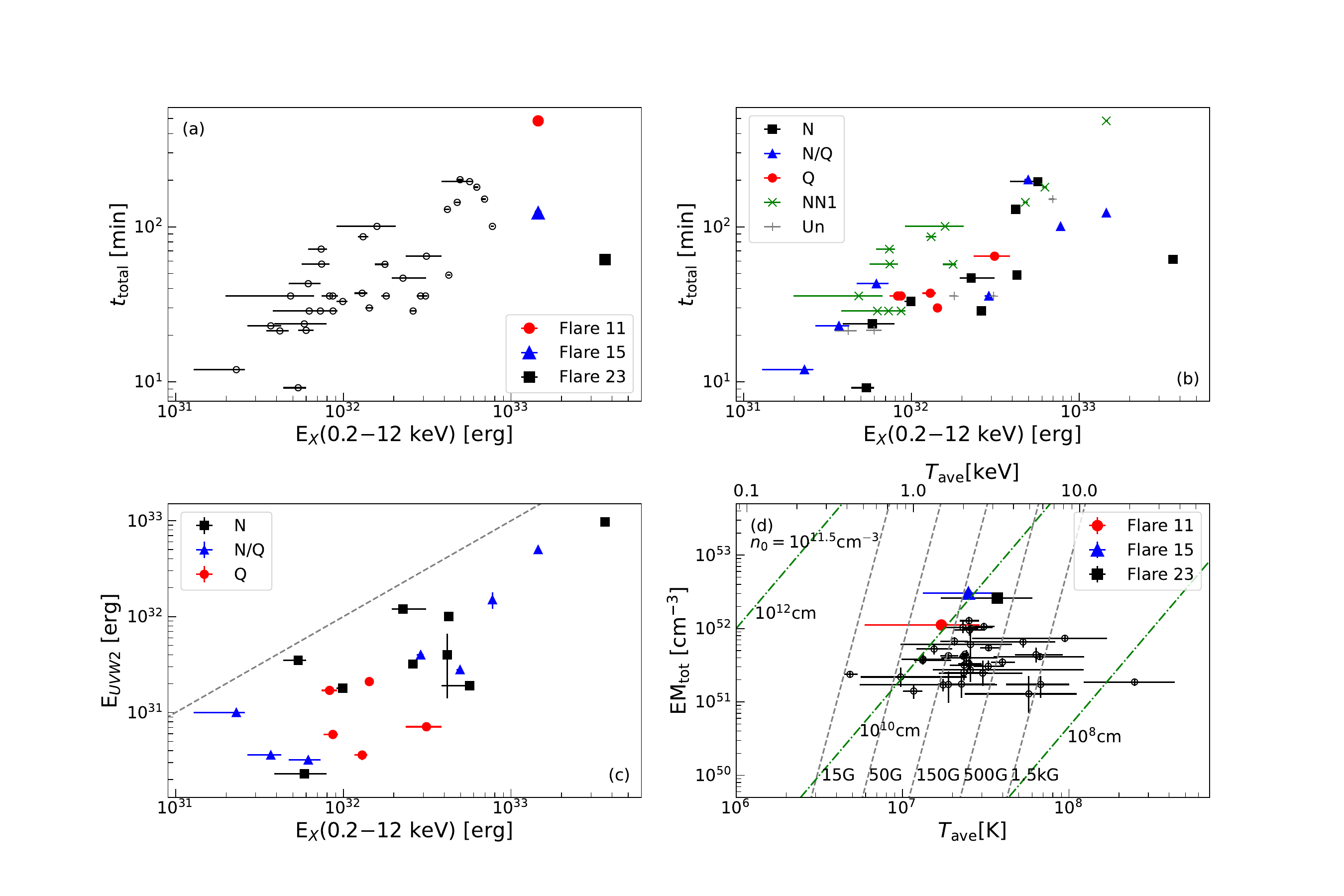}{1.0\textwidth}{\vspace{0mm}}
    }
     \vspace{-5mm}
     \caption{
(a) The X-ray flare energy in \color{black} the \color{black} 0.2 -- 12 keV range ($E_{\rm{X}}$) vs. 
flare total duration ($t_{\rm{total}}$) from Table \ref{table:X-ray_flarefit_timeave}.
The three large remarkable flares (Flares 11, 15, and 23), which are analyzed in Sections \ref{subsec:ana_flare_23} \& \ref{subsec:ana_flare_11_and_15}, are highlighted with the red circle, blue triangle, and black square.
(b) Same as (a), but the data points are plotted with Neupert classifications in T23 (also listed in Table \ref{table:X-ray_flarefit_timeave}).
(c) The X-ray flare energy in 0.2 -- 12 keV range ($E_{\rm{X}}$) vs. the UVW2-band NUV flare energy $E_{\rm{UVW2}}$ from Table \ref{table:X-ray_flarefit_timeave}. The data points are plotted with the Neupert classifications as in (b). The line
showing $E_{\rm{X}}=E_{\rm{UVW2}}$ is plotted for reference (dashed line).
(d) The flare X-ray average temperature ($T_{\rm{ave}}$) vs. the flare total emission measure ($EM_{\rm{tot}}$). As in (a), Flare 11, 15, and 23 are highlighted with the red circle, blue triangle, 
and black square.
The gray dashed lines correspond to Equation (\ref{eq:B-EMT}) with $n_{0}=10^{11.5}$ cm$^{-3}$
and $B$=15G, 50G, 150G, 500G, and 1.5kG. 
The green dash dotted lines correspond to Equation (\ref{eq:L-EMT}) with 
$n_{0}=10^{11.5}$ cm$^{-3}$
and $L$=10$^{8}$, 10$^{10}$, and 10$^{12}$cm. 
}
   \label{fig:plot_time_risedecay_flarelist}
   \end{center}
 \end{figure}

As described in Sections \ref{subsec:flare_atlas} \& \ref{subsec:X-ray_specana_time-average}, 
38 flares are identified in the EPIC-pn X-ray lightcurve data 
(Figures \ref{fig:X-ray_Ha_obsid_0822740301_lc} -- \ref{fig:X-ray_Ha_obsid_0822740601_lc}), 
and the analysis results of the time-averaged flare X-ray spectral components 
are listed in Table \ref{table:X-ray_flarefit_timeave}.
The X-ray flare energies in \color{black} the \color{black} 0.2 -- 12 keV range ($E_{\rm{X}}$), 
X-ray flare total durations ($t_{\rm{total}}$), 
and UVW2-band NUV flare energies ($E_{\rm{UVW2}}$)
are plotted in Figure \ref{fig:plot_time_risedecay_flarelist}(a)--(c).
The X-ray energies of the 38 flares are in the range of $E_{\rm{X}}=
2\times 10^{31} - 4\times 10^{33}$ erg.
In Figure \ref{fig:plot_time_risedecay_flarelist}(a), 
there is a rough positive  \color{black} trend \color{black}  with some scatter 
between $E_{\rm{X}}$ and $t_{\rm{total}}$,
and the three remarkable flares (Flares 11, 15, and 23) analyzed 
in Sections \ref{subsec:ana_flare_23} \& \ref{subsec:ana_flare_11_and_15} 
are the most energetic flares (with $E_{\rm{X}}>10^{33}$ erg) among all the 38 flares. The Neupert classification introduced in T23 \color{black} is \color{black} plotted just for 
reference in Figure \ref{fig:plot_time_risedecay_flarelist}(b). 
Figure \ref{fig:plot_time_risedecay_flarelist}(c) suggests 
that there is also a rough positive  \color{black} trend \color{black}  between 
$E_{\rm{X}}$ and $E_{\rm{UVW2}}$. 

The flare energy partition among different wavelengths is
an important topic of stellar flares since this can have constraints
on how flare energy release \color{black} processes \color{black} occur in the different layers of 
\color{black} stellar flaring \color{black} atmosphere from photosphere to corona (e.g., \citealt{Osten+2015,Stelzer+2022_A&A}). 
The detailed multi-wavelength flare 
\color{black} energies \color{black} of 
the remarkable flares analyzed in Section \ref{sec:remarkable_flares} 
(Flares 21, 22, 23, 11, 15, 35, 36, and 65) are  
summarized in Table \ref{table:energy_remarkable_flares}.
The detailed energy partition discussions as in \citet{Osten+2015}
are not directly \color{black} applicable to our study \color{black} since we do not have the NUV/optical broad-band flare
spectra, which is necessary for correct energy \color{black} estimations \color{black}  (we only have $UVW2$, $U$, and $V$-band emissions in this campaign).
However, it is interesting to briefly compare Flares 21, 22, and 23, 
which occurred successively on the same day (cf. Section \ref{subsec:ana_flare_23} and Figure \ref{fig:Oct13_TEM_multi_lc}).
As for Flares 22 and 23, 
more flare energy is emitted in the observed NUV \& optical continuum bands than the X-ray band ($E_{\rm{X}}<E_{\rm{UVW2}}+E_{\rm{U}}+E_{\rm{V}}$. 
Flare 22 shows 
$E_{\rm{X}}=2.27^{+0.83}_{-0.32}\times 10^{32}$ erg
and $E_{\rm{UVW2}}+E_{\rm{U}}+E_{\rm{V}} \approx 5.8\times 10^{32}$erg,
while Flare 23 shows $E_{\rm{X}}=1.45^{+0.02}_{-0.03}\times 10^{33}$ erg
and $E_{\rm{UVW2}}+E_{\rm{U}}+E_{\rm{V}} \approx 3.6\times 10^{33}$erg (Table \ref{table:energy_remarkable_flares}).
In contrast, the opposite result ($E_{\rm{X}}>E_{\rm{UVW2}}+E_{\rm{U}}+E_{\rm{V}}$) can be seen in the case of Flare 21:
$E_{\rm{X}}=3.13^{+1.00}_{-0.84}\times 10^{32}$ erg and 
$E_{\rm{UVW2}}+E_{\rm{U}}+E_{\rm{V}} \approx 1.1\times 10^{32}$erg 
(Table \ref{table:energy_remarkable_flares}).
This example highlights \color{black} a \color{black}  variety of flare energy partitions and Neupert responses among M-dwarf flares (see also T23 for 
\color{black} detailed  \color{black} discussions). As a future study with more multi-wavelength observation data especially including NUV/optical broad-band flare spectroscopy (e.g., \citealt{Hawley+1991,Kowalski_2022_FrASS,Kowalski+2025_ApJ}), 
it is important to update the flare energy partition scaling relations (e.g., \citealt{Osten+2015}) by considering the variety between X-ray (thermal) and optical/NUV (non-thermal) emissions of M-dwarf flares.

The X-ray temperature ($T$) and emission measure (EM) values from
soft X-ray spectral analysis results 
are useful for investigating the physical parameters of the flaring plasma
such as flare size scales (flare loop lengths) and magnetic field strengths 
(e.g., 
\citealt{Reale+1997_A&A,Reale_2007_A+A,Shibata_Yokoyama_1999,Shibata_Yokoyama+2002,Raassen+2007,Osten+2006,Osten+2016,Namekata+2017_PASJ,Pillitteri+2022,Notsu+2024_ApJ}
). \citet{Shibata_Yokoyama_1999,Shibata_Yokoyama+2002} discussed
the scaling relations of $T$ and EM for solar and stellar flares,
which consider the energy balance between conduction cooling and magnetic reconnection heating (see also \citealt{Shibata+2011} for review).
The scaling laws by \citet{Shibata_Yokoyama+2002} show that 
the flare magnetic field ($B$) and characteristic length of the flare loop
($L$) can be expressed in terms of the flare EM ($EM=n^{2}V$, $n^{2}$ is the coronal electron density, and $V$ is the volume), the preflare coronal electron density ($n_{0}$), and flare temperature ($T$):
  \begin{eqnarray}
    B &=& 50\left(\frac{\rm{EM}}{10^{48}\rm{cm}^{-3}}\right)^{-1/5}\left(\frac{n_{0}}{10^{9}\rm{cm}^{-3}}\right)^{3/10}\left(\frac{T}{10^{7}\rm{K}}\right)^{17/10} \rm{G}\ ,  \label{eq:B-EMT}\\
    L &=& 10^{9}\left(\frac{\rm{EM}}{10^{48}\rm{cm}^{-3}}\right)^{3/5}\left(\frac{n_{0}}{10^{9}\rm{cm}^{-3}}\right)^{-2/5}\left(\frac{T}{10^{7}\rm{K}}\right)^{-8/5} \rm{cm}\ .  \label{eq:L-EMT}
  \end{eqnarray}
Here simple order-of-magnitude estimates are used (e.g., $V=L^{3}$).
These simple scaling laws were validated with solar flare observations and can estimate $B$ and $L$ values with an accuracy of a factor of 3 (\citealt{Namekata+2017_PASJ}). 

In Figure \ref{fig:plot_time_risedecay_flarelist}(d),
the average X-ray temperature ($T_{\rm{ave}}$) and total emission measure 
of the 38 X-ray flares from Table \ref{table:X-ray_flarefit_timeave}
are plotted with the lines corresponding to 
Equations (\ref{eq:B-EMT}) \& (\ref{eq:L-EMT}). 
The quiescent electron density estimates from the Ne IX lines 
in Section \ref{subsec:X-ray_specana_quiescent}  
have suggested $n_{0}\lesssim 10^{11.5-11.9}$cm$^{-3}$.
\citet{Osten+2006} measured the quiescent electron densities
of the active M3.5 flare star EV Lac with more transition region and corona lines, 
and have shown at least $n_{0}\geq10^{11.5}$ cm$^{-3}$ around the quiescent coronal temperature of AU Mic ($T_{\rm{ave}}\sim 10^{6.9} - 10^{7.0}$ K).
Considering these two results, \color{black} we assume \color{black} $n_{0}= 10^{11.5}$ cm$^{-3}$
in this study. It is noted that this is a very rough assumption, 
the estimated $B$ and $L$ values can vary by a factor of two or three 
depending on the different assumptions of $n_{0}$ (see also Table 7 of \citealt{Notsu+2024_ApJ} for the difference of $B$ and $L$ values with different 
$n_{0}$ values). 
As seen in Figure \ref{fig:plot_time_risedecay_flarelist}(d),
most of the flares are located in the range of 
$B \sim$ 50 G -- 1.5 kG and $L \sim 3\times10^{8}$ cm -- $2\times 10^{10}$ cm 
= 0.006 -- 0.38 $R_{\rm{star}}$. 
These ranges can be some rough constraints for future modeling studies of the X-ray flare loop emission, 
but it should be noted that the scaling relations in Equations (\ref{eq:B-EMT}) \& (\ref{eq:L-EMT}) were originally derived for the energy balances around the flare peak times 
(\citealt{Shibata_Yokoyama+2002}) with various assumptions as described in the above (e.g., $V=L^{3}$) and the potential uncertainties (e.g., up to order of magnitude) should 
be kept in mind (cf. \citealt{Namekata+2017_PASJ}).

\begin{figure}[ht!]
   \begin{center}
      \gridline{
\fig{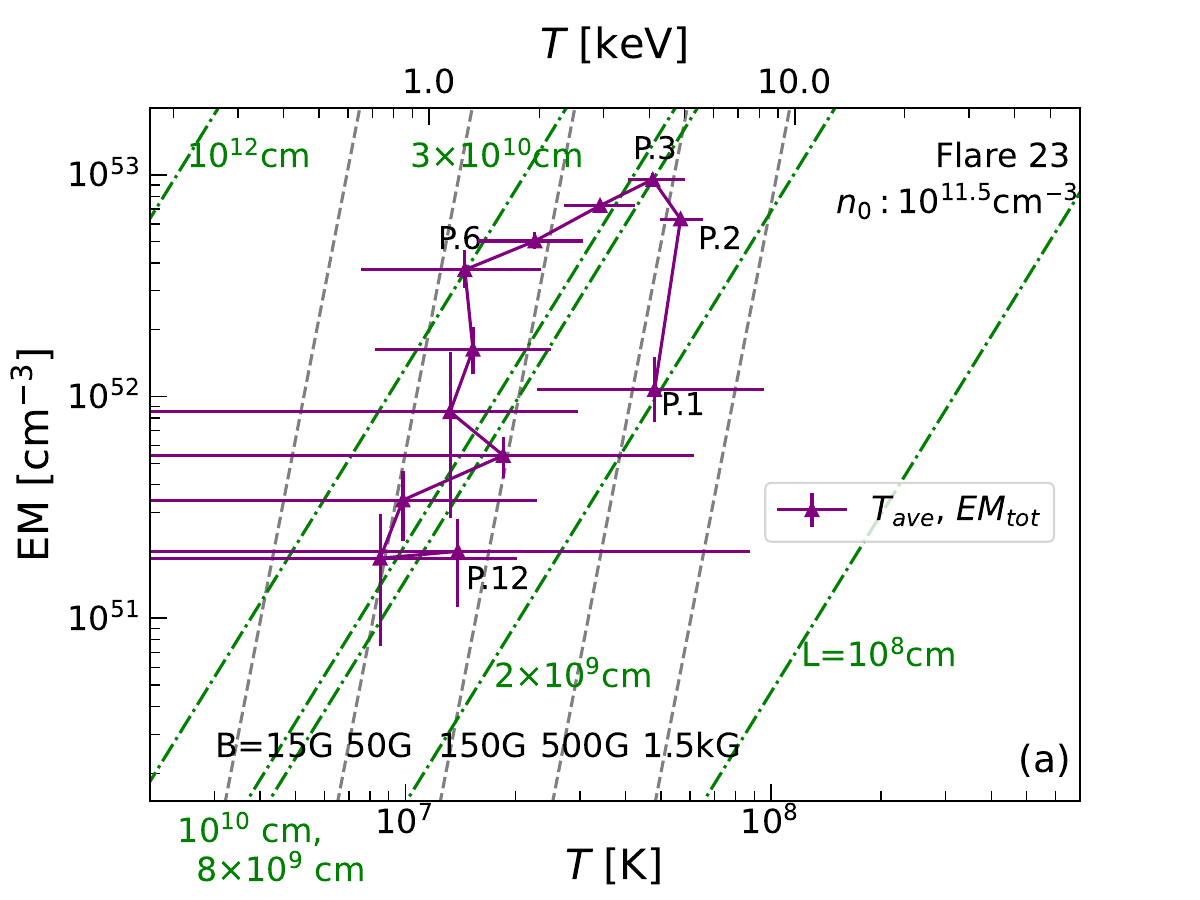}{0.45\textwidth}{\vspace{0mm}}
\fig{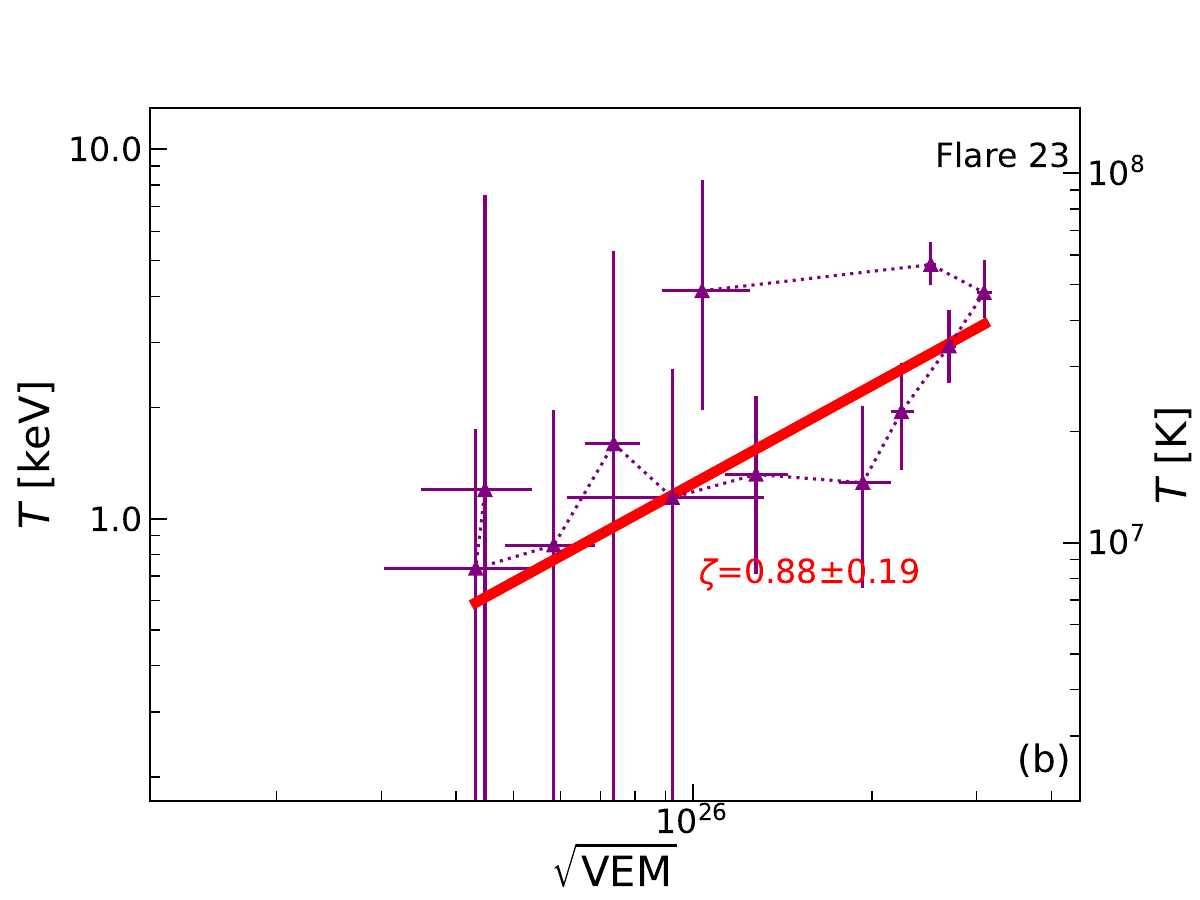}{0.45\textwidth}{\vspace{0mm}}
    }
     \vspace{-10mm}
      \gridline{
\fig{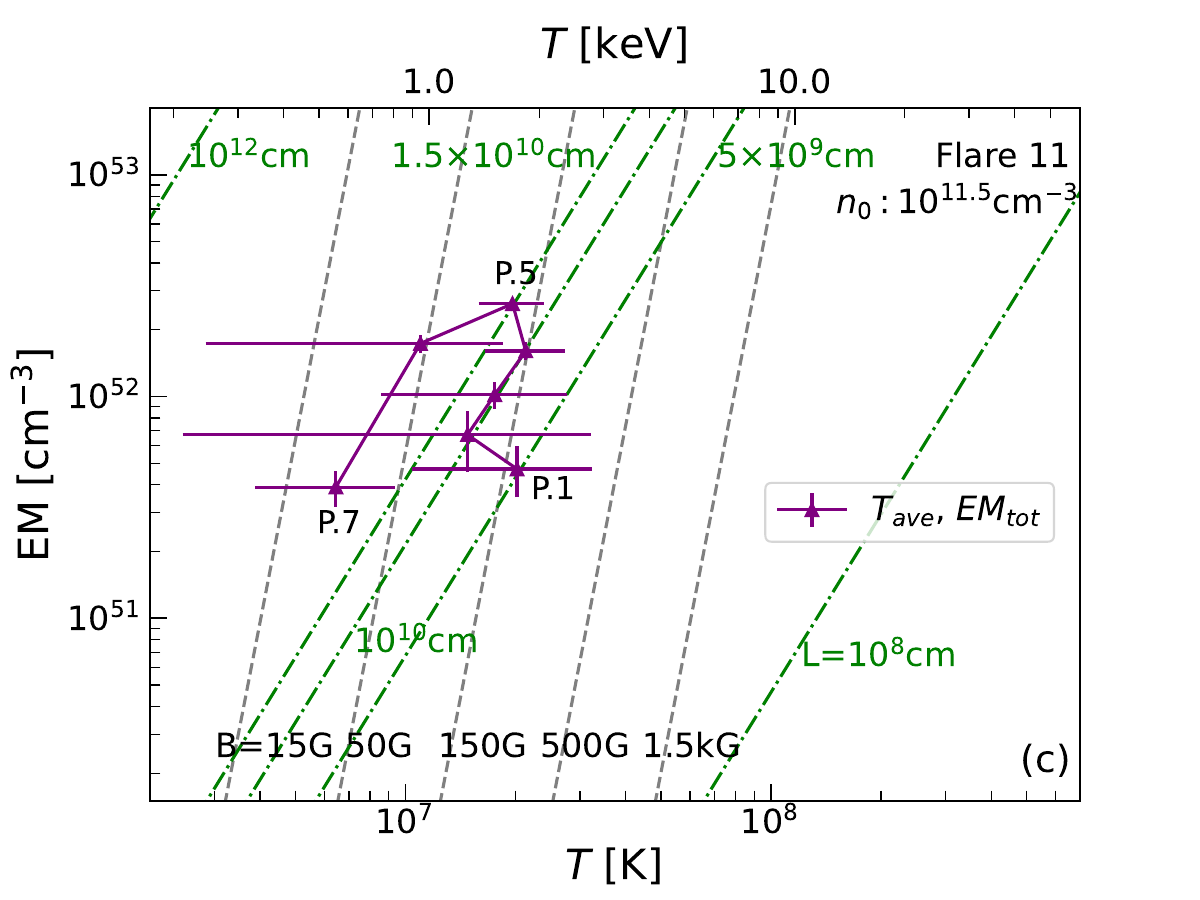}{0.45\textwidth}{\vspace{0mm}}
\fig{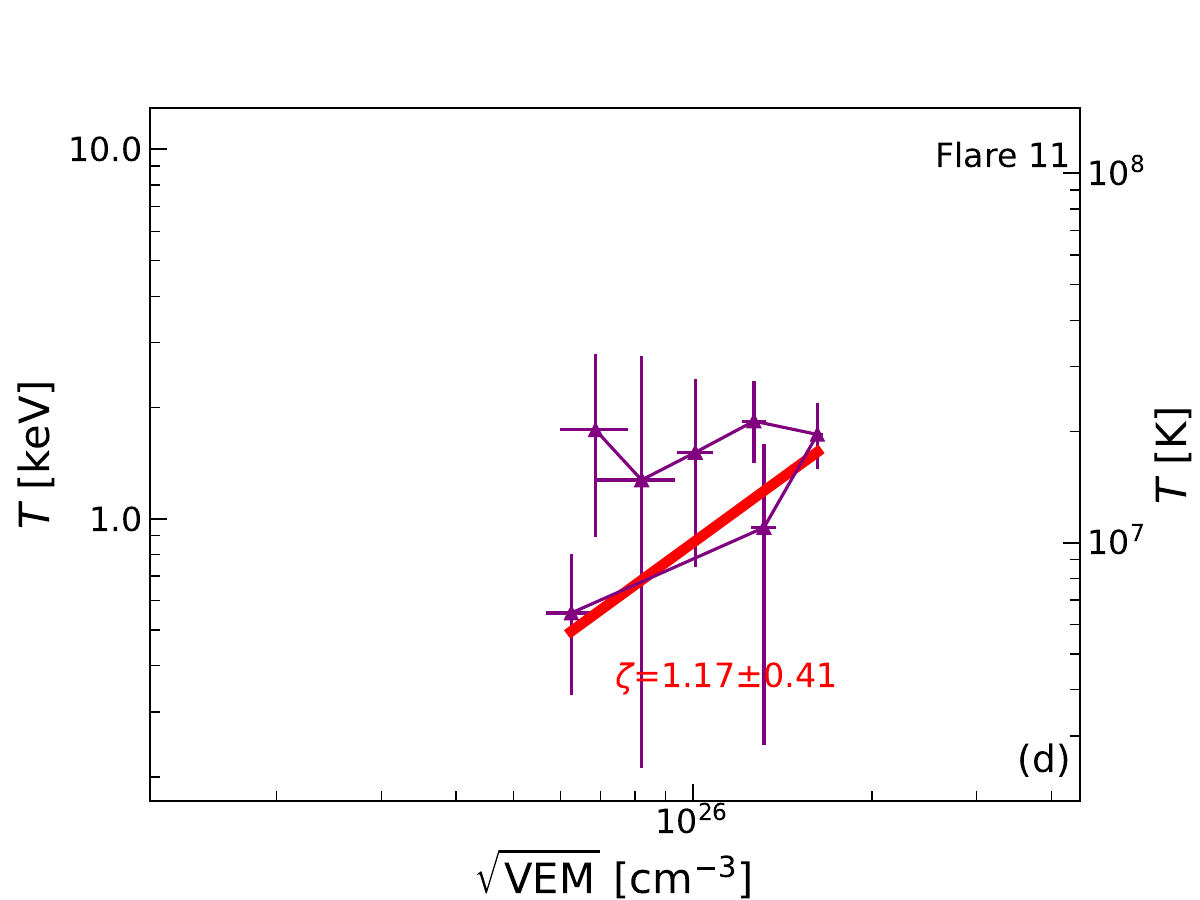}{0.45\textwidth}{\vspace{0mm}}
    }
     \vspace{-10mm}
      \gridline{
\fig{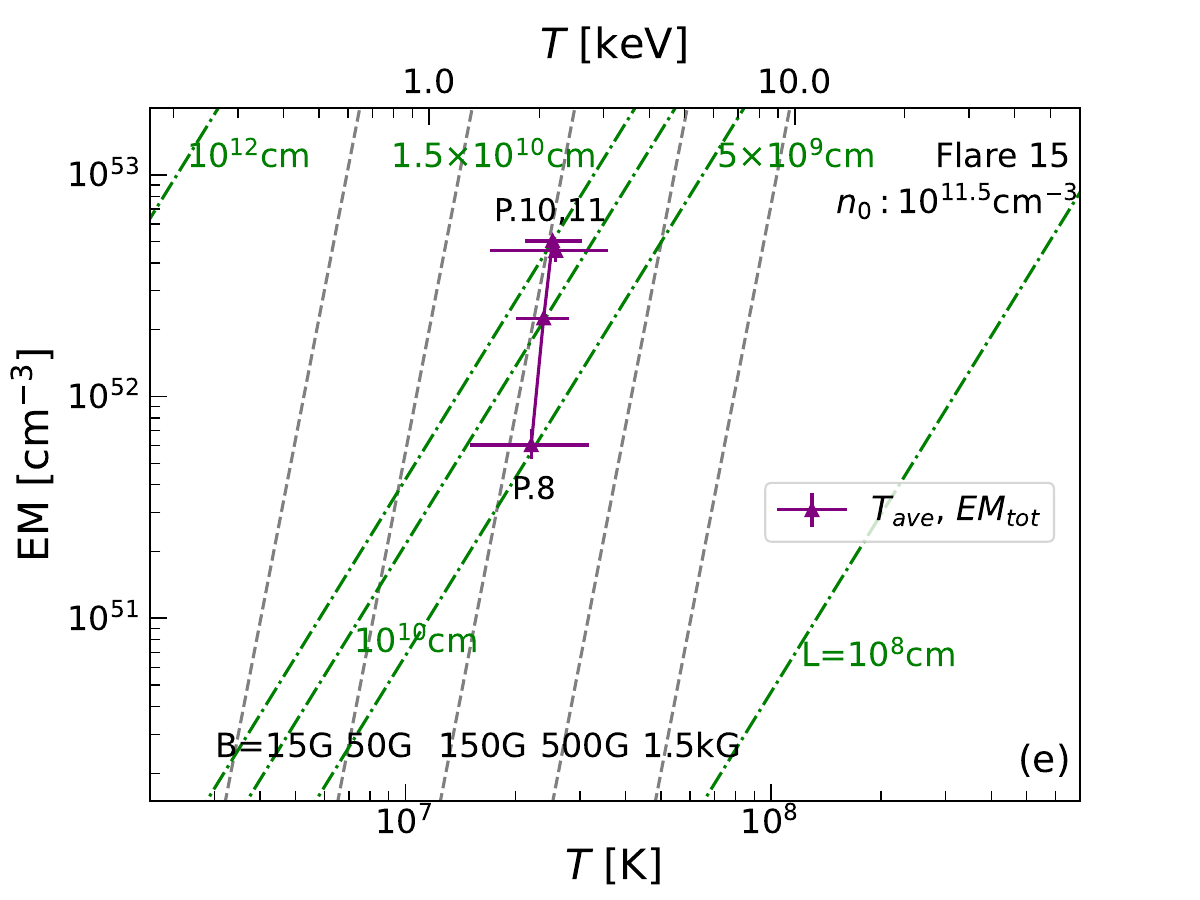}{0.45\textwidth}{\vspace{0mm}}
\fig{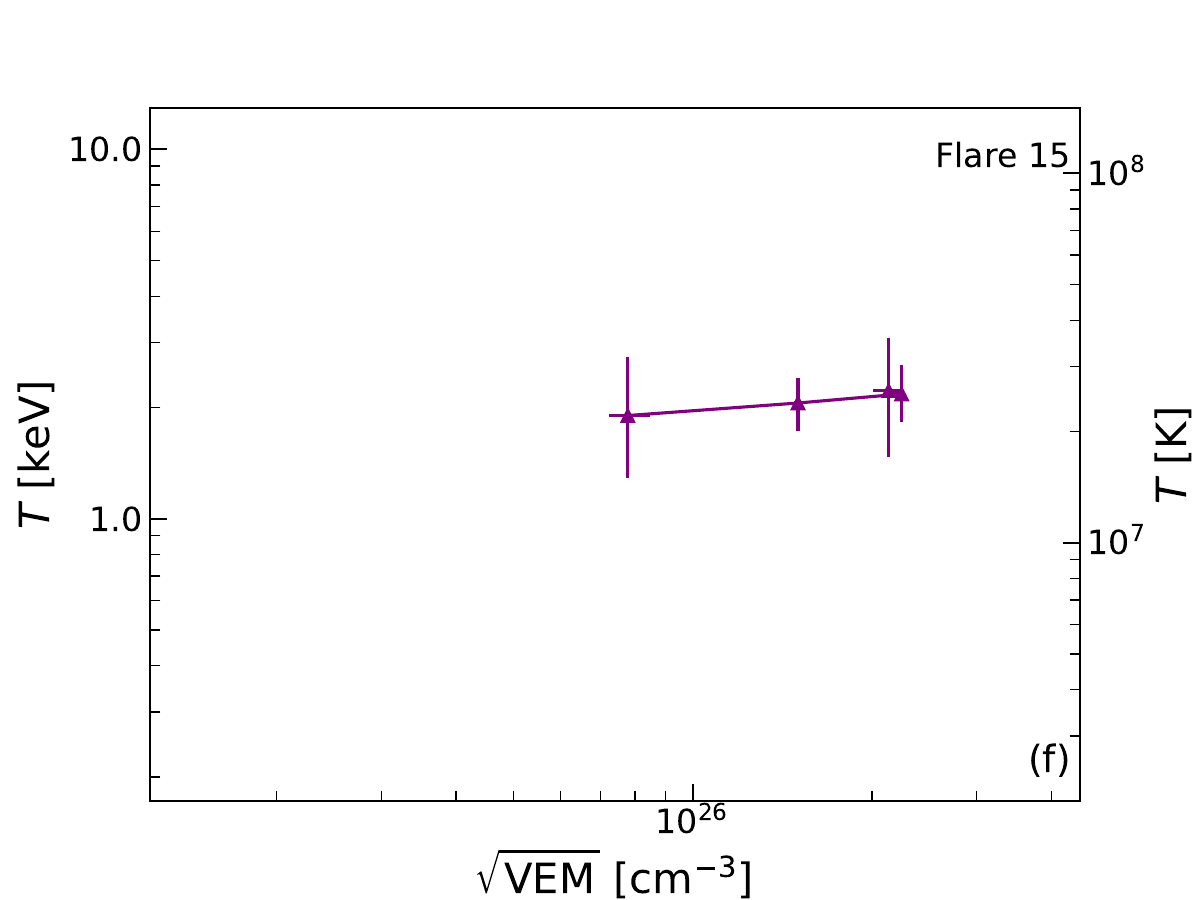}{0.45\textwidth}{\vspace{0mm}}
    }
     \vspace{-5mm}
     \caption{
(a) Time evolution of $T_{\rm{ave}}$ and $E_{\rm{tot}}$ of Flare 23 (cf. Figure \ref{fig:Flare23_TEM_multi_lc} and Table \ref{table:flare23_Xray_fit}). 
The gray dashed lines and green dash dotted lines are plotted with the same way as Figure \ref{fig:plot_time_risedecay_flarelist}(d).
The phase numbers (e.g., ``P.1", ``P.2") plotted in this figure correspond to those in Figure \ref{fig:Flare23_TEM_multi_lc}(e).
(b) The trend of  $T_{\rm{ave}}$ vs. the square root of volume emission measure ($VEM$) derived from each spectrum, for Flare 23, 
along with a determination of
the slope $\zeta$, given by the red solid line.
(c) \& (d) Same as (a) \& (b), but for Flare 11 (cf. Figure \ref{fig:Flare11and15_TEM_multi_lc} and Table \ref{table:flare11_Xray_fit}). 
(e) \& (f) Same as (a) \& (b), but for Flare 15 (cf. Figure \ref{fig:Flare11and15_TEM_multi_lc} and Table \ref{table:flare11_Xray_fit}).
The slope $\zeta$ is not determined for Flare 15 since the flare decay phase is not observed.
}
   \label{fig:TEM_eachflare}
   \end{center}
 \end{figure}

In Section \ref{subsec:ana_flare_23} and \ref{subsec:ana_flare_11_and_15},
we have derived the $T-\mathrm{EM}$ time evolutions of Flares 23, 11, and 15,
as shown in Figures \ref{fig:Flare23_TEM_multi_lc} \&  \ref{fig:Flare11and15_TEM_multi_lc} and listed in 
Tables \ref{table:flare23_Xray_fit} \&
\ref{table:flare11_Xray_fit}. 
Then we plot these $T-\mathrm{EM}$ time evolutions 
of the three flares (Flares 23, 11, and 15) 
with Equations (\ref{eq:B-EMT}) \& (\ref{eq:L-EMT})
in Figure \ref{fig:TEM_eachflare} (a), (c), \& (e).
Comparing Figures \ref{fig:Flare23_TEM_multi_lc} and \ref{fig:TEM_eachflare}(a),
Flare 23 starts already with high temperature: $T_{\rm{ave}}>40$ MK for Phase 1 (``P.1" in Figure \ref{fig:TEM_eachflare}(a)). The temperature is highest at Phase 2 ($T_{\rm{ave}}>50$ MK), and the emission measure ($EM_{\rm{tot}}$) is at the peak at Phase 3. From Phase 3 to 6, both $T_{\rm{ave}}$ and $EM_{\rm{tot}}$ values gradually decrease. 
After Phase 6 until the flare end (Phase 12),
the $T_{\rm{ave}}$ does not change so much 
while the $EM_{\rm{tot}}$ values continue to decrease. 
Comparing these overall evolution properties with Equations (\ref{eq:B-EMT}) \& (\ref{eq:L-EMT}) in Figure \ref{fig:TEM_eachflare}(a), 
we can interpret \color{black} the $T-\mathrm{EM}$ time evolution of Flare 23 as described in the following of this paragraph\color{black}, if we assume that 
the flare heating energy balance considered for these equations 
(cf. \citealt{Shibata_Yokoyama+2002}) 
holds over the most of the flare phase \footnote{
We also assume that the flares occur with subsequent appearances of multiple loops (cf. \citealt{Shibata+2011}).}. 
The X-ray flare emission originated from 
a flare loop with higher magnetic field and smaller loop size ($B\sim600$G and $L\sim2\times10^{9}$cm $\sim0.04R_{\rm{star}}$) at Phase 1, and at the Phase 2 ($T_{\rm{ave}}$ attains a maximum value) the magnetic field does not change ($B\sim600$G) but the X-ray emission comes from the larger expanded loop
($L\sim5\times10^{9}$cm$\sim0.10R_{\rm{star}}$). 
At Phase 3 ($EM_{\rm{tot}}$ takes the peak), the flare loop size becomes bigger or the emission comes from subsequent larger loops ($L\sim8\times10^{9}$cm$\sim0.15R_{\rm{star}}$), 
but the magnetic field starts to become smaller ($B\sim400$G).
As the flare X-ray emission decays from Phase 3 to 6 (``early" decay phase of the flare), 
the size of the dominant flare loops 
still continue to increase (from $L\sim8\times10^{9}$cm to $L\sim3\times10^{10}$cm$\sim0.57R_{\rm{star}}$), 
while the magnetic field becomes smaller ($B\sim$400 G to $B\sim$65 G). 
After Phase 6 to Phase 12 (``late" decay phase), the magnetic field does not change so much, 
while the loop size becomes slightly smaller. 
However, we should be careful about the evolution properties 
in these ``late" decay phase (Phase 6--12), since the error range of spectral analysis results (e.g., $T$ and EM values) are large.
\color{black}
Moreover, as described above, the simple scaling laws applied here can only provide very rough estimates of $B$ and $L$ values.
The scaling relations in Equations (\ref{eq:B-EMT}) \& (\ref{eq:L-EMT}) were originally derived for the energy balances around the flare peak times 
(\citealt{Shibata_Yokoyama+2002}) with various assumptions as described in the above (e.g., $V=L^{3}$, $n_{\rm{e}}$ is fixed to $10^{11.5}$cm$^{-3}$) and such assumptions can cause roughly up to an order-of-magnitude uncertainty on the estimates of $B$ and $L$ values (see also \citealt{Namekata+2017_PASJ}, which suggests a factor of 3 uncertainty based on solar data comparisons, but the real uncertainty when applying the scaling relations into stellar data can becomes larger (i.e. up to an order-of-magnitude as mentioned here) considering the above assumptions). 
This systematic uncertainty can be dominant compared with the uncertainty caused by the errors of the observed $T$ and $EM$ values. For example, in the case of ``Phase 3" data in Figure \ref{fig:TEM_eachflare}(a), the errorbars of the $T$ and $EM$ values suggests much smaller uncertainty of the $L$ value with only a factor of $\sim$0.1--0.2 if we assume that the green dash dotted lines from Equation (\ref{eq:L-EMT}) do not include any systematic uncertainty mentioned above.
\color{black}
In spite of these limitations, the qualitative discussions how flare evolves in the $T-\mathrm{EM}$ diagram can provide insights on the flare evolution around the flare peak, as discussed in the following.

The $T-\mathrm{EM}$ evolution tracks similar to Flare 23 have been reported 
in several stellar X-ray flares, as summarized in Figure 2 of \citet{Aschwanden+2008_ApJ}, on the basis of earlier stellar flare X-ray observational studies (e.g., \citealt{Stern+1992_ApJ,Tsuru+1989_PASJ,Maggio+2000_A+A,Reale+2004_A+A}). \citet{Shibata_Yokoyama+2002} discussed the flare $T-\mathrm{EM}$ diagram as a ``flare H-R diagram", based on $n-T$ diagram originally obtained from earlier one-dimensional hydrodynamic numerical simulations of solar flare loop evolutions 
(e.g., \citealt{Jakimiec+1992_A+A}, see also \citealt{Sylwester_1996_SSRv,Guedel2004_A&ARv} for reviews). 
In this case, when a sudden flare energy release occurs, the temperature of flare plasma increases until the energy \color{black} reaches \color{black}  the lower atmosphere (chromosphere and photosphere) (see ``pre-evaporation" \& ``reconnection heating" phase in Figures 6 \& 7 of \citealt{Shibata_Yokoyama+2002}). 
We may interpret that this is roughly consistent with the observed evolution of Flare 23.
The flare temperature already increased at Phase 1 (Figure \ref{fig:TEM_eachflare}(a)), when the UVW2 emission (non-thermal heating proxy in the case of this observation) already takes the peak before the X-ray emission increase 
as seen in Figure \ref{fig:Flare23_TEM_multi_lc} (see also Neupert discussions in T23). The non-thermal energy injection of this flare, which correspond to the NUV and optical emissions, concentrates in and before the X-ray emission rise phase.
After that, the temperature becomes constant, and the evaporation starts with the flare loop density gradually increasing owing to evaporation as long as the flare
heating continues (see ``evaporation" phase in Figures 6 \& 7 of \citealt{Shibata_Yokoyama+2002}). 
We may interpret that this is also 
roughly consistent with 
the observed evolution of Flare 23.
The temperature keeps the high value from Phase 1 to 3, while the EM values dramatically increase (Figure \ref{fig:TEM_eachflare}(a)).
The late-phase evolution (from Phase 3 through Phase 6 and further to Phase 12) might also be similarly interpreted as the decay phase track suggested in Figures 6 of \citet{Shibata_Yokoyama+2002}.

On the other hand, these evolutions should be observed as a combination of multi loop evolutions (e.g., multithread hydrodynamic modeling studies by \citealt{Hori+1997,Warren2006}). 
Larger loops with smaller magnetic loops appear in the standard model of flares (cf. \citealt{Shibata+2011}).
For example, the size scales of the well-known Bastille-day solar flare, which was measured as the expansion of UV ribbons, expand by several times ($\sim$ factor of 4) during the rise phase of the flare \color{black}(\citealt{Qiu+2010_ApJ}). \color{black}
It is interesting to see the similar trend \color{black} in \color{black} the size scale of Flare 23 \color{black} which \color{black} becomes several times ($\sim$ factor of 4) larger in the rise phase (from P1 to P3: from 2$\times$10$^{9}$ cm to 8$\times$10$^{9}$cm in Figure \ref{fig:TEM_eachflare}(a)), although \color{black} the \color{black}
uncertainty of the determined $L$ \color{black} value \color{black} 
is large as mentioned \color{black} above \color{black}. 
As a result, the X-ray time evolution, especially $T-\mathrm{EM}$ evolution of the clear Neupert-type Flare 23 with the simultaneous multi-wavelength observations \color{black} is \color{black} consistent with \color{black} the \color{black} 
flare evolution track described in \citet{Shibata_Yokoyama+2002}, 
which follows the standard model (and basically the same as the so-called CSHKP model) of solar and stellar flares  (see  \citealt{Shibata+2011} for the details)  \footnote{It can be also interesting note that recent solar observations (e.g., \citealt{Hudson+2021_MNRAS}) have reported ``hot onsets" of some solar X-ray flares: Hard ``hot onset" interval occurs during the initial soft X-ray increase and definitely before any detectable hard X-ray emission (non-thermal emission). It can be interesting to note that Flare 23 already started with high temperature at Phase 1 (Figure \ref{fig:TEM_eachflare}) in relation to this, but in the case of AU Mic's Flare 23, we can see that the NUV (UVW2) emission (= non-thermal emission proxy) has the peak value just before (or at least almost simultaneously with) the soft X-ray emission starting to increase (Figure \ref{fig:Flare23_TEM_multi_lc}). 
}. 

It is noted that the peak temperature and emission measure values of Flare 23 are larger than the normal solar flare ones. Following \citet{Shibata_Yokoyama+2002}, this can be interpreted within the standard model framework that the energetic stellar flares are associated with stronger magnetic field and larger size scales, but more detailed flare X-ray loop modeling studies in the case of active M-dwarfs \color{black} would be required \color{black} for further quantitative discussions (e.g., how much hot X-ray emission components can be generated, compared with normal solar flares ? cf. \citealt{Osten+2016,Kowalski+2024_LRSP}).

The $T-\mathrm{EM}$ evolutions of Flares 11 and 15, are shown in Figure \ref{fig:TEM_eachflare}(c) \& (e). It is noted that the decay phase was not observed for Flare 15 (cf. Figure \ref{fig:Flare11and15_TEM_multi_lc}).
The X-ray lightcurve shapes of these two flares are in contrast to Flare 23 (a clear Neupert flare) from the viewpoint that the flare X-ray emissions rise more gradually and the flare X-ray durations are longer (see Figures \ref{fig:Flare23_TEM_multi_lc} and \ref{fig:Flare11and15_TEM_multi_lc}).
The flare peak temperatures of Flare 11 and 15 are smaller, the magnetic fields are smaller, and loop sizes at the flare peak times (at Phases 5 and 10 in Figure \ref{fig:TEM_eachflare}(c) \& (e)) are larger, compared with Flare 23 in Figure \ref{fig:TEM_eachflare}(a). For example, the loop size scales at the flare peaks of Flare 11 and 15 (Phases 5 and 10) are both $L\sim 1.5\times10^{10}\rm{cm}~\sim 0.29 R_{\rm{star}}$, and these are about twice 
larger than the value at the flare peak of Flare 23 
($L\sim 8\times10^{9}\rm{cm}~\sim 0.15 R_{\rm{star}}$).
These differences may reflect \color{black} the
relation between flare impulsiveness 
and flare structures/properties.  
For example, in the case of solar flares, there are impulsive flares and long-durational-event (LDE) flares whose X-ray durations and flare structures are different 
(see \citealt{Shibata+2011} for review).
However, \color{black} statistical characterizations with more flare events are necessary for further discussions.

In addition to the above $T-\mathrm{EM}$ method based on \citet{Shibata_Yokoyama+2002} (Equations (\ref{eq:B-EMT}) \& (\ref{eq:L-EMT})), the coronal loop length of stellar flares
have been also often estimated by using the method 
of \citet{Reale+1997_A&A}, from the decay phase X-ray lightcurves (e.g., \citealt{Reale_2007_A+A,Osten+2010_ApJ,Osten+2016,Guarcello+2019}) \footnote{It can be helpful to note that \citet{Namekata+2017_PASJ} compared the flare loop estimation methods of \citet{Shibata_Yokoyama+2002} and \citet{Reale+1997_A&A} by using the same solar flare data. The estimated values obtained with both scaling laws are roughly consistent with the true values from the spatially resolved solar observations, 
although the method of \citet{Reale+1997_A&A} had  larger scatter. They argued that the method of \citet{Shibata_Yokoyama+2002} might be more robust especially in case that the time evolution of the flare is not fully observed (as in the case of Flare 15 in our study).}. As described in \citet{Reale_2007_A+A} and \citet{Osten+2016},
the thermodynamic loop decay time ($\tau_{\rm{th}}$) can be expressed as 
  \begin{eqnarray}
  \tau_{\rm{th}}=\alpha l / \sqrt{T_{\rm{max}}}
      \label{eq:tau_th}
  \end{eqnarray}
where $\alpha=3.7\times10^{-4}$ cm$^{-1}$ s$^{-1}$ K$^{1/2}$,
$l$ is the loop half length (cm),
and $T_{\rm{max}}$ is the flare maximum temperature (K),
  \begin{eqnarray}
T_{\rm{obs}}=\xi T_{\rm{obs}}^{\eta} ,
      \label{eq:TmaxTobs_Osten2016}
  \end{eqnarray}
and $T_{\rm{obs}}$ is the maximum best-fit temperature derived from temperature fitting of the data. The ratio of the observed
exponential light curve decay time $\tau_{\rm{LC}}$ to $\tau_{\rm{th}}$ can be
written with the slope $\zeta$ of the flare decay in the $\log(\sqrt{VEM} - T)$ plane ($VEM$ is the volume emission measure):
  \begin{eqnarray}
\tau_{\rm{LC}}/\tau_{\rm{th}}=\frac{c_{\rm{a}}}{\zeta - \zeta_{\rm{a}}} + q_{\rm{a}} 
= F(\zeta) .
      \label{eq:taulc_tauth}
  \end{eqnarray}
The parameters in Equations (\ref{eq:TmaxTobs_Osten2016}) and (\ref{eq:taulc_tauth}) are determined 
as $c_{\rm{a}}=0.51$, $\zeta_{\rm{a}}=0.35$, $q_{\rm{a}}=1.36$, $\xi=0.130$,
and $\eta=1.16$ for the XMM EPIC data (Table A.1 of \citealt{Reale_2007_A+A}).
From these expressions, the loop flare length is expressed as
  \begin{eqnarray}
l=\frac{\tau_{\rm{LC}}\sqrt{T_{\rm{max}}}}{\alpha F(\zeta)} ,
      \label{eq:l_decay}
  \end{eqnarray}
which is valid for $0.35 \leq \zeta \leq 1.6$ (Table A.1 of \citealt{Reale_2007_A+A}).
In this study,  we apply that $T_{\rm{ave}}$ and $EM_{\rm{tot}}$ in Tables \ref{table:flare23_Xray_fit} and \ref{table:flare11_Xray_fit} values correspond to $T_{\rm{obs}}$ and $VEM$ in these equations.

Figure \ref{fig:TEM_eachflare}(b), (d), and (f) show $\sqrt{VEM} - T$ diagrams
for Flares 23, 11, and 15. 
The $\zeta$ values of Flares 23 and 11 are determined as $\zeta=0.88\pm0.19$ and 
$\zeta=1.17\pm0.41$, respectively (Flare 15 is not analyzed since the decay phase is not observed). The exponential decay time $\tau_{LC}$ are determined 
by fitting the EPIC-pn lightcurves 
(Figures \ref{fig:Flare23_TEM_multi_lc}(e) and \ref{fig:Flare11and15_TEM_multi_lc}(b)): $\tau_{\rm{lc}}= 674 \pm 11$ s (Flare 23) and $\tau_{\rm{lc}}= 4818 \pm 69 $ s (Flare 11). It is noted that the X-ray flare decay curve of Flare 11 
is somewhat different from the exponential function (Figure \ref{fig:Flare11and15_TEM_multi_lc}), compared with Flare 23 (Figure \ref{fig:Flare23_TEM_multi_lc}), 
so this can cause larger error of $\tau_{\rm{lc}}$ that is not included in the systematic 1-$\sigma$ uncertainty value shown here.
$T_{\rm{obs}}$ values are 
from $T_{\rm{ave}}$ values at the flare flux peaks: 
$T_{\rm{ave}}=4.09^{+0.92}_{-0.60}$ keV $=4.75^{+1.07}_{-0.70}\times10^{7}$ K 
for Phase 3 of Flare 23 (Table \ref{table:flare23_Xray_fit}),
and $T_{\rm{ave}}=1.69^{+0.37}_{-0.33}$ keV $=1.96^{+0.43}_{-0.38}\times10^{7}$ K 
for Phase 5 of Flare 11
(Table \ref{table:flare11_Xray_fit}).
Applying these parameters into Equation (\ref{eq:l_decay}), 
the loop semi-length $l$ values of Flares 23 and 11 are estimated to be 
$l=8.0^{+2.3}_{-2.2} \times 10^{9}$ cm $= 0.15^{+0.05}_{-0.05} R_{\rm{star}}$,
and $l=4.0^{+1.1}_{-1.4} \times 10^{10}$ cm $= 0.77^{+0.21}_{-0.26} R_{\rm{star}}$, respectively. 
Interestingly, this $l$ value of Flare 23 ($l= 0.15^{+0.05}_{-0.05} R_{\rm{star}}$) is very consistent with $L\sim8\times10^{9}$cm$\sim0.15R_{\rm{star}}$ of Flare 23 at the flare flux peak (Phase 3) on the basis of $T-\mathrm{EM}$ diagram (Figure \ref{fig:TEM_eachflare}(a)) in the above. 
On the other hand, the $l$ value of Flare 11 ($l= 0.77^{+0.21}_{-0.26} R_{\rm{star}}$) is bigger than $L\sim 1.5\times10^{10}$cm$\sim 0.29 R_{\rm{star}}$ of Flare 11 from $T-\mathrm{EM}$ diagram (Figure \ref{fig:TEM_eachflare}(c)), but the difference is still within the range of a factor of 2 or 3. 
It must be noted again \color{black} that both \color{black} estimation methods include various assumptions
\color{black} 
causing up to an order-of-magnitude uncertainty, as described in the above part of this subsection. \color{black}
As a result, we can suggest  \color{black} that both \color{black} estimation methods can provide similar rough estimates of the flare peak loop size scales, and that 
Flare 11, which has longer gradual durations in X-ray, can have larger
loop size than Flare 23 ($L\sim l \sim0.15^{+0.05}_{-0.05} R_{\rm{star}}$).

\subsection{Flare emissions in chromospheric lines: no clear blue/red wing asymmetries}
\label{sec:discuss_no_asymmetries}

As summarized in Section \ref{sec:intro}, blue/red wing asymmetries  
of optical chromospheric emission line profiles
(e.g., H$\alpha$, H$\beta$, Ca II lines) during M-dwarf flares 
have been intensively investigated \color{black} for years (e.g., 
\citealt{Houdebine+1990,Eason+1992,Gunn+1994,Crespo-Chacon+2006,Fuhrmeister+2008,Fuhrmeister+2011_A+A,Fuhrmeister+2018,Vida+2016,Vida+2019,Honda+2018,Muheki+2020_EVLac,Maehara+2021,Notsu+2024_ApJ,Inoue+2024_PASJ,Kajikiya+2025_ApJ_PaperII,Kajikiya+2025_ApJ_PaperI}), \color{black} and these are often discussed from the viewpoint of prominence/filament eruptions,
which can be related with stellar CMEs.
As described in Sections \ref{subsec:flare_atlas} and \ref{sec:remarkable_flares}, 
17 flares are identified in 
the H$\alpha$ and H$\beta$ lightcurves of the whole campaign (Figures \ref{fig:allEW_AUMic_XMM_opt}--\ref{fig:lc_HaHb_only_flares_Oct24-25}, see also Table 6 of T23), 
but none of these flares show clear blue/red wing asymmetries as seen 
in \color{black} the above previous studies of M-dwarf line asymmetries  \color{black}. 
As reported in Sections \ref{subsec:ana_flare_23} \& \ref{subsec:ana_flare_65} (and discussed in the following Section \ref{sec:discuss_line_broadening}),
Flares 22, 23, and 65 show remarkable (multi-wavelength) flare emissions with clear symmetric broadenings of various chromospheric lines, 
but none of these three flares show clear
\color{black} blue/red \color{black} wing asymmetries during their flare time evolutions. 
As mentioned in Section \ref{subsec:ana_flare_23}, 
Figure \ref{fig:flare23_HaSpec}(d) suggest the H$\alpha$ line profile 
at around the flare peak (Time [3]) may have slightly larger flux in the blue wing than in the red wing, but this potential blue wing asymmetry is too small to conclude that there is a clear blue wing enhancement as seen in the previous studies.

As another notable point, \citet{Veronig+2021} reported the potential X-ray dimming event just before two flares (Flares 35 and 36). 
No clear blue/red wing asymmetries of H$\alpha$ and H$\beta$ lines 
as reported in Section \ref{subsec:ana_flare_35_and_36}. 
This result suggests that signatures of the filament/prominence eruptions \color{black}
(cf. \citealt{Notsu+2024_ApJ,Kajikiya+2025_ApJ_PaperI} and the footnote in Section \ref{subsec:ana_flare_35_and_36})
\color{black}
are at least not clearly confirmed with this potential X-ray dimming event, which has been discussed as a stellar CME candidate \color{black} (\citealt{Veronig+2021,Veronig+2025_LRSP}).\color{black}

The recently found Neptune sized exoplanets have been making the AU Mic 
system as an important target for studying atmospheric loss on exoplanets (\citealt{Feinstein+2022_AJ,do_Amaral_2025_ApJ}), 
especially the effects of stellar coronal mass ejections (e.g., \citealt{Alvarado-Gome+2022_ApJ}). From this point,
the non-detection of clear blue/red wing asymmetries among the 17 H$\alpha$ \& H$\beta$ flares \color{black} can be \color{black} a notable result, but we cannot necessarily conclude that the AU Mic 
does not cause any prominence/filament eruptions only from the observation data of this study.
First, \citet{Vida+2019} reported at least one slow blue wing enhancement event from AU Mic.
Second, 
if AU Mic flares tend to occur in high latitude or stellar limb regions, 
the observed blue-shifted/red-shifted velocities could be smaller and 
could not be observed as distinguishable line asymmetries. 
\color{black} There are no established observational evidence of the flare locations 
of AU Mic and this possibility is only an hypothesis, which is very roughly
assumed by analogy with the flare observations of a different very active star (\citealt{Ilin+2021_MNRAS}), which reported the possibility of high latitude flares.
However, it is interesting to note \color{black} that 
starspot modeling results in \citet{Ikuta+2023_ApJ}  have at least 
suggested the possibility of high latitude active regions, although the spot modeling results include various uncertainties.
Third, according to several recent studies with the H$\alpha$ flare monitoring observations (\citealt{Notsu+2024_ApJ,Kajikiya+2025_ApJ_PaperI,Kajikiya+2025_ApJ_PaperII}), only $\sim$10--20\% of the flares show clear blue/red wing asymmetries. 
We might be able to speculate that the observations of only 17 H$\alpha$ flares in this study cannot be statistically significant, especially if we take into account the following potential difficulties.
Fourth, the flare line profile symmetric broadenings very clearly appear during some remarkable flares (see Flares 22, 23, and 65 reported in Sections \ref{subsec:ana_flare_23} \& \ref{subsec:ana_flare_65}), but these evident symmetric broadenings caused by flare emissions can make it difficult to detect weak asymmetric components 
of the H$\alpha$ and H$\beta$ lines.
\color{black} Fifth\color{black}, 
most of the H$\alpha$ monitoring observations have been mainly conducted 
for mid/late M-dwarfs \color{black} (e.g., \citealt{Vida+2016,Vida+2019,Fuhrmeister+2018,Muheki+2020_EVLac,Notsu+2024_ApJ,Kajikiya+2025_ApJ_PaperI})\footnote{\color{black} As also cited in Section \ref{subsec:discuss_quiescent},  \citet{Odert+2025_MNRAS} also conducted a flare spectroscopy monitoring program of AU Mic including the H$\alpha$ line, but they wrote that spectral line asymmetry will be analyzed in their next separate study. Future comparisons on line asymmetries between the results of our study and their next study are important. Also there are several recent studies  investigating line asymmetries of flares on G-dwarfs although the number of detected flares are still small 
(\citealt{Namekata+2022_NatAst,Namekata+2024_ApJ,Leitzinger+2020,Leitzinger+2024_MNRAS}). In future, comparisons with such G-dwarf studies would be also helpful how spectral types affect the existence of line asymmetries.}, 
\color{black} but we might speculate that these cooler M-dwarfs can have a larger H$\alpha$ emission contrast between the quiescent and blue/red-shifted components, compared with early M-dwarfs like AU Mic. 

\color{black} 
Here, we list five possibilities on why these 17 flares 
do not show any clear blue/red asymmetries.
However, in order to provide final conclusions, 
it is important to conduct more detailed radiation modeling of potential erupting prominences/filaments for early M-dwarfs like AU Mic 
(cf. modeling examples of mid M-dwarfs in \citealt{Leitzinger+2022}).
More detailed studies on the above listed points incorporating radiation modeling are necessary to clarify the potential stellar CME impacts on the AU Mic exoplanet system (cf. \citealt{Alvarado-Gome+2022_ApJ}).
Future more coronal dimming observations (e.g., ESCAPE mission; \citealt{France+2022_JATIS,Mason+2025_ApJ}) and simultaneous observations of 
chromospheric line blue wing asymmetries are also important.
\color{black}

\subsection{Flare emissions in chromospheric lines: multi-wavelength time evolutions and line profile broadenings}
\label{sec:discuss_line_broadening}

As described in Sections \ref{subsec:ana_flare_23} and \ref{sec:discuss_Xray_flare}, 
Flare 23 is the largest X-ray amplitude flare with the largest multi-wavelength coverage in this campaign, and this flare is categorized as a Neupert flare by T23. 
There are clear differences of the flare durations among the chromospheric lines 
(Figures \ref{fig:Flare23_TEM_multi_lc}).
The H$\alpha$ duration of Flare 23 is longer than that of the H$\beta$ line, 
and the similar trend is seen for Flare 65 in Section \ref{subsec:ana_flare_65}. 
This can be relevant/consistent with the results of mid M-dwarf flare observations in \citet{Kowalski+2013}, which have reported
the ``time decrement" trend of Balmer lines (see also \citealt{Kowalski+2019_ApJ_HST}). 
\citet{Kowalski+2013} also reported Ca II K flare emissions can have the delayed peaks and longer durations than those of the H$\alpha$ line (see also \citealt{Hawley+1991}), but those trends are not clearly seen in the Ca II 8542\AA~line emission of Flares 23 and 65 in this study: The durations of Ca II 8542\AA~line emission is shorter or comparable to those of the H$\alpha$ line.
Only from the current datasets, 
it is difficult to clarify whether this difference 
is caused by some difference in the
line formation process of these two lines during flares generally (Ca II K and Ca II 8542 \AA~lines), 
or we are seeing  differences between various flares.
The Na I D1\&D2 emission decays much faster than H$\alpha$, H$\beta$, and Ca II 8542\AA~lines during  both Flares 23 and 65, but in the case of He I D3 5876\AA~ lines, there is a difference between Flares 23 and 65: the He I D3 emission decays similarly with the Na I D1\&D2 emission during Flare 23, while the He I D3 emission lasts longer and also evolve similarly with the H$\beta$ line during Flare 65.
There is another notable difference between the H$\alpha$ and X-ray time evolutions of Flare 23. The flare emission peaks of the H$\alpha$ and the other chromospheric lines \color{black}
occur \color{black} earlier than that of the soft X-ray emission, but the H$\alpha$ and H$\beta$ emission decay times last longer than the soft X-ray emission (Figure \ref{fig:Flare23_TEM_multi_lc}(a)\&(c)). 
In addition, the H$\alpha$ and  H$\beta$ emission already gradually started to increase even before the NUV (UVW2) and optical emissions (at Time $<$3.3h in Figure \ref{fig:Flare23_TEM_multi_lc}(a)).
These time evolution differences of flare emissions among the different 
chromospheric lines, X-ray emission and NUV/optical continuum emissions \color{black} (e.g., why different chromospheric lines have diffrent peak times as described above) \color{black} could lead to important clues to establish self-consistent physical models of stellar flares, but there are not yet any comprehensive physical models that also explain the absolute and relative timescales of 
the various optical spectral variations 
(cf. Section 7.1 of \citealt{Kowalski+2024_LRSP}).
The flare chromospheric line emissions originate not only from the flare ribbon regions, but also from late-phase flare cool flare loops 
\color{black} have been recently discussed (e.g.,\citealt{Heinzel+2018,Wollmann+2023_A+A,Yang_Kai+2023_ApJ,Otsu+2024_ApJ,Odert+2025_MNRAS,Ichihara+2025_PASJ}) \color{black}, 
and the data of multiple chromospheric lines in this study 
could help further studies on this topic, 
including comparisons \color{black} 
with radiation modeling results and solar flare data analysis
(e.g., \citealt{Wollmann+2023_A+A,Otsu+2024_ApJ,Odert+2025_MNRAS}). \color{black}

During Flare 23 in Section \ref{subsec:ana_flare_23}, 
we found that 
accompanied with the NUV/optical continuum (e.g., UVW2, U-band) emissions, 
the H$\alpha$ and H$\beta$ lines show symmetric broadenings with $\sim\pm$400 km s$^{-1}$ and $\sim\pm$600 km s$^{-1}$, respectively, and these broadenings decay quickly as the NUV \& optical continuum emissions decay (Figures \ref{fig:maps_flare23}--\ref{fig:flare23_HbSpec}). 
The He I D3 5876\AA~and Na I D1 \& D2 lines also show symmetric broadenings with $\sim\pm 200-250$ km s$^{-1}$, while no notable broadenings are seen in the Ca II 8542 line (Figures \ref{fig:maps_flare23}, \ref{fig:flare23_HeNaSpec}, \& \ref{fig:flare23_Ca8542Spec}).
Flare 22 in Section \ref{subsec:ana_flare_23}, which is a smaller flare in terms of the flare peak fluxes and durations than Flare 23 (Figure \ref{fig:Oct13_TEM_multi_lc}), shows similar H$\alpha$ \& H$\beta$ broadenings but with smaller velocities, and they are similarly synchronized with the NUV/optical continuum emission increase (Figures \ref{fig:maps_flare23}--\ref{fig:flare23_HbSpec}). 
Flare 65 in Section \ref{subsec:ana_flare_65} also showed symmetric line broadenings in  the H$\alpha$ and H$\beta$ lines around the emission peaks like Flare 23 (Figures \ref{fig:maps_flare65} \& \ref{fig:spec_HaHb_flare65}), although the exact broadening velocities are slightly smaller than Flare 23 in the above.
The Na I D1\&D2 lines \color{black} of Flare 65 \color{black} show broadenings \color{black} as in the case of Flare 23\color{black}, but no broadening is confirmed for the He I D3 line (Figures \ref{fig:maps_flare65} \& \ref{fig:spec_HeNaCa8542_flare65}).
Unfortunately, there are no other multi-wavelength observation data for Flare 65 (e.g., optical/NUV continuum, X-ray emissions). In the following part of this subsection, we discuss these broadening properties especially for Flare 23 as a representative large flare  with the multi-wavelength coverage and wider line broadening velocities.

\begin{figure}[ht!]
   \begin{center}
      \gridline{
\fig{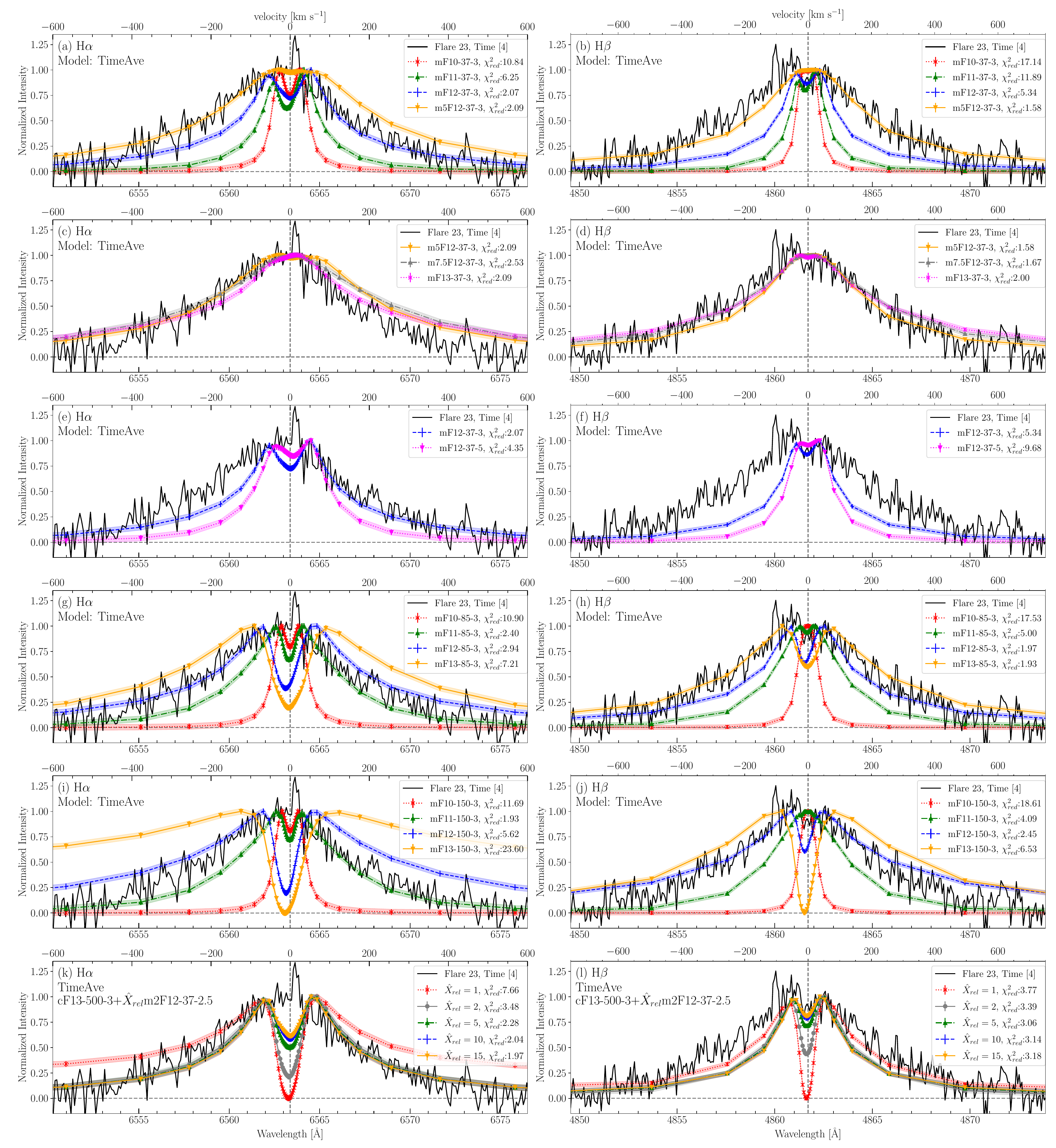}{1.0\textwidth}{\vspace{0mm}}
    }
     \vspace{-5mm}
     \caption{
Comparisons of the H$\alpha$ and H$\beta$ spectra between the observation 
(black solid lines, the quiescent phase component subtracted spectra at Time [4] of Flare 23 taken from Figures \ref{fig:flare23_HaSpec} \& \ref{fig:flare23_HbSpec}) and modeling results (colored).  
The observed spectra correspond to the peak of line broadenings 
during Flare 23 (cf. Figure \ref{fig:maps_flare23}), and 
the normalizations of the line center intensities are done with the Gaussian fitting around the line center components 
(for the range of $\pm$150 km s$^{-1}$ for H$\alpha$ and $\pm$200 km s$^{-1}$ for H$\beta$, respectively). 
The modeling results, which are normalized where the line emission is largest considering the existence of the central reversal components, 
correspond to the time average spectra from the M-dwarf RHD calculation results provided in 
\citet{Kowalski+2024_ApJ} (see the main text of Section \ref{sec:discuss_line_broadening} for more details). 
The error bars and shaded ranges correspond to the 1$\sigma$ error of the above Gaussian fittings to the observed line peak intensities around the line centers.
The modeling cases of \texttt{5F12-37-3} are plotted twice in (a) -- (d), so that the comparisons of different model cases become easier.
}
   \label{fig:HaHb_Obs_Model_TimeAve}
   \end{center}
 \end{figure}

The symmetric (or nearly symmetric) wing broadening of hydrogen Balmer
lines is a remarkable property of stellar flares (\citealt{Hawley+1991,Namekata+2020_PASJ,Kowalski_2022_FrASS}).
The H$\alpha$ and H$\beta$ broadenings of Flare 23 
synchronized with the evolution of optical/NUV continuum emissions 
can be interpreted as the results of intense heating 
by high-energy particles (e.g., electrons) in the flare impulsive phase \color{black} 
(e.g., ``Stark broadening", or electric pressure broadening of hydrogen,
by ambient charged particles (electrons, protons, ions) in the flaring chromosphere)\color{black} following previous theoretical/modeling investigations (see Section 10 of \citealt{Kowalski+2024_LRSP} for review). 
The assumption of the intense heating by high-energy particles  \color{black} is \color{black} consistent with the intense NUV/optical emissions 
and high-frequency gyrosynchrotron radio emissions during this flare (cf. T23, T25).
In order to investigate 
how flare heating parameters affect the observed line profile broadenings, we compare our observed line broadening profiles of Flare 23
with the results of a one-dimensional non-local thermodynamic equilibrium (NLTE) radiative-hydrodynamic (RHD) flare modeling using the \texttt{RADYN} code 
(\citealt{Carlsson_Stein_1992_ApJL,Carlsson_Stein_1995_ApJL,Carlsson_Stein_1997_ApJL,Carlsson_Stein_2002_ApJL,Allred+2015_ApJ,Carlsson+2023_A+A,Kowalski+2024_ApJ}). 
\citet{Kowalski+2024_ApJ} recently provided the time-dependent M-dwarf flare model grid results with various high-energy electron beam parameters \footnote{
The model grid outputs, which we also use in this study, are publicly available 
\color{black} at the Zenodo repository (\citealt{Kowalski+2024_Zenodo}).
}, which include line profile 
evolution results of chromospheric lines (e.g., H$\alpha$, H$\beta$, H$\gamma$ lines) 
with \texttt{TB09+HM88} opacity line profile calculations (\citealt{Tremblay+2009_ApJ}).
In Figure \ref{fig:HaHb_Obs_Model_TimeAve}, the modeling results are compared with 
the observed H$\alpha$ and H$\beta$ line profiles 
(the quiescent component subtracted spectra) at
Time [4] of Flare 23 (Figures \ref{fig:maps_flare23} -- \ref{fig:flare23_HbSpec}) when both the line broadenings and E.W. values of the H$\alpha$ and H$\beta$ lines are the largest.
The model IDs in Figure \ref{fig:HaHb_Obs_Model_TimeAve} correspond to those in
\citet{Kowalski+2024_ApJ}. For example, \texttt{mF10-37-3} is in the main (\texttt{m}) model group of \citet{Kowalski+2024_ApJ}, and has the injected energy flux density of $10^{10}$ erg cm$^{-2}$ s$^{-1}$ (\texttt{F10}) with the low-energy cut off of $E_{c}=37$ keV and the power-law index of $\delta=3$. The modeled profiles of the H$\alpha$ and H$\beta$ lines plotted in Figure \ref{fig:HaHb_Obs_Model_TimeAve} correspond to 
the time-averaged data over the total 10-sec modeling time of \citet{Kowalski+2024_ApJ},
in order to see rough overall trends how the line profile broadenings change depending on 
the input high energy electron parameters. 
We note that the time-averaged model result 
over 10-sec do not incorporate evolution of the particle acceleration properties over a longer flare duration.
Because of this, 
the modeling results presented here are just expected to simply represent
the broadening widths caused by the most dominant flare loop 
at around the peak time of H$\alpha$ and H$\beta$ line broadenings, and only the ``normalized" line profiles are compared in Figure \ref{fig:HaHb_Obs_Model_TimeAve} 
(the absolute line flux values are not discussed).
Future modeling studies considering multiple loop contributions can provide more realistic results also with the effects of time-evolution.

In Figure \ref{fig:HaHb_Obs_Model_TimeAve}, 
panels (a)--(f) show the modeling results with the low energy cutoff of 
$E_{c}=37$ keV, while panels (g)--(j) show those with higher energy cutoff values $E_{c}=$85 and 150 keV. 
Panels (a)--(d) show that the line broadening widths of both H$\alpha$ and H$\beta$ lines become larger as the energy flux density ($F$) becomes larger.
$\delta=3$ and $\delta=5$ cases with the same $E_{c}$ value and $F=1\times10^{12}$ erg cm$^{-2}$ are compared in panels (e) \& (f) 
\footnote{
\citet{Kowalski+2024_ApJ} mostly investigated the cases with $\delta=3$. \texttt{mF12-37-5} and \texttt{m5F10-37-5} are only available cases with $\delta\neq3$, and only the \texttt{mF12-37-5} case is used in the panels (e) \& (f) since \texttt{mF12-37-3} relatively match the observed profile in panels (a) \& (b).
}, and $\delta$ value also affects the broadening.
This is consistent with the result of \citet{Namekata+2020_PASJ}, which conducted the H$\alpha$ line modeling results of \texttt{mF10-37-5}, \texttt{mF12-37-5}, and 
\texttt{mF12-37-3}, but we show more modeling parameter cases and H$\beta$ results in this study. We can see the line profiles with 
$F=5\times10^{12}-1\times10^{13}$ erg cm$^{-2}$ s$^{-1}$ 
can roughly reproduce the observed 
wide wing broadenings of H$\alpha$ and H$\beta$ lines 
($\sim\pm$400 km s$^{-1}$ and $\sim\pm$600 km s$^{-1}$, respectively), 
although a single model cannot match ``both" of the H$\alpha$ and H$\beta$ line profiles. 
For example, the \texttt{mF12-37-3} model result can roughly match well with
the red wing of the H$\alpha$ line (panel (a)), but this model shows a much narrower profile
than the observed H$\beta$ profile (panel (b)). 
The higher energy cutoff cases ($E_{c}=$85 and 150 keV), 
which are often used to reproduce hot temperature components
of the flare optical 
continuum emissions (cf. \citealt{Kowalski+2024_ApJ}), show similar line broadenings with $F\sim10^{11}-10^{13}$ erg cm$^{-2}$ s$^{-1}$ cases, 
but the line center absorption components (central reversal components, cf. \citealt{Cram_Woods_1982_ApJ,Kowalski+2024_ApJ}) 
evidently seen in the modeling results are not consistent 
with the observed profiles of the H$\alpha$ and H$\beta$ lines. 

It is noted that there are slight line center absorption components in the original H$\alpha$ and H$\beta$ line profiles at Time [4], but they are not clearly seen if the quiescent component is subtracted (see Figures \ref{fig:flare23_HaSpec} \& \ref{fig:flare23_HbSpec}).
In addition, the H$\beta$ line at Time [3] of Flare 65 \color{black} shows \color{black} a potential ``dip" around the line center 
in the quiescent subtracted component (Figure \ref{fig:spec_HaHb_flare65}(h)). 
While it is possible that this potential dip of Flare 65 is similar to the models, 
further quantitative assessment of the continuum level increase at the line wavelengths is not possible without optical broadband photometry measurements 
(Unfortunately, Flare 65 has no photometry observation data of the NUV/optical continuum emissions).

Recently, \citet{Kowalski+2025_ApJ} reported superflares 
with largest NUV flare luminosity observed to date from an M star and their rising NUV flare spectra 
are not well represented by a single blackbody / modeling component. 
They discussed that these rising NUV spectra 
can be reproduced with the RADYN modeling results from the combination 
of \texttt{cF13-500-3} (larger energy 
flux with very large low-energy cutoff) and \texttt{m2F12-37-5} 
(smaller energy flux with smaller low-energy cutoff).
This newly introduced model case 
of \texttt{cF13-500-3}, which 
explains very bright FUV/NUV continuum components (e.g., \citealt{Kowalski+2025_ApJ}), is now 
used for the fitting the broadband continuum data of this Flare 23 of AU Mic (Kowalski et al. in prep.).
Then, the modeling cases (\texttt{cF13-500-3} $+\ \hat{X}_{rel}$ \texttt{m2F12-37-5})\footnote{\color{black} $\hat{X}_{rel}$ is the ratio of best-fit filling factors of the two models (see \citealt{Kowalski+2025_ApJ} for details)} are 
also plotted in panels (k) and (l) of Figure \ref{fig:HaHb_Obs_Model_TimeAve},
All the cases with $\hat{X}_{rel}\geq2$ roughly reproduce the observed H$\alpha$ line broadenings at the wings but existence of the central reversals in the modeled profiles are very different from the observed profiles, as also seen in the above cases of $E_{c}=$85 and 150 keV (panels (g)--(j)).
The central reversal properties should be associated 
with the formation height differences between the line wing and line center (cf. \citealt{Kowalski+2024_ApJ}) and can be a subject of the future more detailed modeling papers.

These comparison results have suggested some of the electron beam parameters that have been 
used to reproduce M-dwarf NUV/optical continuum emissions (e.g., \citealt{Kowalski+2024_ApJ,Kowalski+2025_ApJ}) can roughly reproduce existence of 
very large line broadenings of the H$\alpha$ and H$\beta$ line profiles seen around the peak of a superflare on AU Mic (Flare 23). 
It is noted that these electron beam flux densities ($F\sim5\times10^{12}-10^{13}$ erg cm$^{-2}$ s$^{-1}$) are much stronger than those assumed for normal solar flares ($F\sim10^{10}-10^{11}$ erg cm$^{-2}$ s$^{-1}$), and this is considered to be associated with the strong magnetic field flux densities of energetic M-dwarf flares (cf. \citealt{Kowalski+2024_LRSP}).
Although the line broadenings are roughly reproduced, 
there are many future research topics for further quantitative investigations \footnote{\color{black} It is also noted that the number of modeling cases plotted in Figure \ref{fig:HaHb_Obs_Model_TimeAve} have limitations since we simply used the modeling cases already provided in \citet{Kowalski+2025_ApJ}. 
 After the future topics raised here are resolved, it is also interesting to increase the number of modeling case results with slightly different input parameters.
.}
The line-shape parameters of \texttt{TB09+HM88} are used in the modeling results 
from \citet{Kowalski+2024_ApJ}, but this should be updated with newer parameters 
(cf. \citealt{Cho+2022_ApJ}) and will be the subject of future work (A. F. Kowalski \& T. A. Gomez, in preparation). 
The absolute flare peak flux values in UVW2, $U$, and $V$ bands
can provide another constraint to the electron beam flux parameters.
This will be also the subject of our upcoming paper of the AU Mic project papers (after T23, T25, and this Paper III), and the results should be compared with the input parameters used in Figure \ref{fig:HaHb_Obs_Model_TimeAve}.

\section{Summary and Conclusions} \label{sec:summary}

A unique 7-day multiwavelength flare campaign of AU Mic, a young active M1-dwarf with exoplanets, was conducted to investigate energetic flares across the electromagnetic spectrum.
In this Paper III, we presented soft X-ray spectral analysis results from the XMM-Newton data, and chromospheric line profile evolutions from the SMARTS/CHIRON optical high-dispersion spectroscopic data. In total, 38 flares are identified in the X-ray data (Figures \ref{fig:allEW_AUMic_XMM_opt}--\ref{fig:X-ray_Ha_obsid_0822740601_lc}), while 17 flares are identified in H$\alpha$ and H$\beta$ lines (Figures \ref{fig:allEW_AUMic_XMM_opt} -- \ref{fig:lc_HaHb_only_flares_Oct24-25}).
Among these flares, 
the four types of flares are highlighted as remarkable flares in this study.

\begin{itemize}

\item Flare 23: the large Neupert-type flare with the largest multi-wavelength coverage in this campaign and clear symmetric (or nearly symmetric) broadenings of chromospheric lines (Section \ref{subsec:ana_flare_23}). 
\item Flares 11\&15: The two long duration X-ray flares
(Section \ref{subsec:ana_flare_11_and_15}).
\item Flares 35\&36: Two flares preceding the potential X-ray dimming event, originally reported in \citet{Veronig+2021} (Section \ref{subsec:ana_flare_35_and_36}).
\item Flare 65: Another energetic flare showing clear broadenings of chromospheric lines (Section \ref{subsec:ana_flare_65}).

\end{itemize}
The key findings of this study are summarized in the following.

\begin{enumerate} 
\renewcommand{\labelenumi}{(\arabic{enumi})}

\item 
The quiescent (non-flare) H$\alpha$ and H$\beta$ emission components show quasi-periodic 
variations, which roughly correspond to the rotational period of AU Mic ($\sim$4.863 days) (Figure \ref{fig:TESS_comparison_rotation}). The quiescent X-ray lightcurve also show some variability but the periodicity is not clear within the limit of the 7-day data of this study. More simultaneous high-precision optical photometry (e.g., TESS data) with H$\alpha$, H$\beta$, and X-ray observations are needed for characterization on correlations/anti-correlations of the rotational modulations in these different wavelengths (see Section \ref{subsec:discuss_quiescent}).

\item
Spectral analysis of quiescent component X-ray data are conducted in Section \ref{subsec:X-ray_specana_quiescent}. The abundances of the quiescent phase spectra tend to show the 
so-called Inverse FIP effect (
\color{black}
Figures \ref{fig:specfit_QuieALL1_0822740301}(j) \& 
\ref{fig:specfit_QuieALL1_0822740401} (j) -- 
\ref{fig:specfit_QuieALL1_0822740601} (j)
\color{black}
). The quiescent phase $T-\mathrm{EM}$ relation is discussed 
with the theoretical scaling laws proposed by \citet{Takasao+2020_ApJ}, which incorporated 
the effects of the size distribution of magnetic features on the stellar surface (Figure \ref{fig:quie_T_EM}). As a result, it may be suggested that X-ray active stars like AU Mic have much more (e.g., $\sim$100$\times$) active regions for a given size than the Sun,
though there are some limitations and uncertainty of the scaling-law assumptions (see Section \ref{subsec:discuss_quiescent_X-ray}). 
\color{black} 
This potential confirmation of a large number of active regions on the early M-dwarf AU Mic 
is consistent with the results of \citet{Takasao+2020_ApJ}, which suggested the same conclusions for active G-dwarfs.
\color{black} 

\item
The time-averaged spectral analysis of the X-ray flares are conducted by using the PN and MOS1 spectra of XMM-Newton in Section \ref{subsec:X-ray_specana_time-average}. The flare RGS spectra will be discussed in a future paper. The X-ray energy (0.2 -- 12 keV range) of the 38 flares are in the range of 
$E_{\rm{X}}= 2\times 10^{31} - 4\times 10^{33}$ erg (Figure \ref{fig:plot_time_risedecay_flarelist}). 
Variety of flare energy partitions and Neupert responses are seen (as discussed in T23).
For example, there are both flares with $E_{\rm{X}}>E_{\rm{UVW2}}+E_{\rm{U}}+E_{\rm{V}}$ 
and $E_{\rm{X}}<E_{\rm{UVW2}}+E_{\rm{U}}+E_{\rm{V}}$. 
By applying the flare $T-\mathrm{EM}$ scaling relations from \citet{Shibata_Yokoyama+2002},
the average X-ray temperature ($T_{\rm{ave}}$) and total emission measure values
of the 38 X-ray flares suggest that most of the flares locate in the range of 
$B \sim$ 50G  -- 1.5 kG and $L \sim 3\times10^{8}$ cm -- $2\times 10^{10}$ cm 
= 0.006 -- 0.38 $R_{\rm{star}}$, although large uncertainty (e.g., up to 
order-of-magnitude error) should be noted (Figure \ref{fig:plot_time_risedecay_flarelist}).

\item
Assuming the scaling relations of \citet{Shibata_Yokoyama+2002},
the $T-\mathrm{EM}$ evolution of flares 
can also provide the information of magnetic field and flare loop size scale evolutions, although large uncertainty should be noted.
In the case of Flare 23,
the flare starts with $B\sim600$G and $L\sim2\times10^{9}$cm $\sim0.04R_{\rm{star}}$,
and then the flare loop size becomes bigger at the flare peak 
(e.g., $B\sim400$G and $L\sim8\times10^{9}$cm$\sim0.15R_{\rm{star}}$).
As the flare X-ray emission decays after the peak (``early" decay phase of the flare), 
sizes of the dominant flare loops still continue to expand (e.g., from $L\sim8\times10^{9}$cm to $L\sim3\times10^{10}$cm$\sim0.57R_{\rm{star}}$) 
while the magnetic field becomes smaller ($B\sim$400 G to $B\sim$65 G) (Figure \ref{fig:TEM_eachflare}). 
The overall $T-\mathrm{EM}$ evolution track of this clear Neupert-flare Flare 23 
\color{black} is \color{black} consistent with the flare evolution track described as 
``Flare H-R diagram" in \citet{Shibata_Yokoyama+2002}, 
which follows the standard model of solar and stellar flares (see Section \ref{sec:discuss_Xray_flare}). 

\item
Flares 11 and 15 show X-ray lightcurves 
more gradual and have longer durations than Flare 23. 
The $T-\mathrm{EM}$ relations of these two flares show smaller peak temperature,
the smaller magnetic fields, and larger size scales (flare loops), compared with Flare 23 (Figure \ref{fig:TEM_eachflare}). 
These differences may reflect the relations between flare impulsiveness 
and flare structures/properties, although statistical characterization with more flare events is necessary for further discussions.

\item 
The coronal loop sizes of Flares 23 and 11 are also estimated by
using the method of \citet{Reale+1997_A&A}, from the decay phase X-ray lightcurves.
As a result, both methods (\citealt{Shibata_Yokoyama+2002} and \citealt{Reale+1997_A&A}) 
\color{black} provide \color{black} consistent 
rough estimates of the flare peak loop sizes, and it becomes more probable that Flare 11, which has longer gradual durations in X-ray, can have larger loop size than Flare 23  (see Section \ref{sec:discuss_Xray_flare}).

\item
None of the 17 H$\alpha$ \& H$\beta$ flares show clear blue/red 
wing asymmetries during their flare time evolutions.
As a notable point, \citet{Veronig+2021} reported the potential X-ray dimming event just before two remarkable flares (Flares 35 and 36). 
The analysis result in Section \ref{subsec:ana_flare_35_and_36} suggests that signatures of the filament/prominence eruptions (cf. \citealt{Notsu+2024_ApJ,Kajikiya+2025_ApJ_PaperI} and references therein) are at least not clearly confirmed with this potential X-ray dimming event, which has been discussed as stellar CME candidates (\citealt{Veronig+2021}).
In spite of these results, we cannot necessarily conclude that the AU Mic 
does not cause any prominence eruptions or filament eruptions only from the observation data of this study, 
and further observations and \color{black} comprehensive \color{black} modeling studies are important \color{black} 
to clarify the real occurrence frequency of prominence/filament eruptions 
\color{black}
and to understand potential stellar CME impacts on the AU Mic exoplanet system (See Section \ref{sec:discuss_no_asymmetries}).

\item 
The multi-wavelength time evolutions of Flares 23 (Figure \ref{fig:Flare23_TEM_multi_lc}) and 65 (Figure \ref{fig:lc1_flare65}) show differences among different chromospheric lines and soft X-ray emission. For example the H$\alpha$ emission last longer than other chromospheric lines and soft X-ray emission. The time evolution differences of multi-wavelength flare emissions could lead to important clues to establish comprehensive physical models of stellar flares, while there are not yet any comprehensive physical models that explain the absolute and relative timescales of the various optical spectral variations (see Section \ref{sec:discuss_line_broadening}).

\item 
During Flare 23 in Section \ref{subsec:ana_flare_23}, 
we found that accompanied with the NUV/optical continuum (e.g., UVW2, U-band) emissions, 
the H$\alpha$ and H$\beta$ lines show symmetric broadenings with $\pm$400 km s$^{-1}$ and $\pm$600 km s$^{-1}$, respectively, and these broadenings decay quickly as the NUV/optical continuum emissions decay (Figures \ref{fig:maps_flare23}--\ref{fig:flare23_HbSpec}). 
The He I D3 5876\AA~and Na D1 \& D2 lines also show symmetric broadenings with $\sim200-250$ km s$^{-1}$ around the flare peak, while no notable broadenings are seen in Ca II 8542 line (Figures \ref{fig:maps_flare23}, \ref{fig:flare23_HeNaSpec}, \& \ref{fig:flare23_Ca8542Spec}). Flare 65 showed similar line broadenings, although the exact broadening velocities are slightly smaller than Flare 23 (Figures \ref{fig:maps_flare65} -- \ref{fig:spec_HeNaCa8542_flare65}).

\item
The H$\alpha$ and H$\beta$ broadenings of Flare 23 
synchronized with the optical/NUV continuum emission evolution 
can be interpreted as the results of intense heating 
by high-energy particles (e.g., electrons) in the flare impulsive phase. 
We compared the observed H$\alpha$ \& H$\beta$ line profiles with 
the M-dwarf RHD flare model grid results from \citet{Kowalski+2024_ApJ}.
As a result, some of the electron beam heating parameters 
that have been used to reproduce M-dwarf NUV/optical continuum emissions 
can roughly reproduce the existence of these large line broadenings of the H$\alpha$ and H$\beta$ line profiles seen around the peak of a superflare on AU Mic (Flare 23).  These electron beam flux densities ($F\sim5\times10^{12}-10^{13}$ erg cm$^{-2}$ s$^{-1}$) are much stronger than those assumed for normal solar flares ($F\sim10^{10}-10^{11}$ erg cm$^{-2}$ s$^{-1}$), and this is considered to be associated with the strong magnetic field flux densities of energetic M-dwarf flares.
There are many future research topics for further quantitative investigations beyond this paper (e.g., broadening of other lines, updates of line shape parameters, electron beam parameters to reproduce the flare NUV/optical continuum emission, multiple loop effects).

\end{enumerate}

%% Also note that the akcnowlodgment environment does not support long amounts of text. If you have a lot of people and institutions to acknowledge, do not use this command. Instead, create a new \section{Acknowledgments}.
\begin{acknowledgments}
We would like to thank Drs. Brad D. Carter, John P. Wisniewski , Eliot H. Vrijmoet , Graeme L. White , Todd J. Henry, Rodrigo H. Hinojosa, Wei-Chun Jao, Jamie R. Lomax, James E. Neff, Leonardo A. Paredes, and Jack Soutter
for their contributions during early parts of the 2018 AU Mic Campaign.
We also would like to appreciate Mr. Shun Inoue and Dr. Teruaki Enoto for the discussions on X-ray spectral analysis, and Dr. Kazunari Shibata for the discussions on X-ray flare loop evolutions.
Y.N. also would like to acknowledge the the relevant discussions in the International Space Science Institute (ISSI)
Workshop ``Stellar Magnetism and its Impact on (Exo)Planets (\url{https://workshops.issibern.ch/stellar-magnetism/})" (held on June 2-6, 2025), which helped to revise this paper.

We acknowledge funding support through NASA ADAP award Number 80NSSC21K0632, NASA XMM-Newton Guest Observer AO-17 Award 80NSSC19K0665, and 
NSF/AGS Award 1916509.
Y.N. also acknowledges funding support from NASA TESS Cycle 6 Program 80NSSC24K0493, 
NASA TESS Cycle 7 Program 80NSSC25K7906,
NASA NICER Cycle 6 Program 80NSSC24K1194, 
NASA Swift Cycle 20 Program 80NSSC25K7378, 
and STScI HST GO Program 17464. 

This work is based on observations obtained with XMM-Newton, an ESA science mission with instruments and contributions directly funded by ESA Member States and NASA.
This research also used data from the SMARTS 1.5m\&0.9m telescopes at Cerro Tololo Inter-American Observatory (CTIO), and this telescope is operated as part of the SMARTS Consortium.  This research makes use of observations from the Las Cumbres Observatory (LCO) global telescope network. We used the LCO observation time allocated to the University of Colorado.  
We used the data from the VLA, which is a facility of the National Science Foundation operated under cooperative agreement by Associated Universities, Inc.
This paper also includes data collected with the TESS mission, obtained from the
Mikulski Archive for Space Telescopes (MAST) at the Space
Telescope Science Institute (STScI). 
We sincerely appreciate all the consortium/observatory scientist/staff members for their large contributions in carrying out our observations.

\end{acknowledgments}

\facilities{XMM,CTIO:1.5m,CTIO:0.9m,LCOGT,VLA}

\appendix

\section{Additional figures and tables of the X-ray spectral analysis 
and CHIRON optical spectra}
\label{appen_sec:additional_figures}

\begin{figure}[ht!]
   \begin{center}
      \gridline{
\fig{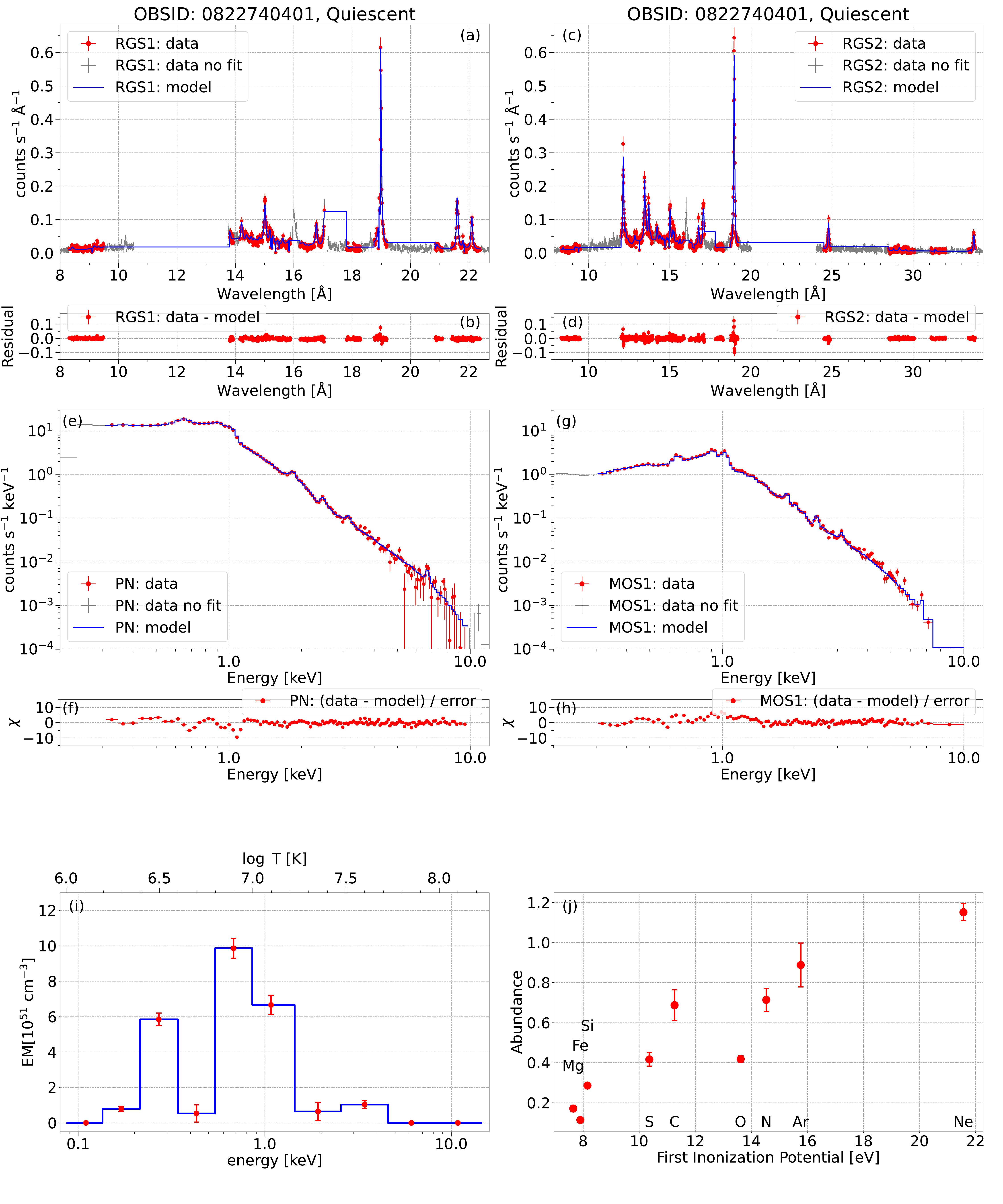}{1.0\textwidth}{\vspace{0mm}}
    }
     \vspace{-5mm}
     \caption{
    Same as Figure \ref{fig:specfit_QuieALL1_0822740301}, but for Obs-ID 0822740401 (cf. Figure \ref{fig:X-ray_Ha_obsid_0822740401_lc}).}
   \label{fig:specfit_QuieALL1_0822740401}
   \end{center}
 \end{figure}

\begin{figure}[ht!]
   \begin{center}
      \gridline{
\fig{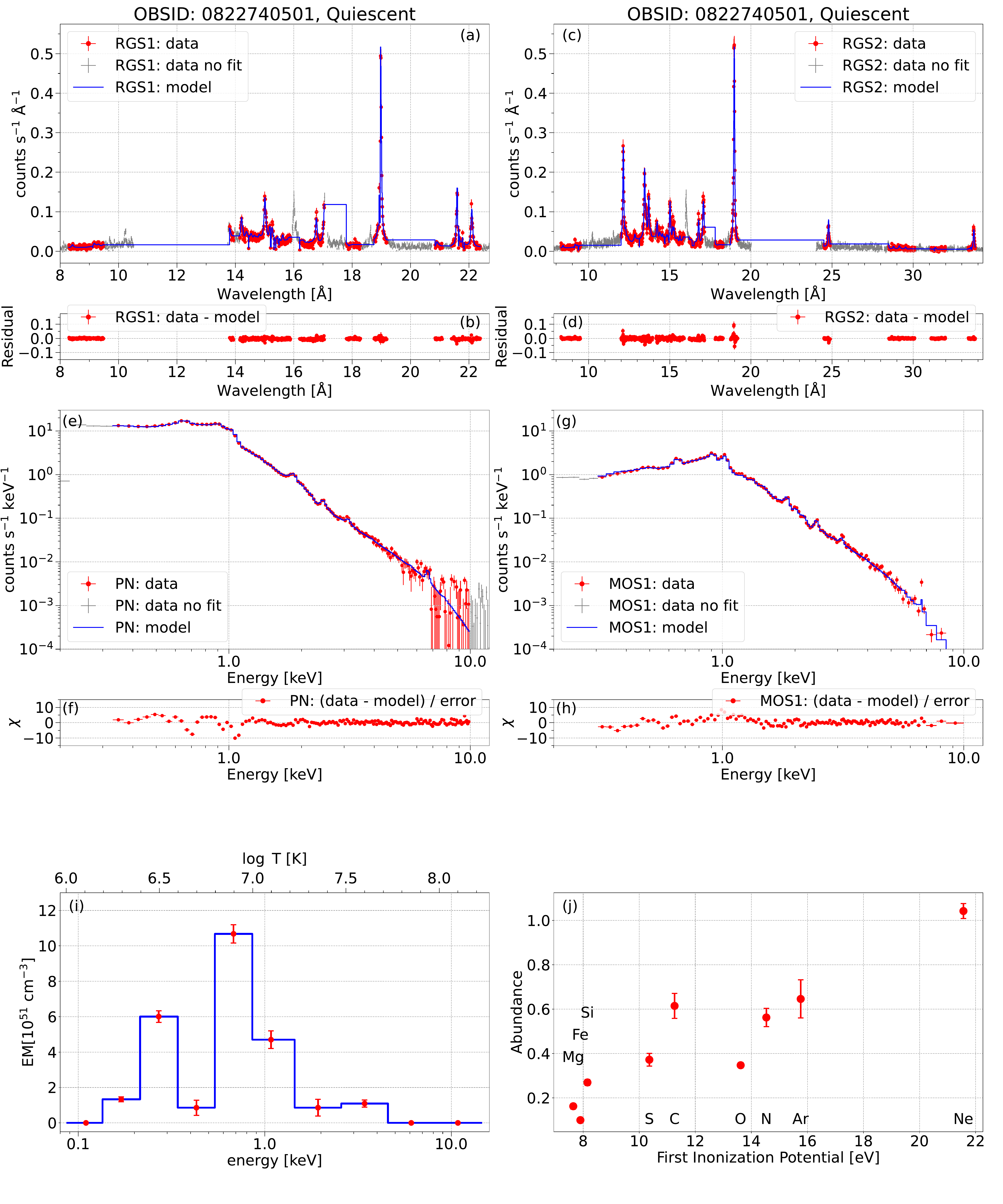}{1.0\textwidth}{\vspace{0mm}}
    }
     \vspace{-5mm}
     \caption{
    Same as Figure \ref{fig:specfit_QuieALL1_0822740301}, but for Obs-ID 0822740501 (cf. Figure \ref{fig:X-ray_Ha_obsid_0822740501_lc}).}
   \label{fig:specfit_QuieALL1_0822740501}
   \end{center}
 \end{figure}

\begin{figure}[ht!]
   \begin{center}
      \gridline{
\fig{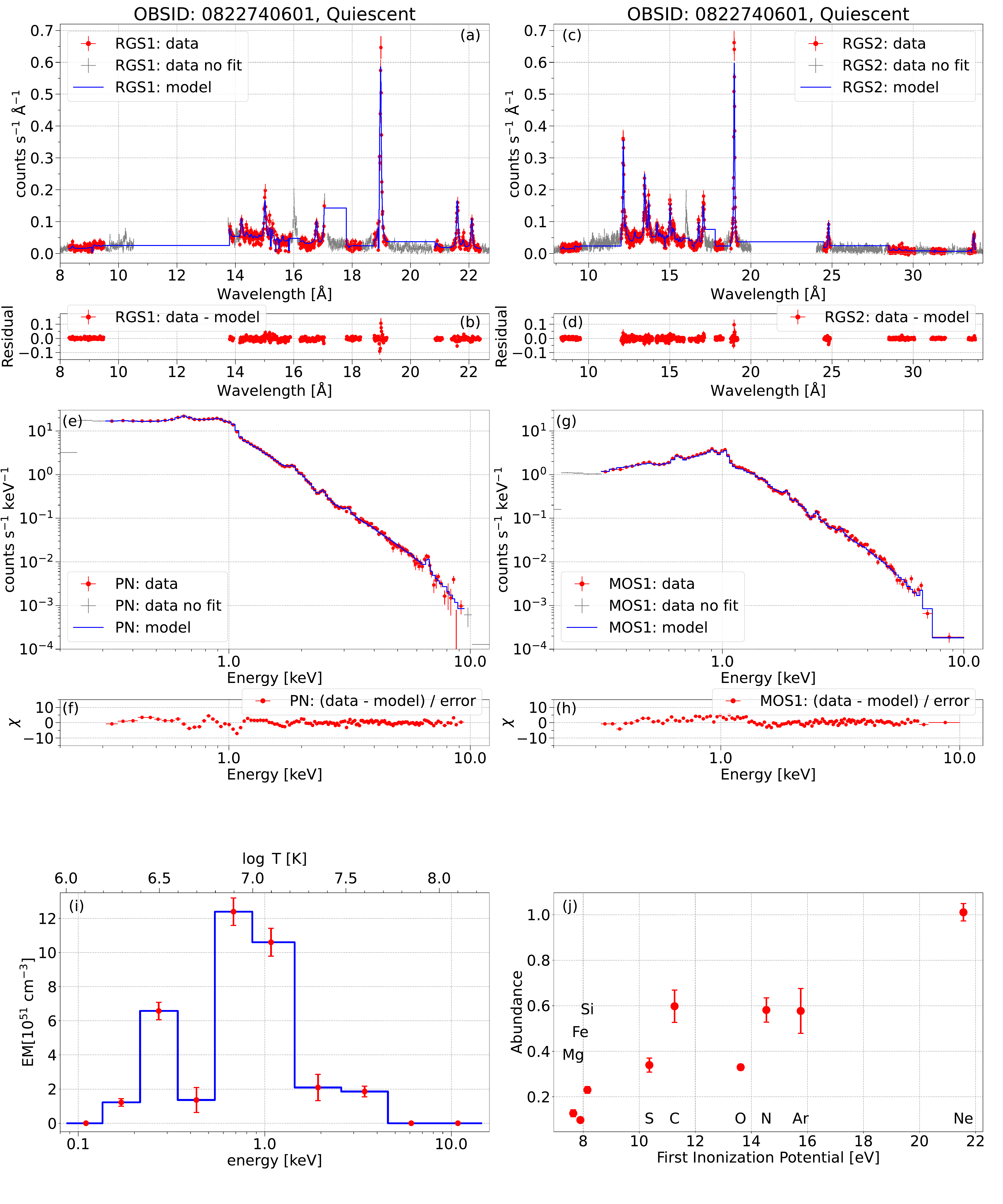}{1.0\textwidth}{\vspace{0mm}}
    }
     \vspace{-5mm}
     \caption{
    Same as Figure \ref{fig:specfit_QuieALL1_0822740601}, but for Obs-ID 0822740601 (cf. Figure \ref{fig:X-ray_Ha_obsid_0822740601_lc}).}
   \label{fig:specfit_QuieALL1_0822740601}
   \end{center}
 \end{figure}

\begin{figure}[ht!]
   \begin{center}
      \gridline{
\fig{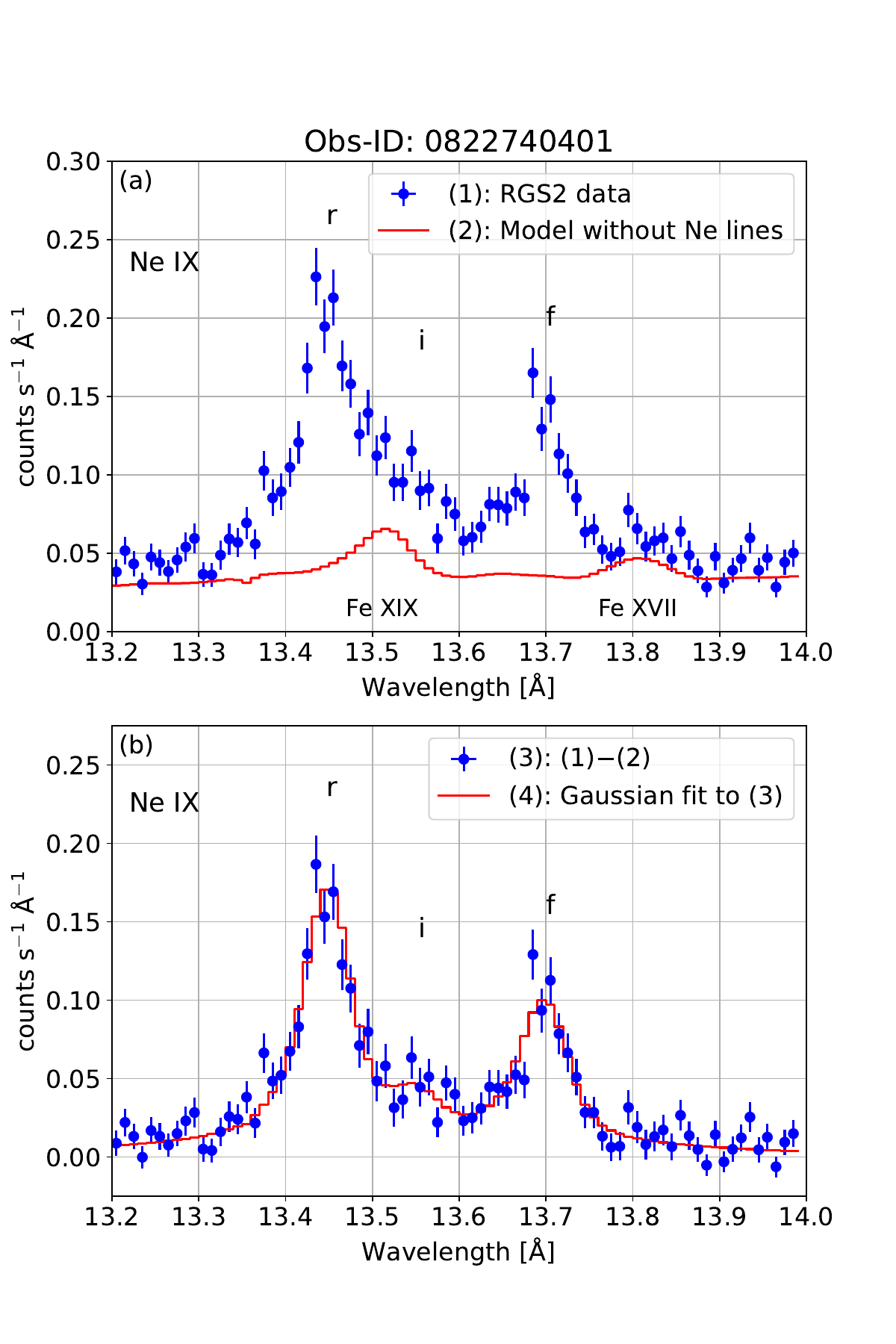}{0.5\textwidth}{\vspace{0mm}}
\fig{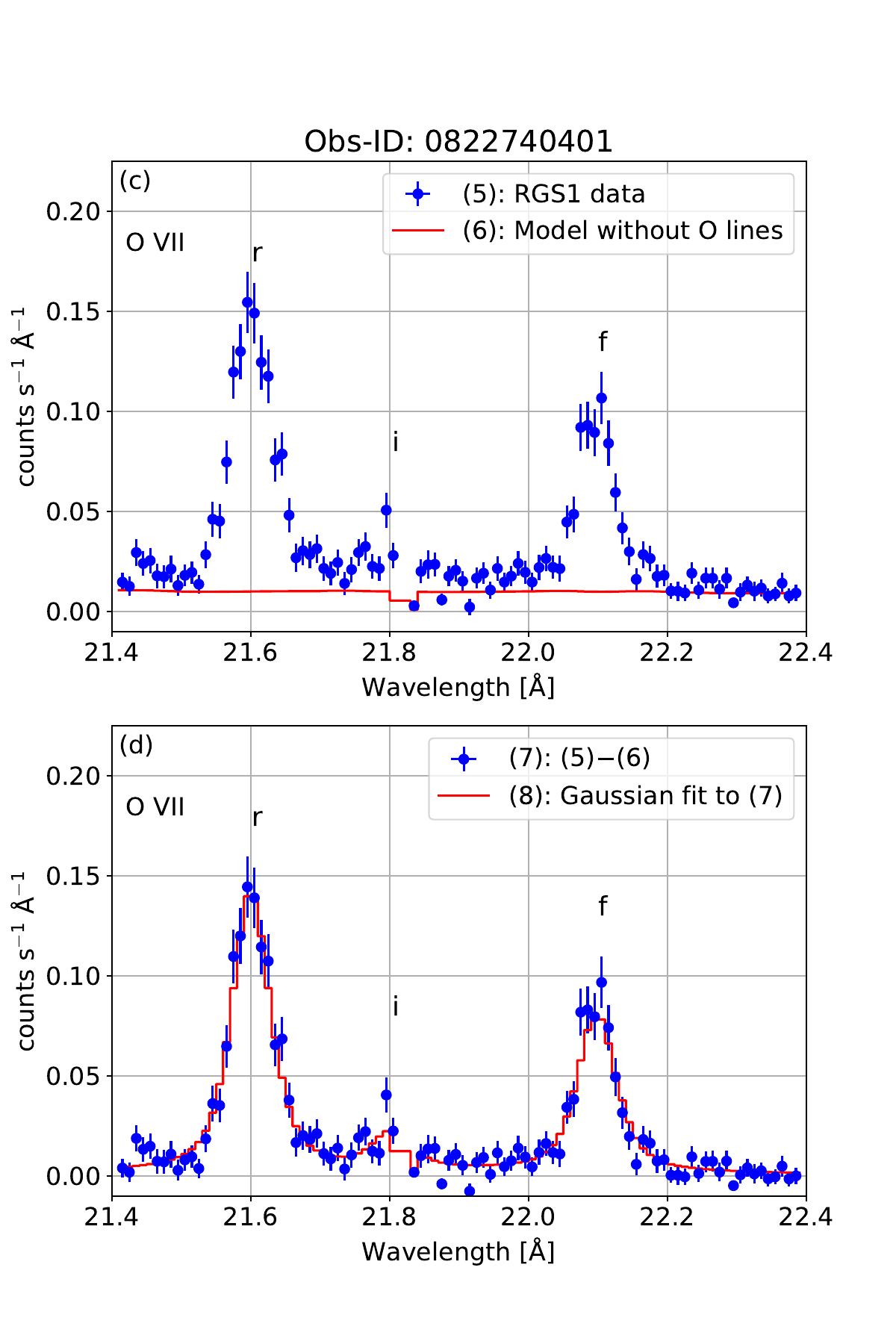}{0.5\textwidth}{\vspace{0mm}}
    }
     \vspace{-5mm}
     \caption{
Same as Figure \ref{fig:quieden_0822740301} but for Obs-ID 0822740401.
}
   \label{fig:quieden_0822740401}
   \end{center}
 \end{figure}

\begin{figure}[ht!]
   \begin{center}
      \gridline{
\fig{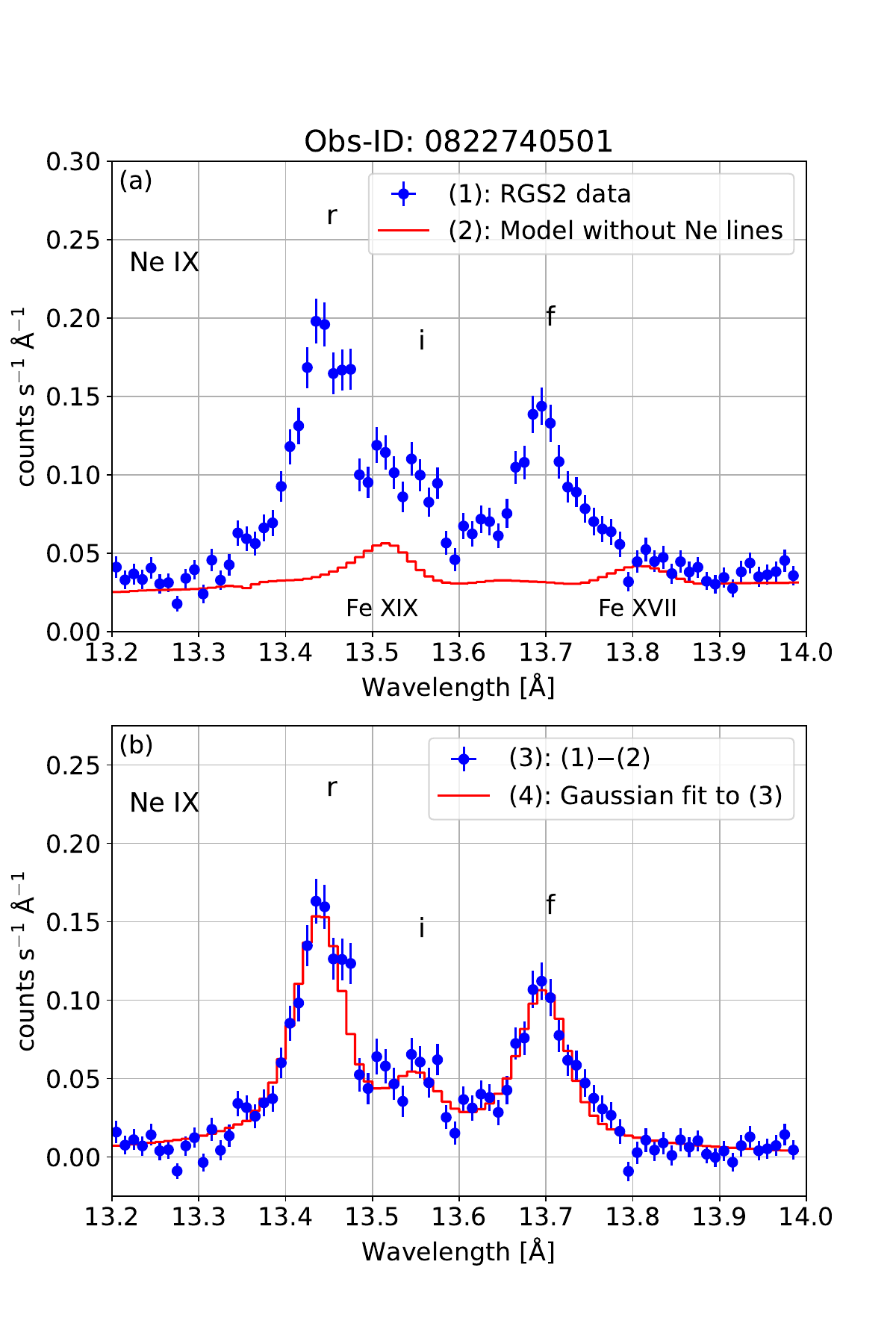}{0.5\textwidth}{\vspace{0mm}}
\fig{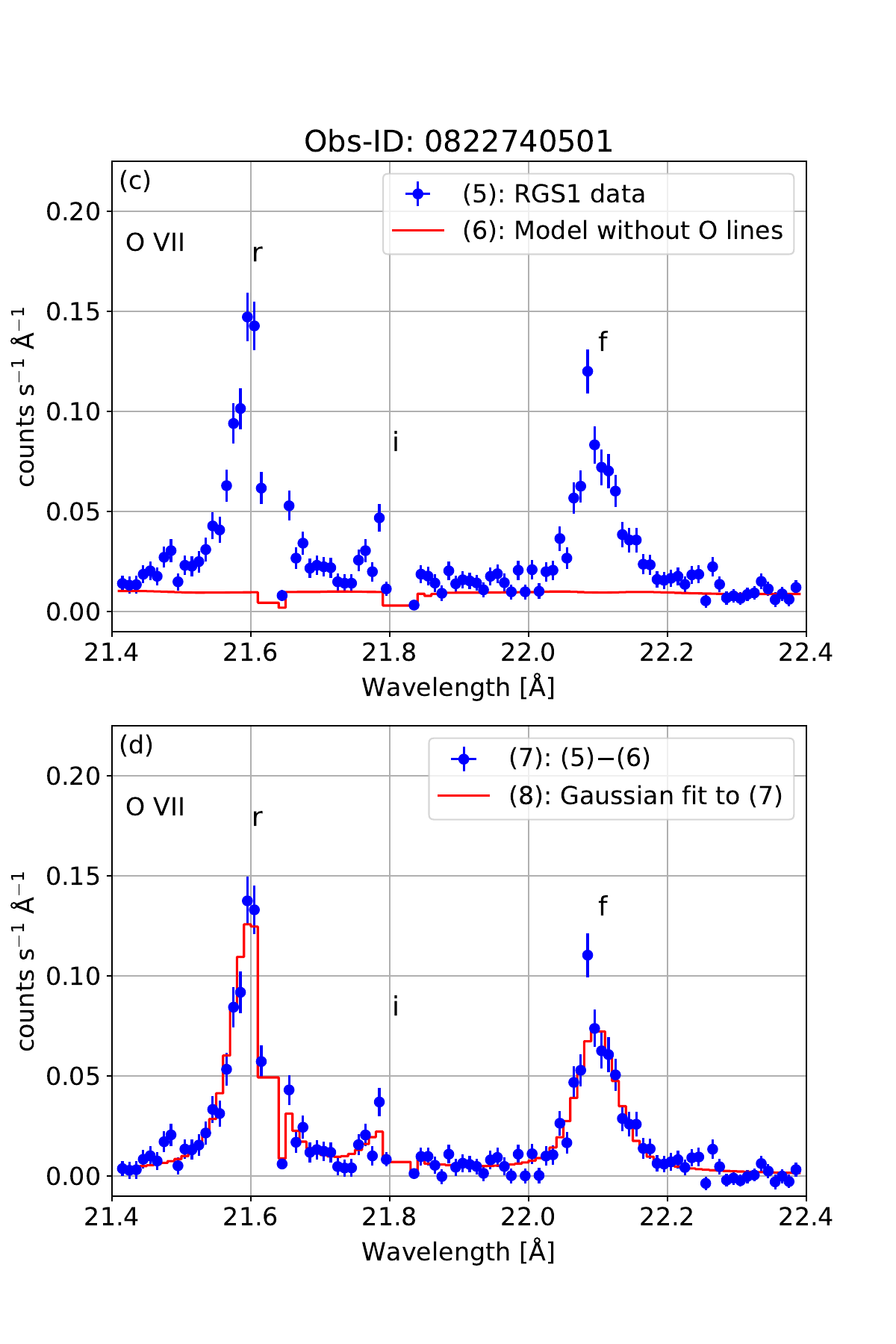}{0.5\textwidth}{\vspace{0mm}}
    }
     \vspace{-5mm}
     \caption{
Same as Figure \ref{fig:quieden_0822740301} but for Obs-ID 0822740501.
}
   \label{fig:quieden_0822740501}
   \end{center}
 \end{figure}

\begin{figure}[ht!]
   \begin{center}
      \gridline{
\fig{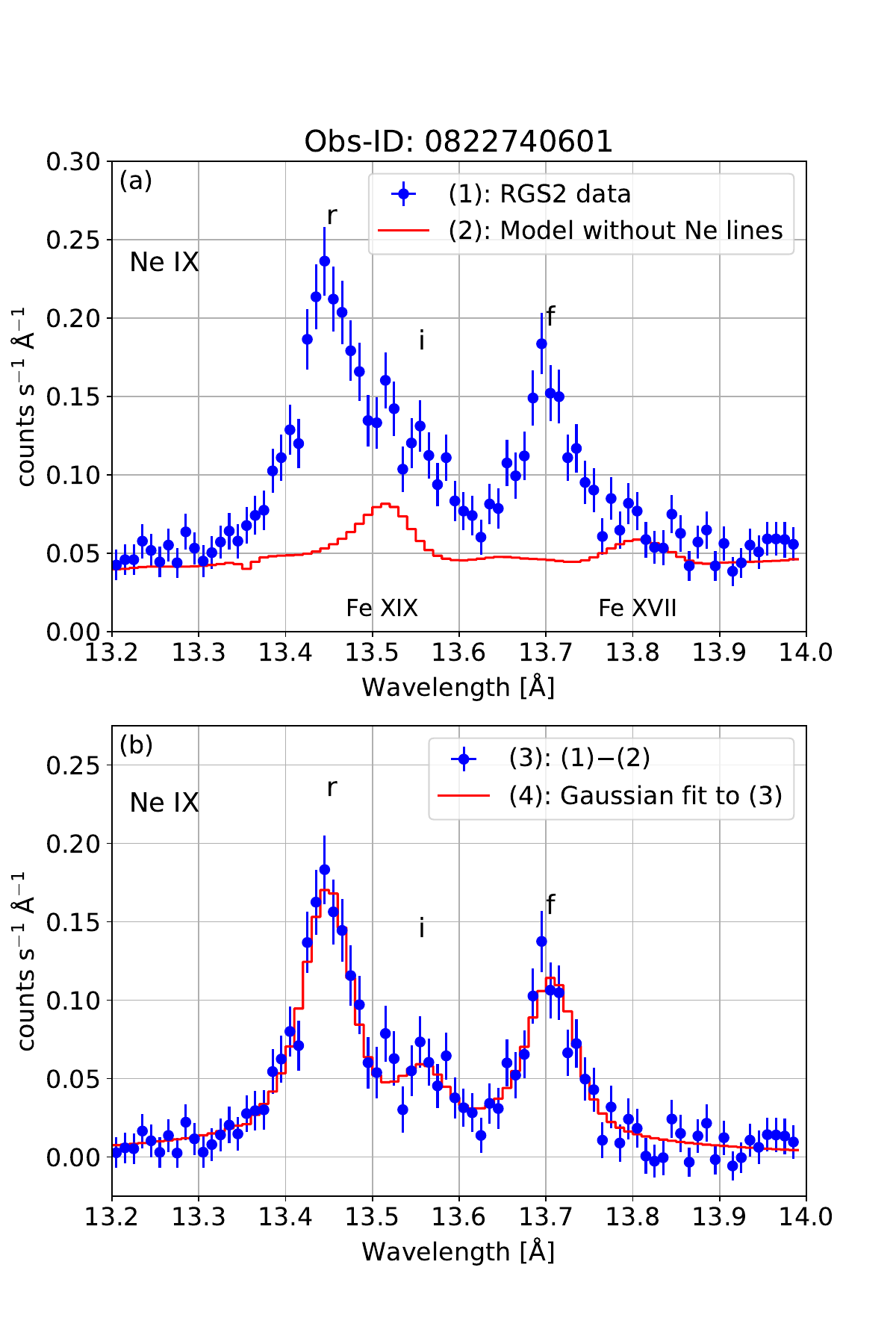}{0.5\textwidth}{\vspace{0mm}}
\fig{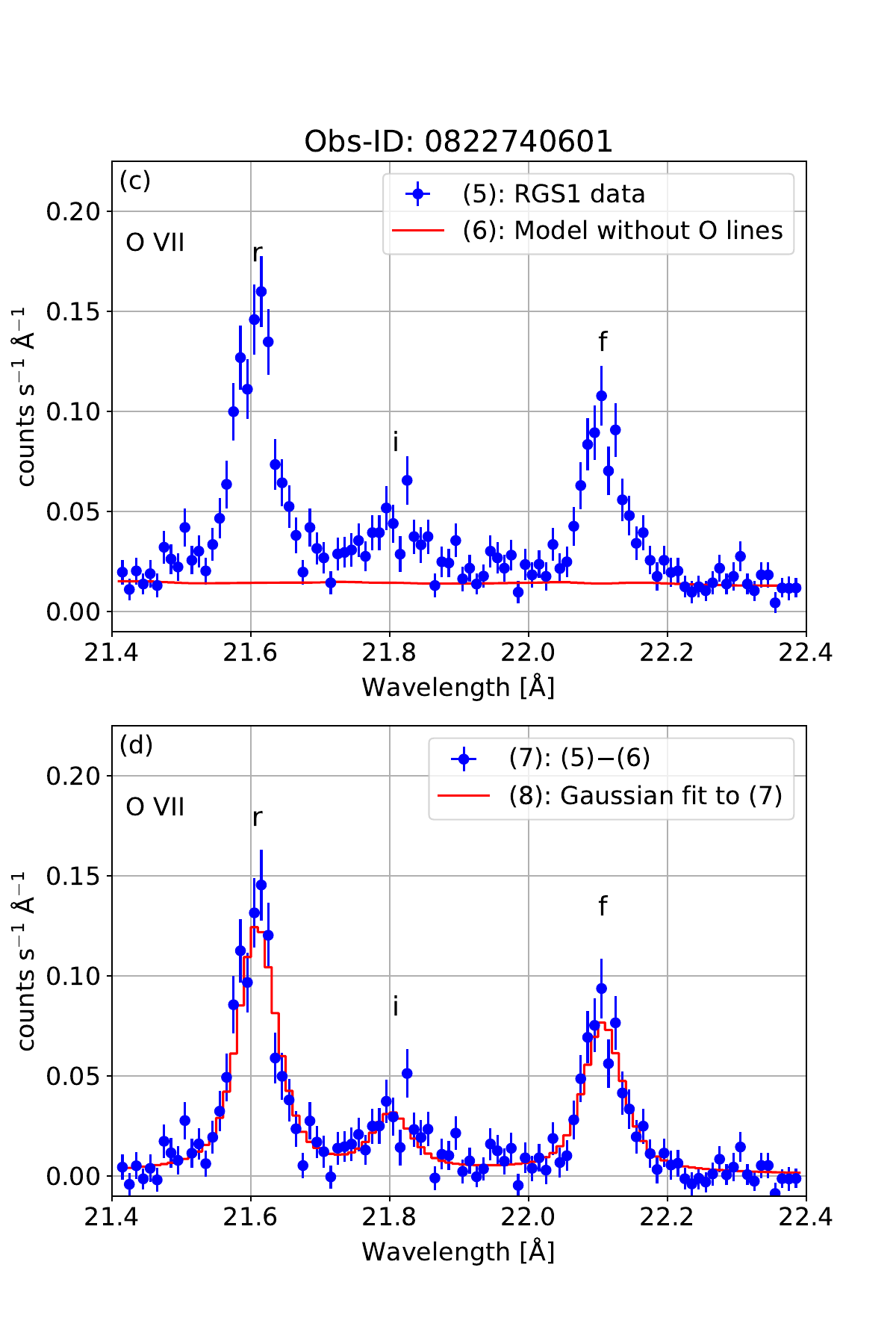}{0.5\textwidth}{\vspace{0mm}}
    }
     \vspace{-5mm}
     \caption{
Same as Figure \ref{fig:quieden_0822740301} but for Obs-ID 0822740601.
}
   \label{fig:quieden_0822740601}
   \end{center}
 \end{figure}

\begin{figure}[ht!]
   \begin{center}
      \gridline{
\fig{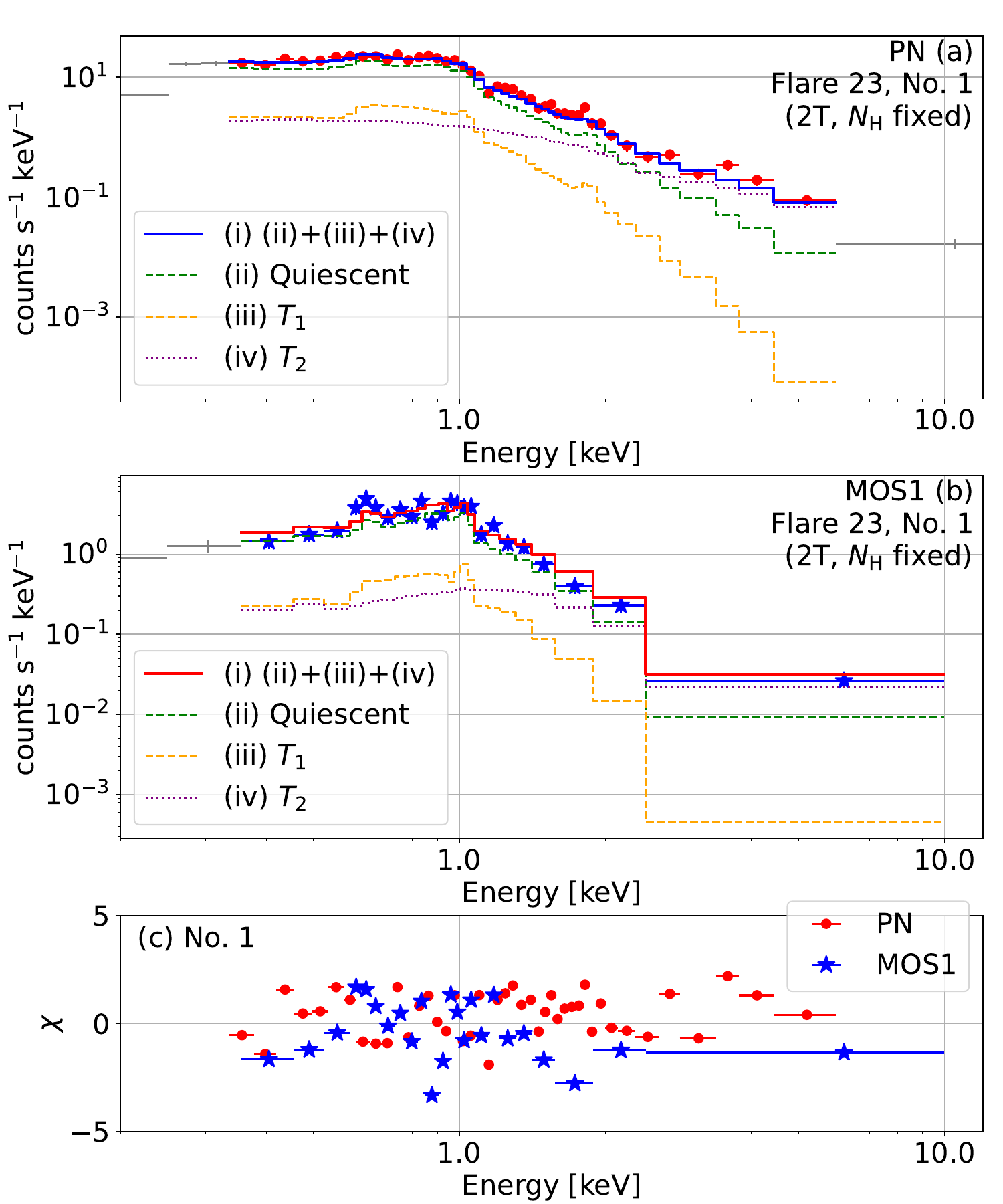}{0.5\textwidth}{\vspace{0mm}}
\fig{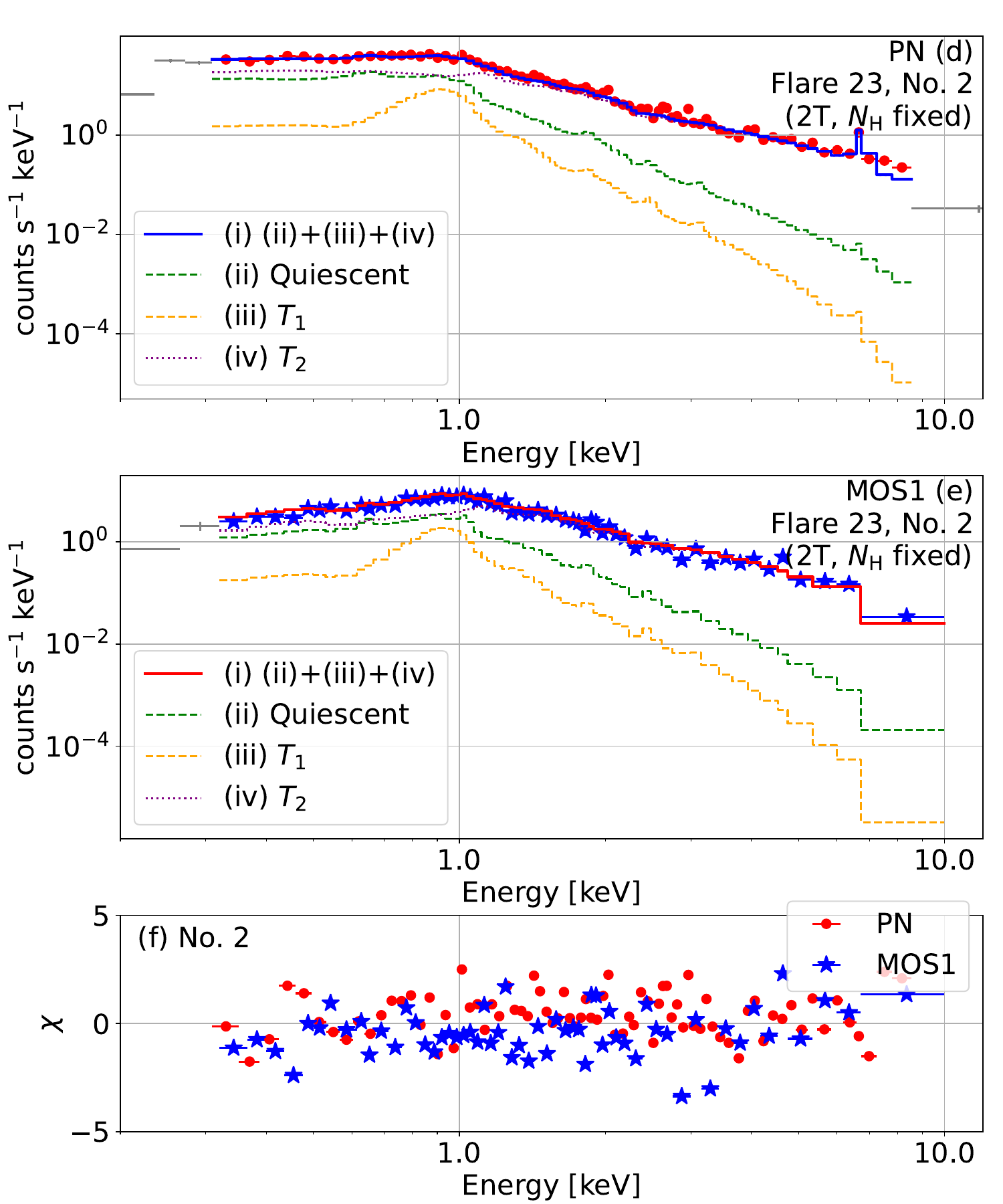}{0.5\textwidth}{\vspace{0mm}}
    }
     \vspace{-5mm}
      \gridline{
\fig{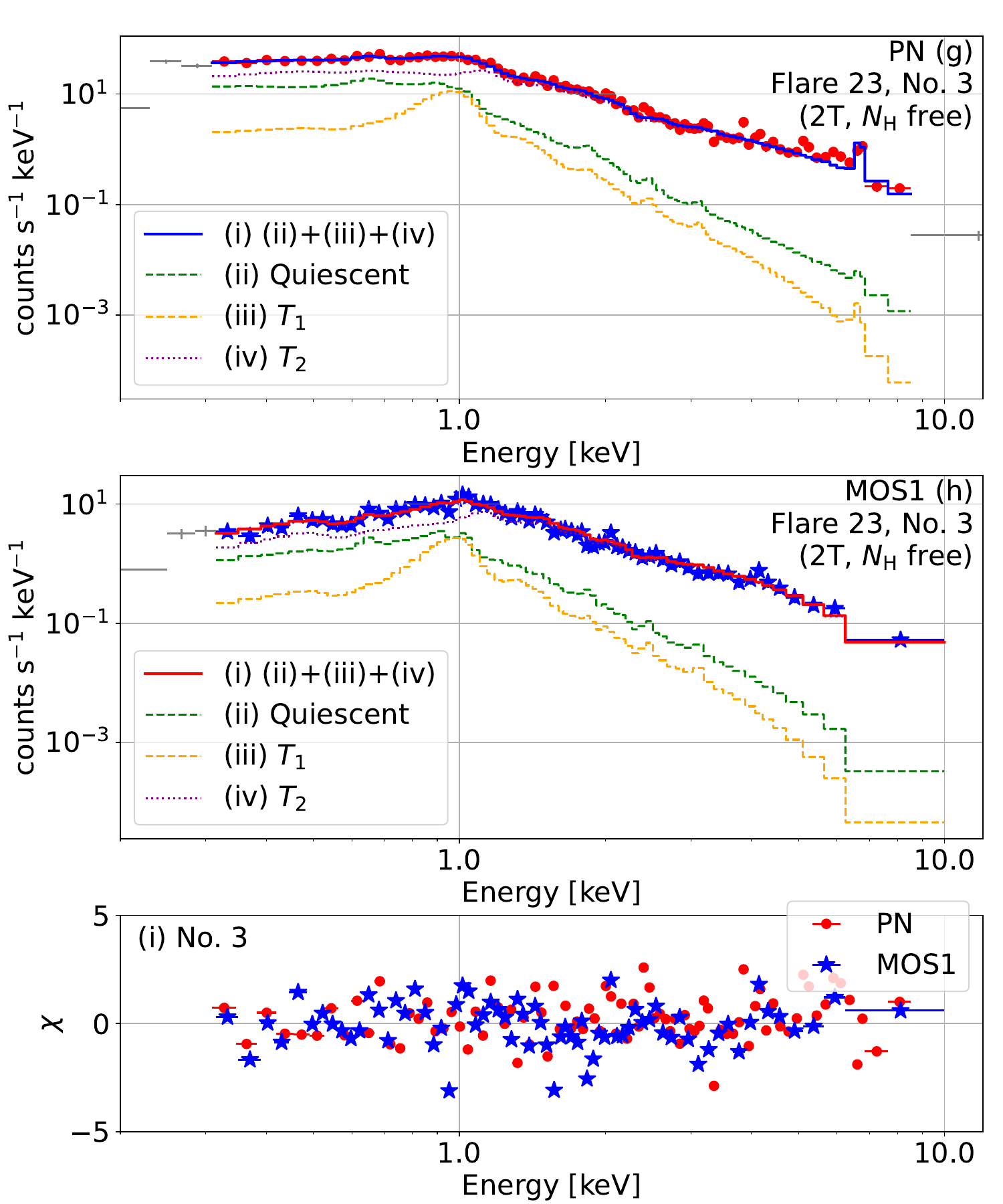}{0.5\textwidth}{\vspace{0mm}}
\fig{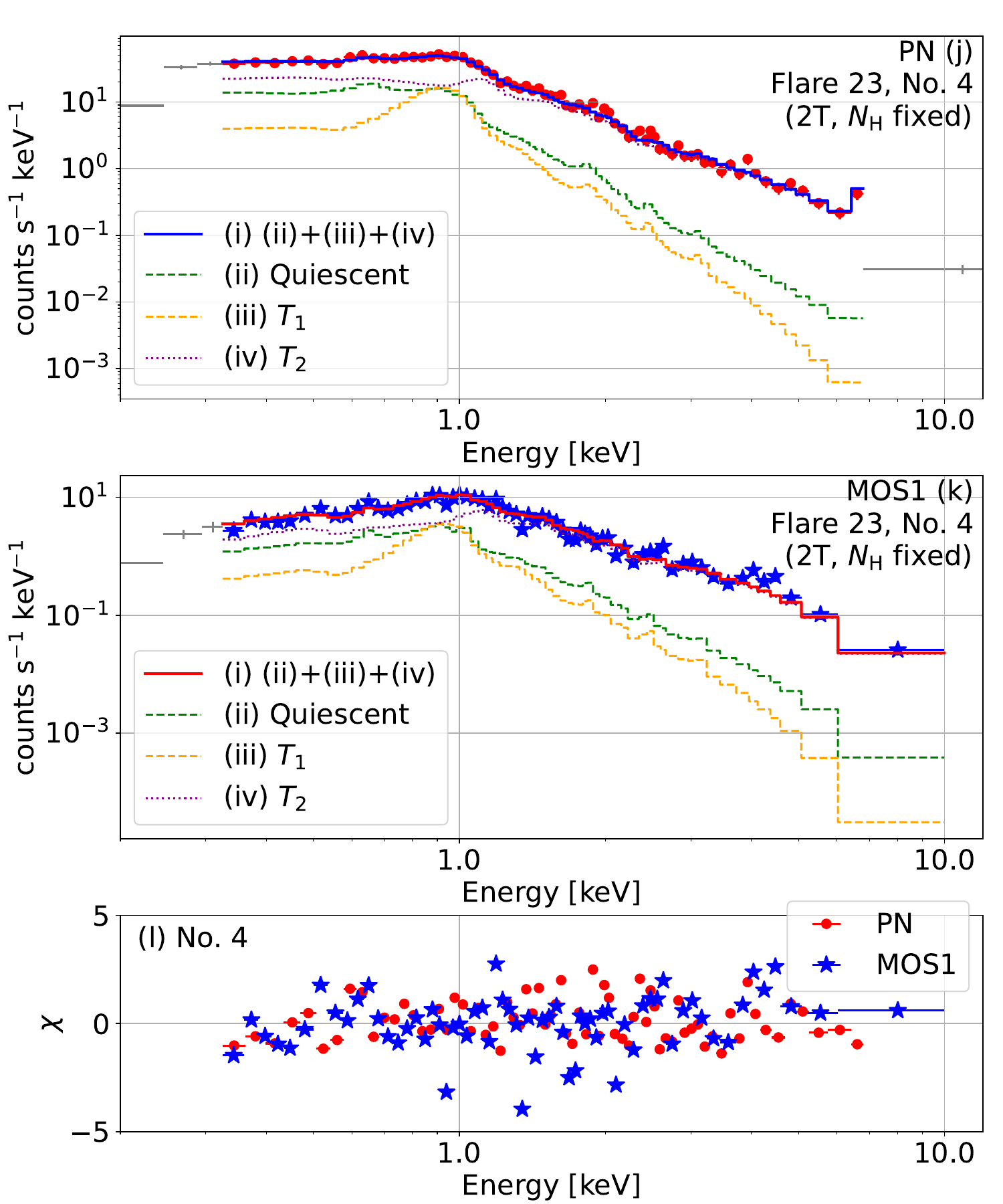}{0.5\textwidth}{\vspace{0mm}}
    }
     \vspace{-5mm}
     \caption{
The PN and MOS1 spectra and best-fit model result for the Phases No. 1 -- No. 4 (cf. Figure \ref{fig:Flare23_TEM_multi_lc}) of Flare 23.
The data and fit results are plotted with the same way as Figure \ref{fig:RiseDecayFit_fig1_Flare22}.
}
   \label{fig:specfig_Flare23_TEM_quie_each_No.1-No.4}
   \end{center}
 \end{figure}

\begin{figure}[ht!]
   \begin{center}
      \gridline{
\fig{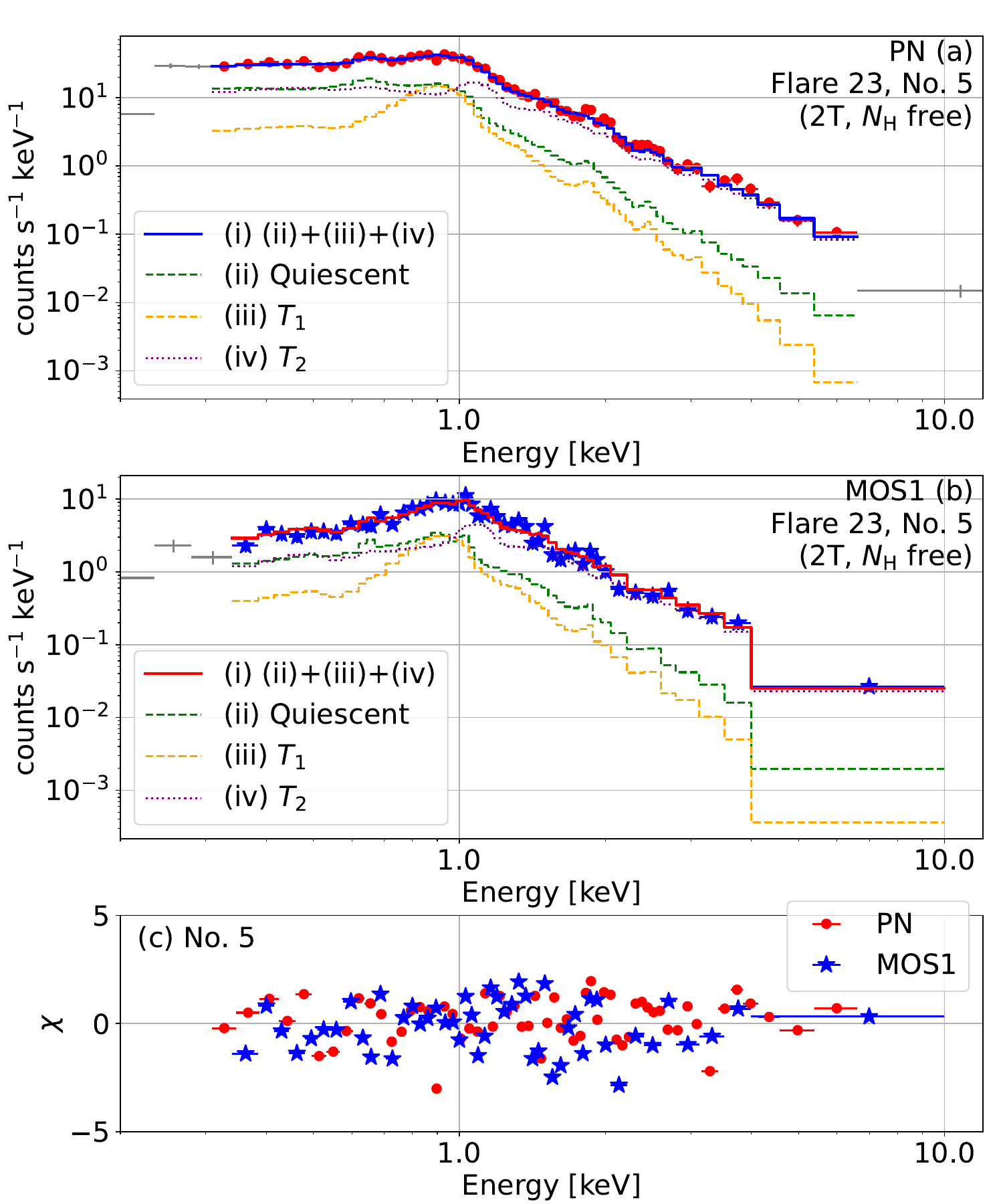}{0.5\textwidth}{\vspace{0mm}}
\fig{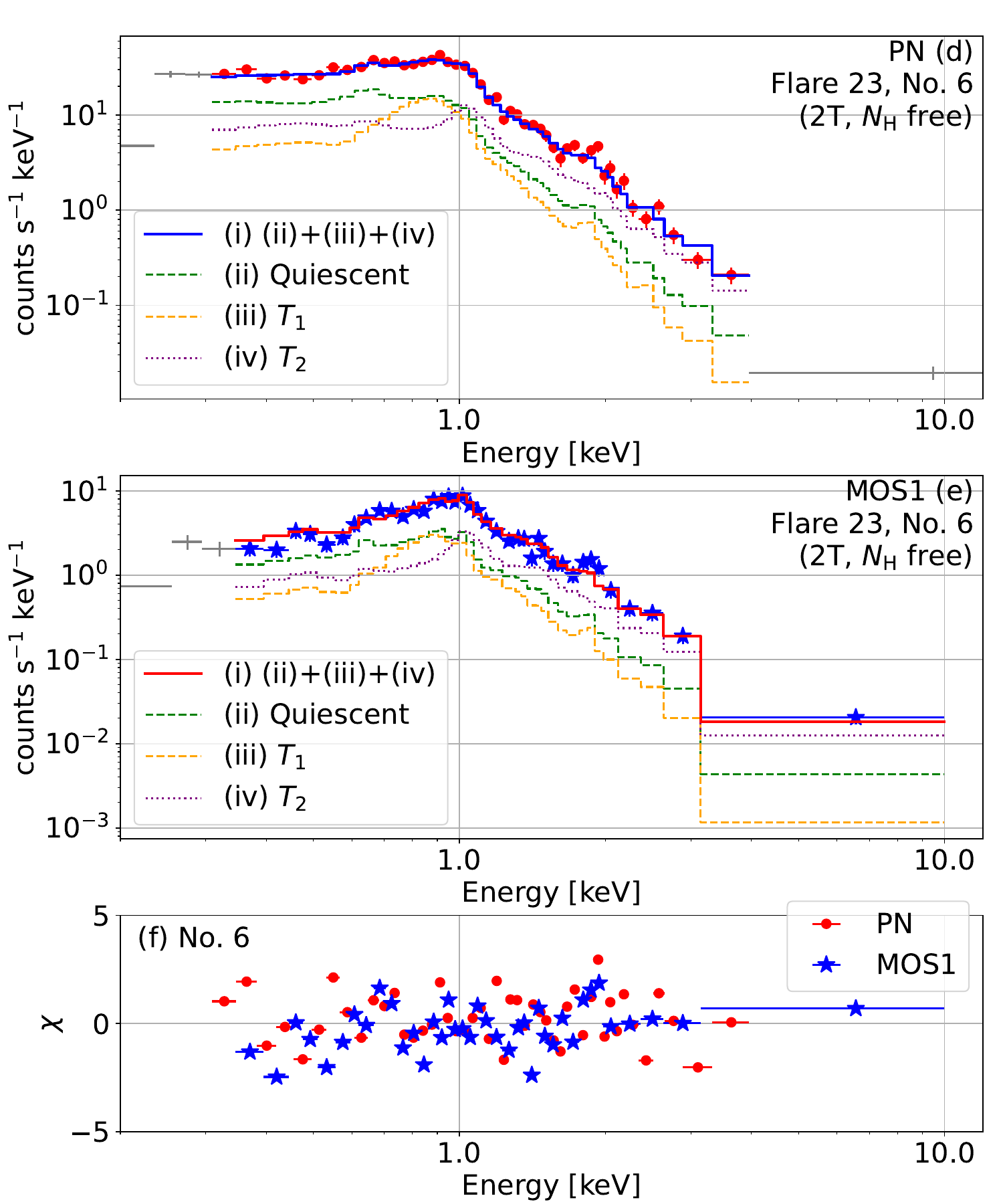}{0.5\textwidth}{\vspace{0mm}}
    }
     \vspace{-5mm}
      \gridline{
\fig{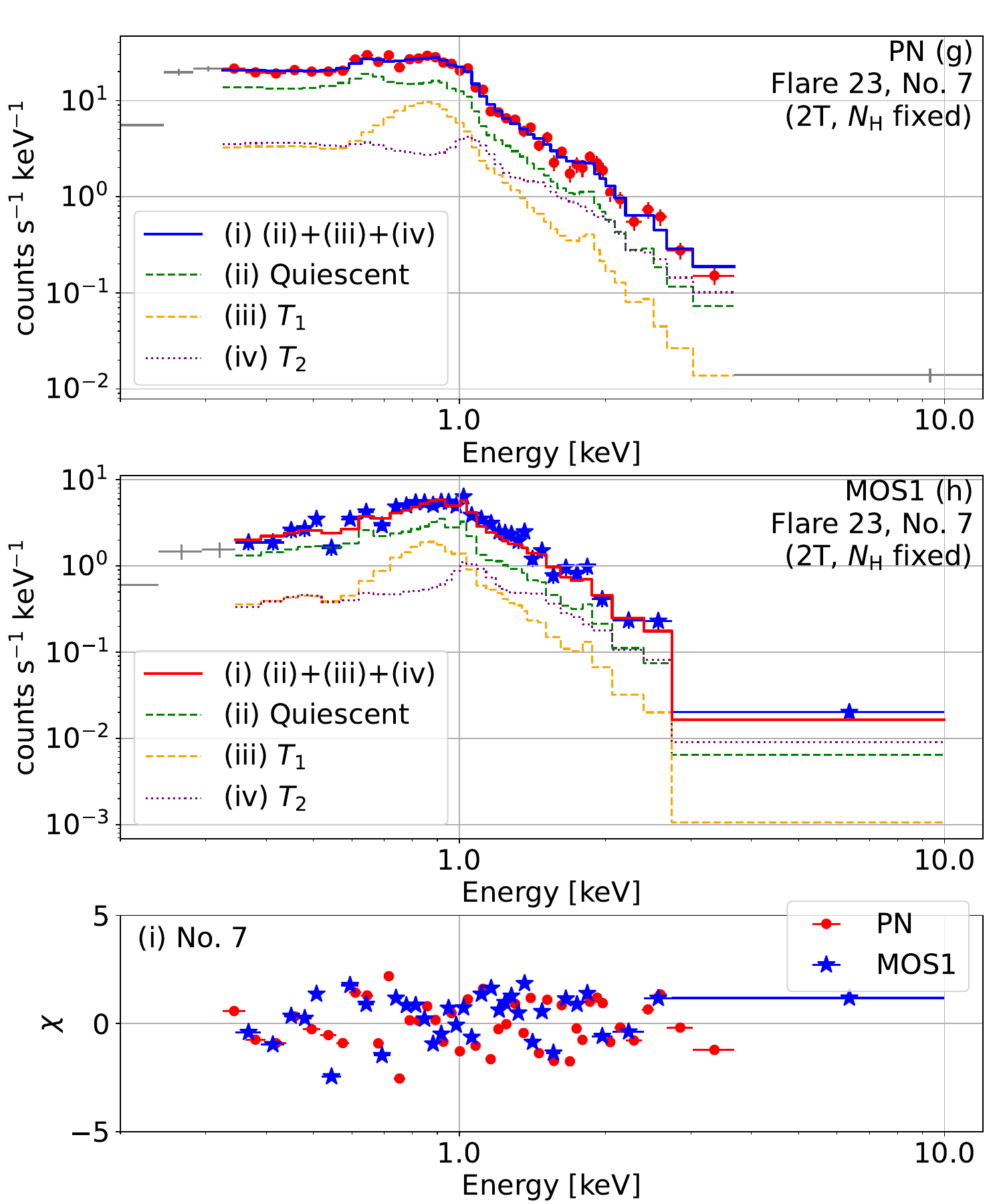}{0.5\textwidth}{\vspace{0mm}}
\fig{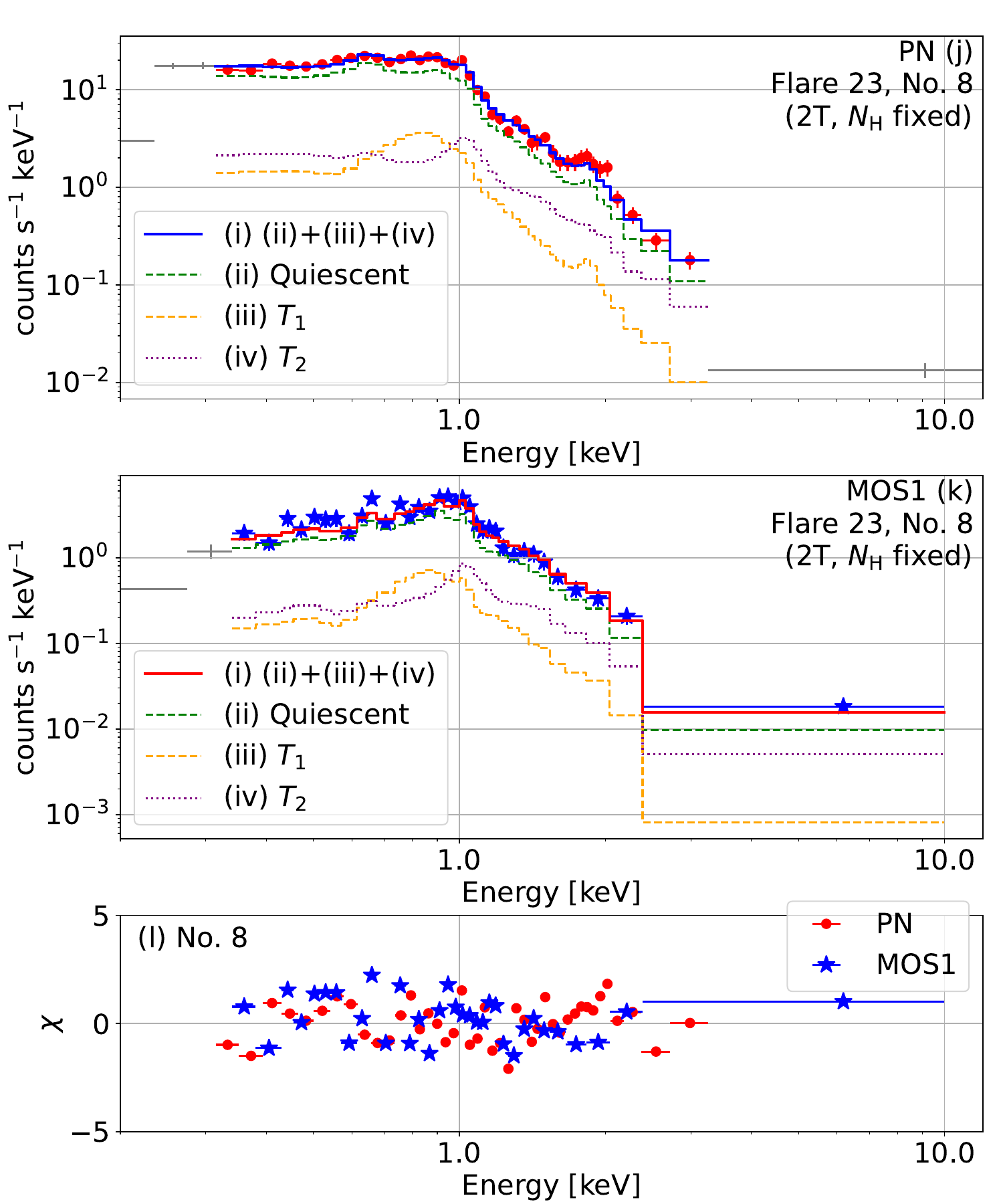}{0.5\textwidth}{\vspace{0mm}}
    }
     \vspace{-5mm}
     \caption{
Same as Figure \ref{fig:specfig_Flare23_TEM_quie_each_No.1-No.4},
but the Phases No. 5 -- No. 8 (cf. Figure \ref{fig:Flare23_TEM_multi_lc}).
}
   \label{fig:specfig_Flare23_TEM_quie_each_No.5-No.8}
   \end{center}
 \end{figure}

\begin{figure}[ht!]
   \begin{center}
      \gridline{
\fig{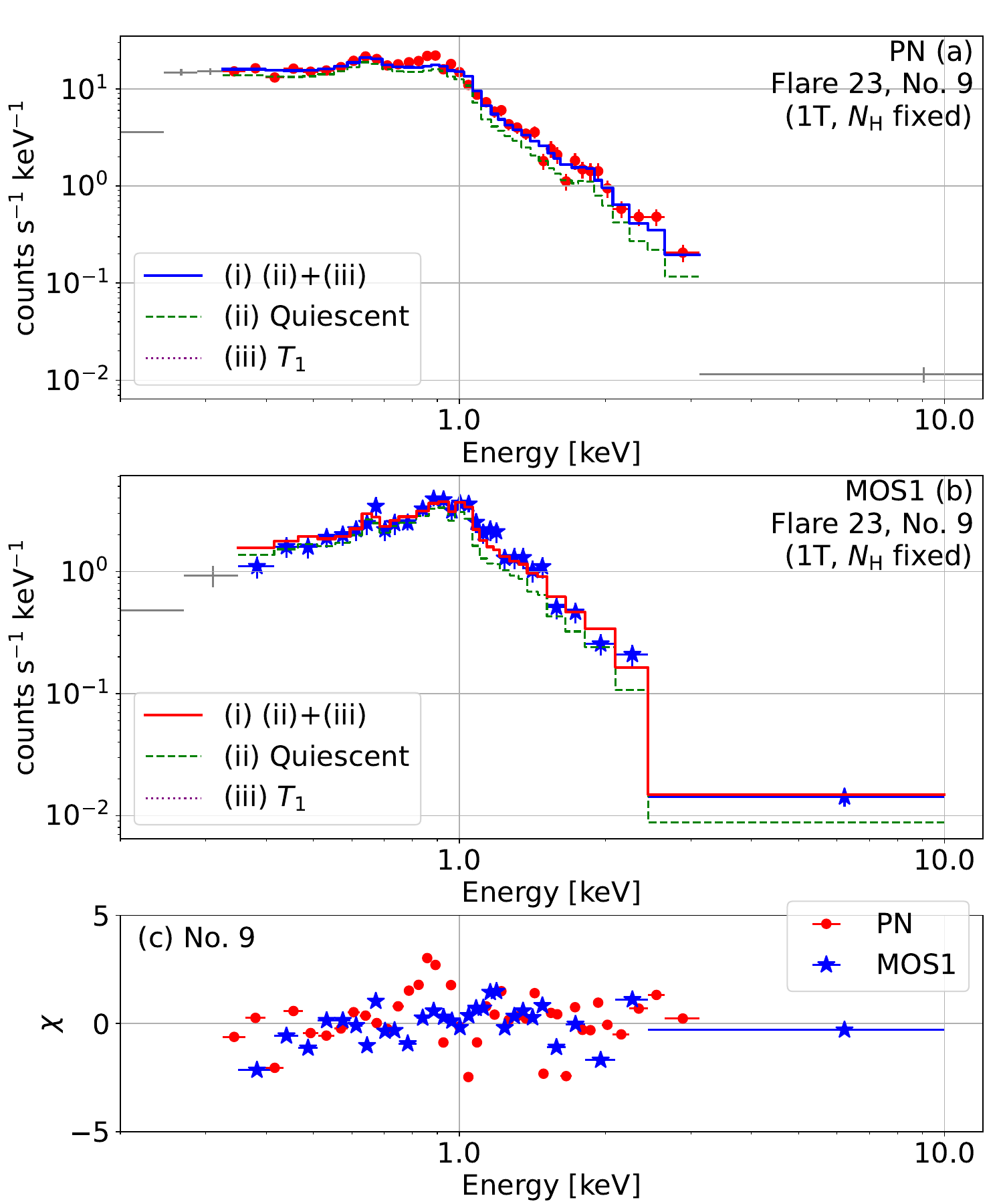}{0.5\textwidth}{\vspace{0mm}}
\fig{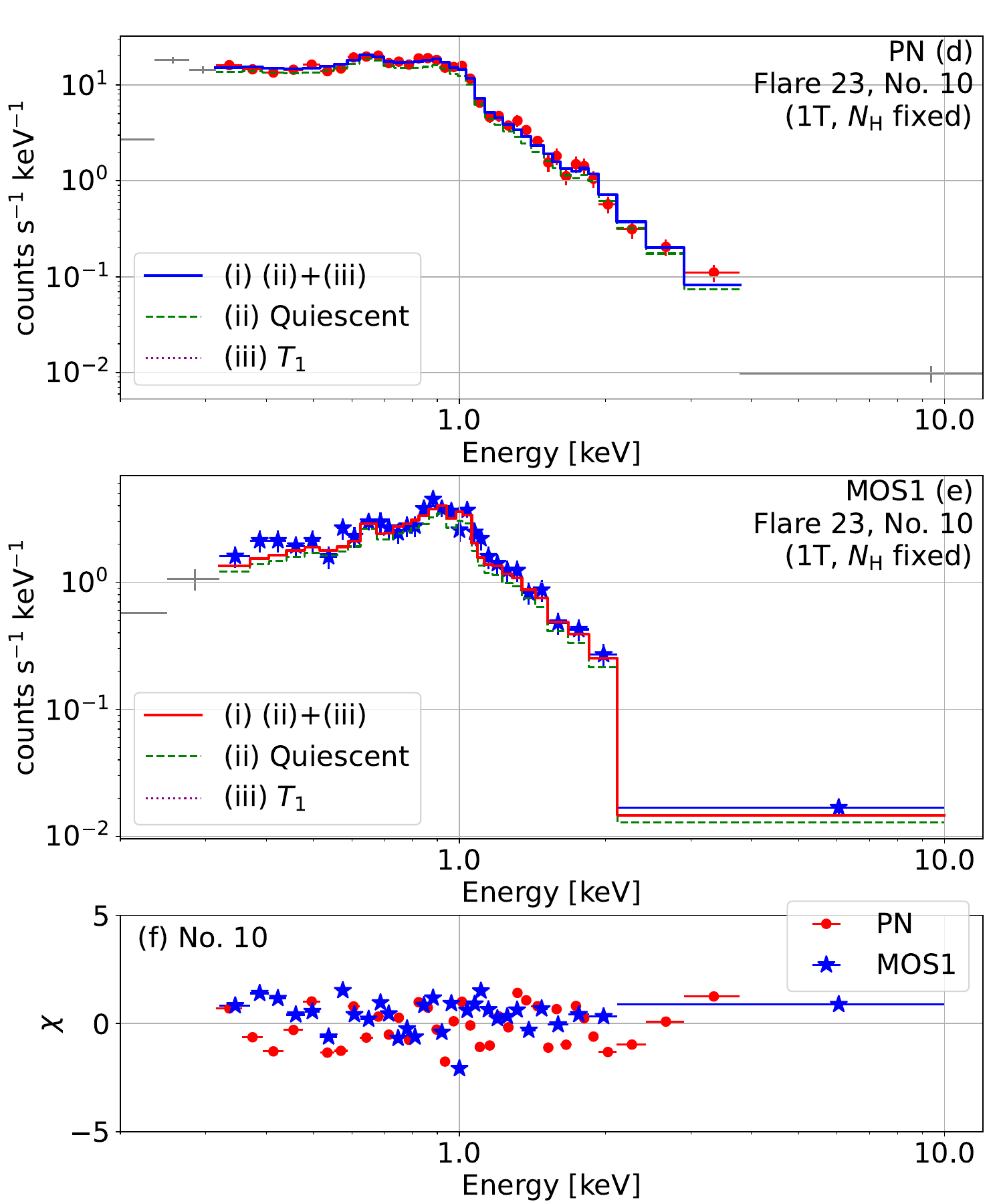}{0.5\textwidth}{\vspace{0mm}}
    }
     \vspace{-5mm}
      \gridline{
\fig{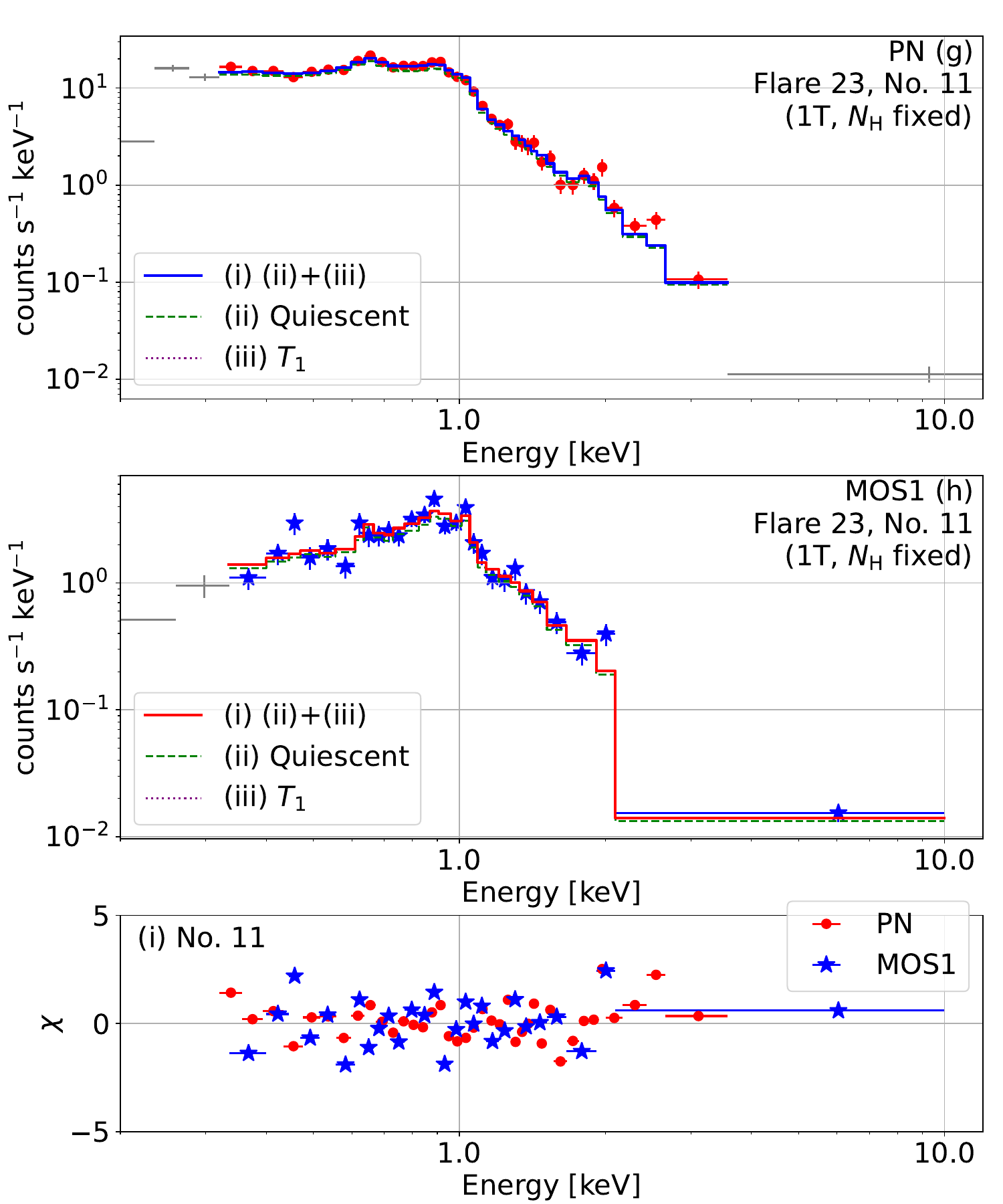}{0.5\textwidth}{\vspace{0mm}}
\fig{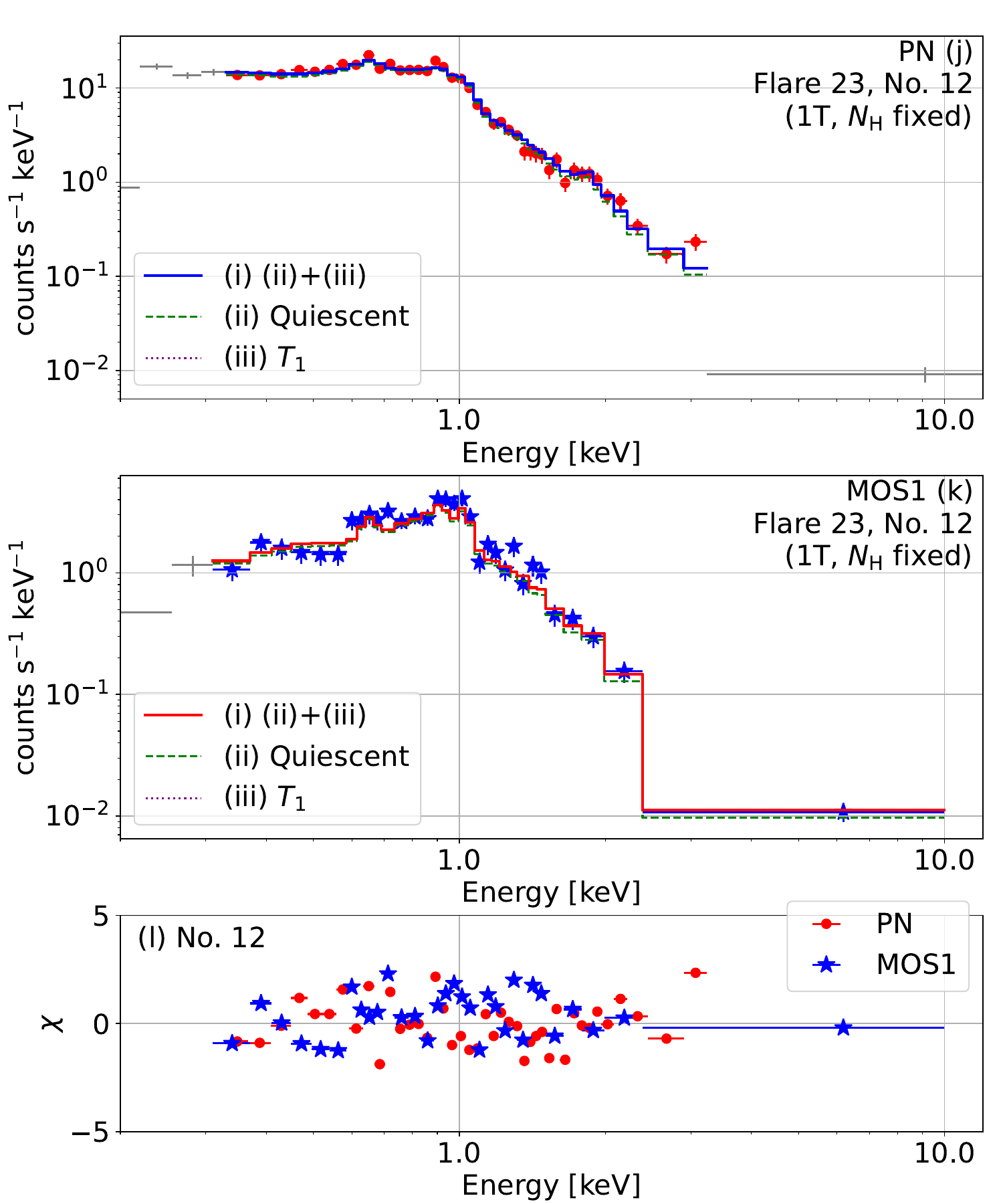}{0.5\textwidth}{\vspace{0mm}}
    }
     \vspace{-5mm}
     \caption{
Same as Figure \ref{fig:specfig_Flare23_TEM_quie_each_No.1-No.4},
but the Phases No. 9 -- No. 12 (cf. Figure \ref{fig:Flare23_TEM_multi_lc}).
}
   \label{fig:specfig_Flare23_TEM_quie_each_No.9-No.12}
   \end{center}
 \end{figure}

\begin{figure}[ht!]
   \begin{center}
      \gridline{
\fig{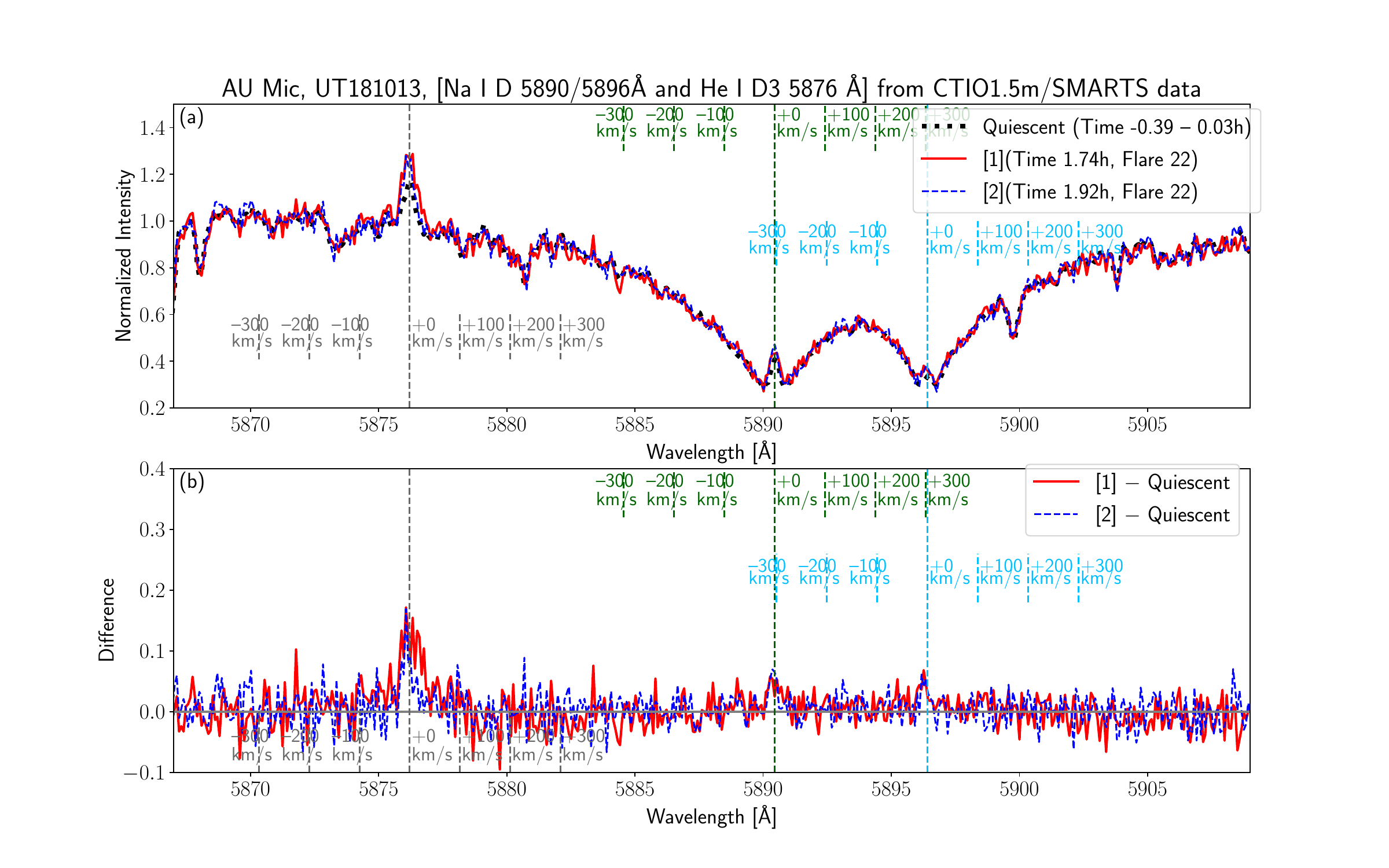}{0.5\textwidth}{\vspace{0mm}}
\fig{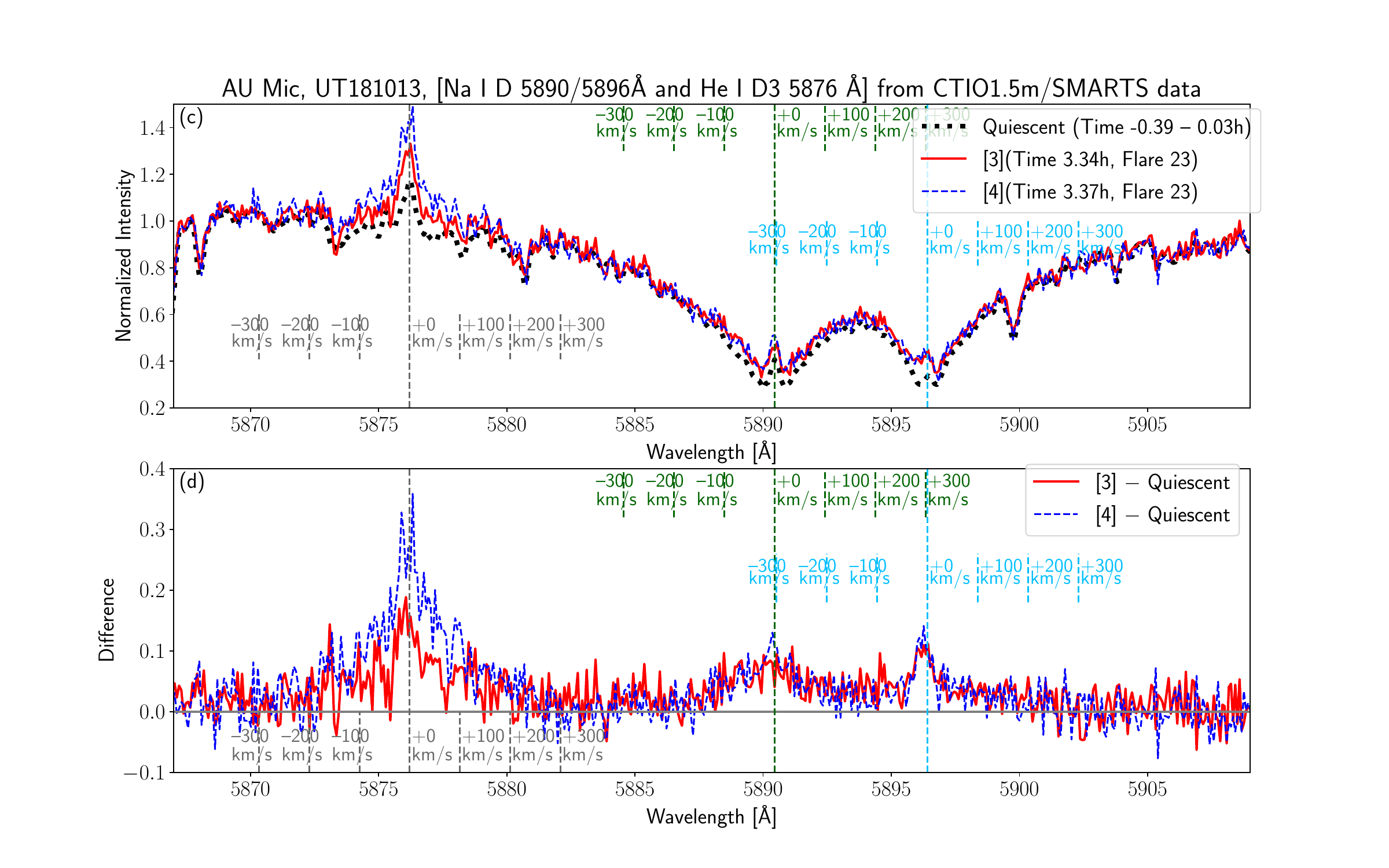}{0.5\textwidth}{\vspace{0mm}}
    }
     \vspace{-5mm}
      \gridline{
\fig{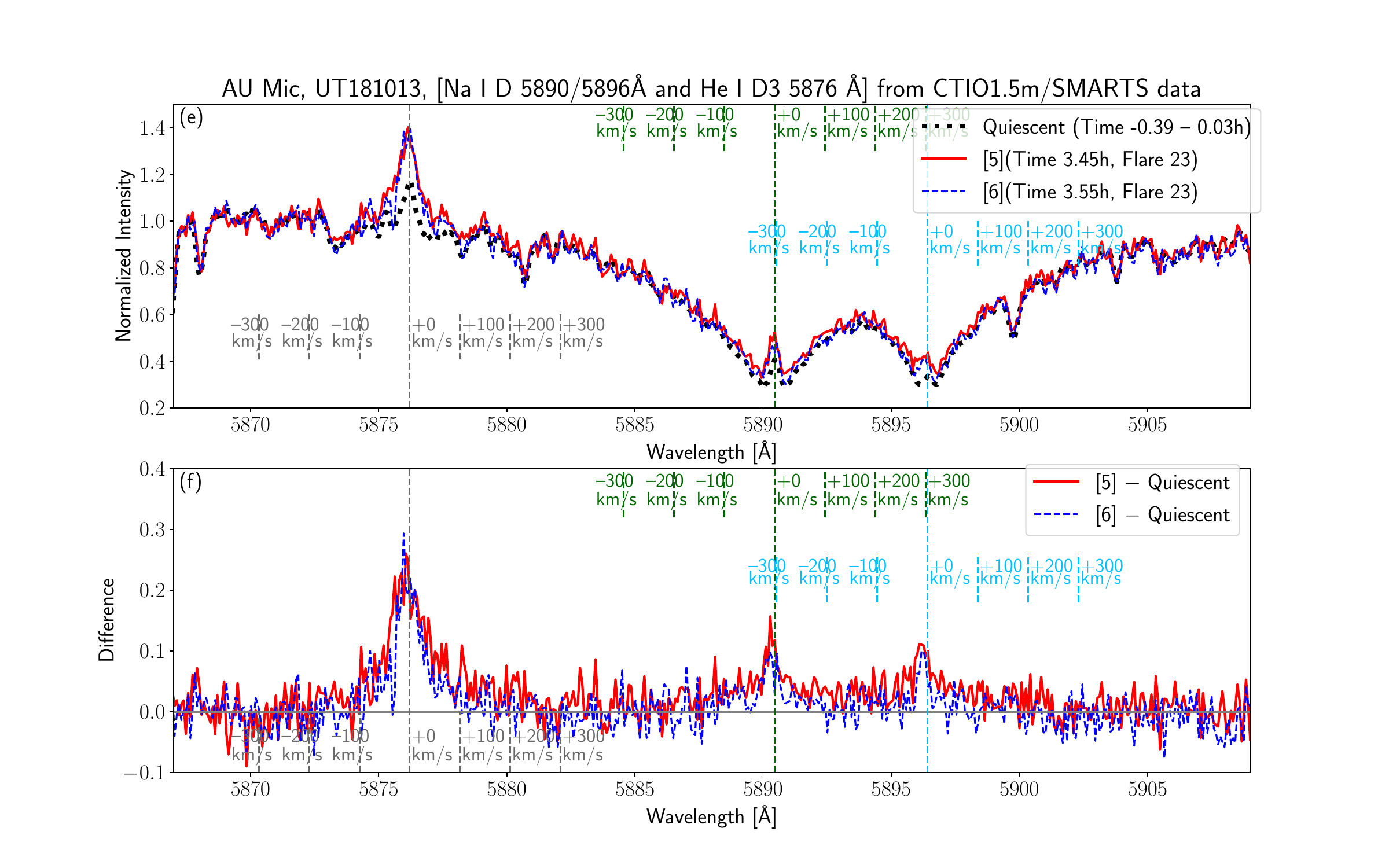}{0.5\textwidth}{\vspace{0mm}}
\fig{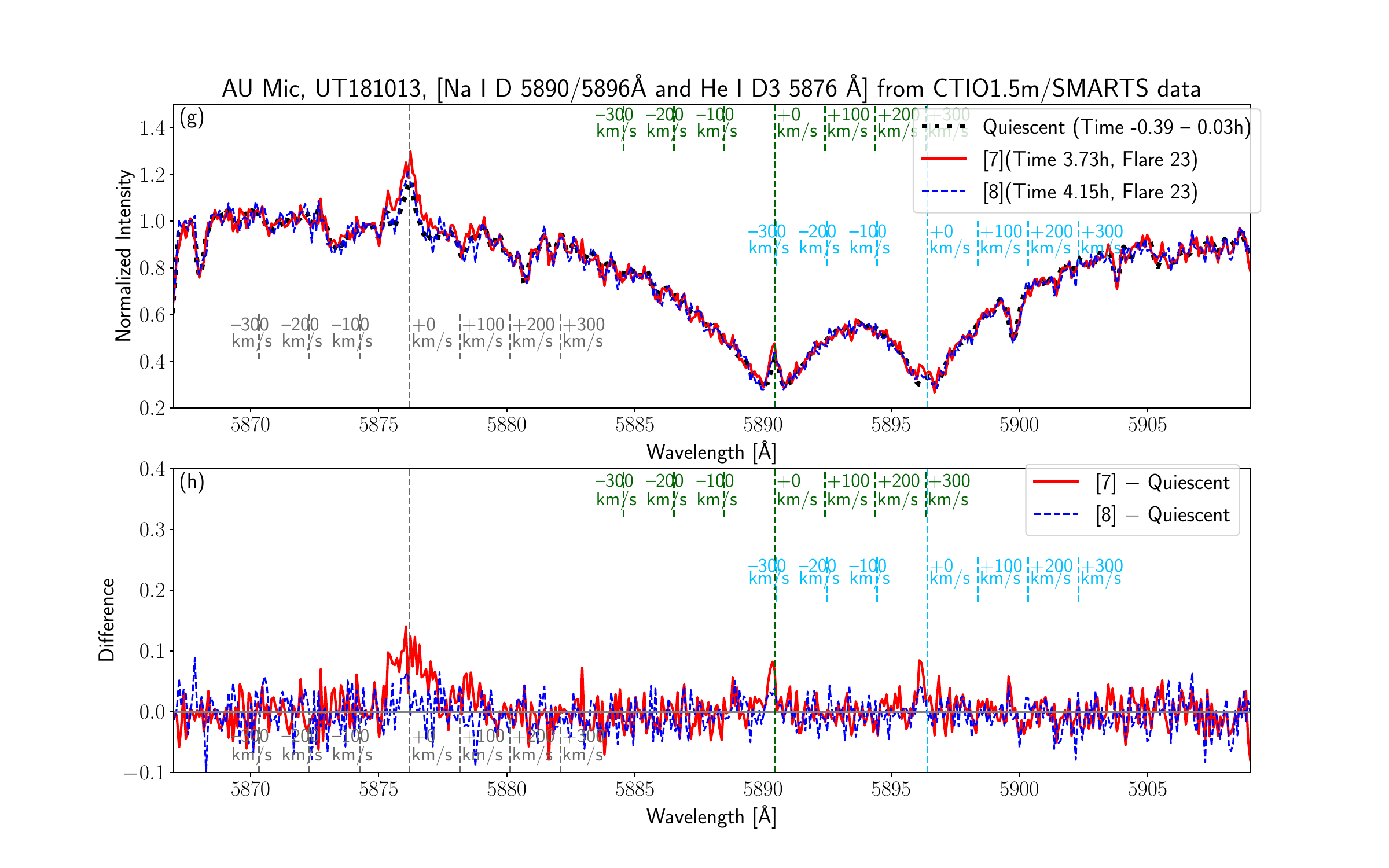}{0.5\textwidth}{\vspace{0mm}}
    }
     \caption{
Same as Figure \ref{fig:flare23_HaSpec}, the He I D3 5876\AA~and Na D1 \& D2 lines.
}
   \label{fig:flare23_HeNaSpec}
   \end{center}
 \end{figure}

\begin{figure}[ht!]
   \begin{center}
      \gridline{
\fig{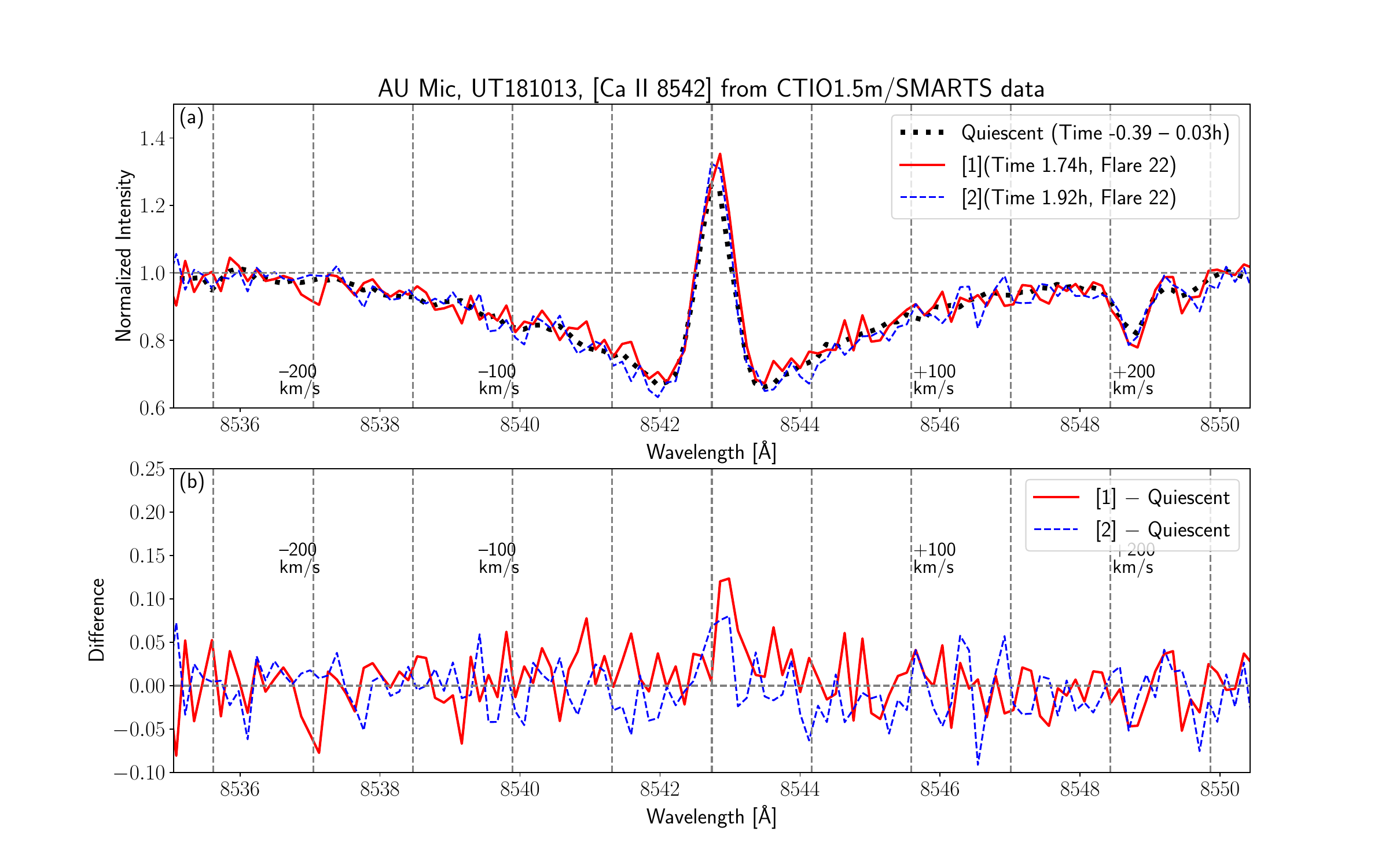}{0.5\textwidth}{\vspace{0mm}}
\fig{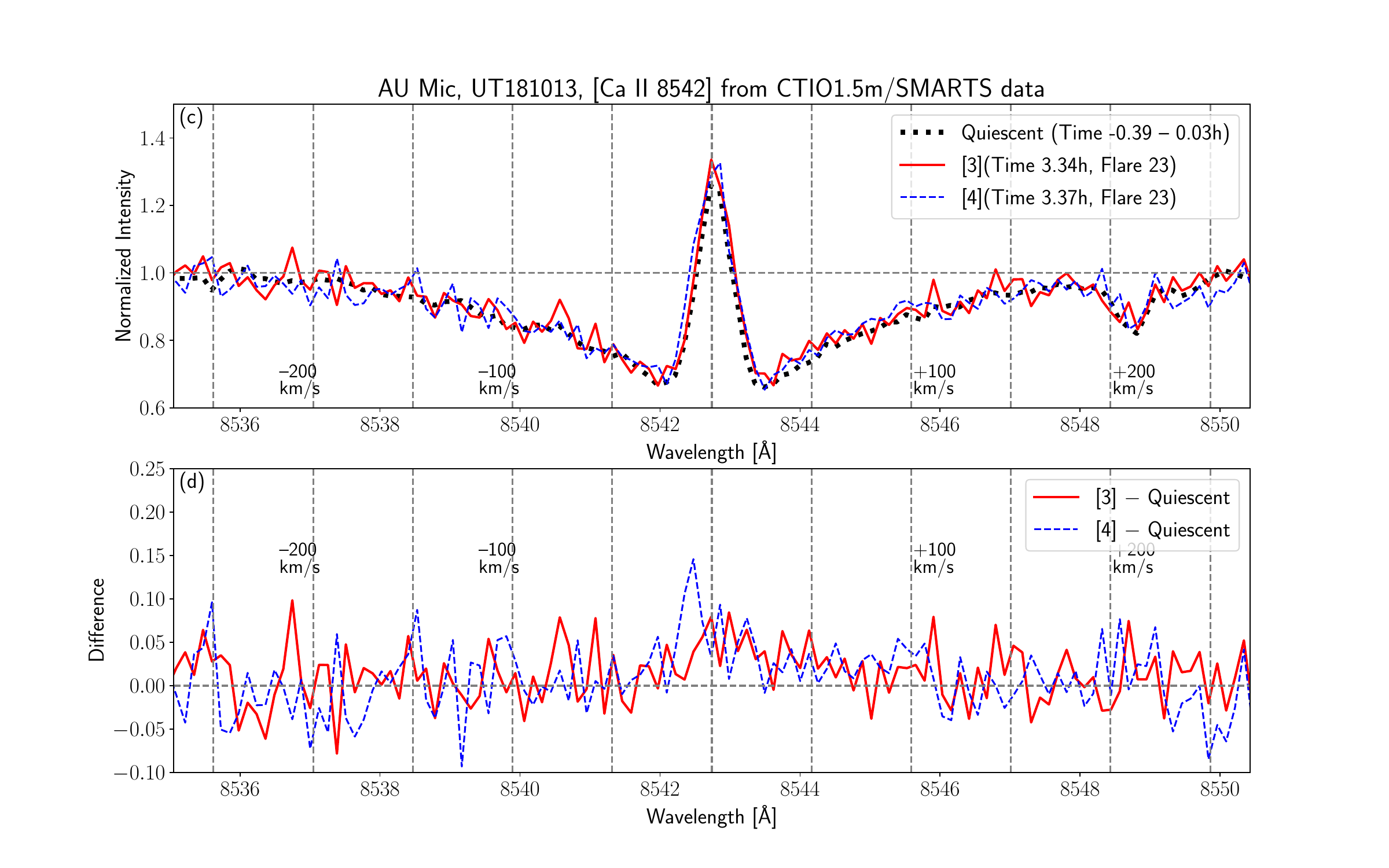}{0.5\textwidth}{\vspace{0mm}}
    }
     \vspace{-5mm}
      \gridline{
\fig{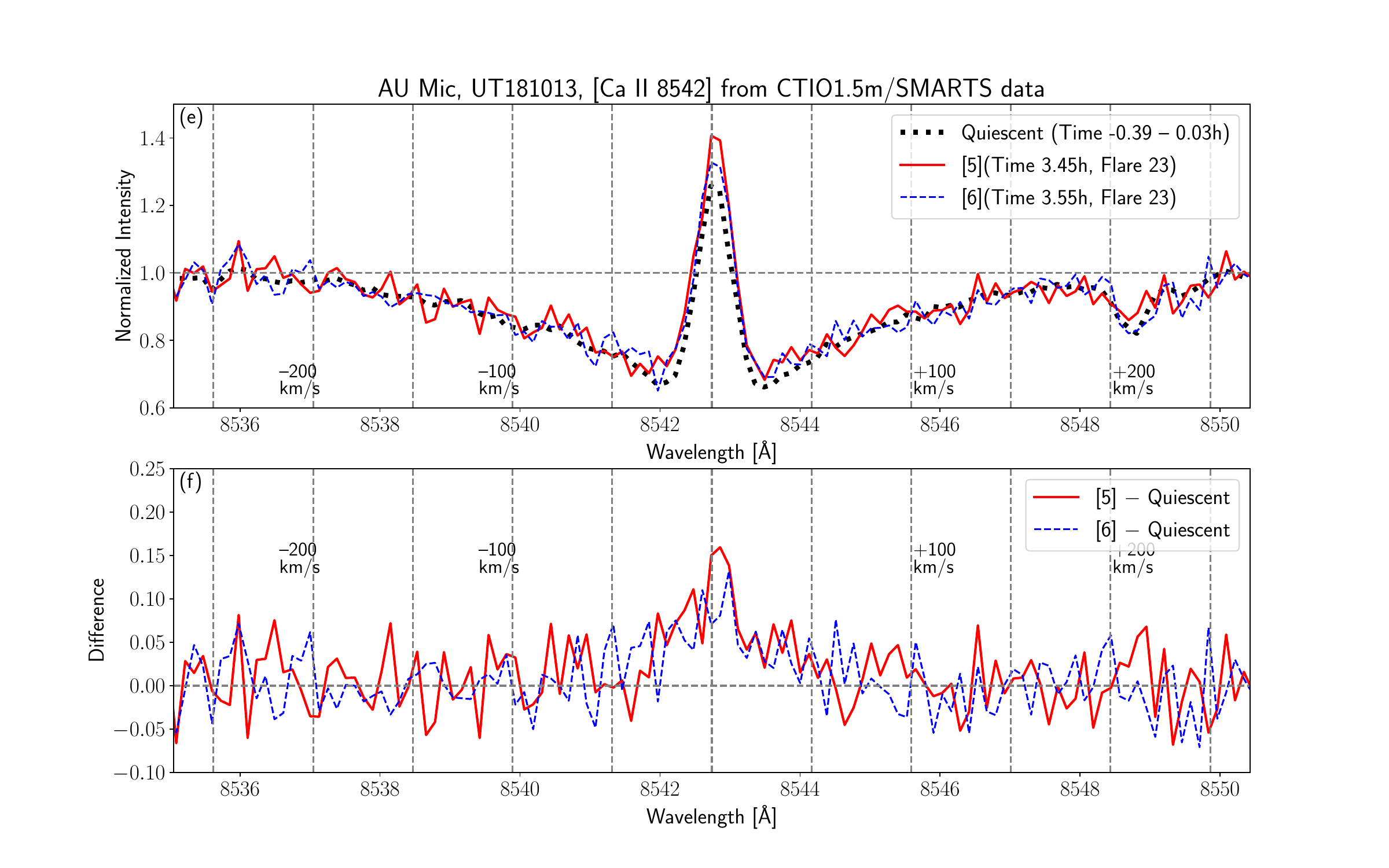}{0.5\textwidth}{\vspace{0mm}}
\fig{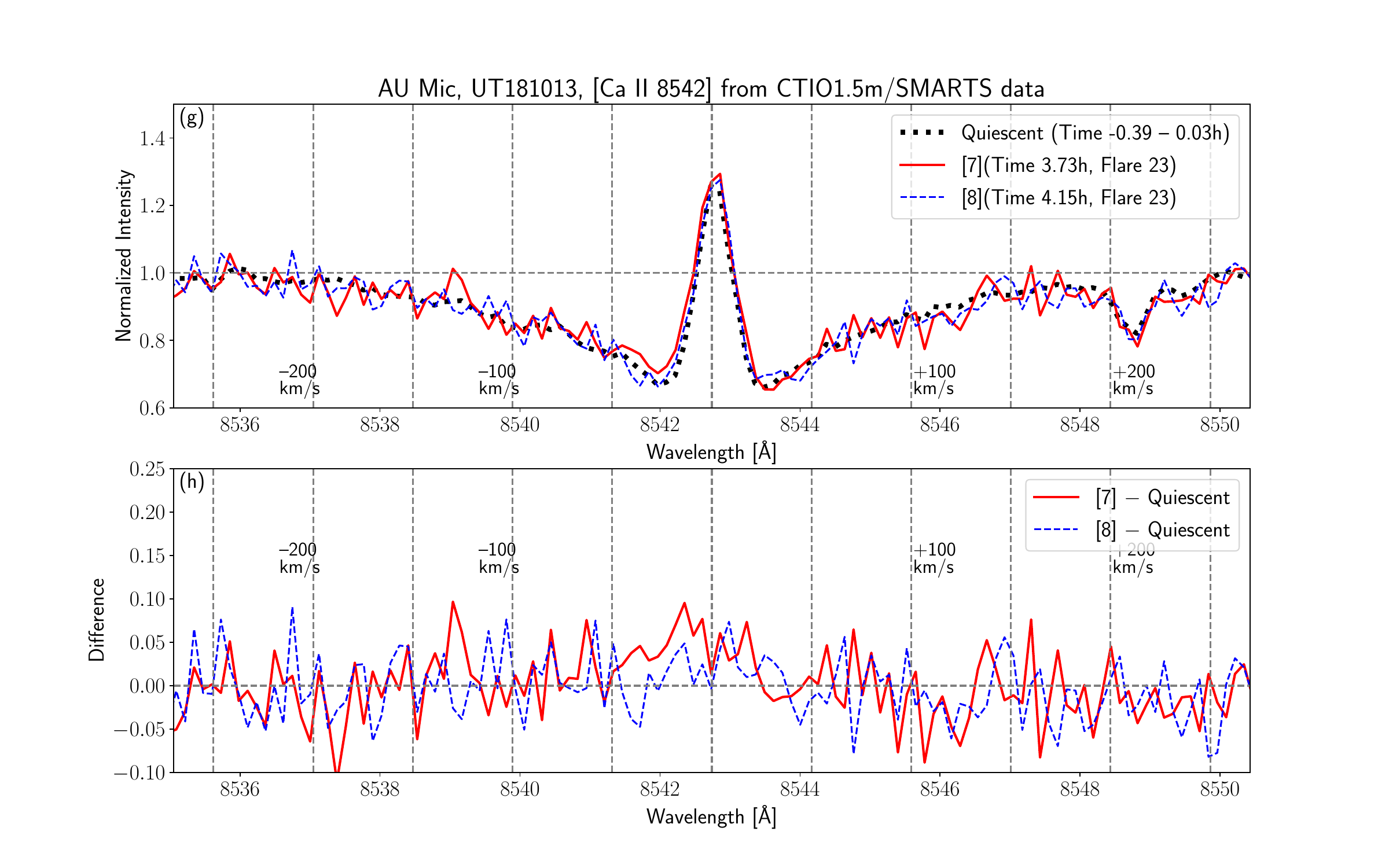}{0.5\textwidth}{\vspace{0mm}}
    }
     \caption{
Same as Figure \ref{fig:flare23_HaSpec}, the Ca II 8542\AA~line.
}
   \label{fig:flare23_Ca8542Spec}
   \end{center}
 \end{figure}

\begin{figure}[ht!]
   \begin{center}
      \gridline{
\fig{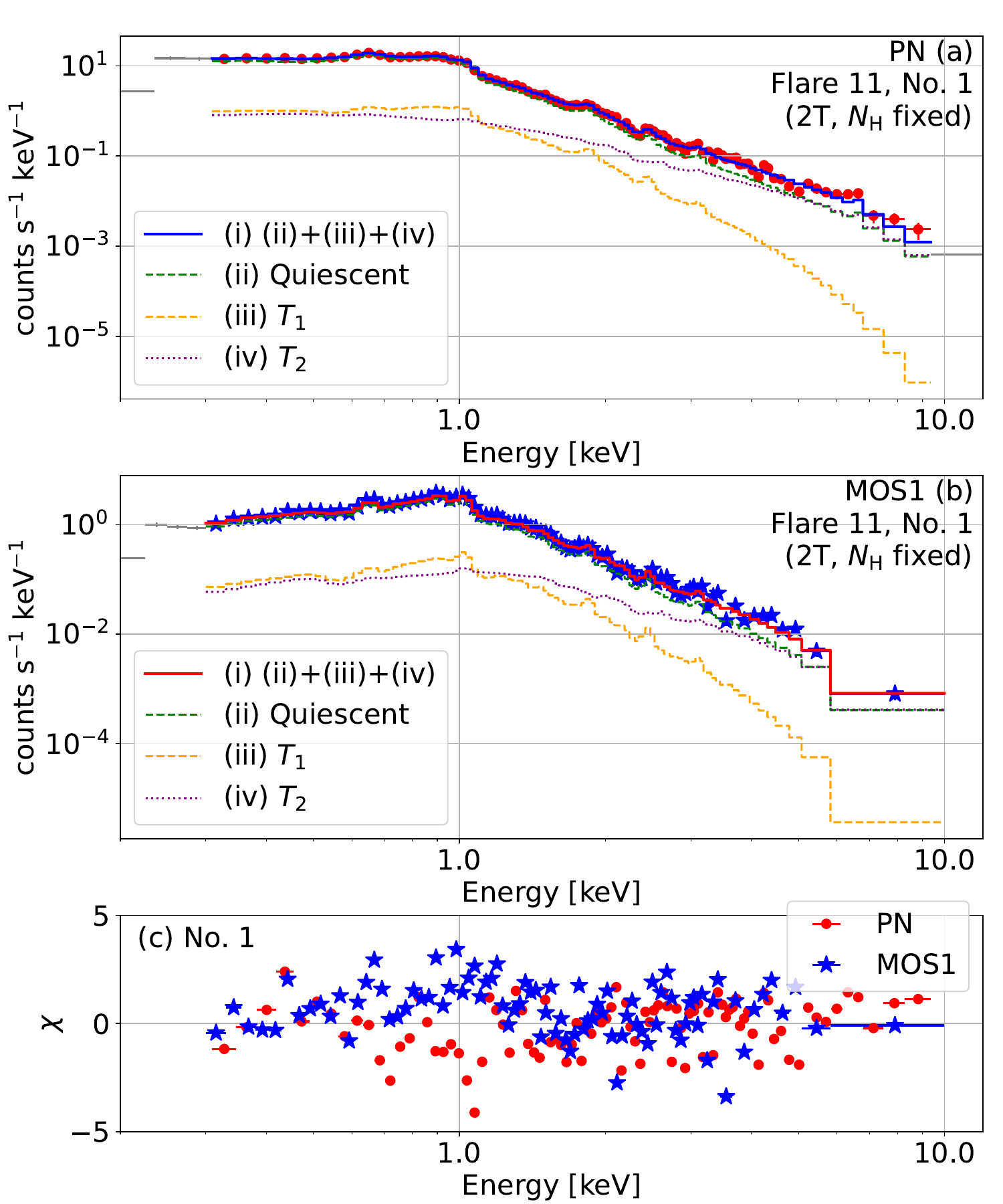}{0.5\textwidth}{\vspace{0mm}}
\fig{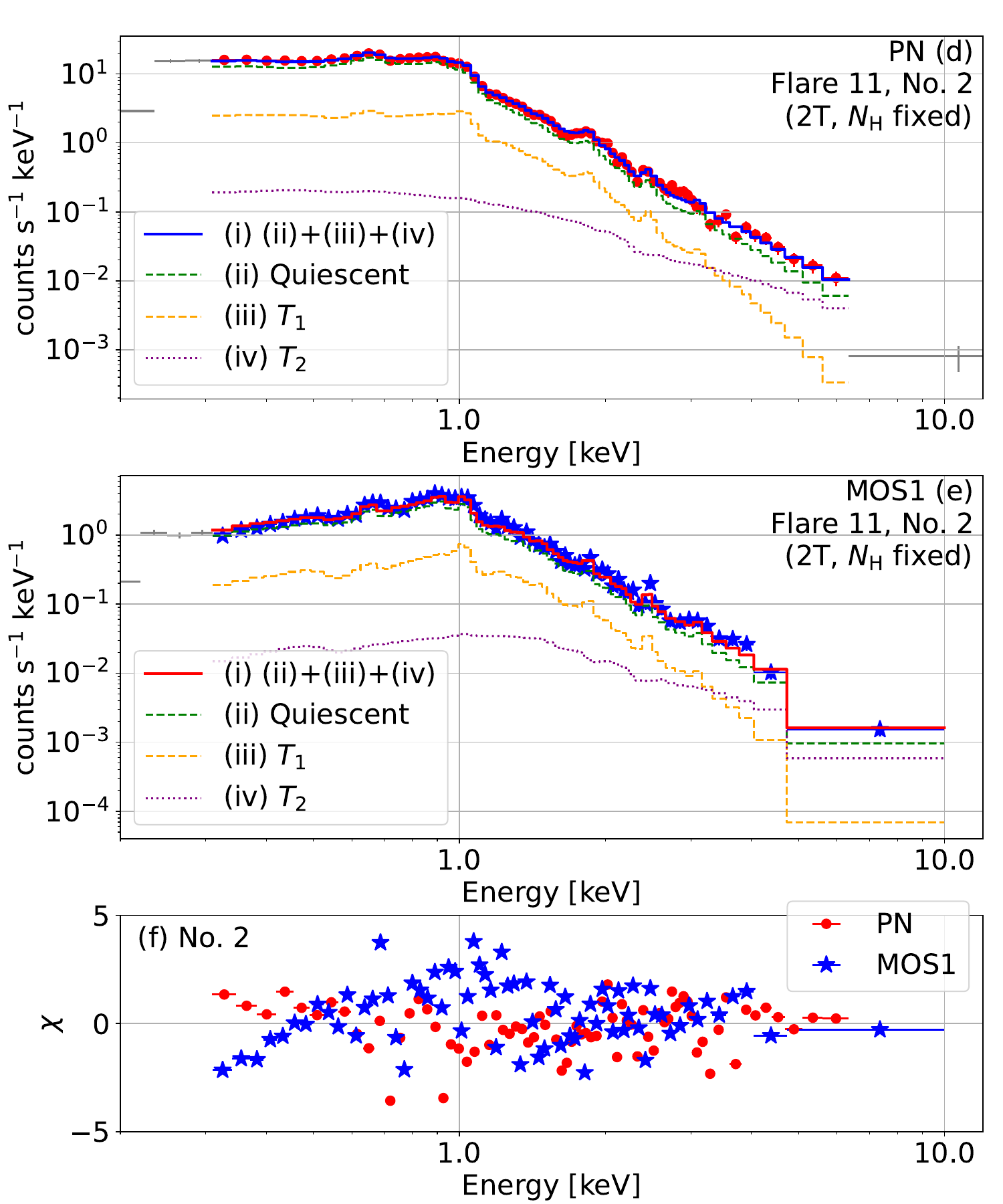}{0.5\textwidth}{\vspace{0mm}}
    }
     \vspace{-5mm}
      \gridline{
\fig{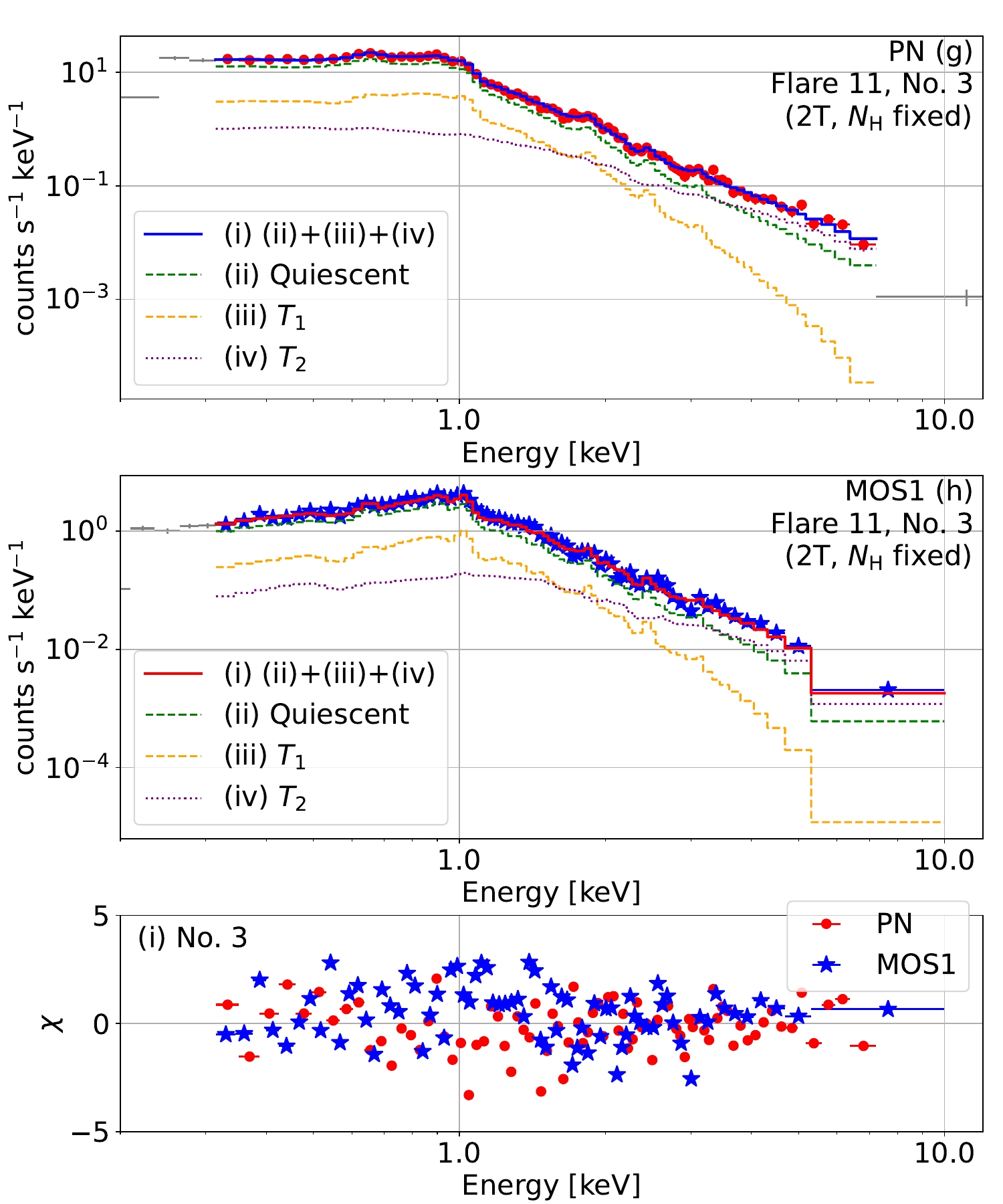}{0.5\textwidth}{\vspace{0mm}}
\fig{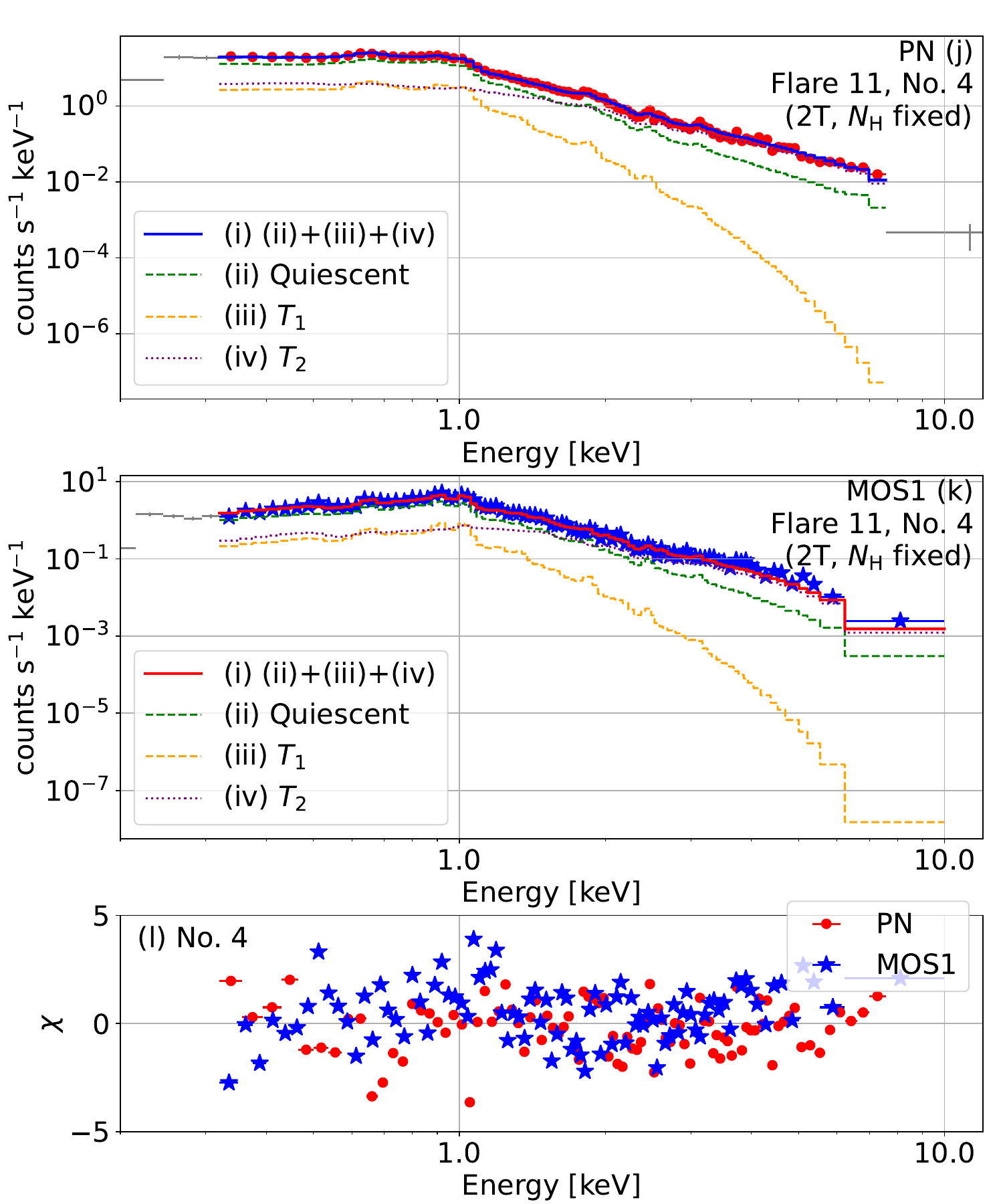}{0.5\textwidth}{\vspace{0mm}}
    }
     \vspace{-5mm}
     \caption{
The PN and MOS1 spectra and best-fit model result for the Phases No. 1 -- No. 4 (cf. Figure \ref{fig:Flare11and15_TEM_multi_lc}) of Flares 11.
The data and fit results are plotted with the same way as Figures  \ref{fig:RiseDecayFit_fig1_Flare22} and 
\ref{fig:specfig_Flare23_TEM_quie_each_No.1-No.4}.
}
   \label{fig:specfig_Flare11_TEM_quie_each_No.1-No.4}
   \end{center}
 \end{figure}

\begin{figure}[ht!]
   \begin{center}
      \gridline{
\fig{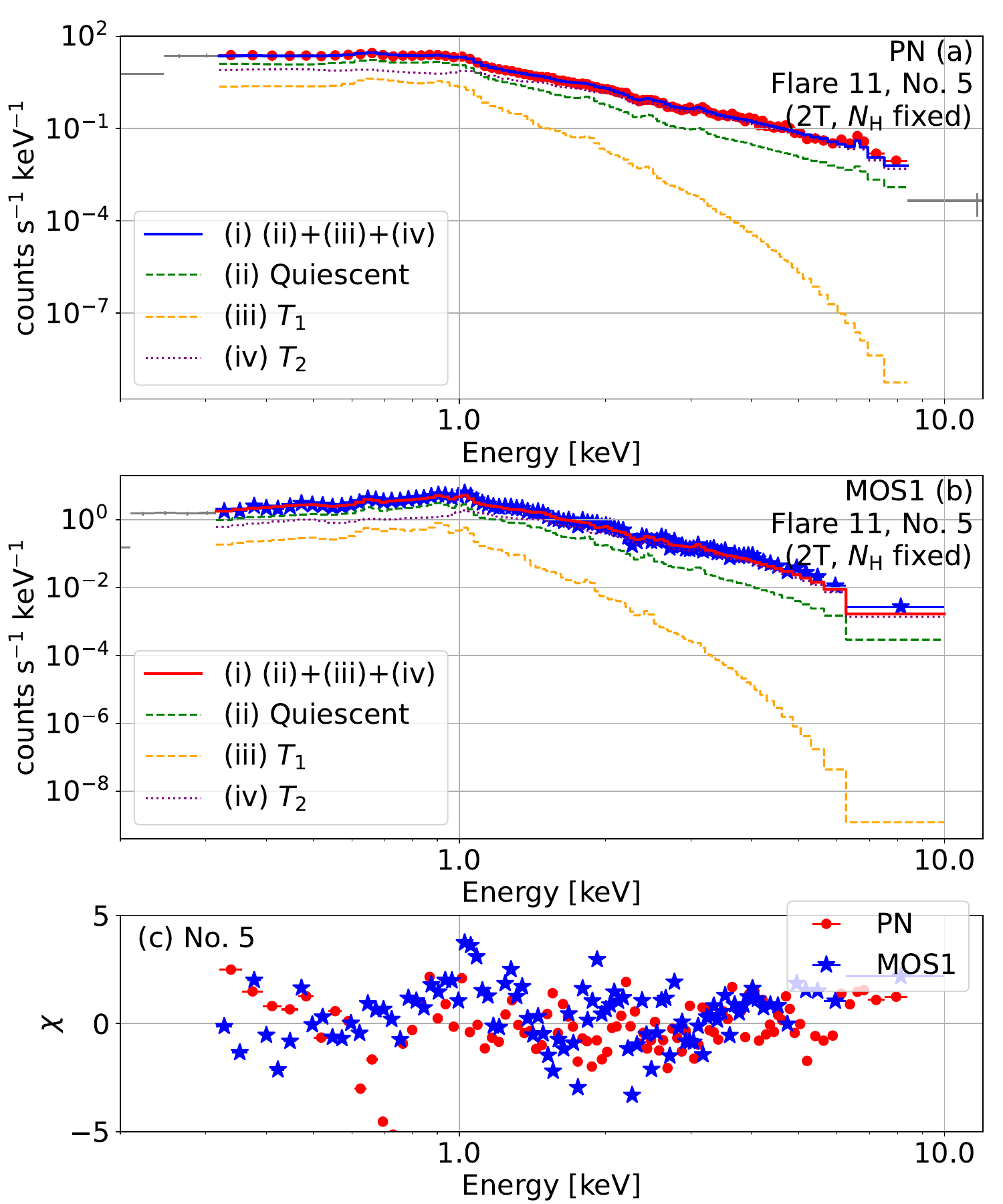}{0.5\textwidth}{\vspace{0mm}}
\fig{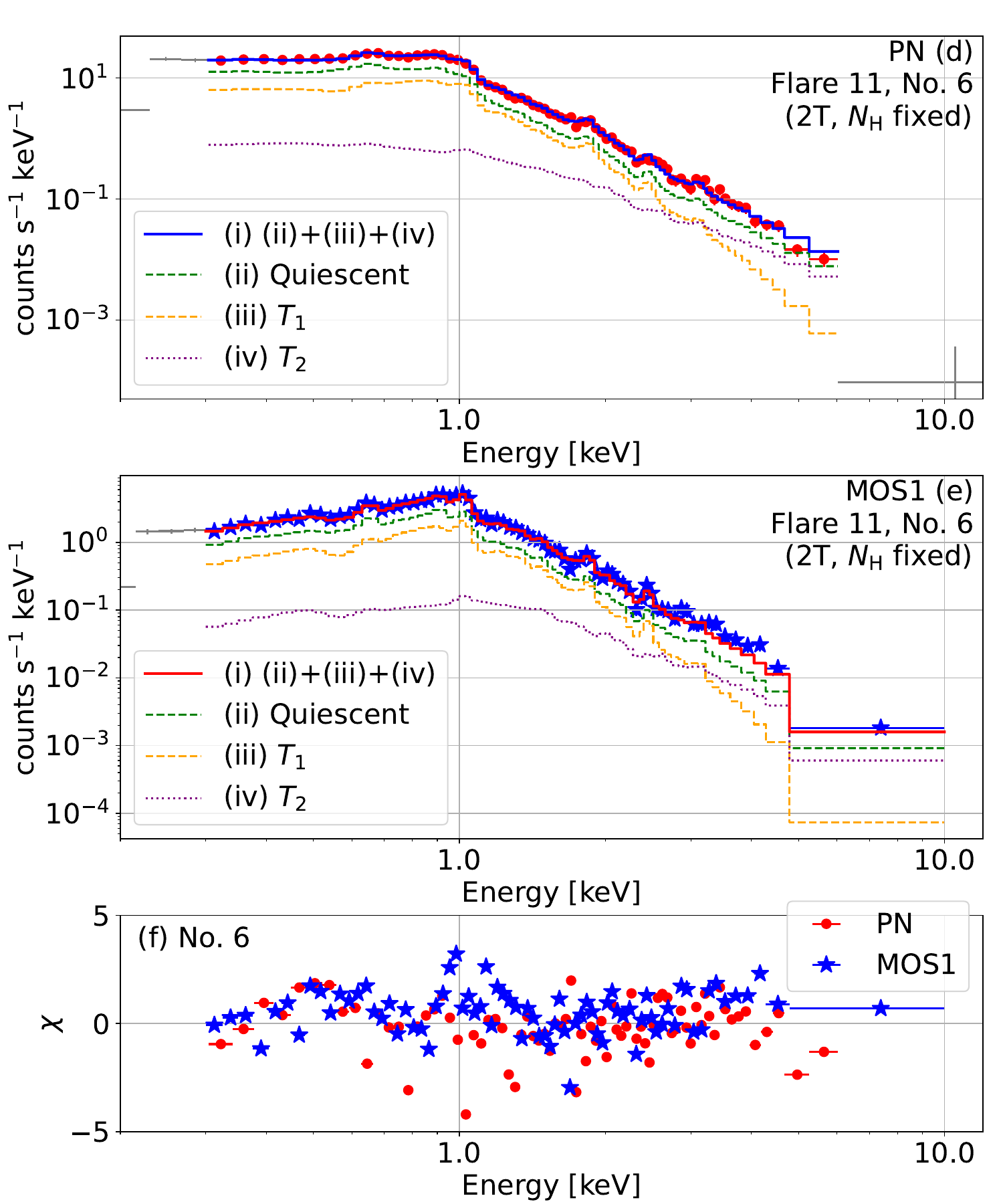}{0.5\textwidth}{\vspace{0mm}}
    }
     \vspace{-5mm}
      \gridline{
\fig{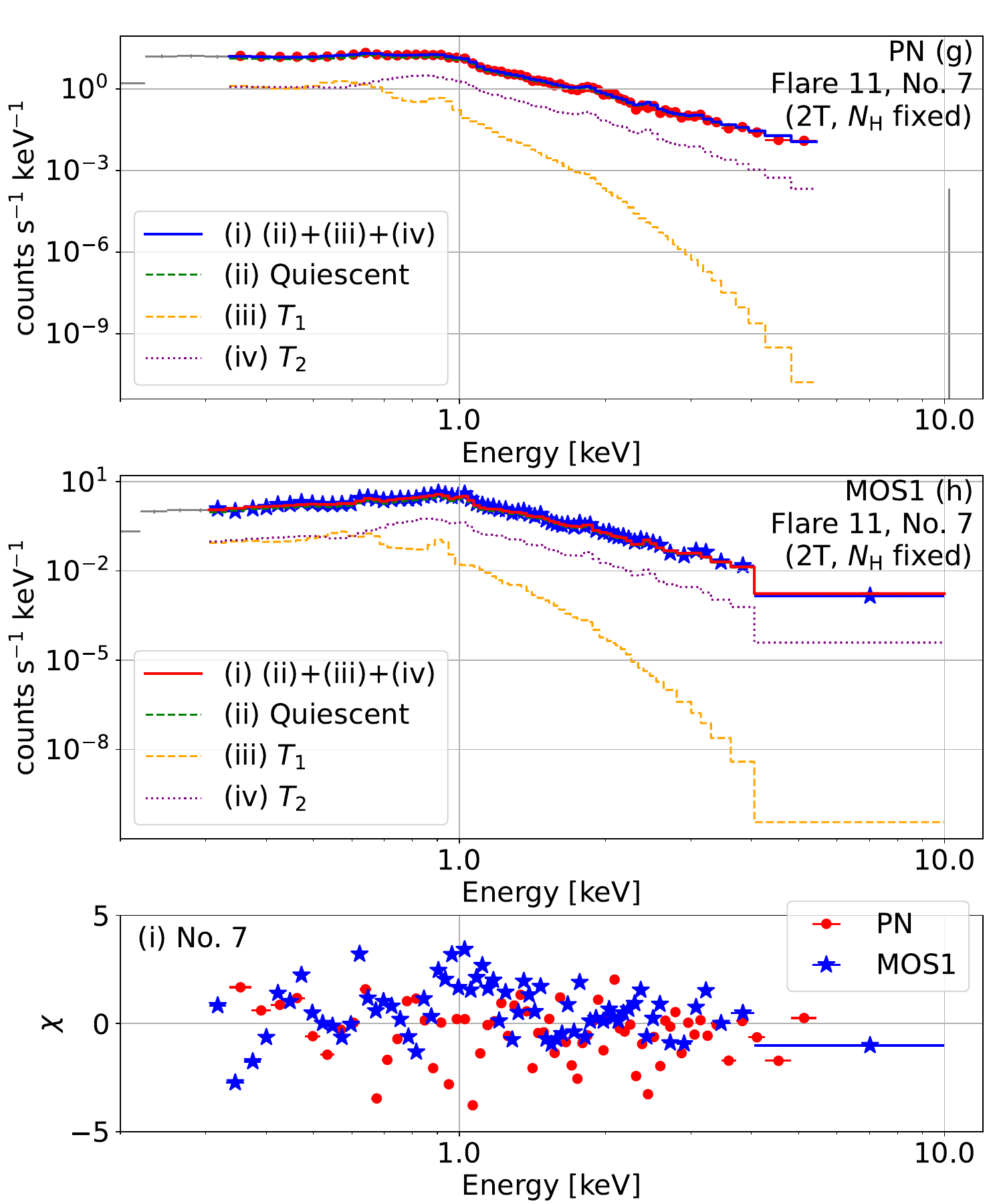}{0.5\textwidth}{\vspace{0mm}}
    }
     \vspace{-5mm}
     \caption{
Same as Figure \ref{fig:specfig_Flare11_TEM_quie_each_No.1-No.4},
but the Phases No. 5 -- No. 7 of Flare 11 (cf. Figure \ref{fig:Flare11and15_TEM_multi_lc}).
}
   \label{fig:specfig_Flare11_TEM_quie_each_No.5-No.7}
   \end{center}
 \end{figure}

\begin{figure}[ht!]
   \begin{center}
      \gridline{
\fig{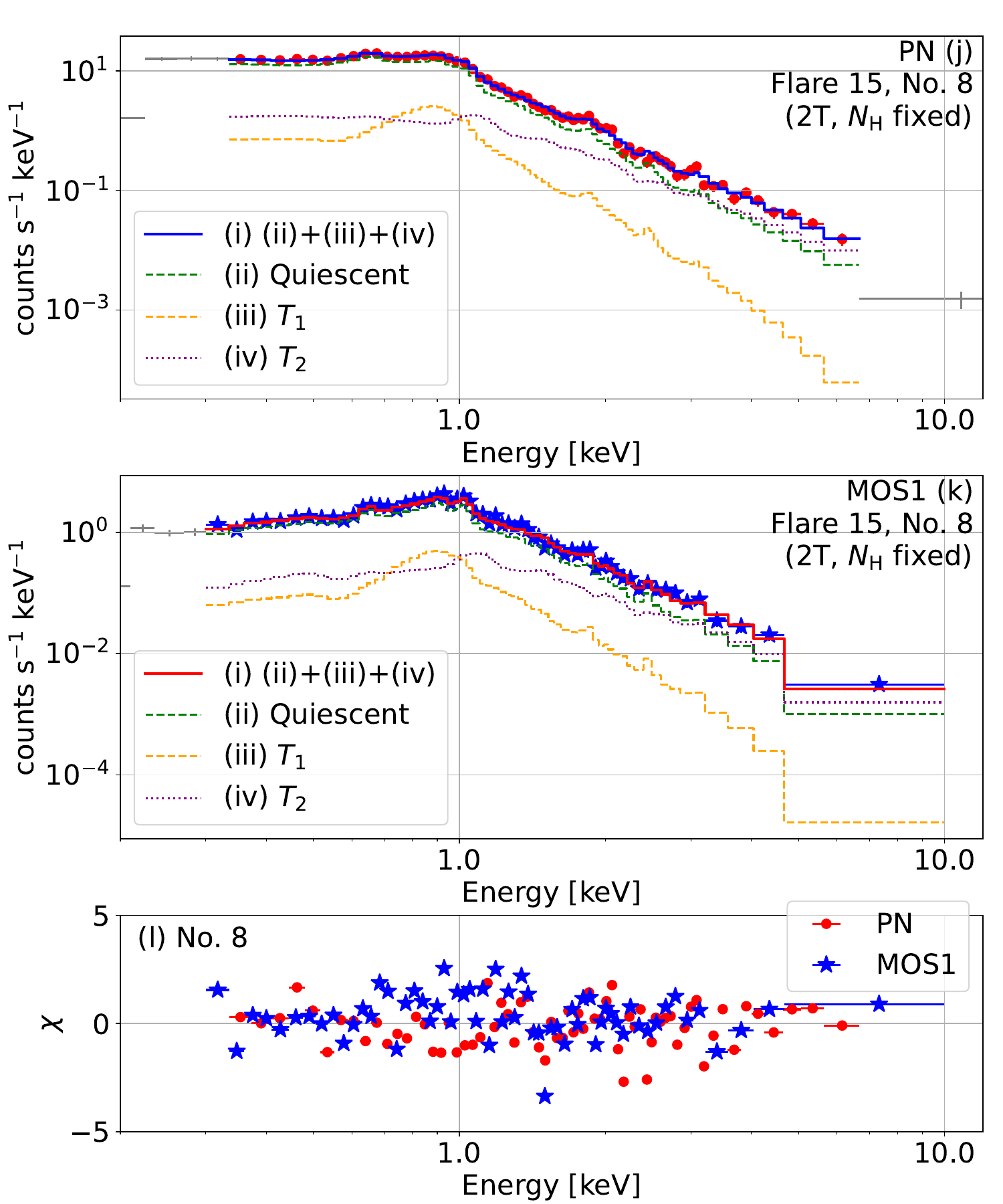}{0.5\textwidth}{\vspace{0mm}}
\fig{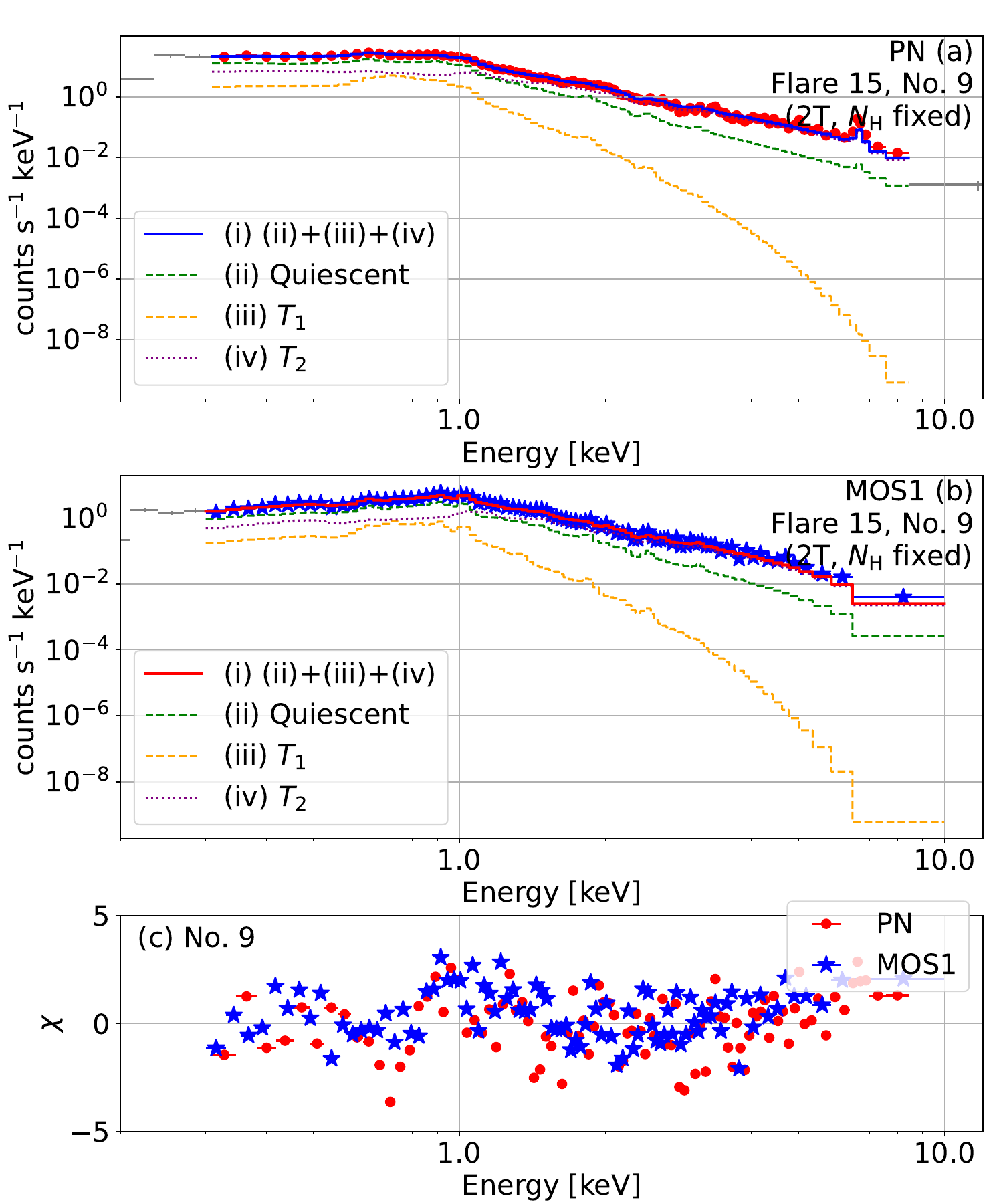}{0.5\textwidth}{\vspace{0mm}}
}
     \vspace{-5mm}
      \gridline{
\fig{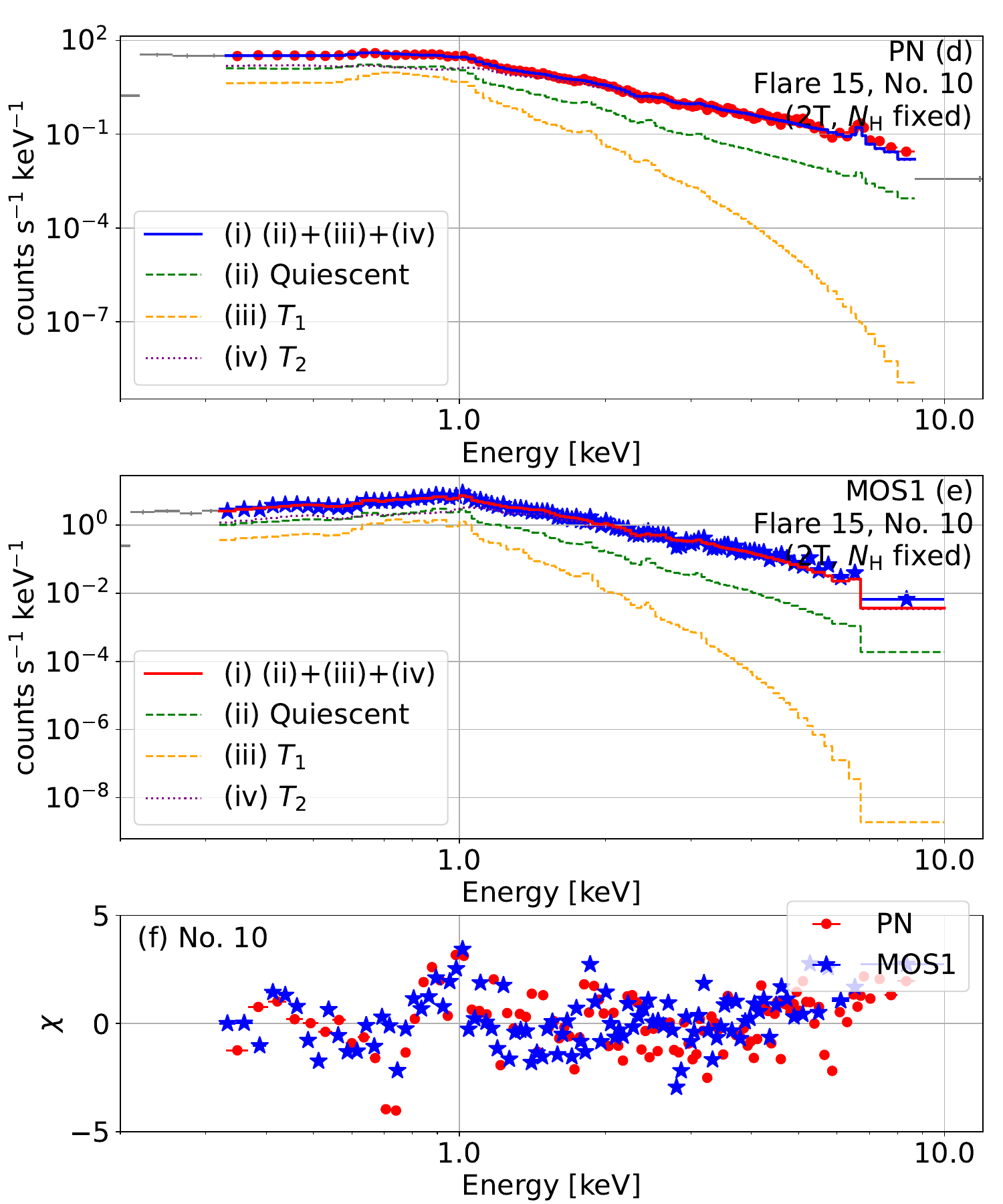}{0.5\textwidth}{\vspace{0mm}}
\fig{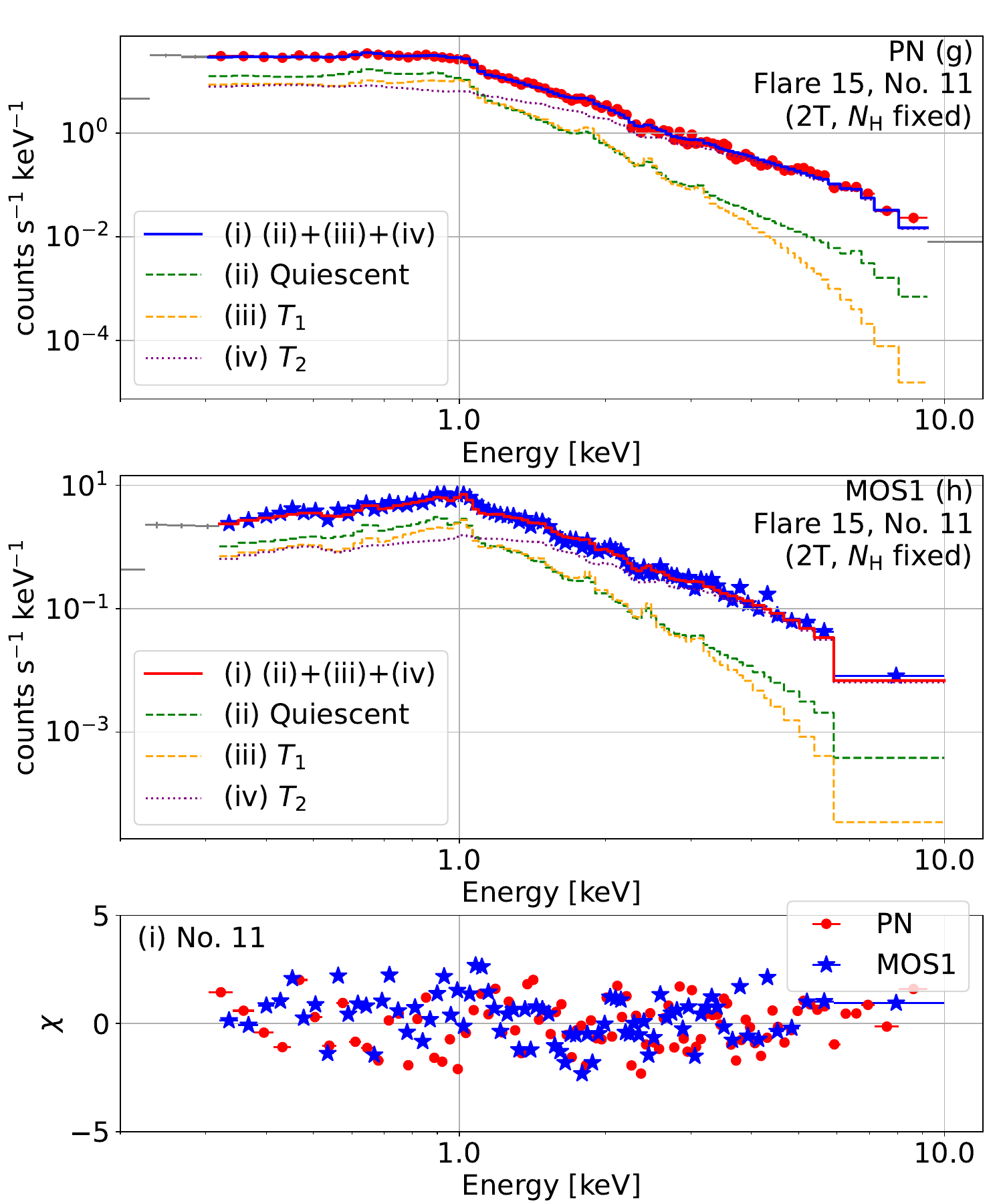}{0.5\textwidth}{\vspace{0mm}}
    }
     \vspace{-5mm}
     \caption{
Same as Figure \ref{fig:specfig_Flare11_TEM_quie_each_No.1-No.4}, but for the Phases No. 8 -- No. 11 (cf. Figure \ref{fig:Flare11and15_TEM_multi_lc}) of Flares 15.
}
   \label{fig:specfig_Flare15_TEM_quie_each_No.8-No.11}
   \end{center}
 \end{figure}

\begin{figure}[ht!]
   \begin{center}
      \gridline{
\fig{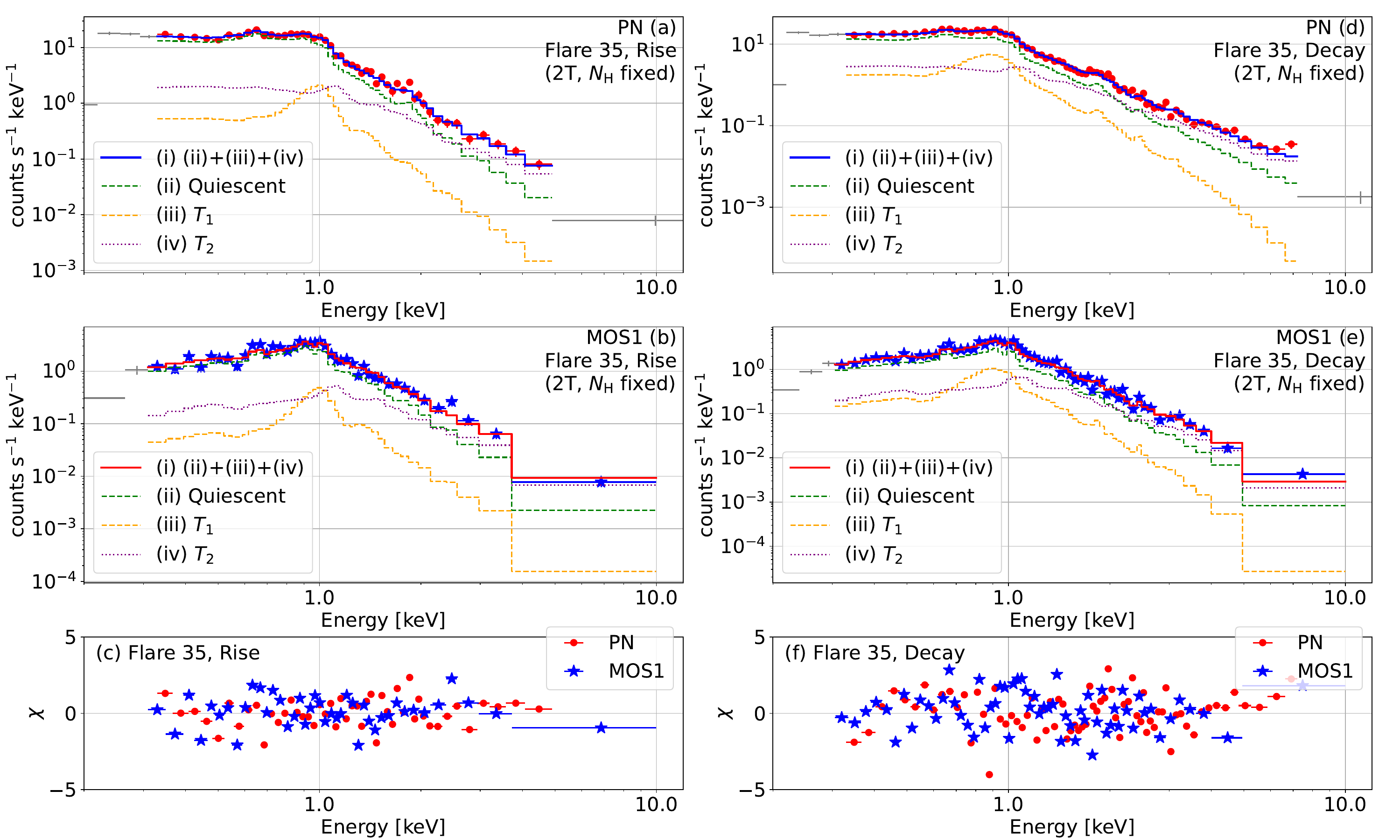}{0.95\textwidth}{\vspace{0mm}}
    }
     \vspace{-5mm}
      \gridline{
\fig{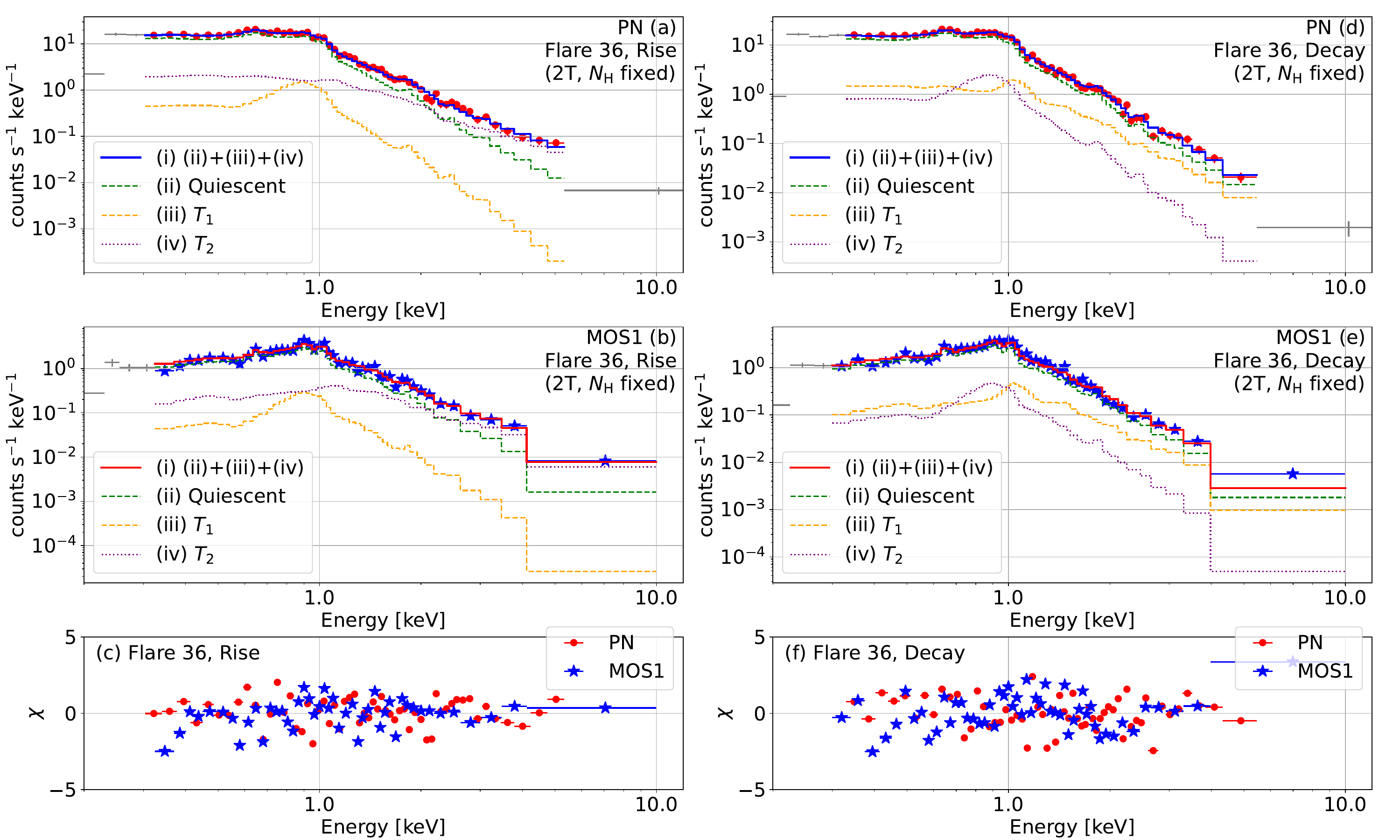}{0.95\textwidth}{\vspace{0mm}}
    }
     \vspace{-5mm}
     \caption{
The PN and MOS1 spectra and best-fit model result for rise and decay phase data of Flares 35 and 36 (cf. Figure \ref{fig:Flare35and36_TEM_multi_lc1}), which are plotted with the same way as Figure \ref{fig:RiseDecayFit_fig1_Flare22}.
}
   \label{fig:RiseDecayFit_fig1_Flare35and36}
   \end{center}
 \end{figure}

\begin{figure}[ht!]
   \begin{center}
      \gridline{
\fig{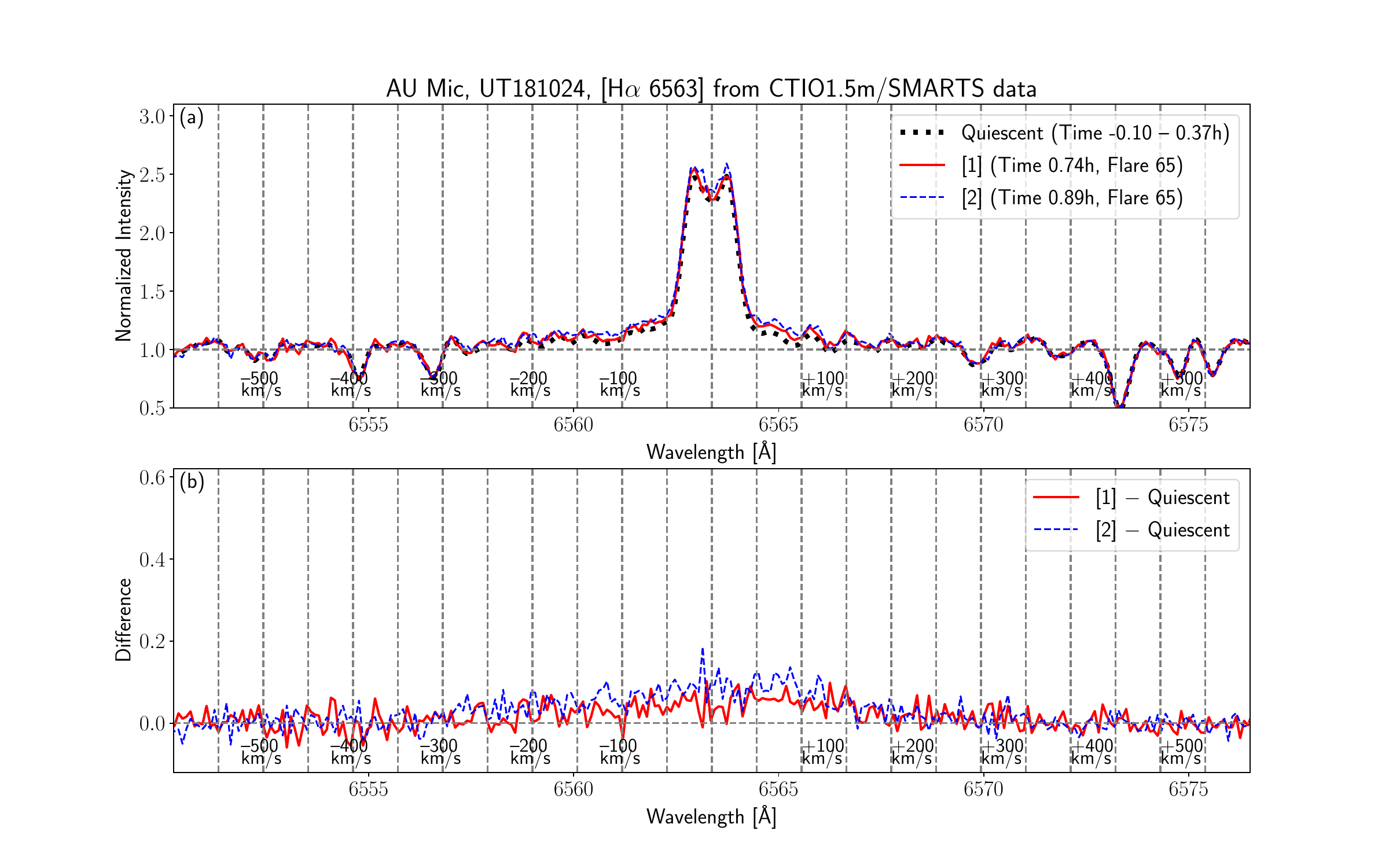}{0.5\textwidth}{\vspace{0mm}}
\fig{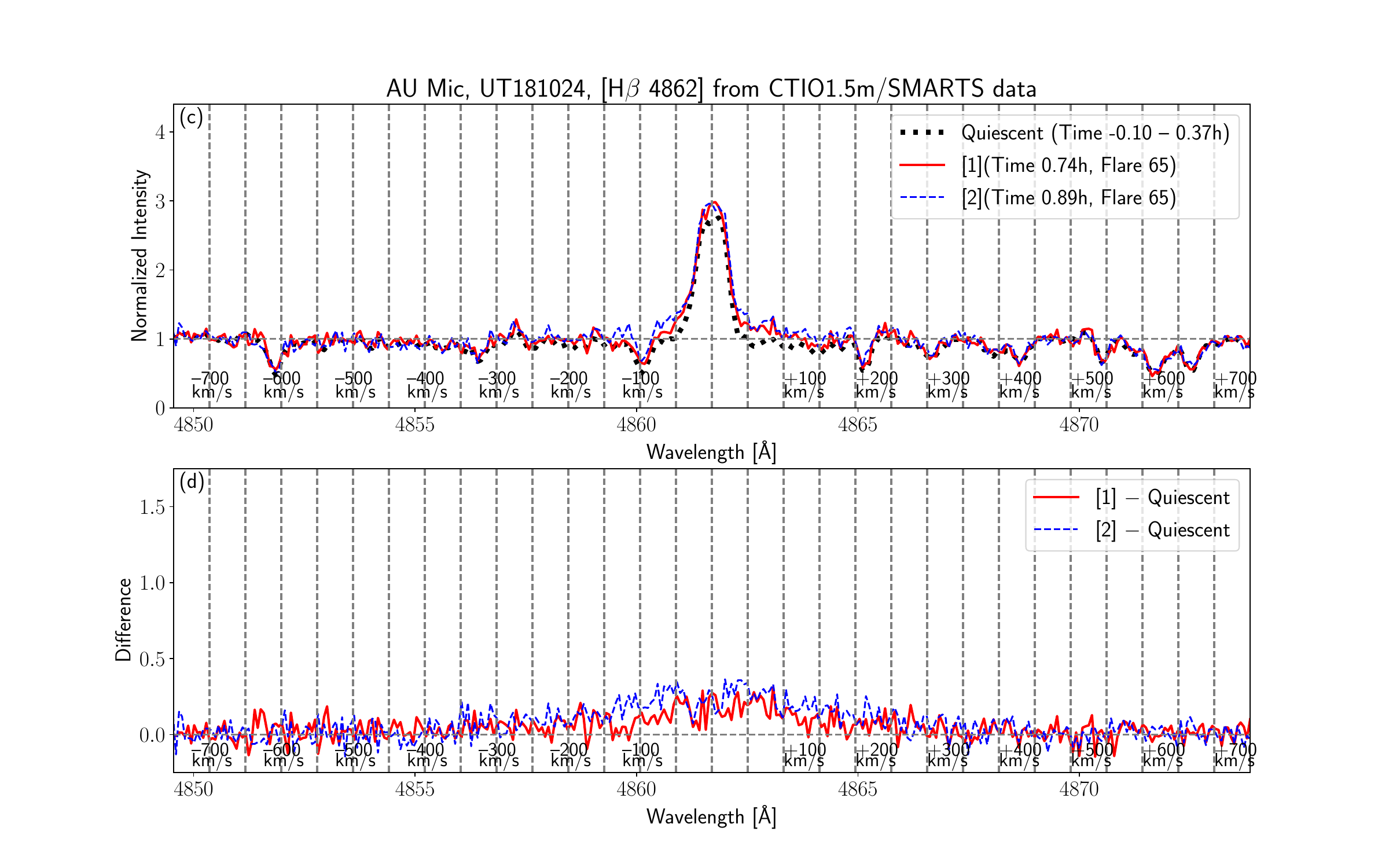}{0.5\textwidth}{\vspace{0mm}}
  }
     \vspace{-5mm}
      \gridline{
\fig{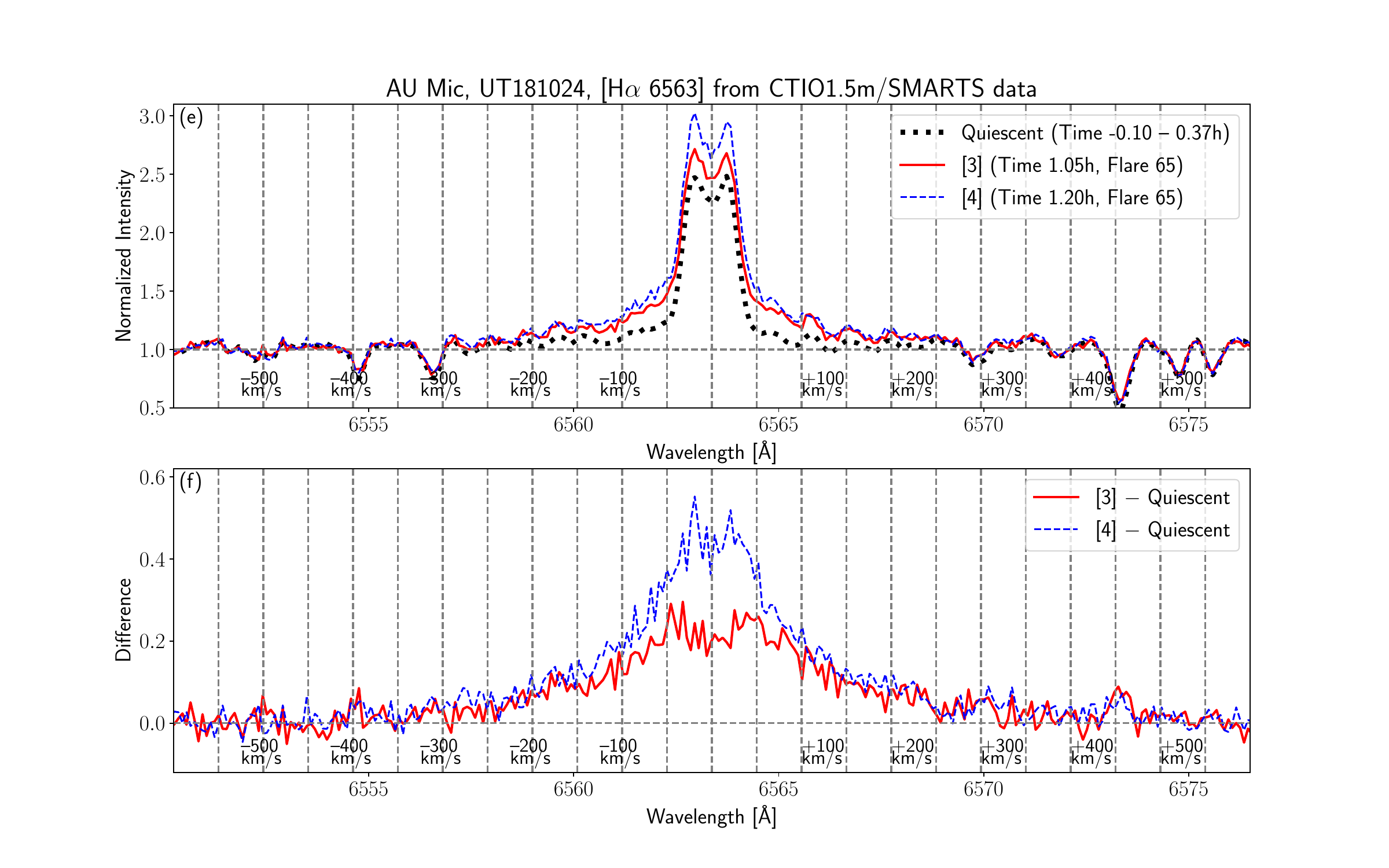}{0.5\textwidth}
{\vspace{0mm}}
\fig{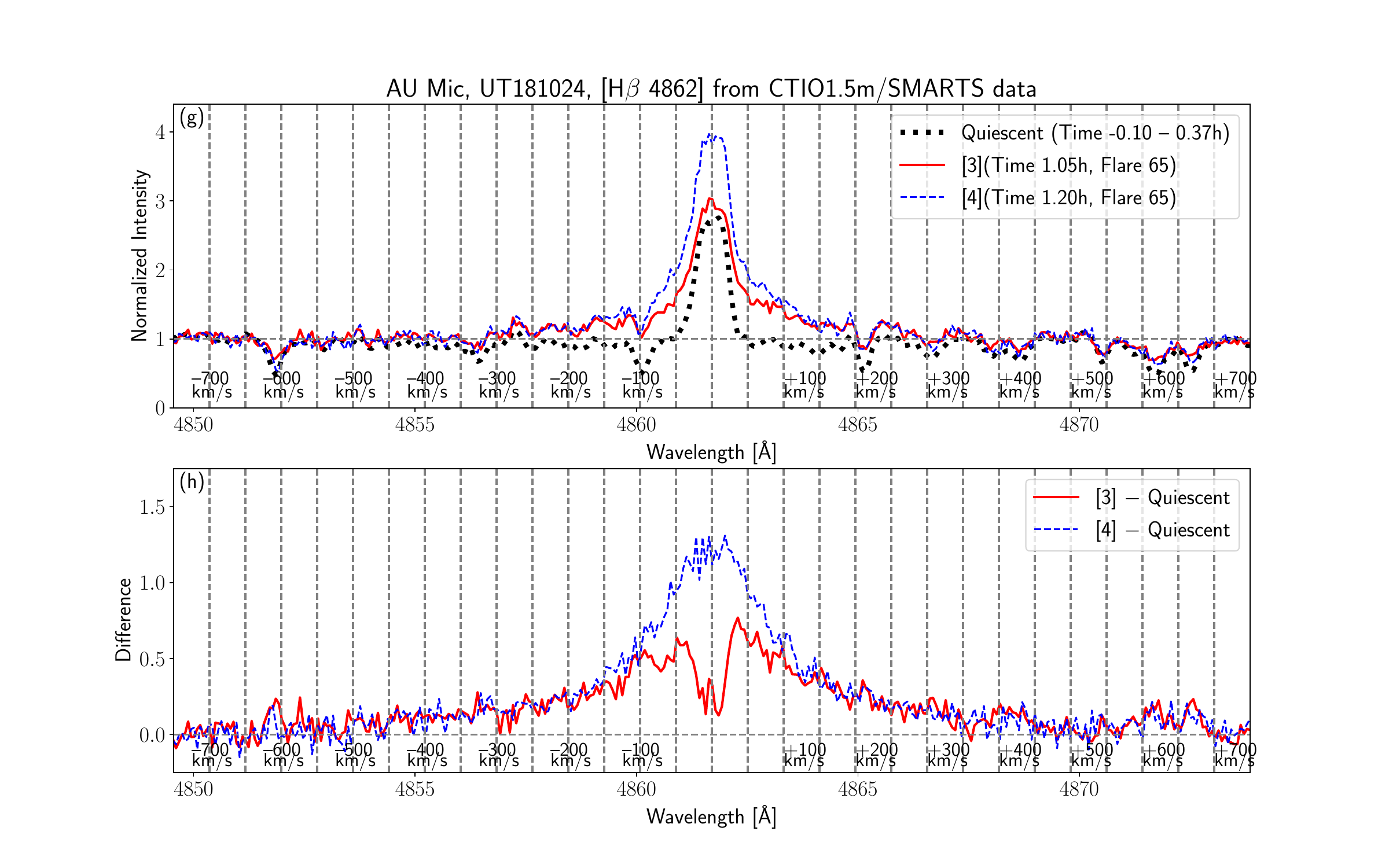}{0.5\textwidth}
{\vspace{0mm}}
}
     \vspace{-5mm}
      \gridline{
\fig{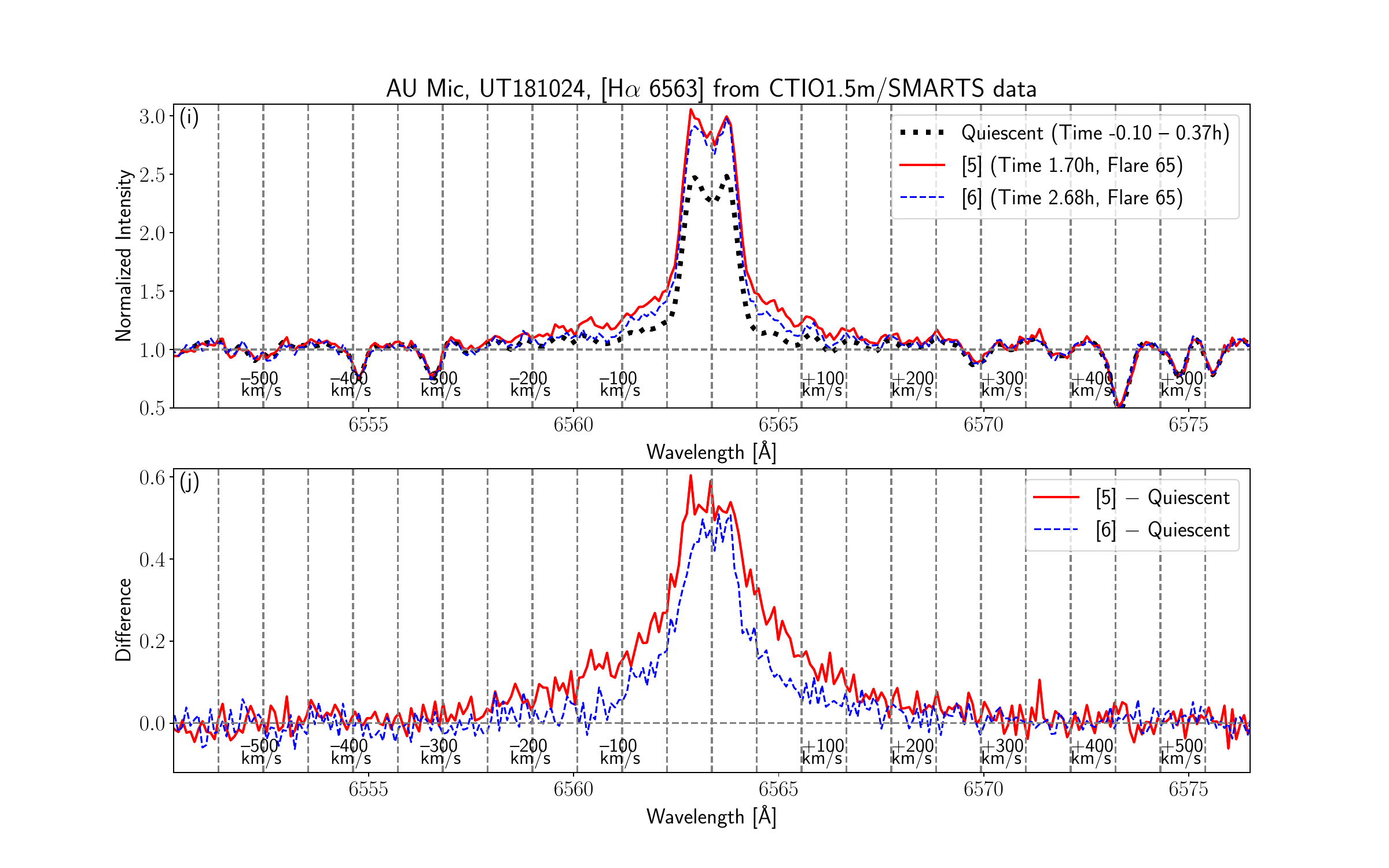}{0.5\textwidth}{\vspace{0mm}}
\fig{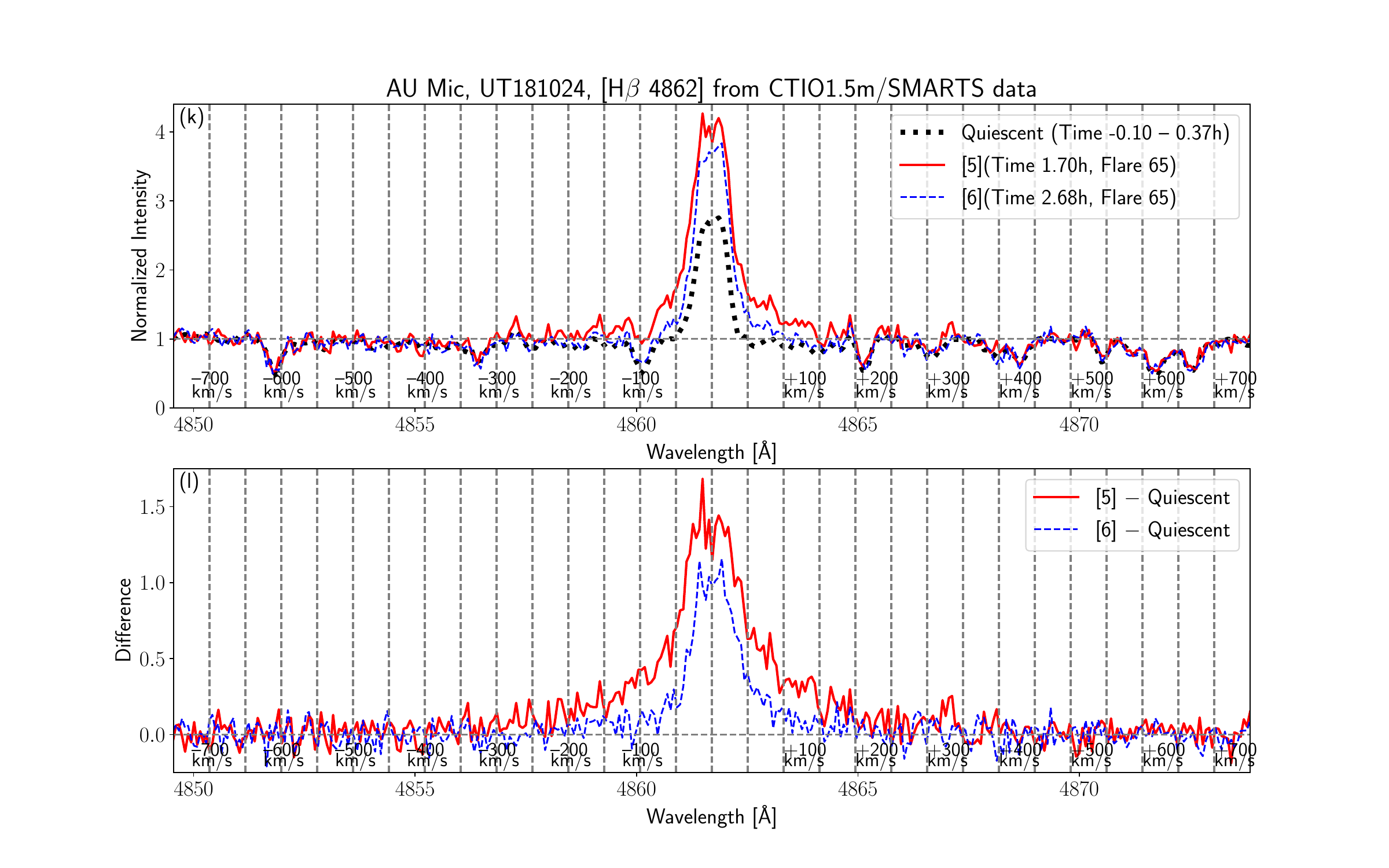}{0.5\textwidth}{\vspace{0mm}}
    }
     \vspace{-5mm}
     \caption{
     (a) Line profiles of the H$\alpha$ emission line during Flare 65 on 2018 
October 24 from the CHIRON spectra. 
The horizontal and vertical axes represent the wavelength and flux normalized by the continuum. 
The gray vertical dashed lines with velocity values represent the Doppler velocities from the H$\alpha$ line center.
The red solid and blue dashed lines indicate the line profiles at the time [1] and \color{black} time \color{black}  [2], respectively, which are indicated in 
Figure \ref{fig:maps_flare65} (a) \& (b) (lightcurve and intensity map)
and are during Flare 65. 
The black dotted line indicates the line profiles in quiescent phase, which are the average profile during -0.10-- +0.37 hr on this date (see Figure \ref{fig:maps_flare65}(a)).
(e) \& (i) Same as panel (a), but the line profiles at time [3] -- \color{black} time \color{black} [6] during Flare 65.
(b), (f), \& (j) Same as panels (a), (e), \& (i), respectively,
but the line profile differences from the quiescent phase.
(c), (d), (g), (h), (k), \& (l) Same as panels (a), (b), (e), (f), (i), \& (j), respectively, but for the H$\beta$ line.
}
   \label{fig:spec_HaHb_flare65}
   \end{center}
 \end{figure}

\begin{figure}[ht!]
   \begin{center}
      \gridline{
\fig{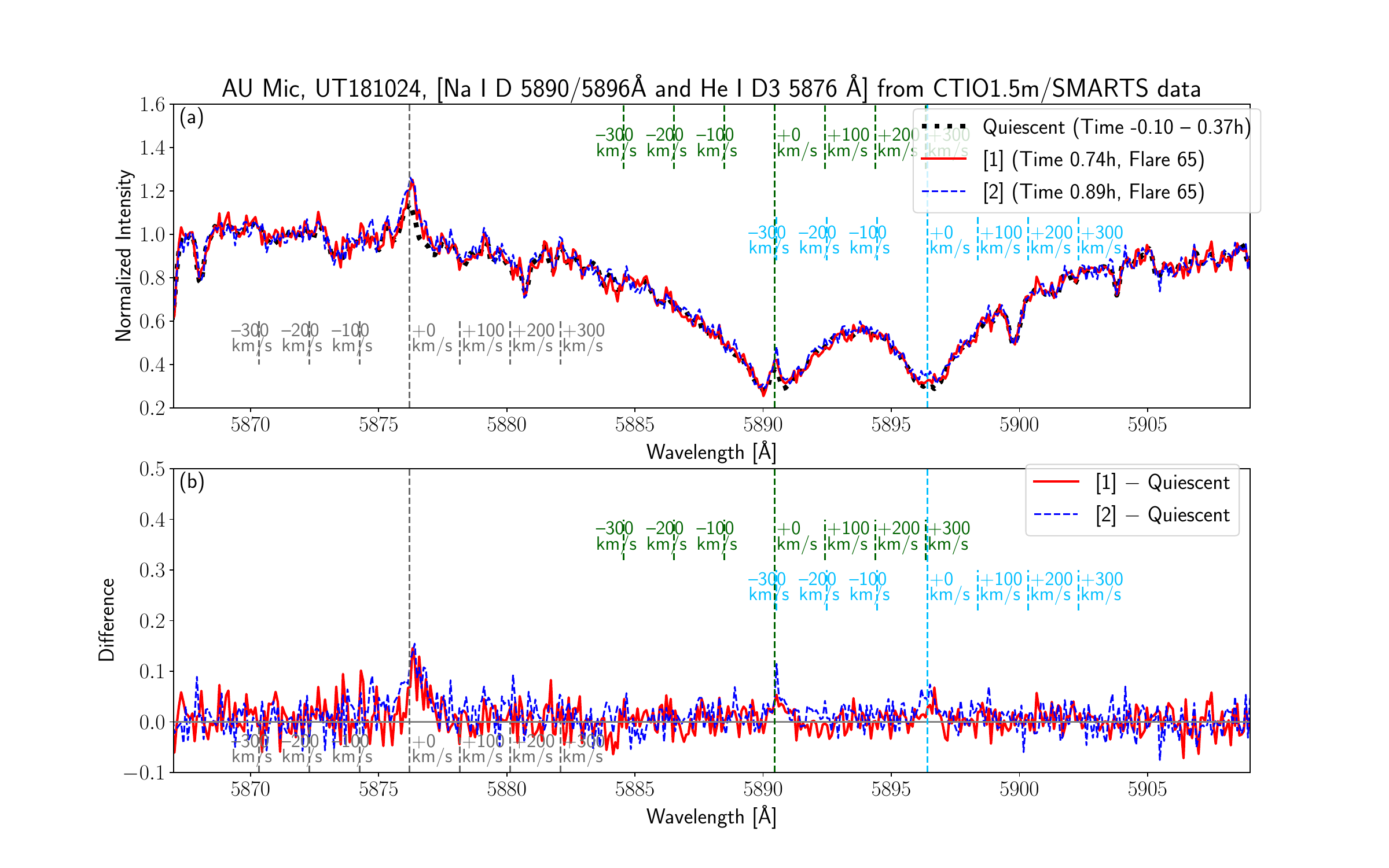}{0.5\textwidth}{\vspace{0mm}}
\fig{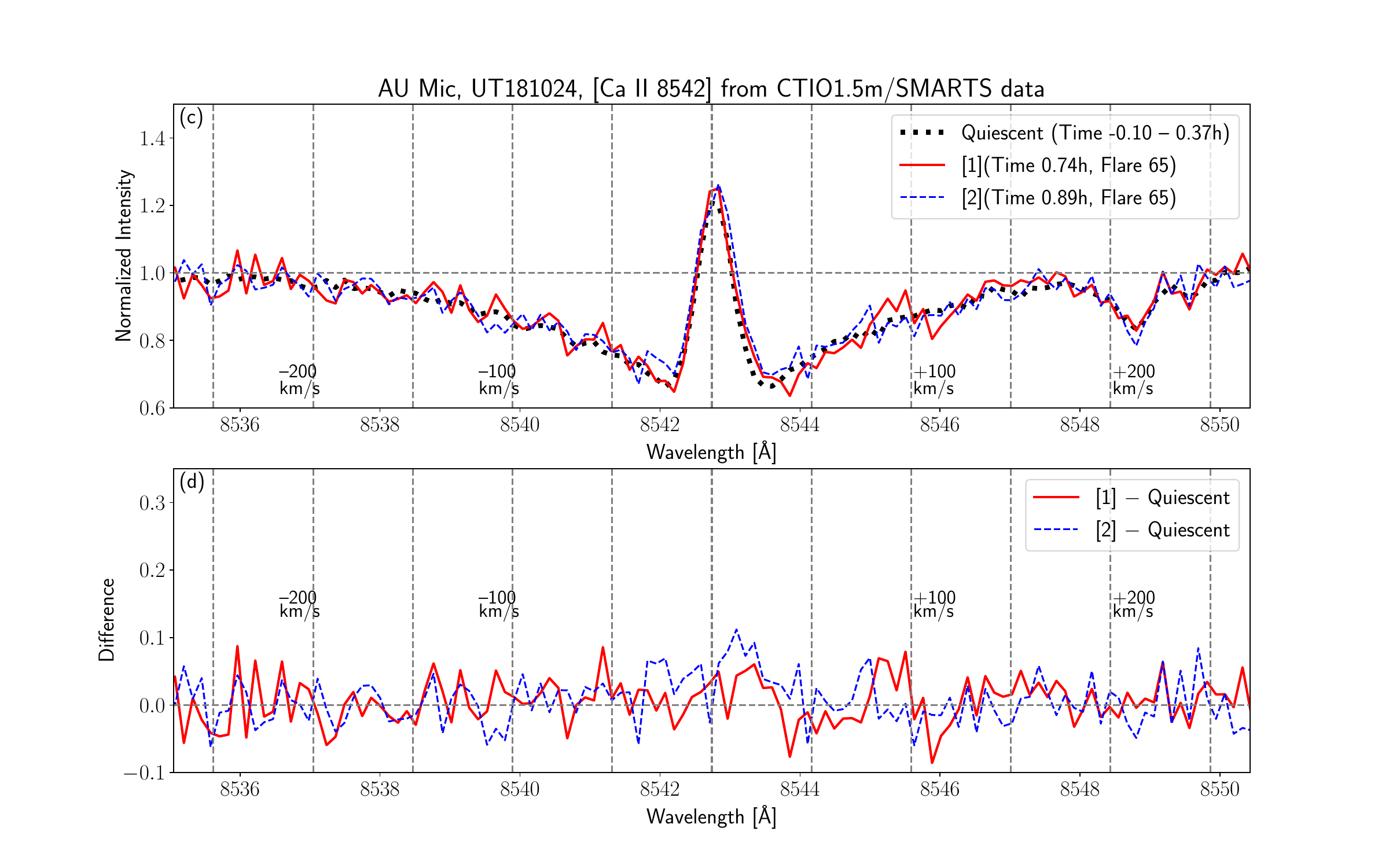}{0.5\textwidth}{\vspace{0mm}}
  }
     \vspace{-5mm}
      \gridline{
\fig{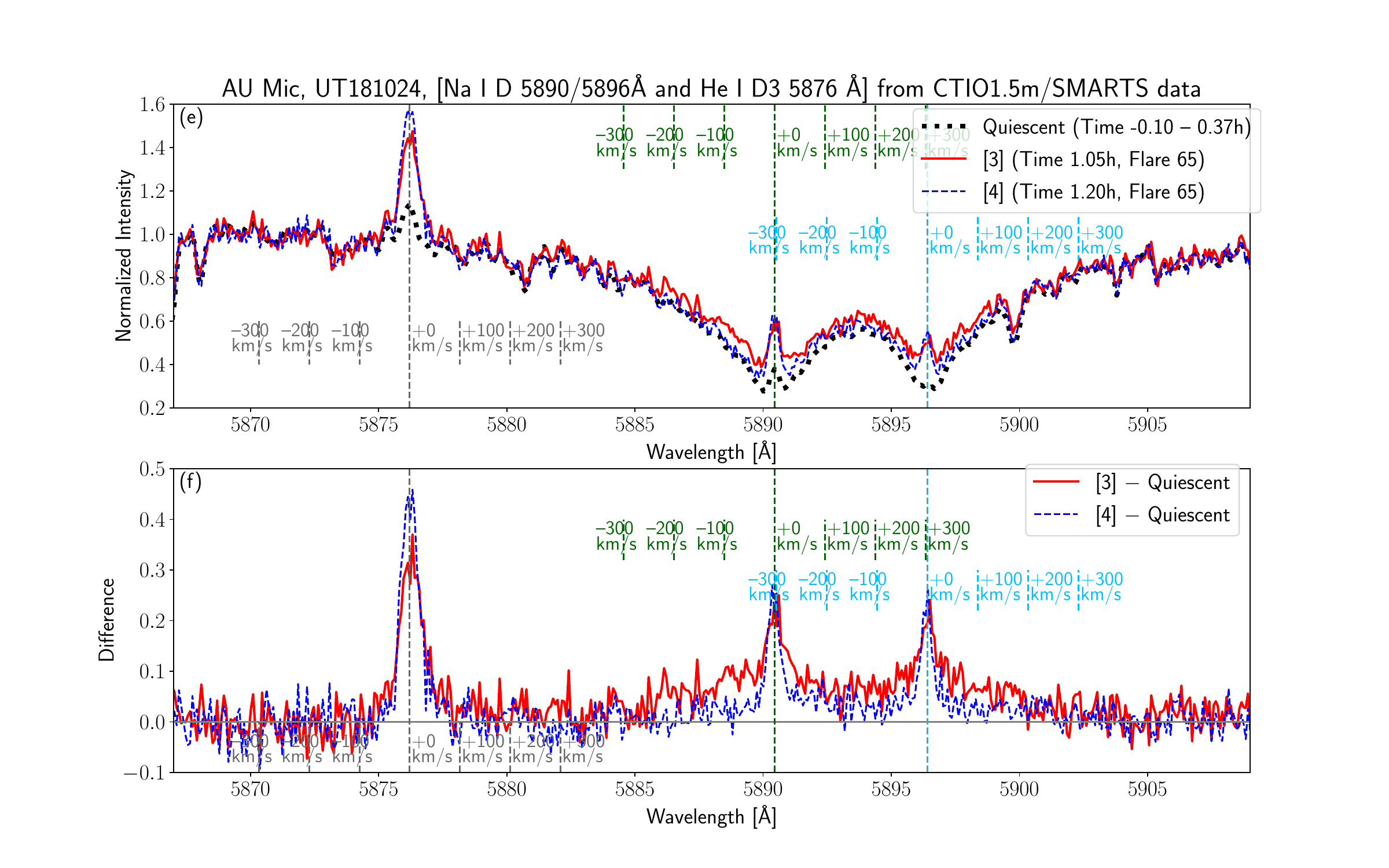}{0.5\textwidth}
{\vspace{0mm}}
\fig{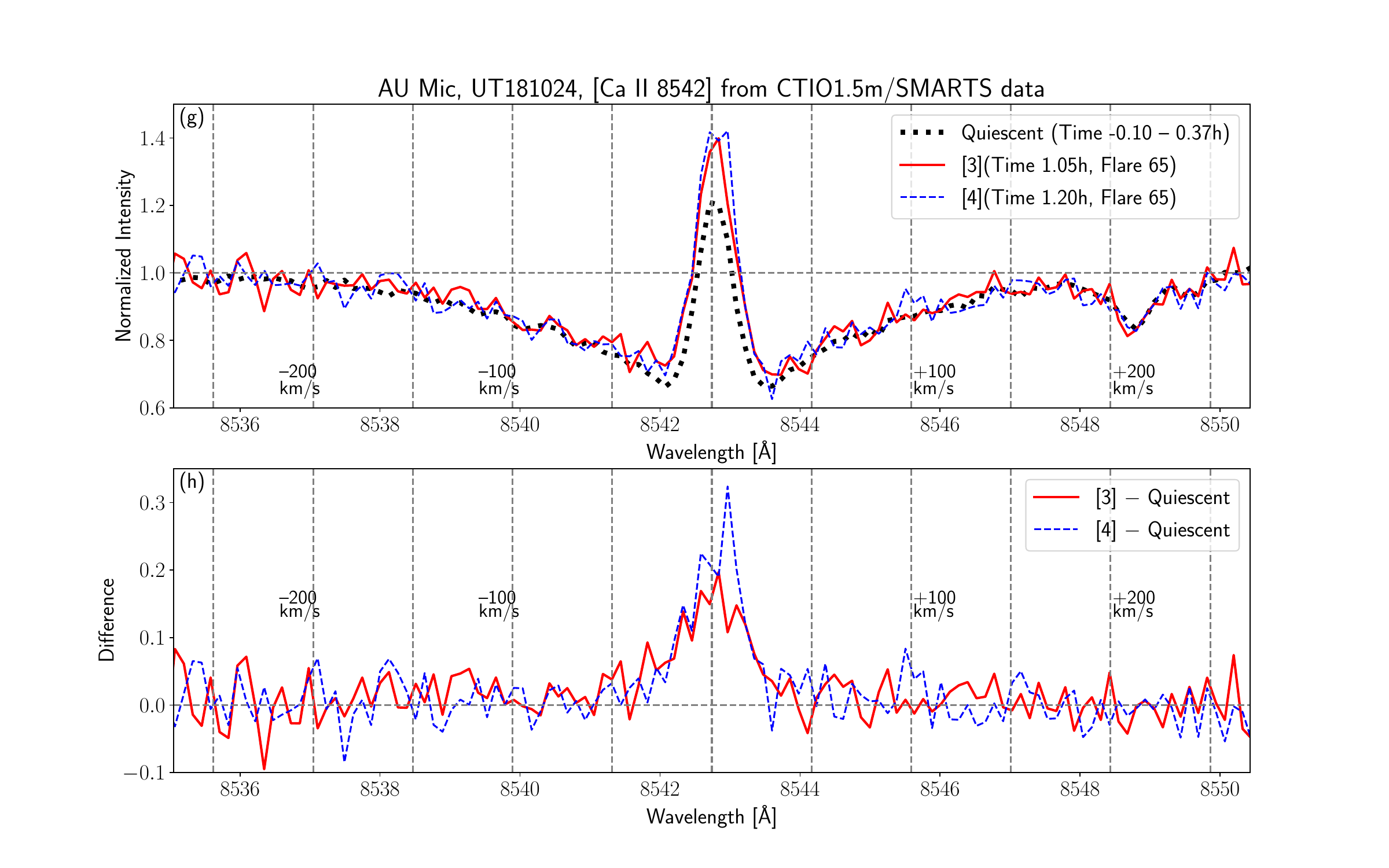}{0.5\textwidth}
{\vspace{0mm}}
}
     \vspace{-5mm}
      \gridline{
\fig{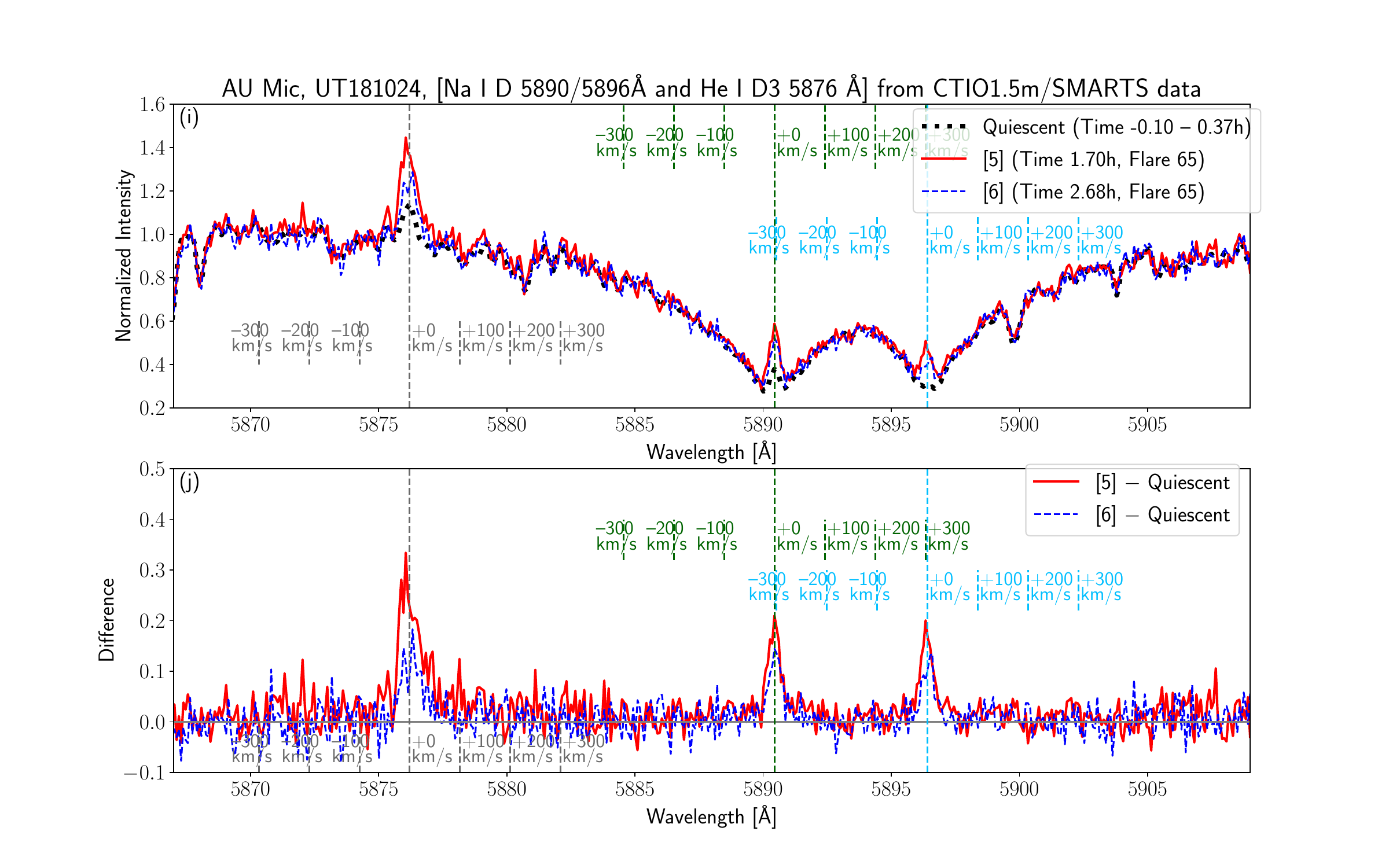}{0.5\textwidth}{\vspace{0mm}}
\fig{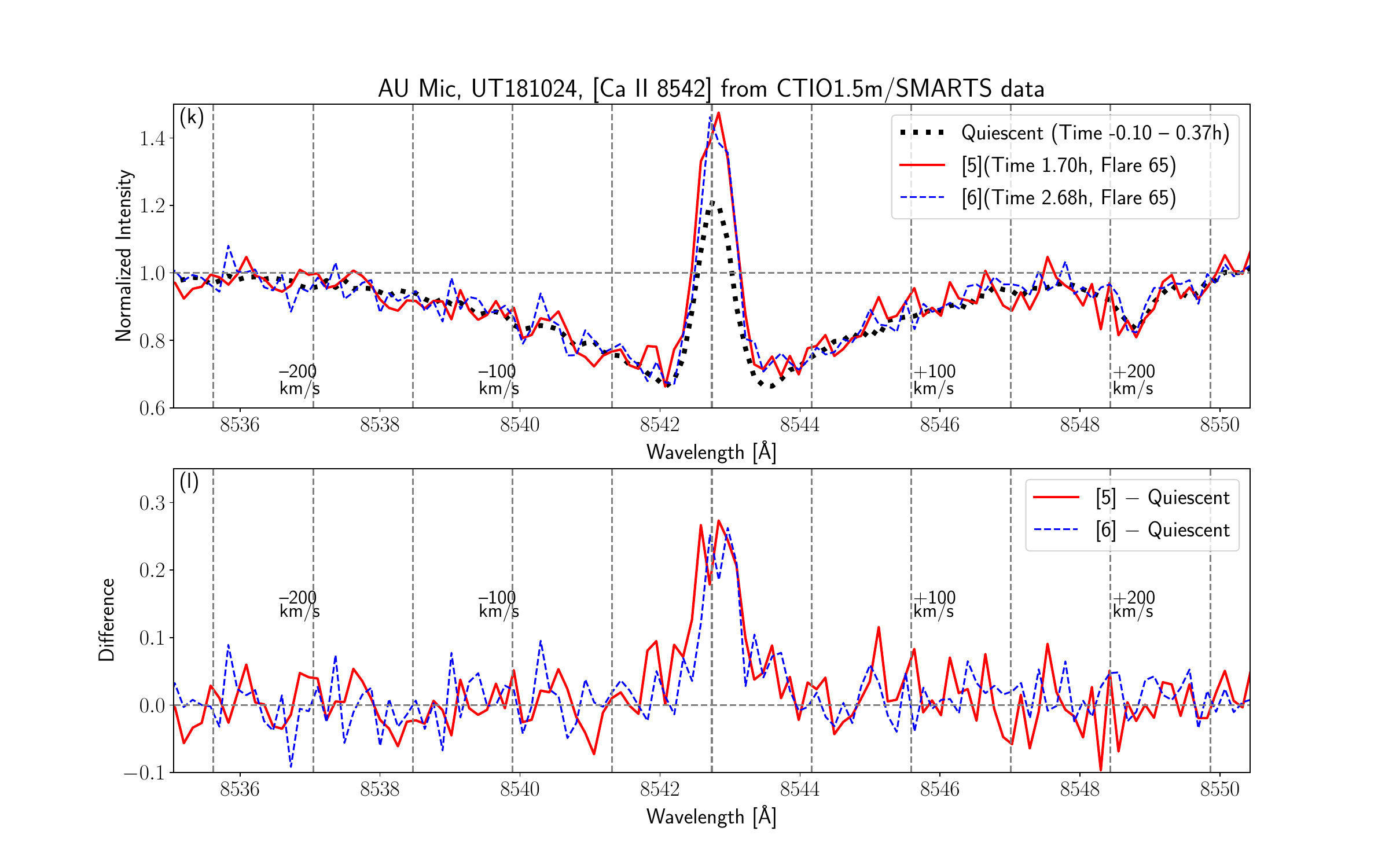}{0.5\textwidth}{\vspace{0mm}}
    }
     \vspace{-5mm}
     \caption{
Same as Figure \ref{fig:spec_HaHb_flare65}, the He I D3 5876\AA, Na D1 \& D2, and Ca II 8542 lines.
}
   \label{fig:spec_HeNaCa8542_flare65}
   \end{center}
 \end{figure}

\color{black}
In Section \ref{subsec:X-ray_specana_quiescent}, 
the quiescent phase X-ray spectra of Obs-ID 0822740301 with fitting results 
are shown in Figure \ref{fig:specfit_QuieALL1_0822740301},
while the RGS spectra focusing on Ne IX and OVII lines are shown in Figure 
\ref{fig:quieden_0822740301}. 
The same figures for Obs-IDs 0822740401,  0822740501, and  0822740601 
are shown Figures \ref{fig:specfit_QuieALL1_0822740401} -- \ref{fig:quieden_0822740601} in this appendix section.
The spectral 
fitting results of the time-averaged X-ray flare spectra in Section 
\ref{subsec:X-ray_specana_time-average} are listed in Table \ref{table:X-ray_flarefit_timeave}.
The time-resolved PN and MOS1 X-ray spectra and the best-fit model results for Flare 23 (Section \ref{subsec:ana_flare_23}) are shown in 
Figures \ref{fig:specfig_Flare23_TEM_quie_each_No.1-No.4}, \ref{fig:specfig_Flare23_TEM_quie_each_No.5-No.8} and, \ref{fig:specfig_Flare23_TEM_quie_each_No.9-No.12}.
Figures  \ref{fig:flare23_HeNaSpec} and \ref{fig:flare23_Ca8542Spec}
show the line profiles of the He I D3 5876\AA, Na D1 \& D2, and Ca II 8542\AA~emission lines during Flares 22 and 23 from the CHIRON spectra (Section \ref{subsec:ana_flare_23}). 
The time-resolved PN and MOS1 X-ray spectra and the best-fit model results for Flares 11 and 15 (Section \ref{subsec:ana_flare_11_and_15}) are shown in 
Figures \ref{fig:specfig_Flare11_TEM_quie_each_No.1-No.4}, 
\ref{fig:specfig_Flare11_TEM_quie_each_No.5-No.7}, and 
\ref{fig:specfig_Flare15_TEM_quie_each_No.8-No.11}.
The PN and MOS1 X-ray spectra and the best-fit model results for rise and decay phase data of Flares 35 and 36 (Section \ref{subsec:ana_flare_35_and_36}) are
shown in Figure \ref{fig:RiseDecayFit_fig1_Flare35and36}.
Figures \ref{fig:spec_HaHb_flare65} and \ref{fig:spec_HeNaCa8542_flare65} 
the line profiles of the H$\alpha$, H$\beta$, He I D3 5876\AA, Na D1 \& D2, and Ca II 8542\AA~emission lines during Flare 65 from the CHIRON spectra (Section \ref{subsec:ana_flare_65}).

\color{black}

\clearpage

\begin{longrotatetable}
\begin{deluxetable*}{cclcccccccccc}
\tablecaption{Best fitting parameters for the time-averaged rise and decay phase component spectra of X-ray flares. Statistical
90\% confidence region errors are shown for the X-ray fitting parameters.}
\tablewidth{0pt}
   \tablehead{
   \colhead{Flare ID} &    
\colhead{$QC_{\rm{PN}}^{\rm{Flare}}$ \tablenotemark{\rm \ddag}} & 
\colhead{Rise:} &
\colhead{$t_{\rm{rise}}$\tablenotemark{\rm *}} &
\colhead{$N_{\rm{H}}$ \tablenotemark{\rm \S}} &
\colhead{$T_{1}$} &
\colhead{$EM_{1}$} &
\colhead{$T_{2}$} &
\colhead{$EM_{2}$} &
\colhead{$Z_{\rm{Fe}}$} &
\colhead{$\chi^{2}/\rm{d.o.f}$} &
\colhead{$\chi^{2}_{\rm{black}}$} & 
\colhead{$L_{\rm{X}}$ \tablenotemark{$\S$}} 
\\
\colhead{} & 
\colhead{(cps)} & 
\colhead{}&
\colhead{(sec)} & 
\colhead{(10$^{20}$cm$^{-2}$)}&
\colhead{(keV)} &
\colhead{(10$^{51}$cm$^{-3}$)} &
\colhead{(keV)} &
\colhead{(10$^{51}$cm$^{-3}$)} & 
\colhead{($Z_{\rm{Fe,}\odot}$)}& 
\colhead{}&
\colhead{}&
\colhead{(10$^{29}$ erg s$^{-1}$)}
\\
\cline{3-13} 
\colhead{(Obs-ID\tablenotemark{\rm \dag}})&  
\colhead{}& 
\colhead{Decay:} &
\colhead{$t_{\rm{decay}}$\tablenotemark{\rm *}} &
\colhead{$N_{\rm{H}}$ \tablenotemark{\rm \S}} &
\colhead{$T_{1}$} &
\colhead{$EM_{1}$} &
\colhead{$T_{2}$} &
\colhead{$EM_{2}$} &
\colhead{$Z_{\rm{Fe}}$} &
\colhead{$\chi^{2}/\rm{d.o.f}$} &
\colhead{$\chi^{2}_{\rm{black}}$} & 
\colhead{$L_{\rm{X}}$ \tablenotemark{$\S$}}
\\
\colhead{}&  
\colhead{}& 
\colhead{}&
\colhead{(sec)} & 
\colhead{(10$^{20}$cm$^{-2}$)}&
\colhead{(keV)} &
\colhead{(10$^{51}$cm$^{-3}$)} &
\colhead{(keV)} &
\colhead{(10$^{51}$cm$^{-3}$)} & 
\colhead{($Z_{\rm{Fe,}\odot}$)}& 
\colhead{}& 
\colhead{}&
\colhead{(10$^{29}$ erg s$^{-1}$)}
\\
\cline{3-13} 
\colhead{}&  
\colhead{}& 
\colhead{Total:}&
\colhead{$t_{\rm{total}}$\tablenotemark{\rm *}}&
\colhead{}& 
\colhead{$T_{\rm{ave}}$ \tablenotemark{$\sharp$}}&
\colhead{$EM_{\rm{tot}}$ \tablenotemark{$\sharp$}} &
\colhead{$E_{\rm{X}}$ \tablenotemark{$\sharp$}} &
\colhead{$E_{\rm{UVW2}}$ \tablenotemark{\rm \P}} &
\colhead{Neupert \tablenotemark{\rm \P}} & 
\colhead{}& 
\colhead{}&
\colhead{}
\\
\colhead{}&  
\colhead{}& 
\colhead{}&
\colhead{(sec)}& 
\colhead{}& 
\colhead{(keV)}&
\colhead{(10$^{51}$cm$^{-3}$)} &
\colhead{(10$^{31}$erg)} &
\colhead{(10$^{30}$erg)} &
\colhead{}& 
\colhead{}&
\colhead{}&
\colhead{}
}
   \startdata 
1 & 17.86 & Rise: & 10.33 & fixed & 2.51$_{-0.71}^{+1.15}$ & 2.54$_{-0.72}^{+0.74}$ & \multicolumn{2}{c}{(1 Temperature)} & 1.27$_{-1.27}^{+1.27}$ & 114$/$85 & 1.34 &  3.46$_{-0.08}^{+0.07}$ \\ 
(301) & & Decay: & 22.67 & fixed & 1.42$_{-0.17}^{+0.21}$ & 5.07$_{-0.58}^{+0.59}$ & \multicolumn{2}{c}{(1 Temperature)} & 0.11$_{-0.11}^{+0.11}$ & 158$/$115 & 1.38 &  3.60$_{-0.05}^{+0.03}$ \\ 
\cline{3-13}   
& & Total: & 33.00&  & 1.63$_{-0.18}^{+0.24}$ & 4.28$_{-0.46}^{+0.47}$ & 9.9$_{-0.8}^{+0.6}$ & 18.0$\pm$1.0 & N &  &  &  \\ 
\cline{1-13}   
2 & 18.17 & Rise: & 30.50 & fixed & 3.08$_{-0.46}^{+0.54}$ & 3.00$_{-0.39}^{+0.40}$ & \multicolumn{2}{c}{(1 Temperature)} & 0.81$_{-0.81}^{+0.81}$ & 165$/$127 & 1.30 &  3.56$_{-0.05}^{+0.04}$ \\ 
(301) & & Decay: & 113.33 & fixed & 1.87$_{-0.10}^{+0.11}$ & 4.86$_{-0.24}^{+0.25}$ & \multicolumn{2}{c}{(1 Temperature)} & 0.19$_{-0.19}^{+0.19}$ & 344$/$183 & 1.88 &  3.66$_{-0.02}^{+0.02}$ \\ 
\cline{3-13}   
& & Total: & 143.83&  & 2.05$_{-0.11}^{+0.11}$ & 4.47$_{-0.21}^{+0.21}$ & 47.8$_{-1.6}^{+1.9}$ & -- & NN1 &  &  &  \\ 
\cline{1-13}   
4 & 17.40 & Rise: & 11.50 & fixed & 6.33$_{-3.83}^{+6.33}$ & 0.56$_{-0.56}^{+0.69}$ & \multicolumn{2}{c}{(1 Temperature)} & 8.04$_{-8.04}^{+8.04}$ & 71$/$85 & 0.84 &  3.19$_{-0.22}^{+0.03}$ \\ 
(301) & & Decay: & 46.00 & fixed & 1.17$_{-0.18}^{+0.25}$ & 1.99$_{-0.38}^{+0.38}$ & \multicolumn{2}{c}{(1 Temperature)} & 0.12$_{-0.12}^{+0.12}$ & 153$/$128 & 1.20 &  3.18$_{-0.03}^{+0.03}$ \\ 
\cline{3-13}   
& & Total: & 57.50&  & 1.51$_{-0.22}^{+0.32}$ & 1.71$_{-0.32}^{+0.34}$ & 7.4$_{-1.8}^{+0.9}$ & -- & NN1 &  &  &  \\ 
\cline{1-13}   
5 & 17.40 & Rise: & 18.17 & fixed & 3.78$_{-0.84}^{+1.30}$ & 1.81$_{-0.38}^{+0.42}$ & \multicolumn{2}{c}{(1 Temperature)} & 1.47$_{-1.47}^{+1.47}$ & 180$/$106 & 1.70 &  3.31$_{-0.05}^{+0.05}$ \\ 
(301) & & Decay: & 17.66 & fixed & 0.73$_{-0.17}^{+0.13}$ & 1.80$_{-0.97}^{+1.93}$ & 2.04$_{-0.55}^{+15.63}$ & 1.90$_{-1.37}^{+0.76}$ &  0.21$_{-0.13}^{+0.31}$ & 105$/$97 & 1.08 &  3.44$_{-0.08}^{+0.03}$ \\ 
\cline{3-13}   
& & Total: & 35.83&  & 2.20$_{-0.89}^{+8.40}$ & 2.74$_{-0.85}^{+1.05}$ & 8.7$_{-1.0}^{+0.6}$ & 5.9$\pm$0.3 & Q &  &  &  \\ 
\cline{1-13}   
8 & 17.18 & Rise: & 3.67 & fixed & 8.62$_{-3.90}^{+8.62}$ & 4.11$_{-1.70}^{+2.12}$ & \multicolumn{2}{c}{(1 Temperature)} & 3.59$_{-3.59}^{+3.59}$ & 52$/$53 & 0.97 &  4.09$_{-0.35}^{+0.21}$ \\ 
(301) & & Decay: & 5.50 & fixed & 3.59$_{-1.06}^{+1.41}$ & 4.60$_{-1.08}^{+1.11}$ & \multicolumn{2}{c}{(1 Temperature)} & 1.60$_{-1.60}^{+1.60}$ & 61$/$70 & 0.88 &  3.80$_{-0.20}^{+0.12}$ \\ 
\cline{3-13}   
& & Total: & 9.17&  & 5.47$_{-1.37}^{+2.45}$ & 4.40$_{-0.94}^{+1.08}$ & 5.4$_{-1.0}^{+0.6}$ & 35.0$\pm$0.0 & N &  &  &  \\ 
\cline{1-13}   
11 & 15.54 & Rise: & 331.50 & \multicolumn{9}{c}{(Time-resolved analysis in Section \ref{sec:remarkable_flares})} \\ 
(301) & & Decay: & 150.83 & \multicolumn{9}{c}{(Time-resolved analysis in Section \ref{sec:remarkable_flares})} \\ 
\cline{3-13}   
& & Total: & 482.33&  & 1.47$_{-0.96}^{+1.02}$ & 11.19$_{-0.53}^{+0.54}$ & 363.3$_{-6.8}^{+2.7}$ & -- & NN1 &  &  &  \\ 
\cline{1-13}   
15 & 15.54 & Rise: & 90.17 & \multicolumn{9}{c}{(Time-resolved analysis in Section \ref{sec:remarkable_flares})} \\ 
(301) & & Decay: & 33.00 & \multicolumn{9}{c}{(Time-resolved analysis in Section \ref{sec:remarkable_flares})} \\ 
\cline{3-13}   
& & Total: & 123.17&  & 2.14$_{-1.01}^{+1.09}$ & 30.43$_{-1.10}^{+1.17}$ & 288.1$_{-2.5}^{+2.1}$ & 500.0$\pm$40.0 & N$/$Q &  &  &  \\ 
\cline{1-13}   
16 & 16.51 & Rise: & 24.67 & fixed & 0.29$_{-0.09}^{+0.10}$ & 0.97$_{-0.40}^{+0.53}$ & 2.34$_{-1.08}^{+7.86}$ & 0.86$_{-0.45}^{+0.31}$ & 0.20$_{-0.20}^{+1.92}$ & 104$/$102 & 1.02 &  3.00$_{-0.05}^{+0.06}$ \\ 
(401) & & Decay: & 18.33 & fixed & 0.14$_{-0.04}^{+0.08}$ & 1.36$_{-0.79}^{+1.42}$ & 0.79$_{-0.26}^{+0.59}$ & 1.30$_{-0.78}^{+0.78}$ &  0.07$_{-0.07}^{+0.31}$ & 121$/$94 & 1.29 &  3.04$_{-0.11}^{+0.06}$ \\ 
\cline{3-13}   
& & Total: & 43.00&  & 0.84$_{-0.36}^{+1.25}$ & 2.19$_{-0.59}^{+0.77}$ & 6.2$_{-1.5}^{+1.1}$ & 3.2$\pm$0.3 & N$/$Q &  &  &  \\ 
\cline{1-13}   
17 & 17.23 & Rise: & 38.00 & fixed & 0.85$_{-0.56}^{+0.32}$ & 0.51$_{-0.37}^{+1.12}$ & 2.66$_{-0.42}^{+0.76}$ & 3.45$_{-0.83}^{+0.57}$ & 0.24$_{-0.19}^{+0.29}$ & 155$/$132 & 1.18 &  3.46$_{-0.05}^{+0.03}$ \\ 
(401) & & Decay: & 19.33 & fixed & 0.75$_{-0.35}^{+0.17}$ & 3.29$_{-1.85}^{+1.14}$ & 4.37$_{-2.98}^{+4.37}$ & 0.74$_{-0.56}^{+1.72}$ &  0.08$_{-0.05}^{+0.15}$ & 110$/$100 & 1.10 &  3.39$_{-0.17}^{+0.04}$ \\ 
\cline{3-13}   
& & Total: & 57.33&  & 2.08$_{-0.61}^{+1.12}$ & 3.98$_{-0.88}^{+1.08}$ & 17.7$_{-2.3}^{+0.8}$ & -- & NN1 &  &  &  \\ 
\cline{1-13}   
18 & 17.23 & Rise: & 9.17 & fixed & 1.90$_{-0.93}^{+8.36}$ & 1.53$_{-0.72}^{+0.67}$ & \multicolumn{2}{c}{(1 Temperature)} & 0.00$_{-0.00}^{+0.00}$ & 84$/$78 & 1.07 &  3.09$_{-0.08}^{+0.07}$ \\ 
(401) & & Decay: & 26.66 & fixed & 1.05$_{-0.12}^{+0.14}$ & 4.43$_{-0.55}^{+0.57}$ & \multicolumn{2}{c}{(1 Temperature)} & 0.08$_{-0.08}^{+0.08}$ & 128$/$112 & 1.15 &  3.38$_{-0.05}^{+0.05}$ \\ 
\cline{3-13}   
& & Total: & 35.83&  & 1.14$_{-0.13}^{+0.55}$ & 3.68$_{-0.45}^{+0.46}$ & 8.3$_{-0.9}^{+0.9}$ & 17.0$\pm$1.0 & Q &  &  &  \\ 
\cline{1-13}   
20 & 15.92 & Rise: & 15.33 & fixed & 0.45$_{-0.15}^{+0.54}$ & 2.02$_{-1.06}^{+1.82}$ & 3.78$_{-1.83}^{+3.78}$ & 1.74$_{-1.07}^{+1.03}$ & 0.00$_{-0.00}^{+0.15}$ & 80$/$90 & 0.89 &  3.14$_{-0.10}^{+0.05}$ \\ 
(401) & & Decay: & 186.00 & fixed & 0.68$_{-0.10}^{+0.09}$ & 2.82$_{-0.49}^{+0.52}$ & 2.19$_{-0.45}^{+1.05}$ & 1.02$_{-0.43}^{+0.42}$ &  0.06$_{-0.02}^{+0.02}$ & 367$/$185 & 1.98 &  3.10$_{-0.02}^{+0.01}$ \\ 
\cline{3-13}   
& & Total: & 201.33&  & 1.14$_{-0.30}^{+0.39}$ & 3.83$_{-0.61}^{+0.64}$ & 49.7$_{-2.5}^{+1.6}$ & 28.0$\pm$1.0 & N$/$Q &  &  &  \\ 
\cline{1-13}   
21 & 15.88 & Rise: & 13.50 & fixed & 0.72$_{-0.34}^{+0.22}$ & 4.79$_{-2.99}^{+1.91}$ & 5.42$_{-3.16}^{+5.42}$ & 2.20$_{-1.05}^{+2.34}$ & 0.00$_{-0.00}^{+0.07}$ & 85$/$93 & 0.91 &  3.49$_{-0.39}^{+0.06}$ \\ 
(401) & & Decay: & 51.17 & fixed & 0.79$_{-0.08}^{+0.06}$ & 6.50$_{-0.61}^{+0.44}$ & 68.44$_{-62.05}^{+68.44}$ & 0.97$_{-0.30}^{+0.21}$ &  0.04$_{-0.02}^{+0.02}$ & 167$/$139 & 1.20 &  3.50$_{-0.23}^{+0.24}$ \\ 
\cline{3-13}   
& & Total: & 64.67&  & 8.14$_{-5.89}^{+6.47}$ & 7.37$_{-0.85}^{+0.74}$ & 31.3$_{-7.7}^{+7.3}$ & 7.1$\pm$0.3 & Q &  &  &  \\ 
\cline{1-13}   
22 & 16.74 & Rise: & 19.33 & fixed & 0.73$_{-0.12}^{+0.10}$ & 2.34$_{-0.97}^{+1.44}$ & 3.43$_{-0.64}^{+1.34}$ & 5.27$_{-1.06}^{+0.81}$ & 0.20$_{-0.10}^{+0.18}$ & 105$/$118 & 0.89 &  3.93$_{-0.09}^{+0.04}$ \\ 
(401) & & Decay: & 27.34 & 1.17$_{-0.12}^{+2.98}$ & 0.81$_{-0.11}^{+0.08}$ & 5.36$_{-1.40}^{+1.11}$ & 68.44$_{-66.66}^{+68.44}$ & 0.48$_{-0.32}^{+0.28}$ & 0.09$_{-0.04}^{+0.03}$ & 160$/$110 & 1.45 &  3.44$_{-0.19}^{+0.52}$ \\ 
\cline{3-13}   
& & Total: & 46.67&  & 4.56$_{-2.53}^{+2.56}$ & 6.57$_{-1.03}^{+0.96}$ & 22.7$_{-3.3}^{+8.5}$ & 120.0$\pm$0.0 & N &  &  &  \\ 
\cline{1-13}   
23 & 17.51 & Rise: & 9.83 & 1.15$_{-0.74}^{+0.74}$ & 1.07$_{-0.11}^{+0.16}$ & 3.64$_{-1.20}^{+2.87}$ & 5.03$_{-0.40}^{+0.65}$ & 44.44$_{-2.53}^{+1.83}$ & 0.57$_{-0.13}^{+0.12}$ & 228$/$177 & 1.29 &  11.41$_{-0.24}^{+0.25}$ \\ 
(401) & & Decay: & 51.84 & fixed & 0.89$_{-0.03}^{+0.03}$ & 3.15$_{-0.40}^{+0.47}$ & 2.76$_{-0.11}^{+0.11}$ & 15.56$_{-0.46}^{+0.48}$ &  0.51$_{-0.06}^{+0.06}$ & 269$/$194 & 1.39 &  5.82$_{-0.04}^{+0.05}$ \\ 
\cline{3-13}   
& & Total: & 61.67&  & 3.19$_{-0.14}^{+0.17}$ & 23.40$_{-0.68}^{+0.78}$ & 139.1$_{-1.8}^{+2.1}$ & 970.0$\pm$30.0 & N &  &  &  \\ 
\cline{1-13}   
24 & 17.57 & Rise: & 53.67 & fixed & 0.83$_{-0.15}^{+0.14}$ & 0.62$_{-0.24}^{+0.44}$ & 2.81$_{-0.23}^{+0.27}$ & 5.61$_{-0.41}^{+0.41}$ & 0.37$_{-0.13}^{+0.14}$ & 213$/$159 & 1.34 &  3.88$_{-0.04}^{+0.03}$ \\ 
(401) & & Decay: & 47.00 & fixed & 0.92$_{-0.07}^{+0.09}$ & 3.98$_{-1.22}^{+3.72}$ & 2.38$_{-0.21}^{+0.89}$ & 9.30$_{-3.02}^{+1.07}$ &  0.20$_{-0.09}^{+0.07}$ & 185$/$164 & 1.13 &  4.69$_{-0.05}^{+0.04}$ \\ 
\cline{3-13}   
& & Total: & 100.67&  & 2.18$_{-0.42}^{+0.52}$ & 9.52$_{-1.54}^{+1.83}$ & 77.5$_{-1.9}^{+1.4}$ & 150.0$\pm$30.0 & N$/$Q &  &  &  \\ 
\cline{1-13}   
25 & 19.63 & Rise: & 23.83 & fixed & 3.01$_{-0.36}^{+0.42}$ & 5.80$_{-0.50}^{+0.51}$ & \multicolumn{2}{c}{(1 Temperature)} & 0.50$_{-0.50}^{+0.50}$ & 175$/$128 & 1.36 &  4.19$_{-0.05}^{+0.06}$ \\ 
(401) & & Decay: & 6.17 & fixed & 1.91$_{-0.56}^{+1.55}$ & 4.16$_{-1.17}^{+1.14}$ & \multicolumn{2}{c}{(1 Temperature)} & 0.20$_{-0.20}^{+0.20}$ & 69$/$76 & 0.91 &  3.82$_{-0.09}^{+0.08}$ \\ 
\cline{3-13}   
& & Total: & 30.00&  & 2.83$_{-0.32}^{+0.49}$ & 5.46$_{-0.47}^{+0.47}$ & 14.3$_{-0.8}^{+0.9}$ & 21.0$\pm$0.0 & Q &  &  &  \\ 
\cline{1-13}   
27 & 17.32 & Rise: & 9.17 & fixed & 2.68$_{-0.35}^{+0.42}$ & 11.27$_{-0.96}^{+0.96}$ & \multicolumn{2}{c}{(1 Temperature)} & 0.28$_{-0.28}^{+0.28}$ & 131$/$97 & 1.35 &  4.48$_{-0.07}^{+0.10}$ \\ 
(401) & & Decay: & 39.66 & fixed & 0.91$_{-0.05}^{+0.05}$ & 2.35$_{-0.55}^{+0.97}$ & 2.46$_{-0.19}^{+0.27}$ & 7.45$_{-0.73}^{+0.55}$ &  0.43$_{-0.11}^{+0.11}$ & 148$/$152 & 0.97 &  4.35$_{-0.04}^{+0.03}$ \\ 
\cline{3-13}   
& & Total: & 48.83&  & 2.22$_{-0.21}^{+0.23}$ & 10.07$_{-0.76}^{+0.92}$ & 42.6$_{-1.1}^{+0.9}$ & 100.0$\pm$0.0 & N &  &  &  \\ 
\cline{1-13}   
28 & 15.33 & Rise: & 28.17 & fixed & 2.01$_{-0.47}^{+0.86}$ & 2.68$_{-0.56}^{+0.52}$ & \multicolumn{2}{c}{(1 Temperature)} & 0.23$_{-0.23}^{+0.23}$ & 181$/$114 & 1.58 &  2.91$_{-0.03}^{+0.04}$ \\ 
(401) & & Decay: & 167.50 & fixed & 0.65$_{-0.06}^{+0.09}$ & 3.43$_{-0.30}^{+0.33}$ & 25.78$_{-14.40}^{+25.78}$ & 0.96$_{-0.09}^{+0.17}$ &  0.01$_{-0.01}^{+0.01}$ & 353$/$196 & 1.80 &  3.09$_{-0.18}^{+0.02}$ \\ 
\cline{3-13}   
& & Total: & 195.67&  & 5.76$_{-2.72}^{+4.87}$ & 4.14$_{-0.28}^{+0.32}$ & 56.7$_{-18.1}^{+2.0}$ & 19.0$\pm$1.0 & N &  &  &  \\ 
\cline{1-13}   
30 & 15.33 & Rise: & 38.78 & fixed & 0.48$_{-0.06}^{+0.09}$ & 2.71$_{-0.37}^{+0.38}$ & \multicolumn{2}{c}{(1 Temperature)} & 0.05$_{-0.05}^{+0.05}$ & 153$/$115 & 1.33 &  2.85$_{-0.03}^{+0.03}$ \\ 
(401) & & Decay: & 47.55 & fixed & 0.35$_{-0.03}^{+0.04}$ & 2.11$_{-0.27}^{+0.28}$ & \multicolumn{2}{c}{(1 Temperature)} & 0.14$_{-0.14}^{+0.14}$ & 193$/$116 & 1.66 &  2.82$_{-0.03}^{+0.02}$ \\ 
\cline{3-13}   
& & Total: & 86.33&  & 0.42$_{-0.03}^{+0.05}$ & 2.38$_{-0.22}^{+0.23}$ & 13.1$_{-0.9}^{+1.0}$ & -- & NN1 &  &  &  \\ 
\cline{1-13}   
32 & 14.34 & Rise: & 54.50 & fixed & 0.24$_{-0.04}^{+0.04}$ & 0.91$_{-0.20}^{+0.21}$ & \multicolumn{2}{c}{(1 Temperature)} & 0.59$_{-0.59}^{+0.59}$ & 166$/$122 & 1.36 &  2.56$_{-0.03}^{+0.02}$ \\ 
(401) & & Decay: & 46.17 & fixed & 0.36$_{-0.05}^{+0.06}$ & 1.74$_{-0.28}^{+0.31}$ & 68.43$_{-45.36}^{+68.43}$ & 1.26$_{-0.25}^{+0.16}$ &  0.10$_{-0.07}^{+0.11}$ & 174$/$133 & 1.31 &  2.86$_{-0.24}^{+0.17}$ \\ 
\cline{3-13}   
& & Total: & 100.67&  & 21.37$_{-10.83}^{+15.76}$ & 1.87$_{-0.20}^{+0.20}$ & 15.9$_{-6.7}^{+4.7}$ & -- & NN1 &  &  &  \\ 
\cline{1-13}   
34 & 14.40 & Rise: & 83.83 & fixed & 0.78$_{-0.09}^{+0.07}$ & 4.52$_{-0.83}^{+0.82}$ & 4.03$_{-0.96}^{+1.99}$ & 2.23$_{-0.59}^{+0.65}$ & 0.06$_{-0.02}^{+0.03}$ & 247$/$166 & 1.49 &  3.23$_{-0.04}^{+0.02}$ \\ 
(501) & & Decay: & 96.34 & fixed & 0.71$_{-0.10}^{+0.10}$ & 2.24$_{-0.97}^{+0.87}$ & 0.37$_{-0.08}^{+0.06}$ & 1.72$_{-0.78}^{+1.00}$ &  0.09$_{-0.03}^{+0.04}$ & 262$/$142 & 1.85 &  2.86$_{-0.02}^{+0.02}$ \\ 
\cline{3-13}   
& & Total: & 180.17&  & 1.33$_{-0.29}^{+0.38}$ & 5.26$_{-0.82}^{+0.86}$ & 62.5$_{-2.1}^{+1.6}$ & -- & NN1 &  &  &  \\ 
\cline{1-13}   
35 & 15.93 & Rise: & 8.83 & fixed & 1.14$_{-0.24}^{+0.39}$ & 1.17$_{-0.81}^{+4.04}$ & 3.40$_{-0.81}^{+10.42}$ & 5.61$_{-3.09}^{+1.14}$ & 0.65$_{-0.41}^{+0.63}$ & 78$/$85 & 0.91 &  3.79$_{-0.13}^{+0.09}$ \\ 
(501) & & Decay: & 27.00 & fixed & 0.85$_{-0.05}^{+0.05}$ & 3.83$_{-1.00}^{+2.48}$ & 2.26$_{-0.22}^{+0.57}$ & 7.73$_{-1.81}^{+0.80}$ &  0.30$_{-0.11}^{+0.11}$ & 206$/$129 & 1.60 &  4.24$_{-0.06}^{+0.05}$ \\ 
\cline{3-13}   
& & Total: & 35.83&  & 1.99$_{-0.41}^{+1.01}$ & 10.38$_{-1.75}^{+2.22}$ & 30.9$_{-1.2}^{+0.9}$ & -- & Un &  &  &  \\ 
\cline{1-13}   
36 & 15.93 & Rise: & 13.17 & fixed & 0.87$_{-0.38}^{+0.56}$ & 1.01$_{-0.66}^{+5.00}$ & 3.77$_{-0.80}^{+3.77}$ & 5.97$_{-3.32}^{+0.88}$ & 0.32$_{-0.28}^{+0.49}$ & 82$/$98 & 0.84 &  3.78$_{-0.12}^{+0.05}$ \\ 
(501) & & Decay: & 22.66 & fixed & 1.61$_{-1.26}^{+1.61}$ & 3.76$_{-2.82}^{+1.14}$ & 0.85$_{-0.17}^{+0.23}$ & 1.76$_{-0.80}^{+3.87}$ &  0.28$_{-0.12}^{+0.19}$ & 132$/$105 & 1.25 &  3.39$_{-0.05}^{+0.04}$ \\ 
\cline{3-13}   
& & Total: & 35.83&  & 2.21$_{-1.37}^{+1.71}$ & 6.06$_{-2.23}^{+3.16}$ & 18.0$_{-1.2}^{+0.6}$ & -- & Un &  &  &  \\ 
\cline{1-13}   
37 & 13.76 & Rise: & 13.83 & fixed & 0.81$_{-0.44}^{+0.18}$ & 0.96$_{-0.77}^{+1.75}$ & 4.97$_{-1.83}^{+4.97}$ & 2.24$_{-0.82}^{+0.67}$ & 0.29$_{-0.25}^{+1.45}$ & 85$/$87 & 0.97 &  2.88$_{-0.19}^{+0.04}$ \\ 
(501) & & Decay: & 137.17 & fixed & 0.81$_{-0.06}^{+0.06}$ & 4.12$_{-0.77}^{+0.72}$ & 2.88$_{-0.43}^{+0.58}$ & 2.96$_{-0.57}^{+0.62}$ &  0.06$_{-0.02}^{+0.02}$ & 229$/$186 & 1.23 &  3.14$_{-0.03}^{+0.02}$ \\ 
\cline{3-13}   
& & Total: & 151.00&  & 1.77$_{-0.31}^{+0.37}$ & 6.73$_{-0.88}^{+0.88}$ & 69.5$_{-2.9}^{+1.4}$ & -- & Un &  &  &  \\ 
\cline{1-13}   
38 & 15.59 & Rise: & 10.30 & fixed & 4.27$_{-1.91}^{+4.89}$ & 1.63$_{-0.64}^{+0.61}$ & \multicolumn{2}{c}{(1 Temperature)} & 1.69$_{-1.69}^{+1.69}$ & 108$/$81 & 1.34 &  2.97$_{-0.09}^{+0.07}$ \\ 
(501) & & Decay: & 11.03 & fixed & 0.99$_{-0.28}^{+0.26}$ & 3.34$_{-0.88}^{+1.01}$ & \multicolumn{2}{c}{(1 Temperature)} & 0.06$_{-0.06}^{+0.06}$ & 59$/$79 & 0.75 &  2.97$_{-0.06}^{+0.05}$ \\ 
\cline{3-13}   
& & Total: & 21.33&  & 2.01$_{-0.48}^{+0.81}$ & 2.51$_{-0.55}^{+0.60}$ & 4.2$_{-0.7}^{+0.5}$ & -- & Un &  &  &  \\ 
\cline{1-13}   
39 & 16.19 & Rise: & 12.17 & fixed & 10.47$_{-10.47}^{+10.47}$ & 0.01$_{-0.01}^{+0.70}$ & \multicolumn{2}{c}{(1 Temperature)} & 698.94$_{-698.94}^{+698.94}$ & 90$/$80 & 1.13 &  2.85$_{-0.11}^{+0.03}$ \\ 
(501) & & Decay: & 59.66 & fixed & 1.00$_{-0.14}^{+0.13}$ & 1.69$_{-0.36}^{+0.40}$ & \multicolumn{2}{c}{(1 Temperature)} & 0.13$_{-0.13}^{+0.13}$ & 198$/$131 & 1.51 &  2.93$_{-0.03}^{+0.02}$ \\ 
\cline{3-13}   
& & Total: & 71.83&  & 1.01$_{-0.14}^{+0.13}$ & 1.41$_{-0.30}^{+0.35}$ & 7.4$_{-1.2}^{+0.6}$ & -- & NN1 &  &  &  \\ 
\cline{1-13}   
40 & 17.93 & Rise: & 14.33 & fixed & 3.18$_{-0.71}^{+0.93}$ & 2.68$_{-0.57}^{+0.58}$ & \multicolumn{2}{c}{(1 Temperature)} & 1.28$_{-1.28}^{+1.28}$ & 123$/$95 & 1.30 &  3.50$_{-0.05}^{+0.05}$ \\ 
(501) & & Decay: & 14.34 & fixed & 0.86$_{-0.17}^{+0.17}$ & 0.48$_{-0.27}^{+0.74}$ & 2.81$_{-0.60}^{+0.86}$ & 2.95$_{-0.59}^{+0.68}$ &  0.69$_{-0.43}^{+0.87}$ & 99$/$94 & 1.05 &  3.59$_{-0.09}^{+0.03}$ \\ 
\cline{3-13}   
& & Total: & 28.67&  & 2.82$_{-0.52}^{+0.69}$ & 3.06$_{-0.43}^{+0.58}$ & 8.7$_{-0.9}^{+0.5}$ & -- & NN1 &  &  &  \\ 
\cline{1-13}   
41 & 18.07 & Rise: & 12.17 & fixed & 10.82$_{-7.75}^{+10.82}$ & 0.83$_{-0.83}^{+0.53}$ & \multicolumn{2}{c}{(1 Temperature)} & 0.00$_{-0.00}^{+0.00}$ & 162$/$87 & 1.86 &  3.24$_{-0.17}^{+0.20}$ \\ 
(501) & & Decay: & 23.66 & fixed & 0.99$_{-0.43}^{+0.42}$ & 0.52$_{-0.42}^{+1.24}$ & 4.47$_{-2.25}^{+4.47}$ & 1.01$_{-0.63}^{+0.75}$ &  0.42$_{-0.36}^{+1.72}$ & 120$/$105 & 1.14 &  3.32$_{-0.18}^{+0.08}$ \\ 
\cline{3-13}   
& & Total: & 35.83&  & 4.94$_{-2.91}^{+4.65}$ & 1.29$_{-0.57}^{+0.98}$ & 4.8$_{-2.9}^{+1.9}$ & -- & NN1 &  &  &  \\ 
\cline{1-13}   
42 & 18.07 & Rise: & 22.67 & fixed & 2.16$_{-0.44}^{+0.70}$ & 3.55$_{-0.60}^{+0.57}$ & \multicolumn{2}{c}{(1 Temperature)} & 0.51$_{-0.51}^{+0.51}$ & 132$/$110 & 1.20 &  3.54$_{-0.05}^{+0.05}$ \\ 
(501) & & Decay: & 6.00 & fixed & 1.00$_{-0.23}^{+0.35}$ & 1.78$_{-1.16}^{+1.23}$ & \multicolumn{2}{c}{(1 Temperature)} & 0.32$_{-0.32}^{+0.32}$ & 74$/$62 & 1.19 &  3.32$_{-0.15}^{+0.06}$ \\ 
\cline{3-13}   
& & Total: & 28.67&  & 2.02$_{-0.37}^{+0.58}$ & 3.18$_{-0.53}^{+0.52}$ & 7.3$_{-0.9}^{+0.7}$ & -- & NN1 &  &  &  \\ 
\cline{1-13}   
43 & 18.22 & Rise: & 9.83 & fixed & 11.13$_{-8.58}^{+11.13}$ & 1.15$_{-1.15}^{+0.56}$ & \multicolumn{2}{c}{(1 Temperature)} & 0.00$_{-0.00}^{+0.00}$ & 77$/$80 & 0.96 &  3.33$_{-0.23}^{+0.34}$ \\ 
(501) & & Decay: & 13.84 & fixed & 0.97$_{-0.13}^{+0.15}$ & 0.25$_{-0.16}^{+0.57}$ & 4.18$_{-1.39}^{+2.02}$ & 1.90$_{-0.56}^{+0.57}$ &  2.37$_{-1.62}^{+4.27}$ & 98$/$92 & 1.06 &  3.63$_{-0.17}^{+0.06}$ \\ 
\cline{3-13}   
& & Total: & 23.67&  & 5.83$_{-2.21}^{+2.85}$ & 1.73$_{-0.59}^{+0.53}$ & 5.8$_{-2.0}^{+2.1}$ & 2.3$\pm$0.2 & N &  &  &  \\ 
\cline{1-13}   
44 & 19.17 & Rise: & 10.17 & fixed & 2.85$_{-0.67}^{+0.90}$ & 3.57$_{-0.74}^{+0.77}$ & \multicolumn{2}{c}{(1 Temperature)} & 0.88$_{-0.88}^{+0.88}$ & 85$/$90 & 0.95 &  3.81$_{-0.07}^{+0.09}$ \\ 
(501) & & Decay: & 11.33 & fixed & 1.44$_{-0.21}^{+0.30}$ & 3.09$_{-0.76}^{+0.79}$ & \multicolumn{2}{c}{(1 Temperature)} & 0.40$_{-0.40}^{+0.40}$ & 107$/$87 & 1.23 &  3.64$_{-0.07}^{+0.05}$ \\ 
\cline{3-13}   
& & Total: & 21.50&  & 2.16$_{-0.30}^{+0.41}$ & 3.32$_{-0.53}^{+0.55}$ & 6.0$_{-0.6}^{+0.6}$ & -- & Un &  &  &  \\ 
\cline{1-13}   
46 & 19.67 & Rise: & 8.33 & fixed & 5.50$_{-1.70}^{+2.73}$ & 2.80$_{-0.94}^{+0.89}$ & \multicolumn{2}{c}{(1 Temperature)} & 2.85$_{-2.85}^{+2.85}$ & 123$/$84 & 1.46 &  4.12$_{-0.19}^{+0.12}$ \\ 
(601) & & Decay: & 29.00 & fixed & 2.98$_{-0.49}^{+0.61}$ & 3.69$_{-0.50}^{+0.52}$ & \multicolumn{2}{c}{(1 Temperature)} & 0.56$_{-0.56}^{+0.56}$ & 143$/$127 & 1.13 &  3.97$_{-0.05}^{+0.05}$ \\ 
\cline{3-13}   
& & Total: & 37.33&  & 3.43$_{-0.50}^{+0.65}$ & 3.49$_{-0.44}^{+0.45}$ & 13.0$_{-1.3}^{+1.0}$ & 3.6$\pm$0.4 & Q &  &  &  \\ 
\cline{1-13}   
47 & 20.70 & Rise: & 59.92 & fixed & 2.40$_{-0.17}^{+0.18}$ & 4.29$_{-0.31}^{+0.32}$ & \multicolumn{2}{c}{(1 Temperature)} & 0.65$_{-0.65}^{+0.65}$ & 215$/$163 & 1.32 &  4.21$_{-0.04}^{+0.02}$ \\ 
(601) & & Decay: & 69.58 & fixed & 1.65$_{-0.15}^{+0.15}$ & 4.11$_{-0.34}^{+0.36}$ & \multicolumn{2}{c}{(1 Temperature)} & 0.31$_{-0.31}^{+0.31}$ & 231$/$160 & 1.44 &  4.10$_{-0.03}^{+0.03}$ \\ 
\cline{3-13}   
& & Total: & 129.50&  & 2.00$_{-0.12}^{+0.12}$ & 4.20$_{-0.23}^{+0.24}$ & 41.8$_{-2.0}^{+1.3}$ & 40.0$\pm$26.0 & N &  &  &  \\ 
\cline{1-13}   
48 & 20.78 & Rise: & 12.67 & fixed & 2.39$_{-0.99}^{+2.09}$ & 2.11$_{-0.80}^{+0.77}$ & \multicolumn{2}{c}{(1 Temperature)} & 0.73$_{-0.73}^{+0.73}$ & 122$/$96 & 1.27 &  3.91$_{-0.09}^{+0.07}$ \\ 
(601) & & Decay: & 10.33 & fixed & 1.08$_{-0.15}^{+0.22}$ & 1.31$_{-0.96}^{+0.96}$ & \multicolumn{2}{c}{(1 Temperature)} & 0.61$_{-0.61}^{+0.61}$ & 84$/$85 & 0.99 &  3.84$_{-0.12}^{+0.04}$ \\ 
\cline{3-13}   
& & Total: & 23.00&  & 1.95$_{-0.54}^{+1.14}$ & 1.75$_{-0.62}^{+0.61}$ & 3.7$_{-1.0}^{+0.6}$ & 3.6$\pm$0.3 & N$/$Q &  &  &  \\ 
\cline{1-13}   
49 & 22.20 & Rise: & 13.83 & fixed & 2.81$_{-0.90}^{+1.34}$ & 2.11$_{-0.66}^{+0.71}$ & \multicolumn{2}{c}{(1 Temperature)} & 1.20$_{-1.20}^{+1.20}$ & 133$/$101 & 1.31 &  4.21$_{-0.08}^{+0.06}$ \\ 
(601) & & Decay: & 14.84 & fixed & 1.01$_{-0.57}^{+0.31}$ & 1.48$_{-1.10}^{+1.66}$ & 4.09$_{-2.24}^{+4.09}$ & 1.33$_{-0.95}^{+1.45}$ &  0.20$_{-0.20}^{+0.64}$ & 92$/$101 & 0.91 &  4.26$_{-0.26}^{+0.06}$ \\ 
\cline{3-13}   
& & Total: & 28.67&  & 2.61$_{-1.19}^{+1.92}$ & 2.47$_{-0.82}^{+1.19}$ & 6.3$_{-2.5}^{+0.7}$ & -- & NN1 &  &  &  \\ 
\cline{1-13}   
50 & 24.31 & Rise: & 8.00 & fixed & 1.05$_{-0.20}^{+0.29}$ & 0.95$_{-0.65}^{+2.12}$ & 4.14$_{-0.85}^{+1.64}$ & 8.79$_{-1.56}^{+1.32}$ & 0.61$_{-0.39}^{+0.67}$ & 105$/$98 & 1.07 &  5.87$_{-0.18}^{+0.10}$ \\ 
(601) & & Decay: & 27.83 & 4.34$_{-2.23}^{+2.60}$ & 0.88$_{-0.14}^{+0.15}$ & 1.36$_{-0.60}^{+1.31}$ & 2.56$_{-0.26}^{+0.32}$ & 9.52$_{-0.99}^{+1.01}$ & 0.35$_{-0.15}^{+0.17}$ & 207$/$146 & 1.42 &  5.46$_{-0.08}^{+0.06}$ \\ 
\cline{3-13}   
& & Total: & 35.83&  & 2.66$_{-0.33}^{+0.41}$ & 10.63$_{-0.98}^{+1.40}$ & 28.9$_{-1.6}^{+1.1}$ & 40.0$\pm$1.0 & N$/$Q &  &  &  \\ 
\cline{1-13}   
51 & 24.31 & Rise: & 9.00 & fixed & 2.71$_{-0.31}^{+0.38}$ & 14.78$_{-1.14}^{+1.13}$ & \multicolumn{2}{c}{(1 Temperature)} & 0.36$_{-0.36}^{+0.36}$ & 104$/$110 & 0.94 &  6.24$_{-0.12}^{+0.12}$ \\ 
(601) & & Decay: & 19.67 & 4.37$_{-2.70}^{+3.07}$ & 0.84$_{-0.11}^{+0.11}$ & 2.64$_{-1.07}^{+2.17}$ & 2.14$_{-0.28}^{+0.38}$ & 9.24$_{-1.34}^{+1.32}$ & 0.32$_{-0.14}^{+0.20}$ & 166$/$127 & 1.30 &  5.48$_{-0.09}^{+0.07}$ \\ 
\cline{3-13}   
& & Total: & 28.67&  & 2.16$_{-0.26}^{+0.33}$ & 12.79$_{-1.23}^{+1.78}$ & 26.1$_{-1.3}^{+1.0}$ & 32.0$\pm$1.0 & N &  &  &  \\ 
\cline{1-13}   
52 & 21.00 & Rise: & 5.33 & fixed & 1.50$_{-1.50}^{+1.50}$ & 0.00$_{-0.00}^{+1.16}$ & \multicolumn{2}{c}{(1 Temperature)} & 0.10$_{-0.10}^{+0.10}$ & 90$/$61 & 1.47 &  3.65$_{-0.01}^{+0.01}$ \\ 
(601) & & Decay: & 6.67 & fixed & 0.23$_{-0.06}^{+0.13}$ & 1.38$_{-0.71}^{+0.77}$ & 2.73$_{-0.96}^{+2.35}$ & 1.74$_{-1.16}^{+0.99}$ &  2.16$_{-1.75}^{+20.04}$ & 56$/$73 & 0.76 &  4.22$_{-0.25}^{+0.07}$ \\ 
\cline{3-13}   
& & Total: & 12.00&  & 1.62$_{-1.15}^{+1.57}$ & 1.73$_{-0.75}^{+0.87}$ & 2.3$_{-1.0}^{+0.3}$ & 10.0$\pm$1.0 & N$/$Q &  &  &  \\ 
\cline{1-13}   
   \enddata
 \tablenotetext{\rm \dag}{
XMM-Newton Obs-IDs. 301: 0822740301, 401: 0822740401, 
501: 0822740501, and 601: 0822740601. 
}
 \tablenotetext{\rm \ddag}{
Average count rate of PN lightcurve (0.2 -- 12 keV) of the quiescent component around each flare (cf. $QC_{\rm{PN}}^{\rm{OBSID}}$ in Table \ref{table:X-ray_quie_fit_results}).
}
 \tablenotetext{\rm *}{
 Durations of rise, decay, and total (=rise$+$decay) phase of the flare from Table 6 of T23.
 }
 \tablenotetext{\rm \S}{
$N_{\rm{H}}$ = ``fixed" means that it is fixed to the literature value 
2.29$\times$10$^{18}$ cm$^{-2}$ from \citet{Wood+2005_ApJS}.
$L_{\rm{X}}$ is the X-ray luminosity in the 0.2 -- 12 keV range.
 }
  \tablenotetext{\sharp}{
$E_{\rm{X}}$ is the total flare energy in the 0.2 -- 12 keV range.
$T_{\rm{ave}}$, $EM_{\rm{tot}}$, and $E_{\rm{X}}$ values of Flare 11, 15, and 23 listed in the ``Total" rows are from the results of the time-resolved analyses in Section \ref{sec:remarkable_flares} (Tables \ref{table:flare23_Xray_fit} \& \ref{table:flare11_Xray_fit}). 
As for Flare 23, the $T_{\rm{ave}}$, $EM_{\rm{tot}}$, and $E_{\rm{X}}$ values from the time-averaged fitting results (Section \ref{subsec:X-ray_specana_time-average}) are also listed 
in the ``Rise$+$Decay" row for comparison.
}
 \tablenotetext{\rm \P}{
The NUV flare energies in UVW2 band and Neupert classifications from T23. N: Neupert flare, N/Q Neupert flare (passes one
criterion), Q: Quasi-Neupert flare, NN1: Non-Neupert Type I (there is a SXR response but no UVW2 response), and Un: Undetermined (cf. Table 5 of T23). 
}
   \label{table:X-ray_flarefit_timeave}
 \end{deluxetable*}
\end{longrotatetable}

\clearpage

\bibliography{rev1_ynotsu_AUMic_XrayHa}
\bibliographystyle{aasjournal}

\end{document}